\documentclass[aps,a4,prx,reprint,superscriptaddress,notitlepage,floatfix,nolongbibliography]{revtex4-2}

\usepackage[english]{babel}

\usepackage{etoolbox}
\usepackage{titletoc}   
\AtBeginEnvironment{appendix}{\onecolumn}

\usepackage{amsmath}
\usepackage{graphicx}
\usepackage{csquotes}
\usepackage[dvipsnames]{xcolor}
\usepackage[colorlinks=true, allcolors=OliveGreen]{hyperref}
\usepackage{algorithm}
\usepackage{algorithmic}
\usepackage[capitalize]{cleveref}

\usepackage{amsfonts}
\usepackage{bm}

\newcommand{\DD}{\text{D}}

\def\acursed#1{#1}

\usepackage{acronym}

\newacro{HN}{Hopfield network\acused{HN-ad}}
\newacroindefinite{HN}{a}{a}
\newacro{HN-ad}[HN]{Hopfield-network\acused{HN}}
\newacroindefinite{HN-ad}{a}{a}

\newacro{MI}{mutual information\acused{MI-ad}}
\newacroindefinite{MI}{an}{a}
\newacro{MI-ad}[MI]{mutual-information\acused{MI}}
\newacroindefinite{MI-ad}{an}{a}

\def\figurename{Fig.}
\makeatletter
\def\fnum@figure{\textbf{\figurename~\thefigure}}
\makeatother

\usepackage{hyperref}
\usepackage{bookmark}
\usepackage{xr-hyper}
\begin{document}

\title{Inertial Asynchronous Computation}

\author{Doruk Efe Gökmen}
\affiliation{NSF-Simons National Institute for Theory and Mathematics in Biology, Chicago, IL 60611, USA}
\affiliation{James Franck Institute, The University of Chicago, Chicago, IL 60637, USA}

\author{Michel Fruchart}
\affiliation{Gulliver, ESPCI Paris, Université PSL, CNRS, 75005 Paris, France}

\author{Dmitrii Zendrikov}
\affiliation{Institute of Neuroinformatics, University of Zurich and ETH Zurich, Zurich, Switzerland}

\author{Giacomo Indiveri}
\affiliation{Institute of Neuroinformatics, University of Zurich and ETH Zurich, Zurich, Switzerland}

\author{Giulio Biroli}
\affiliation{Laboratoire de Physique Statistique, École normale supérieure,
PSL Research University, 24 rue Lhomond, 75005 Paris, France}

\author{Vincenzo Vitelli}
\affiliation{Leinweber Institute for Theoretical Physics, The University of Chicago, Chicago, IL 60637, USA}
\affiliation{James Franck Institute, The University of Chicago, Chicago, IL 60637, USA}

\begin{abstract}
Computation is the controlled evolution of a state.
Asynchronous evolutions, where all parts of the state change in their own time without stopping each other, put this control in jeopardy.
It is in fact a mystery how natural processes perform asynchronous computations using many units with no global orchestration.
Here we demonstrate how collective computational abilities can emerge in asynchronous many-body systems.
The key insight is to split the physical \enquote{hardware} underlying the computation into two asymmetrically coupled parts, analogous to position and momentum in a harmonic oscillator.
The resulting inertia nudges the evolution of the state so that the asynchronous computation proceeds in the right order.
By treating our inertial asynchronous computer as a nonequilibrium material, we map out its phase diagram numerically and analytically using a framework we dub loop dynamical mean-field theory.
We experimentally demonstrate our approach using analog spiking neuromorphic chips designed to mimic actual neurons in the brain.
In addition, we construct software that can run on asynchronous hardware: we denoise movies whose clean versions were never seen during training, an instantiation of the generalization transition underlying modern machine learning.
Our results point to a general strategy for reliable, decentralized computation in energy-constrained settings from dynamics self-assembly to cell differentiation.
\end{abstract}

\maketitle

Computation, in its essence, is the controlled evolution of a state~\cite{Church1936,Turing1937}.
In contrast with the hardware of our \acursed{phones} or laptops, where computations are globally orchestrated by a central clock~\cite{vonNeumann1967}, natural systems 
compute in a distributed way. 
Examples range from spin systems and molecular self-assembly to economic decision-making and cellular differentiation~\cite{Zdeborova2016,Mezard2009,Mezard2002,Wolfram1983, hopfield82,amari,LITTLE1978281,Yampolskaya2026,Watson2016,Bouchaud2003,Hu2022}.
These processes are intrinsically asynchronous: individual units change their state in their own time and in random order without blocking each other.
Such asynchronous computing raises the prospect of devising more efficient hardware, by triggering computations only where and when they are needed. 
However, asynchrony also raises fundamental engineering and conceptual challenges: different parts of the computation may fall out of order, potentially leading to chaotic dynamics, deadlocks, race conditions, or wildly incorrect results~\cite{async_ca_universality,acdbook,Manohar_2006}.
To avoid these pitfalls, carefully designed architectures have been proposed, effectively implementing a distributed clock by blocking events using memory buffers and delays~\cite{sompolinsky_sequential,Riedel1988,async_ca_universality,acdbook,Manohar_2006,Nakamura1974,Sompolinsky1988,vanVreeswijk1996} 
to keep the sequence of operations in the right order. 

\begin{figure*}[h!]
    \centering
    \includegraphics[width=0.85\linewidth]{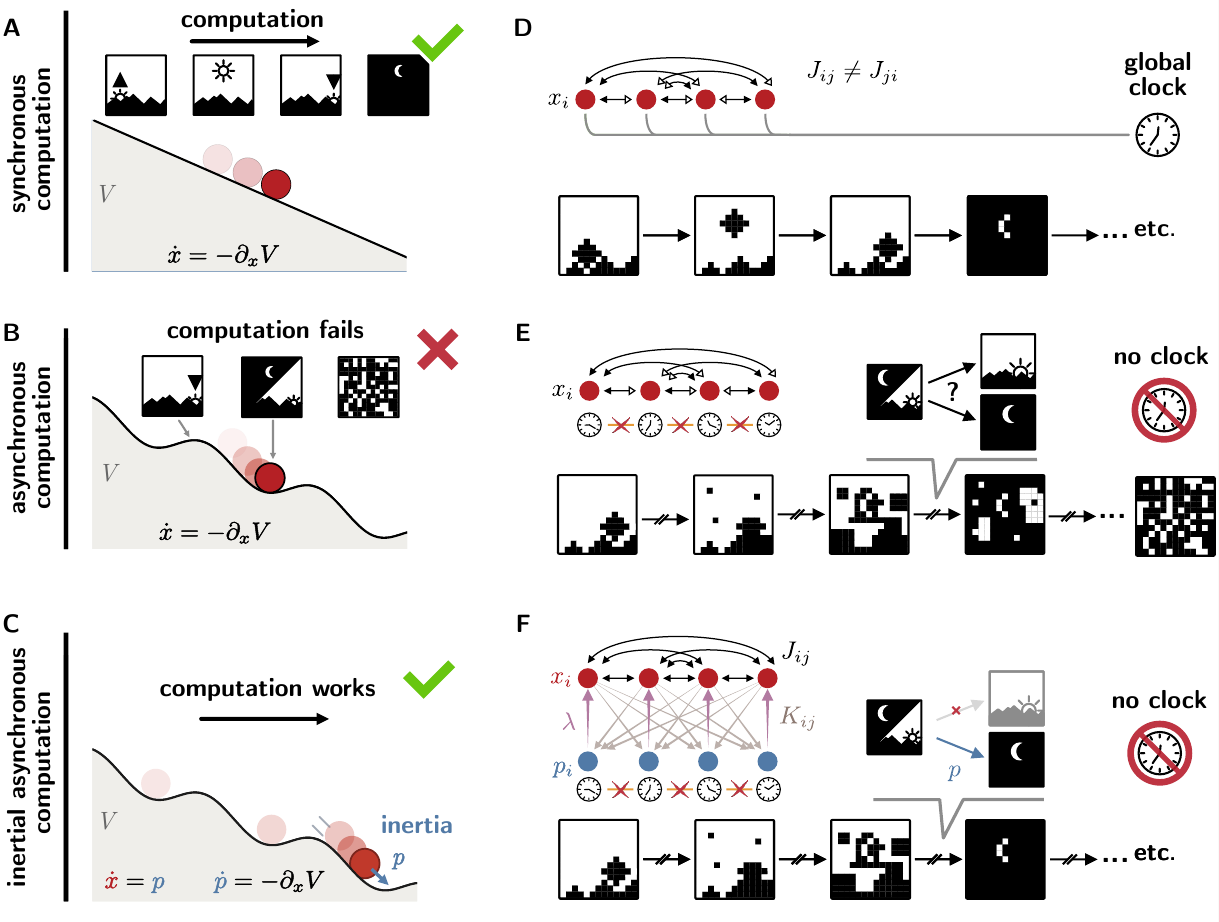}
    \caption{{\bf Saving asynchronous computations from chaos: inertia to the rescue.}
    As an example computation, we consider the retrieval of a sequence of patterns $\bm{\xi}^1 \to \bm{\xi}^2 \to \cdots \bm{\xi}^P \to \bm{\xi}^1 \to \cdots$ of length $P$.
    Here, these are represented by pictures of day, sunset, night, sunrise (going back to day, and so on, so here $P=4$), as shown in panel {\bf A}.
    (Note that in the Hopfield model we describe in the main text, $\bm{\xi}^\mu$ are uncorrelated i.i.d Rademacher random patterns, see Methods for the modern Hopfield version where actual pictures can be encoded.)
    {\bf A-C.} Computation is pictured as a ball (in red) rolling on a hill (in grey).
    In panels {\bf A}-{\bf B}, the ball's motion is overdamped (like in honey),  i.e. its position $x$ follows the first order equation $\dot{x} = -\partial_x V$ in a potential $V(x)$, to represent that the next step in the computation is fully determined by the current one.
    A synchronous computation is a smooth hill where the computation runs (panel {\bf A}), while an asynchronous computation is a bumpy hill where the ball gets stuck (panel {\bf B}).
    Adding inertia to the asynchronous computation, which is encoded in Newton's law of motion $m\ddot{x}=-\partial_x V$ where $m$ is the mass, allows the computation to overcome problematic steps where it may get stuck and to eventually succeed.
    This picture is made concrete and precise in panels {\bf D}-{\bf F} (corresponding to panels {\bf A}-{\bf C}, respectively): there, our goal is to store a cyclic temporal sequence of patterns in an associative (aka \enquote{content-addressable}) memory, from which the sequence can be retrieved \emph{in order} starting from a noisy version of one of the patterns.
    {\bf D.} A Hopfield network with synchronous updates (all spins at a time) can successfully retrieve a sequence encoded in its asymmetric couplings $J_{ij} \neq J_{ji}$. 
    {\bf E.} The retrieval of a sequence fails with asynchronous updates (one spin at a time). This is because the network tends to reach \enquote{mixed states} that have significant overlap with multiple patterns (here represented by a linear superposition of day and night) which eventually devolves into a random state with no overlap with any of the encoded patterns.
    (The mixed states do not happen with synchronous updates because all spins are updated in one go.)
    {\bf F.} Our model associative memory solves the problem of asynchronously retrieving temporal sequences.
    It is a generalized version of the Hopfield model composed of two classes of neurons $x_i(t) = \pm 1$ (red) and $p_i(t) = \pm 1$ (black), with $i=1,\dots, N$.
    These are wired in a specific way specified in Eq.~\eqref{eq:model}: (i) the $x_i$ are connected to each other with a symmetric $J_{ij}$ memorizing each individual pattern (black lines), (ii) the $p_i$ are influenced by the $x_j$ through an asymmetric $K_{ij}$ encoding the temporality of the pattern sequence (beige arrows), and (iii) each $x_i$ is influenced by $p_i$ (but not the other $y_j$) through a parameter $\lambda$ (purple arrows).
    Crucially, there is no global clock: the neurons are asynchronously updated according to Eq.~\eqref{eq:model}.
    \label{fig1}
    }
\end{figure*}

Here, we take an alternative view based on nonequilibrium many-body physics, with the complementary goals of formulating plausible minimal models of natural phenomena and engineering strategies. 
To get a gist of the challenges at play and of the solution we propose, start by picturing computation as a ball rolling on a hill in honey (Fig.~\ref{fig1}A-B).
The ball's motion represents the evolution of the high-dimensional state of the system over time: in our model computation, 
it should cycle through a particular sequence of patterns represented by sunrise, noon, sunset, and midnight.
Synchronous computation is analogous to a smooth hill (Fig.~\ref{fig1}A and left panel of Supplementary Movie \href{https://youtu.be/KiTco9FMmmw}{1}): 
the ball rolls down, meaning that the computation completes. 
Asynchronous computation, however, turns out to be a bumpy hill (Fig.~\ref{fig1}B and right panel of Supplementary Movie \href{https://youtu.be/KiTco9FMmmw}{1}), 
where the ball frequently gets stuck in local minima, representing random mixtures of different patterns. 
To unstuck the ball, we can endow it with inertia: this allows the ball (and the asynchronous computation) to move on and run to completion (Fig.~\ref{fig1}C and and Supplementary Movie \href{https://youtu.be/QX8elACcRv0}{2}). 
The overdamped motion of the ball in honey encodes that the next step of the computation is entirely determined by the current one. 
In contrast, the inertia of the ball (or the computation) represents a memory of its past state.

As we shall see, this simple intuition can be directly transposed to a dynamical model of collective computation known as the Hopfield network~\cite{amari, LITTLE1978281, hopfield82}. 
In this model, the features put by hand in Fig.~\ref{fig1}A-C emerge from interactions between many physical neurons (Fig.~\ref{fig1}D-F).
In particular, the potential $V$ that provides a directionality to the computation in Fig.~\ref{fig1}A-C is a proxy for asymmetric couplings between the neurons.
In our mechanical metaphor, these would correspond to non-reciprocal forces between many particles~\cite{fruchart2026nonreciprocalmanybodyphysics}, effectively endowing the system with self-propulsion and guiding it along the sequence of patterns to retrieve.
Crucially, the addition of inertia to this active many-body setting enables collective asynchronous computations without the need of pausing the computation, cloning the states, or storing them in memory buffers.

\medskip
\noindent{\bf Associative memories as toy computations.}
To concretely illustrate how to overcome the challenges raised by asynchrony, consider one of the most elementary examples of computation: the retrieval of a temporal sequence in an associative memory. These could be sequences of instructions in a computer program, frames of your favourite movie or simply alphabetical letters.

In associative (or content-addressable) memories,  patterns are dynamically retrieved from partial or noisy cues.
This is captured by the celebrated Hopfield network (HN)\acused{HN}\acused{HN-ad} model~\cite{amari, LITTLE1978281, hopfield82}, where  memories are stored in the couplings $J_{ij}$ between $N$ binary variables $x_i = \pm 1$.
These couplings are taken to be a sum of $P$ patterns of the form $J_{ij}=\sum_{\mu}\xi_i^\mu \xi^\mu_j$.
The patterns $\bm{\xi}^\mu$
are then retrieved through the discrete-time spin dynamics
\begin{equation}
        \label{hopfield}
        x_i(t+\Delta t) = {\rm sign}\left(J_{ij} x_j(t) + \eta_i(t) \right)
\end{equation}
where a sum over $j$ is implied, $\eta_i$ is a random white noise, and $\Delta t$ the time between two updates.
When $J_{ij} = J_{ji}$, the dynamics in Eq.~\eqref{hopfield} can be seen as a discrete variant of a stochastic gradient descent, which is directly implemented by stochastic physical processes (see Methods).
As a result, the state $x(t)$ of the system ends up retrieving a static pattern when $x(t) \simeq \xi^\mu$  for some $\mu$.

The retrieval of static memory described above is robust to timing of updates and works both synchronously and asynchronously because it essentially entails solving an equilibrium problem. However, this is not the case for the retrieval of temporal sequences, as illustrated in Fig.~\ref{fig1}D-E.
Such a non-equilibrium functionality requires asymmetric connections ($J_{ij} \neq J_{ji}$), i.e.  $J_{ij}=\sum_{\mu}\xi_i^{\mu+1} \xi^\mu_j$, to drive the memory flow $\mu = 1 \to 2 \to \cdots$~\cite{hopfield_sync_async, hopfield82, sompolinsky_sequential, Amit_1989, gillett_ahm}.

\begin{figure*}[hbt!]
    \centering
    \includegraphics[width=0.99\linewidth]{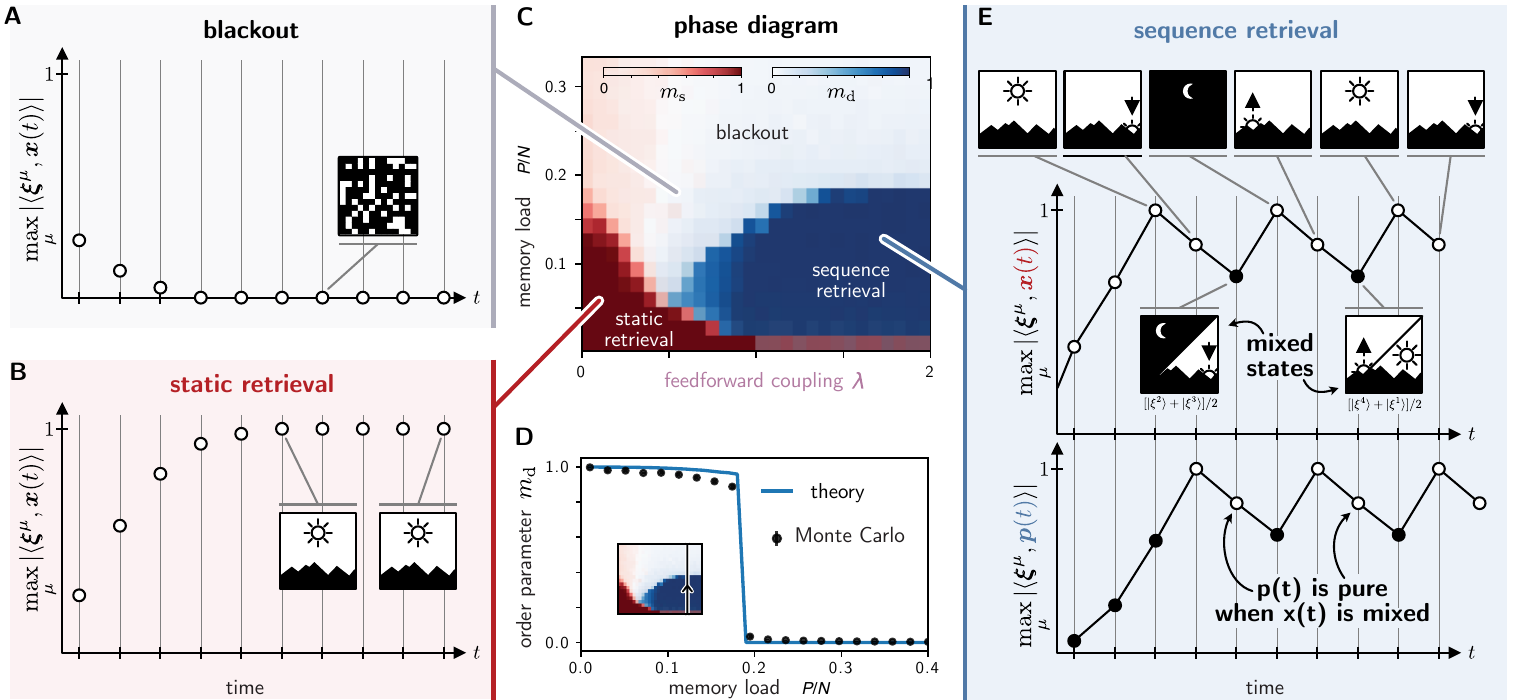}
    \caption{{\bf
    The phase diagram of an asynchronous computer.
    }
    The phase diagram obtained from Monte-Carlo simulations ({\bf C}; see Methods for details) shows three behaviors, that we interpret as thermodynamic phases in the limit where both the number $N$ of neurons and the number of stored patterns $P$ go to infinity, at fixed memory load $P/N$.
    These are: {\bf A} a blackout phase where no stored pattern $\xi^\mu$ is retrieved, so all overlaps $\langle \xi^\mu, x(t)\rangle$ vanish, {\bf B} a static retrieval phase where a single stored pattern (here, day) is retrieved and it does not change in time (Supplementary Movie \href{https://youtu.be/bX1PCPaca9Y}{3}),
    and {\bf E} a sequence retrieval phase where the ordered sequence of patterns is retrieved in order, in a loop, forever.
    We observe that even though both species $x$ and $y$ exhibit mixed states in this phase, it is never happening at the same time, which prevents the system from devolving into chaos (Supplementary Movie \href{https://youtu.be/QX8elACcRv0}{2}).
    The state $\bm{x}(t)$ of the system at any given time can be projected onto the stored patterns $\bm{\xi}^\mu$. When a single pattern (or none at all) is dominant, we say that $\bm{x}(t)$ is a pure state. 
    When multiple patterns have significant overlaps with $\bm{x}(t)$, it is said to be in a mixed state. 
    {\bf D} The system achieves the maximal critical sequential memory capacity of $P/N~0.17$ at large $\lambda$ which can be estimated analytically using loop-space dynamic mean-field theory.
    \label{fig2}
    }
\end{figure*}

In our mechanical metaphor, these asymmetric connections are akin to forces that violate Newton's third law~\cite{fruchart2026nonreciprocalmanybodyphysics} and stir the system along the stored sequence.
In Hopfield networks with unstructured asymmetry~\cite{amari, LITTLE1978281, hopfield82}, asynchrony destabilizes memory flow: early-updating neurons chaotically race ahead of the late ones, fracturing the state into incoherent mixtures of multiple patterns~\cite{PhysRevA.36.4922, PhysRevA.37.4865}, as illustrated in Fig.~\ref{fig1}E (and pictured as dips on the hill in Fig.~\ref{fig1}B).

\medskip
\noindent{\bf Inertia keeps computations in order.}
Inspired by the mechanical analogy in Fig.~\ref{fig1}A-C, we consider a hardware partitioned  into two asymmetrically wired sets of neurons $x_i$ and $p_i$ that play the role of position and momentum respectively and follow the dynamics (Fig.~\ref{fig1}F)
\begin{subequations}
\label{eq:model}
    \begin{align}
        x_i(t+\Delta t) & = {\rm sign}\left( J_{ij} x_j(t) + \lambda p_i(t) + \eta^x_i(t) \right)
        \label{eq:model_a}
        \\
        p_i(t+\Delta t) & = {\rm sign}\left( K_{ij} x_j(t) + \eta^p_i(t) \right).
        \label{eq:model_b}
    \end{align}
\end{subequations}
where $\lambda$ is an asymmetric coupling between the two neuron species,
while $J_{ij} = \frac{1}{N} \sum_\mu \xi_i^\mu \xi^\mu_j$,
and $\eta_i^{x,p}$ are independent white noises.
In addition, the asymmetric ($K_{ij}\neq K_{ji}$) coupling $K_{ij}=\frac{1}{N} \sum_\mu \xi_i^{\mu+1} \xi^\mu_j$, 
makes the neurons $\bm{p}$ predict the target memory based on the current state of $\bm{x}$ (but not conversely). In this respect,
equations \eqref{eq:model_a} and \eqref{eq:model_b} resemble Hamilton equations $\dot{x}=p/m$ and $\dot{p}=-\partial_x V$, respectively, with $\lambda$ being analogous to an inverse inertial mass.
In the Methods, we discuss further how the $p_i$ neurons in Eq.~\eqref{eq:model} induce inertia in the computation (very much like momentum does in classical mechanics, but for the discrete variables $x_i$ and $p_i$) and compare this strategy to explicit time delays. 
Unlike the single-body dynamics in Fig.~\ref{fig1}A-C, our collective computations are non-variational dynamical systems because $K_{ij}\neq K_{ji}$. As a result, mixed states are vestiges of the incipient chaotic dynamics, not associated with extrema of an energy landscape.

The asynchronous Hopfield network described by Eqs.~\eqref{eq:model_a} and \eqref{eq:model_b} lies at the heart of the model asynchronous computer architecture that we investigate in this study.  The computation is asynchronous in the sense that all neurons are updated exactly once in a random order during each update sweep.
As we will show, it is the interplay between the asymmetry of $K_{ij} \neq K_{ji}$ and the asymmetry between $K_{ij}$ and $\lambda$ that leads to
efficient recall of temporal sequences bypassing mixed states of successive patterns (Fig.~\ref{fig1}E) that would otherwise impede it (note that sequences can be retrieved even when $J_{ij}=0$ at the price of missing static retrieval).
Indeed, when $\lambda=0$, Eq.~\eqref{eq:model_a} is a standard Hopfield model with symmetric couplings, which tends to recall one of the patterns, while Eq.~\eqref{eq:model_b} reads the pattern recalled in the $\bm{x}$ neurons and outputs the next pattern in the sequence of Fig.~\ref{fig1}C (with an order specified by $K_{ij}$).
The term $\lambda p_i(t)$ in Eq.~\eqref{eq:model_a} therefore acts as an inertial bias which nudges the $\bm{x}$ neurons Hopfield network towards the next pattern.

As we shall see, this general strategy, while formalised here through Hopfield networks, is readily applicable in more realistic spiking neural networks, which we experimentally demonstrate on a mixed-signal asynchronous neuromorphic processor~\cite{dynapse}.
More broadly, it could be applied to dynamic versions of phenomena ranging from molecular self-assembly to epigenetic memory, which have been modelled using Hopfield networks~\cite{Yampolskaya2026}.

\medskip
\noindent{\bf From blackout to asynchronous computation.}
Numerical simulations of the asynchronous Hopfield network \eqref{eq:model} reveal the existence of three phases, shown in Fig.~\ref{fig2}.
First, a {blackout phase} (in gray in Fig.~\ref{fig2}) corresponds to the case where $\bm{x}(t)$ never significantly overlaps with any pattern $\bm{\xi}^\mu$: the pattern overlaps $m^\mu(t) \equiv \langle\bm{\xi}^\mu, \bm{x}(t)\rangle$ vanish for all $\mu$.
In this case, the network fails at recalling any memory, like a student during a tough oral examination.

Second, a {static retrieval phase} is shown in red in Fig.~\ref{fig2}, where a static memory is successfully retrieved by the Hopfield network ($\bm{x}(t) \simeq \bm{\xi}^{1}$, so $m^\mu$ is of order one).
The system enters this phase when the inertial mass $1/\lambda$ is sufficiently large.
In this phase, the system randomly condenses into one stored pattern, spontaneously breaking the symmetry between the $P$ patterns, analogous to how a cooled magnet spontaneously magnetizes either north or south (see Supplementary Movie \href{https://youtu.be/bX1PCPaca9Y}{3}).
The dynamics transition from the ergodic exploration of configuration space observed in the blackout phase to a confinement within a single region. 
This phase is characterized by a finite value of the order parameter
$m_{\rm s} \equiv \max_{\mu=1\dots P}
m^\mu(t\to\infty)$
which gives the maximum overlap among all patterns stored in the model.

\begin{figure*}[hbt!]
    \centering
    \includegraphics[width=0.9\linewidth]{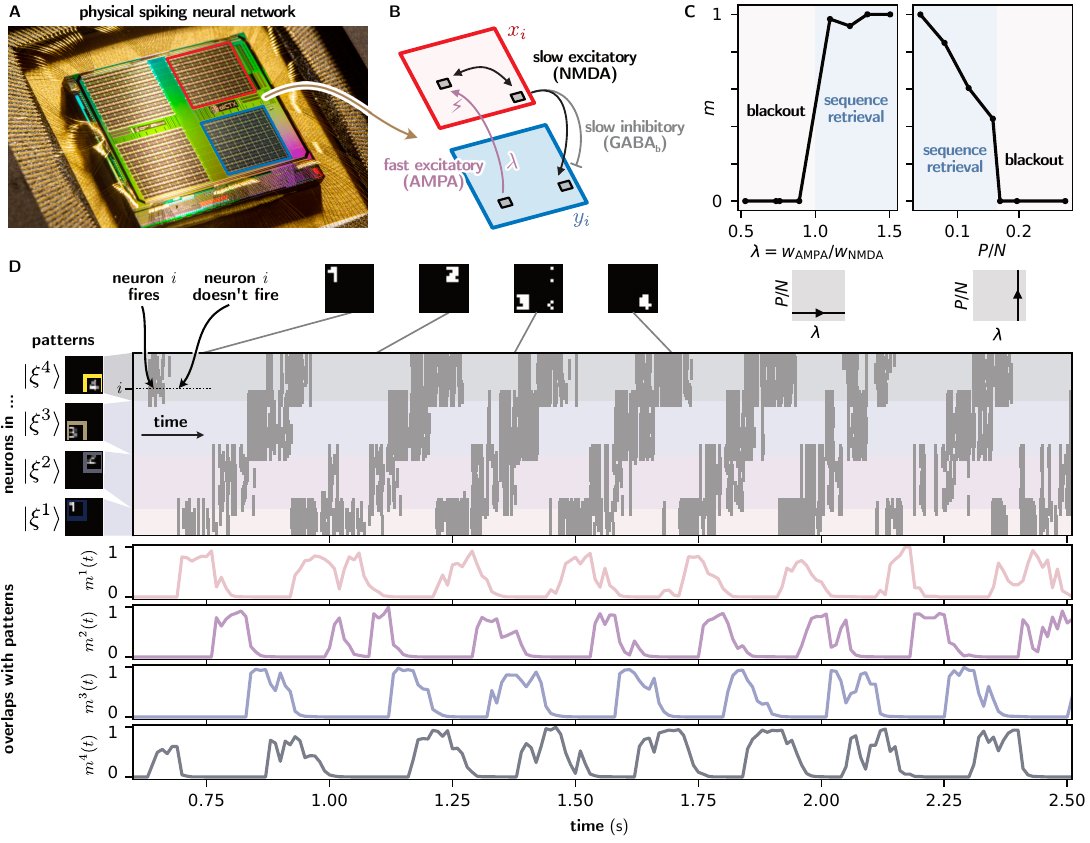}
    \caption{{\bf Experimental demonstration on asynchronous hardware.}
    {\bf A} The neuromorphic chip DYNAP-SE consists of an array of mixed-signal analog/digital circuits that emulate the dynamics of spiking neurons (top right) and synapses (bottom right) in continuous time.
    {\bf B}
    We represent the two neuron species $\mathbf{x}$ and $\mathbf{y}$ on separate cores of the chip, and implement the recurrent Hebbian synapses within each core, while inter-core connections implement the asymmetric coupling between the two species of neurons.
    {\bf C} We map out the phase diagram of this neural system by tuning these synapses (\emph{cf.} Extended Data Fig.~\ref{ed_phase_diagram}D). Left: we observe a transition from static memory retrieval to dynamic sequence retrieval as we increase the AMPA synapse which determines the asymmetric coupling $\lambda$. Right: we can load more memories to the network by rewiring the NMDA and GABAb synapses, whereby we observe a blackout beyond a critical load $P_c/N\approx 0.17$ (\emph{cf.} Extended Data Fig.~\ref{ed_phase_diagram}A).
    {\bf D} Recording of the activity of spiking neurons, in a network loaded with a sequence of patterns containing digits $1$, $2$, $3$ and $4$, respectively. Upon stimulating the network (in the first $0.5$s) with a pulse shaped as $4$, the network then successfully retrieves the sequence.
    See Methods and SI for experimental details.
        \label{fig:hardware_demo}
    }
\end{figure*}

Last, there is a {dynamic retrieval phase} in which the network retrieves a temporal sequence (in blue in Fig.~\ref{fig2}C, also see Supplementary Movie \href{https://youtu.be/QX8elACcRv0}{2}).
The overall picture is that both $\bm{x}(t)$ and $\bm{p}(t)$ follow a sequence $\bm{\xi}^{1} \to \bm{\xi}^{2}\to \cdots \to \bm{\xi}^{P} \to \bm{\xi}^{1}$, but the detailed picture is more subtle (see the blue box in Fig.~\ref{fig2}, where $P=4$).
In particular, simulations revealed that there can be mixed states where the overlaps $m^\mu$ with two or more of the encoded patterns $\bm{\xi}^\mu$ are significant (black dots in the figure), in contrast with the blackout and static retrieval phases which involve pure states where $\bm{x}(t)$ has a significant overlap with at most one of the patterns (white dots in the figure).
These mixed states are at the origin of the instability of sequence retrieval in the unstructured asymmetric Hopfield network \eqref{hopfield}.
As one can see in Fig.~\ref{fig1}F, mixed states  occur periodically (with a period $\varpi=3$ in the figure), but not necessarily with the same period as the sequence (so $\varpi \neq P$ in general; in fact, these can be incommensurate both with each other and with the duration of a single sweep of all neurons, see Methods).
Crucially, our simulations and mathematical analysis in SI show that sequence retrieval self-organises so that $\bm{p}(t)$ is in a pure state whenever $\bm{x}(t)$ is in a mixed state (and conversely): this delay between the two classes of neurons is what avoids a descent to chaos.
The sequence retrieval phase is characterized by an order parameter $m_{\rm d}$ that captures the overlap of the retrieved sequence (excluding mixed states) with the encoded sequence (see Methods), plotted in blue in the numerical phase diagram of Fig.~\ref{fig2}.

\medskip
\noindent{\bf Statistical physics of asynchronous computation.}
The transition from blackout to asynchronous computation (i.e. sequence-retrieval) in Fig.~\ref{fig2} is reminiscent of a first-order phase transition, such as water freezing: they are marked by a discontinuous jump in the order parameter and hysteresis (see Extended Data Fig.~\ref{ed_phase_diagram}A).
Nevertheless, the nonequilibrium nature of the onset of computation in our asynchronous model precludes a standard thermodynamic classification.
We instead turn to the perspective of dynamical systems to assess the nature of the transition by constructing a low-dimensional embedding that effectively endows the pattern space with the topology of the sequence (see Methods).
We then show that the transition from blackout to sequence retrieval is encoded as a Hopf bifurcation (Extended Data Fig.~\ref{ed_phase_diagram}B), a common way to produce a limit cycle in a dynamical system \cite{strogatz2018nonlinear}.
The first-order behavior of the system is encoded in the subcritical nature of this bifurcation.

In order to derive the phase boundary of the many-body dynamics analytically, we have developed a generalization of a method known as dynamic mean-field theory (DMFT) to the space of periodic temporal sequences, that we dub \textit{loop-space} DMFT (see Methods and SI).
The solution of the loop DMFT equations captures the order parameter dynamics in large-$N$ simulations well (Fig.~\ref{fig2}D, Extended Data Fig.~\ref{ed_phase_diagram} in Methods), provides a first-principle derivation of the phase diagram, and captures the critical capacity $P_c/N \approx 0.17$ at large $\lambda$.


\medskip
\noindent{\bf Building asynchronous hardware in the lab.}
We now ask what happens if our architecture (so far analysed in digital models) is built from the analog components available to physics or biology and simply left to run?
As we cannot wire up a brain at will, we instead implement our neural architecture with physical asynchronous dynamics on a neuromorphic device that shares many features and constraints with real biological neural processing systems~\cite{dynapse, Mead_2022, 7159144, 8259423, 8887553, INDIVERI20253311}.
This device emulates spiking neural networks using analog neuron electronic circuits that have biophysically realistic dynamics and integrate biologically plausible synaptic receptors with distinct time constants~\cite{destexhe1998kinetic} (see Fig.~\ref{fig:hardware_demo}A-B and Methods).
Even though the Hopfield-style models with binary neural signals we have considered up to now are at best an idealized representation of such an analog spiking neural network, we now show that a direct translation of their architecture on this biologically plausible computing substrate allows us to retrieve sequences on this asychronous hardware and make a couple of unexpected discoveries stemming from the irregular nature of silicon neurons.

We represent the two neuron populations on separate cores, and use programmable inhibitory and excitory synapses to implement
the asymmetric coupling $\lambda$ and the Hebbian couplings $J_{ij}$ and $K_{ij}$, as shown in Fig.~\ref{fig:hardware_demo}B.
By varying these synaptic weights, we experimentally map out in Fig.~\ref{fig:hardware_demo}C a phase diagram where we observe both a static-to-dynamic transition driven by fast excitatory synapses connecting the two species, and a blackout transition at a similar critical load $P_c/N \approx 0.17$, as in the minimalistic binary model in Eq.~\ref{eq:model} (compare Fig.~\ref{fig:hardware_demo}B with Extended Data Fig.~\ref{ed_phase_diagram}A).

We demonstrate sequence retrieval by loading $P=4$ patterns encoding digits 1 to 4 (playing the same role as the four landscapes in Figs. \ref{fig1} and \ref{fig2}) into a network of 256 spiking neurons per population.
The input initial conditions of the recurrent dynamics can be set by sending a short initial pulse.
As shown in Fig.~\ref{fig:hardware_demo}D, upon such stimulation, the network autonomously retrieves the complete cyclic sequence, with each digit persisting for $\sim$200~ms.
Because $P=4$ lies well below the critical capacity, a \emph{uniform} pulse is already enough to trigger retrieval.
Closer to capacity, in contrast, this spontaneous symmetry breaking no longer happens and a patterned input aligned with one of the stored digits is required, recovering the metastability we observed for the binary model.
In Extended Data
Fig.~\ref{fig:neuromorph_static_dynamic}, we also checked  that when the relative weight of the feedforward AMPA synapses with respect to the slow excitatory synapses (i.e. $\lambda$) is smaller than 1, the system retrieves a memory statically as opposed to cycling through the sequence, just like in the red region in the phase diagram in Fig.~\ref{fig1}.

There is a key feature of this neuromorphic device that makes it more of an open ended experimental platform than a mere demonstration.
The spike timing of the silicon neurons is irregular due to device mismatch of their analog circuits. This stands in sharp contrast with the regular update schedule assumed up tp now in our model, where each neuron is updated exactly once per sweep.
However, if the neurons' firing times were entirely random (Poissonian), we would expect that the retrieval would be strongly degraded or even completely lost (see Methods), but this is not what we observe.
This is not the case because, as their biological counterparts, the silicon neurons have a refractory period which salvages sequence retrieval by preventing neurons from firing twice in a short period.
Extended Data
Fig.~\ref{fig:nmda_refractory}  shows neuron firing kymographs (like in Fig.~\ref{fig:hardware_demo}D) in the neuromorphic hardware for different values the of the ratio $\tau_{\text{ref}}/\tau_s$ where $\tau_{\text{ref}}$ denotes the refractory period and $\tau_s$ the synaptic time constants.
The experimental data demonstrates that the sequence retrieval ability is lost as $\tau_{\text{ref}}/\tau_s$ decreases, consistent with our qualitative explanation.

\begin{figure*}[hbt!]
    \centering
    \includegraphics[width=0.85\linewidth]{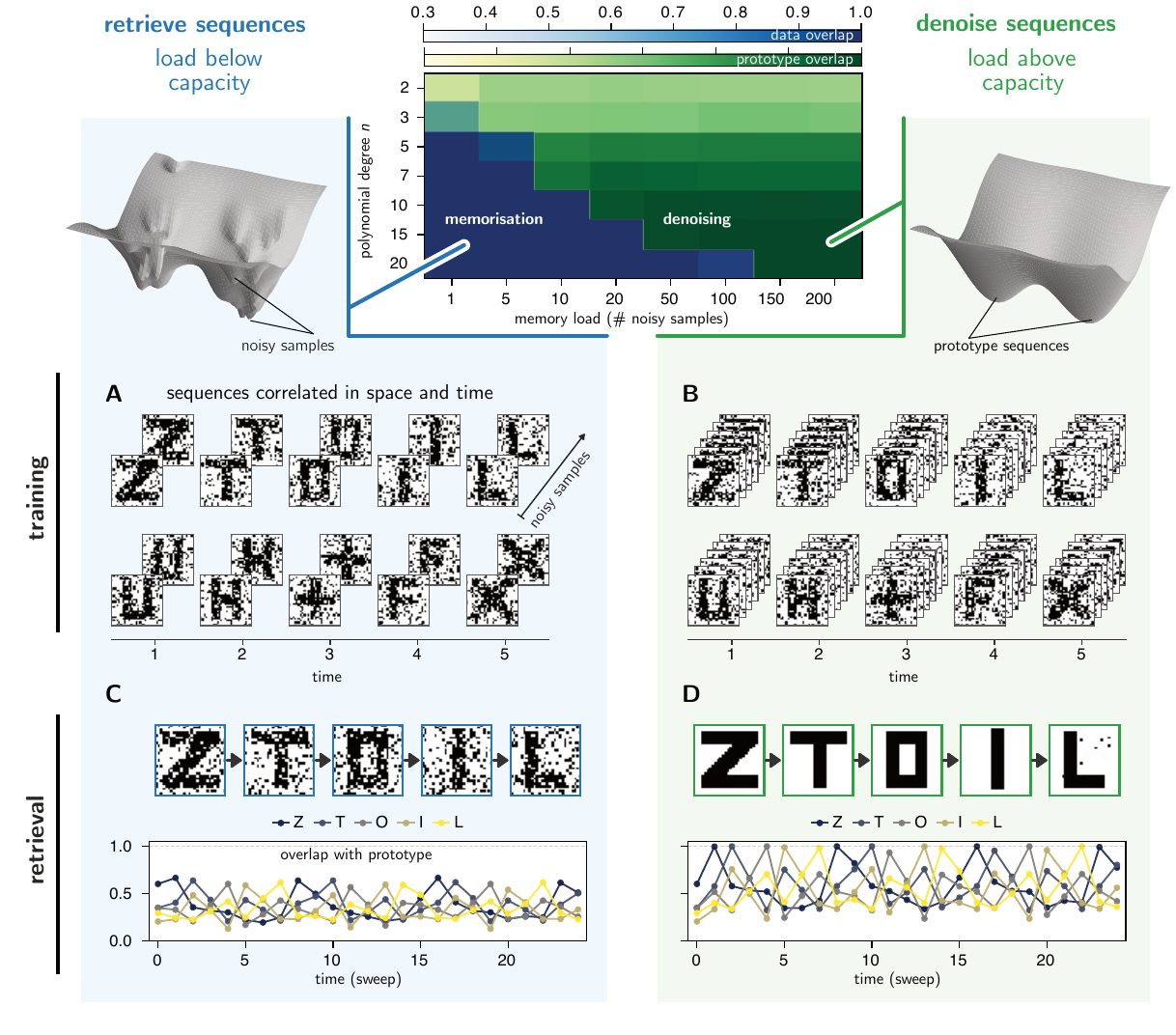}
    \caption{\textbf{Asynchronous software: denoising correlated sequences.
    }
    \textbf{A, B} Training data comprises noisy samples of each frame of different prototype sequences; the clean prototypes themselves are never shown to the network. The network behaves distinctly depending on whether it is loaded with noisy samples below ({\textbf{A}}, memorisation regime) or above ({\textbf{B}}, generalisation regime) its storage capacity.
    \textbf{C, D} Asynchronous retrieval dynamics. In the low-load memorisation regime (\textbf{C}) the retrieved cycle snaps onto whichever stored noisy sample sits closest. Strikingly, well above capacity (\textbf{D}), the same dynamics instead \emph{denoises} the training samples: retrieved frames reconstructs the underlying clean prototype, as revealed by the per-frame prototype overlaps (Z, T, O, I, L) along the trajectory.
    These observations are summarised in a phase diagrams on the $(n, P)$ plane for dense-memory activation $F(m)=m^n$.
    Overlap with the nearest training sample, serves as the memorisation order parameter; it saturates at $m_t\simeq 1$ below a dynamic memory capacity that grows with the nonlinearity $n$. Inset: rugged quasipotential for limit cycles, with inner pits labelling individual memorised noisy samples within each basin.
    At sufficiently large nonlinearity $n$, the prototype overlap is large at high $P$, indicating that the retrieval dynamics successfully remove the noise in the training data. Inset: the quasipotential becomes smooth, with one broad well per prototype sequence.
    \label{fig:denoise}
}
\end{figure*}
\medskip
\noindent{\bf Going nonlinear to compute with correlations.}
Movies you may want to watch display strong correlations between frames.
In fact, the images themselves are also spatially correlated.
In contrast, our analysis so far has been restricted to the case of uncorrelated random patterns, both numerically and with DMFT.
Indeed, spatial and temporal correlations that occur in realistic data severely degrades the retrieval performance of the linear Hebbian Hopfield networks~\cite{Amit_1989, sompolinsky_sequential}.
One way to deal with correlated patterns is to introduce nonlinear couplings between the neurons by making the replacement
\begin{equation}
    J_{ij} \, x_j(t)
    \to
    \sum_{\mu=1}^P \xi_i^{\mu} f\big(\langle\bm{\xi}^\mu, \bm{x}(t)\rangle\big)
\end{equation}
in Eq.~\eqref{hopfield}, where $f$ is a nonlinear activation function (such as a polynomial $f(x) = x^m$).
This construction is known as a modern Hopfield network or as a dense associative memory~\cite{NIPS2016_eaae339c, krotov_modern_hopfield,Ramsauer2020}.
A sufficiently nonlinear $f$ amplifies the separation of partially overlapping memories, and allows retrieving sequences with spatiotemporal correlations.
Note, however, that a standard modern Hopfield network is not enough to asynchronously retrieve sequences.

To deal with asynchrony, we make the following replacement in Eq.~\eqref{eq:model_b}, in addition to the aforementioned replacement in Eq.~\eqref{eq:model_a}
\begin{equation}
    K_{ij} \, x_j(t)
    \to
    \sum_{\mu=1}^P \xi_i^{\mu+1} f\big(\langle\bm{\xi}^\mu, \bm{x}\rangle\big)
\end{equation}
We thus obtain an asynchronous modern Hopfield network with asymmetric couplings capable of encoding correlated sequences \cite{Tahir2024,NEURIPS2023_aa32ebcd} (note the $\mu +1$ index implementing the sequence hopping). 
The resulting dynamics is illustrated in Extended Data Fig.~\ref{fig:modern}A, where a sequence of handwritten digits is retrieved.
As shown in Extended Data Fig.~\ref{fig:modern}B, the memory capacity of our asymmetric dense associative memory scales exponentially with the number of neurons $N$ for uncorrelated sequences, in the same was as the symmetric versions~\cite{krotov_modern_hopfield}.

\medskip
\noindent{\bf Asynchronous software: denoising movies.}
Now that we have established the experimental feasibility of our asynchronous hardware and generalized it to dense memories able to handle correlations, we proceed to illustrate how you would run on this class of hardware a more realistic computation: denoising correlated sequences of images, e.g. your favourite movie. Up to now we have treated the patterns $\bm{\xi}^\mu$ in the training dataset as exact prototypes (see Fig.~\ref{fig:denoise}A) to be stored and faithfully played back. In this view, the network is effectively a content-addressable \emph{lookup table}, i.e. its output can be no cleaner than what was handed to it during training.
Recent results~\cite{PhysRevLett.131.257301, KALAJ2025130946, Pham2025MSG} suggest that this limitation is itself a symptom of under-loading. When a modern Hopfield network is driven far beyond its memory capacity, the spurious attractors that used to be dismissed as retrieval artefacts can reorganise into a smoother manifold, capturing the intrinsic common structure of the training distribution and taking the network from memorising samples to generalising across them.
Here we show that the sequential version of this memorization-to-generalisation transition appears in our model, and that it lets the network \emph{denoise} a whole cyclic sequence asynchronously (see Supplemental Movie \href{https://youtu.be/PxMkNYjKQnU}{4}).

As shown in Fig.~\ref{fig:denoise}, at low load ($P$ small) the network behaves as a standard associative memory: the retrieved sequence snaps onto whichever stored noisy sample sits closest to it, and therefore the prototype overlap is bounded by $1-2\eta_{\rm train}\simeq 0.6$ (Fig.~\ref{fig:denoise}A).
Past a threshold load, the network graduates from reciting a stored movie to a dynamic denoising computation: it infers the structure of the movie from its own noisy recordings and replays the clean version. Each retrieved frame now accurately reconstructs the clean prototype even if this clean version was never seen during training (Fig.~\ref{fig:denoise}B).
As shown in the phase diagram on Fig.~\ref{fig:denoise}, denoising requires the nonlinearity of $f$ to be strong enough ($n \gtrsim 5$), but a larger polynomial degree $n$ also raises the memorisation capacity, so more noisy copies per frame are needed to drive the network past it.

The denoising of sequences that we demonstrated can be seen as a simple form of generalization, where the learned prototype plays the role of the concepts or archetypes emerging from the observation of many examples.
This nontrivial form of computation takes place directly at the level of the physical degrees of freedom (without having to \enquote{encode} the data or \enquote{compile} a program), making it directly available in the physical process where it is implemented, but it is limited in scope.
In the Methods and Extended Data Fig.~\ref{fig:methods_route_B}, we show that universal (Turing complete) cellular automata can also be emulated with our asynchronous architecture by endowing the neural network with additional phase degrees of freedom, coupled with physical aligning interactions, that effectively keep track of local time.

\medskip
\noindent{\bf Outlook.}
Our results provide a first step towards the theoretical understanding of decentralized asynchronous computations as physical processes.
We emphasize that computation can both be the desired outcome of an engineered platform, but also a byproduct of a physical process, or equivalently a way of describing this process.
This is for instance the case for mechanical self-assembly processes (where an object constructs itself from the interactions of its constituents), the assembly of complex ecological communities (made of many species with various abundances), or cellular differentiation (where individual cells have to decide what fate to adopt within an organism).
Even though some of these processes are arguably emerging with no preset goal, it can be fruitful to cast their dynamics as a computation.
In these situations, effective inertia or more complex memory may also emerge from the combination of different nonequilibrium mechanisms.
For instance, in the context of physical self-assembly, cast as a mechanistic associative memory \cite{Murugan2014,Sartori2019,Evans2024,Braz2024}, a delay originating from a nucleation barrier may stabilize nonequilibrium situations where constituents sequentially self-assemble and disassemble into a succession of different structures \cite{Osat2022,Metson2025}. 
Similarly, inertial effects may also arise from error correction mechanisms such as kinetic proofreading in biological assembly processes like DNA replication \cite{Hopfield1974,Boeger2022,Ravasio2026}.

\setcounter{figure}{0}
\renewcommand{\thefigure}{\arabic{figure}}
\renewcommand{\theHfigure}{ED{\arabic{figure}}}

\def\figurename{Extended Data Figure}

\section*{Methods}

\medskip
\noindent{\bf Inertial Hopfield model versus time delays.}
The inertial Hopfield model introduced in the main text is described by the equations (reproduced from Eq.~\eqref{eq:model} in the main text)
\begin{subequations}
    \label{eq:model_methods}
    \begin{align}
        x_i(t+\Delta t) & = {\rm sign}\left( J_{ij} x_j(t) + \lambda p_i(t) + \eta^x_i(t) \right)
        \label{eq:model_a_methods}
        \\
        p_i(t+\Delta t) & = {\rm sign}\left( K_{ij} x_j(t) + \eta^p_i(t) \right).
        \label{eq:model_b_methods}
    \end{align}
\end{subequations}
This model is constructed in analogy with Newton's laws of motion for a particle of mass $m$, position $x$ and linear momentum $p$ subject to a force $F(x)$, written as coupled first-order (Hamilton) equations $\dot{x} = p/m$ and $\dot{p} = F(x)$. This is most apparent when discretizing these equation in time as
\begin{subequations}
\begin{align}
    x(t+\Delta t) &= x(t) + (\Delta t/m) p(t) \\
    p(t+\Delta t) &= p(t) + \Delta t \, F(x(t)).
\end{align}
\end{subequations}
and in the case where $J_{ij}=0$ and $\eta_i^x = 0$ in Eqs.~\eqref{eq:model_methods}, where we see that the neurons $p_i(t)$ play a similar role as momentum in mechanics.
(Note that sequence retrieval is possible even when $J_{ij}=0$.)
Note however significant differences: while classical mechanics deals with continuous quantities $x(t)$ and $p(t) \in \mathbb{R}$, the neurons in our model are still binary variables $x_i(t)$ and $p_i(t) \in \{-1,+1\}$.
In electronic neural networks, where the state is continuous, inertia may for instance be introduced by adding an inductor in the circuit \cite{Babcock1986,Babcock1987,Wheeler1997}.
We emphasize that in our work, the role of inertia is not to accelerate a computation (like in gradient descent with momentum \cite{Polyak1964,Sutskever2013}) but to allow the computation to complete in the first place when run asynchronously.

Extra insights can be gained by considering the second-order equivalent of these equations. 
In the case of classical mechanics, eliminating $p$ from the discrete-time dynamics above yields the second-order difference equation
\begin{equation}
x(t+\Delta t) - 2 x(t) +  x(t-\Delta t) =(\Delta t)^2 F(x(t-\Delta t)) / m
\end{equation}
which is a discrete-time version of the second-order differential equation $\ddot{x} = F(x)/m$ aka Newton's law of motion. 
Similarly, eliminating the $p_i$ neurons leaves a single delayed Hopfield model, second-order update for $x$, namely
\begin{equation}
    \begin{split}
        &x_i(t+\Delta t)  = {\rm sign}\Big(
        J_{ij} x_j(t) 
        \\[-0.15cm]
        &\quad+ \lambda \, {\rm sign}\big( K_{ij} x_j(t-\Delta t) + \eta^p_i(t-\Delta t) \big) + \eta^x_i(t) \Big)
        \label{ours_second_order}
    \end{split}
\end{equation}
Equation \eqref{ours_second_order} is reminiscent of (but different from) the time-delayed dynamics described in Refs.~\cite{sompolinsky_sequential,sompolinsky1988statistical} which read
\begin{equation}
\label{eq:sompolinsky_delay}
\begin{split}
    \!\!\!x_i(t{+}\Delta t)=\text{sign}\big(
    J_{ij}
    \,x_j(t) 
    +\lambda
    K_{ij} x_j(t{-}\tau_{\text{D}}) + \eta^x_i(t)\big)
\end{split}
\end{equation}
when the delay time $\tau_{\text{D}}$ is set to $\tau_{\text{D}} = \Delta t$.
%
%
We emphasize two key differences between these:
\begin{enumerate}
    \item {\it The delays $\tau_{\rm D}$ in Eq.~\ref{eq:sompolinsky_delay} are explicit, and need to be on the order of one full sweep through all $N$ neurons.} That is, to implement it on a real physical system requires copying (without any noise) and storing (long) system histories in a memory buffer. In contrast,  the delays in Eq.~\ref{ours_second_order} arise from physical couplings to the additional momentum-like neurons which have their own noise~$\eta^p$.
    \item {\it The $p$-neurons enter Eq.~\ref{ours_second_order} with a sign function.} Whereas in Eq.~\ref{eq:sompolinsky_delay} the decomposition of the second set of neurons would need to track the integer value of $K_{ij}x_j$. This seemingly small difference results in the critical capacity of the network at intermediate $\lambda$ being higher than that of Eq.~\ref{eq:sompolinsky_delay} (the effect of the additional sign can be seen by comparing Figs.~S8 and S9 for synchronous dynamics for our model and Eq.~\ref{eq:sompolinsky_delay} without delay since it is not needed for synchronous dynamics).
    One can argue that, as opposed to storing state histories, the time delays can alternatively be achieved simply by the using slower synapses for the sequential $K_{ij}$ connections and faster synapses for the auto-associative $J_{ij}$. However, note that for this to work, the two sets of synaptic weights need to be comparable, i.e. $\lambda\sim 1$ in Eq.~\ref{eq:sompolinsky_delay}. 
    Yet in this regime, the critical capacity of the network becomes very small, reducing its usefulness.
    While the sequential capacity is the largest for $\lambda\to \infty$, in this regime the separation of the two time scales essentially disappears since only the fast synapses are relevant in strength.
\end{enumerate}

\medskip
\noindent{\bf Order parameters.}
We characterise the phases by two order parameters. The static-retrieval phase is
marked by a finite
\begin{equation}\label{eq:methods_ms}
    m_{\rm s} \equiv \max_{\mu=1\dots P} m^\mu(t\to\infty),
\end{equation}
i.e. the largest overlap with any single stored pattern, with overlaps $m^\mu(t)=\frac1N\sum_{i=1}^N \xi_i^\mu x_i(t)$.
The dynamic-retrieval phase is
quantified by an order parameter that measures how faithfully the network threads the
stored sequence, evaluated on the pure-state steps and excluding the periodic mixed
states:
\begin{equation}\label{eq:methods_md}
    m_{\rm d} = \frac1P \sum_{t=1}^{P} m^{\,(\mu^\star+(\varpi-1)t)\bmod P}(\varpi\,t),
\end{equation}
where $\varpi$ is the number of internal updates between one condensed pattern and the
next ($\varpi=3$ in Fig.~\ref{fig1}D; the condensed pattern advances
by $\varpi-1=2$ between consecutive pure states) and $\mu^\star$ is the (spontaneously selected)
starting pattern. Sequence retrieval self-organises so that $\bm y(t)$ is pure whenever
$\bm x(t)$ is mixed and conversely; this staggering between the two species is what
averts the descent to chaos.

\begin{figure}[hbt!]
    \centering
    \includegraphics[width=0.95\linewidth]{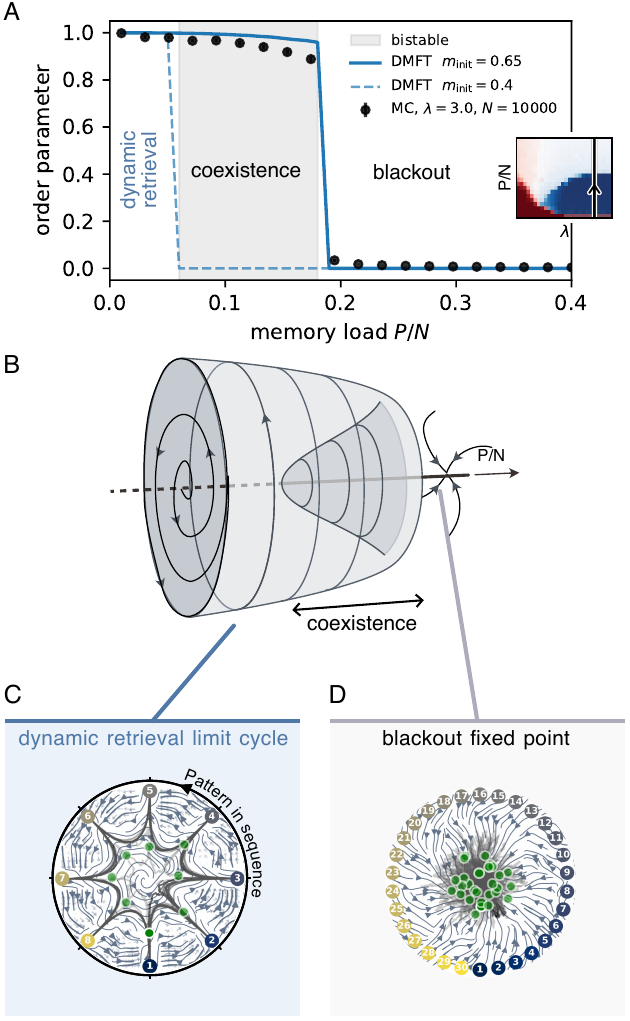}
    \vspace{-0.2cm}
    \caption{{\bf The phase transition between blackout and sequence retrieval.}
    {\bf A}
    A discontinuous transition from dynamic retrieval to blackout is observed as the memory load $P/N$ is increased
    (at fixed $\lambda=3$; inset): the order parameter $\Phi$ for the dynamic retrieval phase from Monte-Carlo simulations (pink markers), and from analytical loop DMFT predictions (purple continuous line) exhibit a sharp jump.
    A previous jump in the order parameter $\Omega$ from loop DMFT (light purple) suggests the presence of a hysteresis region (in grey).
    {\bf B}
    From a dynamical system perspective, this behavior akin to a first-order phase transition can be cast as a subcritical Hopf bifurcation: a stable fixed point (blackout phase; black line) looses stability at a critical $(P/N)_{\text{H}}\simeq 0.06$ and gives rise to an unstable limit cycle (inner surface) which then undergoes a saddle-node of limit cycles at $(P/N)_{\text{SNLC}} \simeq 0.17$, leading to a stable limit cycle (dynamic retrieval phase).
    {\bf C-D}
    This picture is evidenced by low-dimensional embeddings of the high-dimensional dynamics: the state of the ensemble of neurons $x_i(t)$ is embedded into the plane, where the center corresponds to states with vanishing overlap with the stored patterns, while each stored pattern correspond to a point on the circle.
    \label{ed_phase_diagram}
    }
\end{figure}

{\color{black}

\medskip
\noindent{\bf An embedded bifurcation.}
The transition from blackout to retrieval observed in the Monte-Carlo simulations of Fig.~\ref{fig2} is reminiscent of a first-order phase transition, such as water freezing: they are marked by a discontinuous jump in the order parameter ($m_{\text{s}}$ or $m_{\text{d}}$, see Extended Data Fig.~\ref{ed_phase_diagram}A), and are accompanied by other familiar signatures like hysteresis and metastable states, allowing the system to ``superheat'' beyond capacity as patterns are gradually loaded.
In spite of these similarities, the nonequilibrium nature of the model precludes a standard thermodynamic classification, and leads us to turn to the perspective of dynamical systems to assess the nature of the transition.

To do so, we numerically construct a low-dimensional embedding that effectively endows the pattern space with the topology of the sequence, disregarding the natural distance between binary sequences (see below).
This is done by arranging the stored patterns $\bm{\xi}^\mu$ around the unit circle on the plane, and by mapping arbitrary states $\bm{x}(t)$ to the interior based on their overlap with these patterns.


Strikingly, smooth trajectories emerge (Extended Data Figs.~\ref{ed_phase_diagram}C and D), revealing three distinct dynamical regimes as we vary non-reciprocity $\lambda$ and memory load $P/N$.
In the blackout phase, the center of the plane (where the system has vanishing overlap with all patterns) is an attracting fixed point (Extended Data Fig.~\ref{ed_phase_diagram}D).
This fixed point becomes repelling in the retrieval phases: in the static retrieval phase, each stored pattern corresponds to an isolated fixed point, while in the sequence retrieval phase they are all part of a limit cycle connecting them in order (Extended Data Fig.~\ref{ed_phase_diagram}C).

}

To visualise and analyse the high-dimensional dynamics we use a nonlinear circular
embedding. The $P$ stored patterns are placed uniformly on the unit circle,
$\mathbf p_\mu=(\cos\theta_\mu,\sin\theta_\mu)$ with $\theta_\mu=2\pi(\mu-1)/P-\pi/2$, so
they sit in sequential order. A network state $\mathbf x$ is mapped to the plane by its
overlap-weighted pattern positions,
\begin{equation}\label{eq:methods_proj}
    \Pi(\mathbf x)=\sum_{\mu=1}^P w_\mu(\mathbf x)\,\mathbf p_\mu,
\end{equation}
with
\begin{equation*}
    w_\mu(\mathbf x)=\frac{\exp(\tau_d\,m_\mu)}{\sum_\nu \exp(\tau_d\,m_\nu)},
    \qquad m_\mu=\tfrac1N\mathbf x\!\cdot\!\boldsymbol\xi^\mu,
\end{equation*}
where the temperature $\tau_d$ tunes between winner-take-all ($m_\mu\!\approx\!1$ is embedded
near $\mathbf p_\mu$) and soft blending (uniformly small overlaps are embedded near the
origin). The three phases acquire distinct signatures: static retrieval sits at a single
$\mathbf p_\mu$, dynamic retrieval threads the patterns as a closed loop, and blackout
collapses to the origin.

To turn the embedding into a flow field we evolve states and measure the induced 2D
displacement. Rather than inverting the (many-to-one) embedding, we sample actual
trajectories: at each grid point $\mathbf z$ the flow is the distance-weighted average of
the one-step embedded displacements of nearby trajectory points,
\begin{equation}\label{eq:methods_flow}
    \mathbf F(\mathbf z)=\frac{\sum_{k\in\mathcal K}\kappa_k\,[\Pi(\mathbf x^{(k+1)})-\Pi(\mathbf x^{(k)})]}{\sum_{k\in\mathcal K}\kappa_k},
\end{equation}
where
\begin{equation*}
    \kappa_k=\frac{1}{|\Pi(\mathbf x^{(k)})-\mathbf z|+\epsilon}.
\end{equation*}
Under asynchronous updating each step flips at most one spin, so the embedded point
moves by $O(1/N)$ and, as $N\to\infty$, $\mathbf F$ becomes the velocity of a
continuous-time flow $\dot{\mathbf z}=\mathbf F(\mathbf z)$. The relevant stability
criterion is therefore $\mathrm{Re}\,\lambda$ (imaginary axis), not the unit circle.

The origin is a fixed point by symmetry. Linearising via the $2\times2$ Jacobian
\begin{equation}\label{eq:proj_jacobian}
    (\mathbf J_\Pi)_{ab} = \frac{\partial F_a}{\partial z_b}\bigg|_{\mathbf z=\mathbf 0},
\end{equation}
estimated from central finite differences of $\mathbf F$ near the origin, gives a complex-conjugate pair $\lambda_\pm=\mu\pm i\nu$ with
$\mathrm{Re}\,\lambda_\pm=\tfrac12\,\mathrm{tr}\,\mathbf J_\Pi$. Because we reconstruct the
flow in a \emph{neighbourhood} of the origin, we can characterise it even when it repels
(unlike forward iteration of the DMFT map).

\smallskip
\noindent{\it Two scales, one transition.}
The same remembering--forgetting transition is reached from two complementary
constructions at different scales. The loop-space DMFT is a
\emph{macroscopic, discrete-time} map whose stability is set by the unit-circle criterion
$\rho(\mathbf J)<1$; its spectrum approaches the unit circle from within as the
retrieval branch loses robustness, and dynamic retrieval is actually lost at
$P_c/N\approx0.18$ through a fold (saddle-node) with the unstable branch. The circular
embedding (above) is a \emph{microscopic, quasi-continuous} flow whose
stability is set by $\mathrm{Re}\,\lambda<0$; near the origin the
embedded coordinate reduces to the fundamental Fourier mode of the overlap vector, so
the eigenvalue pair is measured directly as the growth and rotation of this mode in
trajectories started at zero overlap, averaged over disorder
(Supplementary Information). Its eigenvalue pair crosses the imaginary
axis at $P/N\approx 0.08$ at $N=400$ (drifting
toward the DMFT Hopf load $\simeq0.06$ with increasing system size), where an oscillatory instability sets in and the zero-overlap state turns from repelling to attracting, rotating at
the sequence-advance rate. The two pictures therefore
consistently diagnose the loss of sequence retrieval, their two
crossings bounding the bistable window of the subcritical bifurcation
(Fig.~\ref{fig2}).


{\color{black}
\medskip
\noindent{\bf Dynamic mean-field theory in loop space.}
The key idea behind standard DMFT is to reduce the very high dimensional dynamics of a many-body system to the dynamics of just a few order parameters \cite{Cugliandolo2024}.
Consider for instance a magnet made of Ising spins $s_i=\pm 1$ all connected to each other with ferromagnetic couplings.
The effect on any given spin of the all the other can be represented by a mean molecular field $h_{\mathrm{eff}}(m)=J m$ which depends on the magnetisation $m=\langle s_i \rangle$ of the magnet. 
After plugging this representation in the equation of motion, the full many-body dynamics then collapses onto a single self-consistent map $m_{t+1}=\langle\tanh(\beta(Jm_t+h))\rangle$.
More generally, DMFT describes the evolution of a macroscopic state vector $\bm{\theta}_t$ containing the relevant order parameters through a self-consistent iterative map of the form $\bm{\theta}_{t+1} = \langle \mathbf{F}(\bm{\theta}_{t}) \rangle$.

Statistical methods including DMFT have been applied to associative memories~\cite{stat_neurodyn,Gardner1987,Horner1989,Clark2026}, but we need to tweak them to handle both asynchrony and the occurrence of limit cycle solutions that spontaneously break time-translation invariance.
Indeed, the synaptic field seen by each neuron is temporally heterogeneous: it mixes contributions from neurons that have already been updated with those still carrying their previous state.
To make this tractable, we first approximate the asynchronous evolution by partitioning the $N$ neurons into many synchronously updated blocks~\cite{Mignacco_2021, lenka_sk}.
We then construct a generic ansatz able to capture limit cycle solutions even when their oscillation period is not commensurate with the duration of a single sweep (i.e., the update of all neurons exactly once).
The resulting loop DMFT allows us to self-consistently predict the period of oscillation from first principles (SI).

To get a glimpse of the phenomenon with lower algebraic complexity, we can use the macroscopic dynamics observed in the Monte-Carlo simulations to guess the size of the sequence, we then obtain generalised DMFT self-consistency equations in loop space with only two blocks, that take the form 
\begin{align}
    \bm{\theta}_{t+1} = \langle \mathbf{F}(\bm{\theta}_{t}, t) \rangle
    \label{loop_dmft}
\end{align}
where $\mathbf{F}(\bm{\theta}, t+\varpi)=\mathbf{F}(\bm{\theta}, t)$, and where the average is taken over random embedded patterns $\bm{\xi}$.
The time periodicity of the function $\bm{F}$ reflects the periodicity of the dynamics (recall that $\varpi$ is the period of occurrence of mixed states, see Fig.~\ref{fig1}).
}

We can now summarise the closed set of DMFT equations governing the macroscopic dynamics, and provide their derivation in the Supplementary Information.
We dub our approach, loop-space DMFT, extending the statistical-neurodynamics approach of Amari--Maginu to capture the non-autonomous macroscopic dynamics induced in our model by an interplay of asynchrony and nonreciprocity.

\smallskip
\noindent{\it Setup.}
In the $\lambda\to\infty$ limit the dynamics reduces to stochastic Glauber updates of
the queue neurons,
\begin{equation}
\begin{split}
    &P\!\left[y_i(T+\omega)\,\big|\,h_i(T)\right]\\
    &\qquad= \tfrac12\!\left[1 + y_i(T+\omega)\tanh\!\big(\beta h_i(T)\big)\right],
\end{split}
\end{equation}
where the macroscopic time $T$ advances by $\omega/N$ per single-neuron update,
so that one full sweep advances $T$ by $\omega$: the constant $\omega$ is the frequency
of the emergent macroscopic cycle in units of sweeps, i.e.\ the dynamics repeats after
$1/\omega = \varpi$ sweeps (here $\varpi = 3$), during which the condensed pattern
advances by two.

\smallskip
\noindent{\it Block-asynchronous approximation.}
Under asynchronous updating the synaptic field at site $i$ mixes updated and
not-yet-updated states,
\begin{equation}\label{eq:methods_field_exact}
    h_i(T) = \frac1N\sum_{\mu=1}^P \xi_i^{\mu+1}
    \!\left[\sum_{j\in U_i}\xi_j^\mu y_j(T) + \sum_{j\notin U_i}\xi_j^\mu y_j(T-\omega)\right],
\end{equation}
with $U_i$ the set of neurons already updated at time $T$. We split the neurons into two
equal blocks $B_a,B_b$ ($|B_I|=N/2$) that update synchronously within a block, with the
inter-block sweep proceeding asynchronously. The block-averaged field then has two
components,
\begin{equation}\label{eq:methods_block_field}
    \mathbf h(T) := \begin{bmatrix} h_a\\ h_b\end{bmatrix}
    = \sum_{\mu=1}^P \xi^{\mu+1}
      \begin{bmatrix} m_b^\mu(T)\\ m_a^\mu(T-\omega)\end{bmatrix},
\end{equation}
where
\begin{equation}
    m_I^\mu = \frac2N\sum_{j\in B_I}\xi_j^\mu y_j(T),\;\; I\in\{a,b\},
\end{equation}
so block $a$ (late updaters) sees block $b$'s current state, while block $b$ (early
updaters) sees block $a$'s state from the previous sweep.

\smallskip
\noindent{\it Three-cyclic mean field.}
Monte Carlo simulations show the macroscopic dynamics is period-3, i.e.\ $\omega=1/3$
(Fig.~\ref{fig2}F). Treating the crosstalk from the $P-1$ uncondensed patterns as
Gaussian noise $\mathbf z$, inspection of the condensed patterns at each step of the
cycle yields the three-cyclic ansatz
\begin{align}
    \mathbf h(T)        &= \xi^{2T+1}\begin{bmatrix} m^{2T}_b(T)\\ m^{2T}_a(T-\omega)\end{bmatrix},\\
    \mathbf h(T+\omega)  &= \begin{bmatrix} \xi^{2T+2}\,m^{2T+1}_b(T+\omega)\\ \xi^{2T+1}\,m^{2T}_a(T)\end{bmatrix},\\
    \mathbf h(T+2\omega) &= \xi^{2T+2}\begin{bmatrix} m^{2T+1}_b(T+2\omega)\\ m^{2T+1}_a(T+\omega)\end{bmatrix}.
\end{align}

\smallskip
\noindent{\it Self-consistent order parameters.}
We now project the stochastic microscopic dynamics onto the condensed overlaps. The
crosstalk from the uncondensed patterns enters as a Gaussian field of variance
$\alpha\mathbf r$, so the projection yields four dynamical order parameters---one per step
of the three-cycle---defined by
\begin{widetext}
\begin{align}
    \mathbf\Theta(T)   &:= \mathbf m^{2T+1}(T+\omega)
        = \Big\langle \xi^{2T+1}\!\int\!\DD\mathbf z\, F_\beta\!\big[\mathbf h(T)+\mathbf z\sqrt{\alpha\mathbf r(T)}\big]\Big\rangle_\xi,\\
    \mathbf\Phi(T)     &:= \mathbf m^{2T+1}(T+2\omega)
        = \Big\langle \xi^{2T+1}\!\int\!\DD\mathbf z\, F_\beta\!\big[\mathbf h(T+\omega)+\mathbf z\sqrt{\alpha\mathbf r(T+\omega)}\big]\Big\rangle_\xi,\\
    \mathbf\Psi(T+1)   &:= \mathbf m^{2T+2}(T+2\omega)
        = \Big\langle \xi^{2T+2}\!\int\!\DD\mathbf z\, F_\beta\!\big[\mathbf h(T+\omega)+\mathbf z\sqrt{\alpha\mathbf r(T+\omega)}\big]\Big\rangle_\xi,\\
    \mathbf\Omega(T+1) &:= \mathbf m^{2T+2}(T+1)
        = \Big\langle \xi^{2T+2}\!\int\!\DD\mathbf z\, F_\beta\!\big[\mathbf h(T+2\omega)+\mathbf z\sqrt{\alpha\mathbf r(T+2\omega)}\big]\Big\rangle_\xi,
\end{align}
\end{widetext}
Here $F_\beta(\cdot)\equiv\tanh(\beta\,\cdot)$, and $\DD\mathbf z=\big[\tfrac{dz_i}{\sqrt{2\pi}}e^{-z_i^2/2}\big]_{i=1}^2$ is the standard normal measure.

\smallskip
\noindent{\it Noise variance and susceptibility.}
The crosstalk variance closes self-consistently with a cross-block coupling,
\begin{subequations}
    \label{eq:dmft_r_eqs}
    \begin{align}
        r_a(T+\omega) = 1 + U_b^2(T+\omega)\,r_b(T)\label{eq:r_a},\\
        r_b(T+\omega) = 1 + U_a^2(T+\omega)\,r_a(T)\label{eq:r_b}
    \end{align}
\end{subequations}
with the corresponding susceptibilities
\begin{equation}
    \mathbf U(T+\omega) = \beta\left\{1 - \left\langle\!\int\!\DD\mathbf z\,F_\beta^2\!\big[\mathbf h(T)+\mathbf z\sqrt{\alpha\mathbf r(T)}\big]\right\rangle_\xi\right\},
\end{equation}
and analogous relations at $T$ and $T+2\omega$. Together these form a closed system of
14 coupled equations (two block components $\times$ four order parameters, plus two
block-resolved variances $\times$ three time steps). 
Numerical solution of the DMFT equations, by integration over the Gaussian noise and iterating to a fixed point, captures the order parameter dynamics in finite-$N$ simulations well (Extended Data Fig.~\ref{ed_phase_diagram}), correctly reproducing both the critical capacity $P_c/N \approx 0.18$ and the discontinuous nature of the remembering-forgetting transition in Fig.~\ref{fig2}.

Note that the non-autonomy of Eq.~\eqref{loop_dmft} (with an explicit (period-$3$) time dependence) is the result of our level of description, where one species has been eliminated; the full loop DMFT equations presented in the SI are autonomous and predict the spontaneous symmetry breaking of time translation observed in the simulations (Supplementary Information). 

\medskip
\noindent{\bf Asynchronous update protocol.}
We study the asynchronous dynamics of the bipartite non-reciprocal model of
Eq.~\eqref{eq:model}. Throughout we use a scheduled sweep in either a fixed or a random-permutation order, in which we update one neuron at a time, one full sweep through all neurons a flip of each neuron in the system is attempted exactly once. 

\begin{figure*}[hbt!]
    \centering
    \includegraphics[width=1\linewidth]{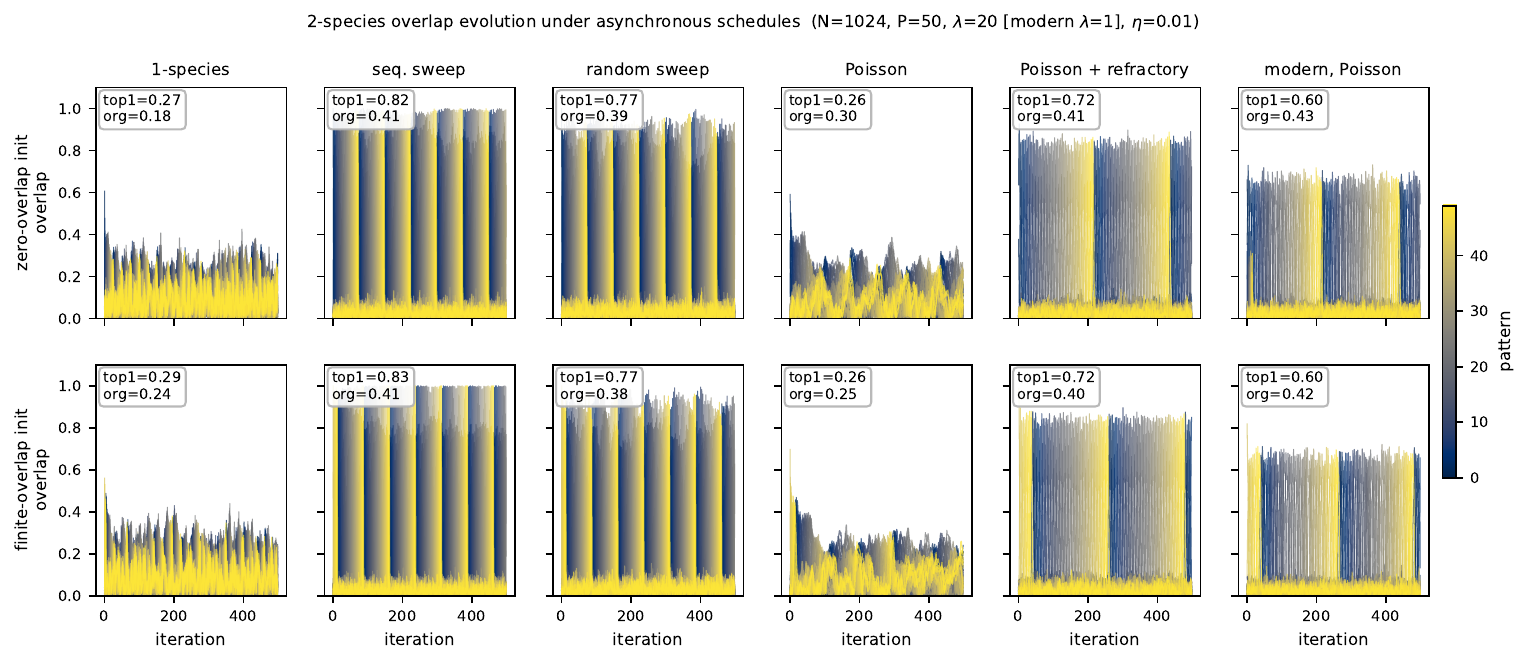}
    \caption{{\bf Asynchronous update schedules and the role of
    refractoriness.}
    Overlap evolution $m^\mu(t)$ with all stored patterns, starting from zero-overlap (top row) and finite-overlap (bottom row) initial conditions.
    From left to right:
    (i)~the single-species sequence network fails to retrieve under any asynchronous
    update protocol at any load;
    (ii, iii)~the bipartite two-species model retrieves the sequence under the
    scheduled (once-per-epoch) asynchronous sweeps used throughout this work, both in
    fixed order (seq.\ sweep) and in random-permutation order (random sweep);
    (iv)~under a fully Poissonian schedule (sites drawn independently with
    replacement) the uneven participation destroys retrieval in the bipartite model;
    (v)~a neuronal refractory period restores sequence retrieval under the Poisson
    schedule;
    (vi)~the modern (dense) version of the two-species model likewise retrieves under
    the Poisson schedule.
    \label{fig:ed_schedules}}
\end{figure*}

If we consider a less regular protocol in which the sites instead drawn independently with replacement (a fully Poissonian schedule),
participation would become uneven---some neurons updating several times per epoch, others
not at all---and the overlaps would drop as the state spreads across consecutive patterns.
A refractory period, which prevents a neuron from updating again too soon, evens out this
participation and restores retrieval (Extended Data
Fig.~\ref{fig:ed_schedules}). Such refractoriness is intrinsic to real neurons
and, together with a finite membrane time constant, to our neuromorphic neurons (below);
the regular updating the protocol uses is therefore not imposed by hand but supplied by
neuronal biophysics itself.

\begin{figure*}[h!]
    \centering
    \includegraphics[width=1\linewidth]{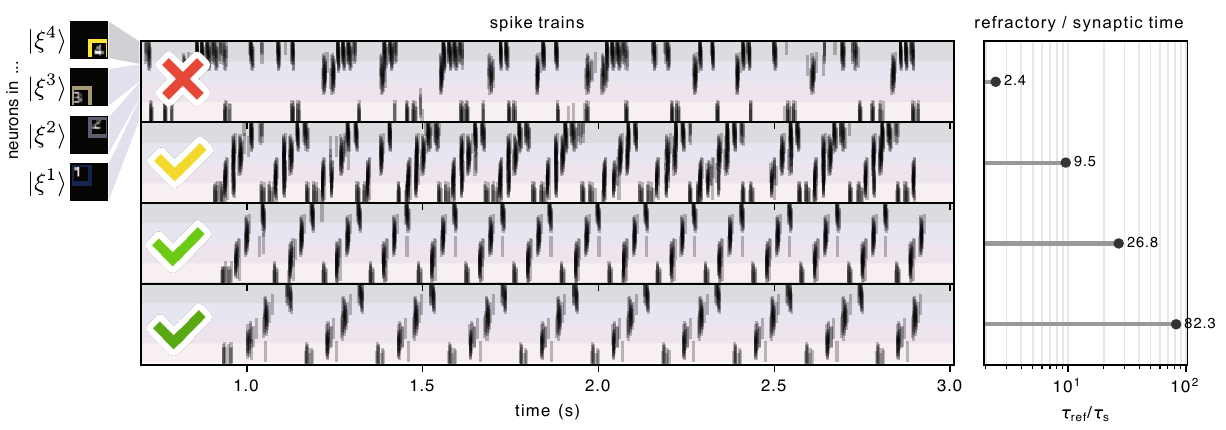}
    \caption{\textbf{Refractory period relative to synaptic time controls sequence
    retrieval in the neuromorphic experiment.} Spike rasters for four NMDA synaptic time constants $\tau_s$ at fixed
    refractory period $\tau_{\rm ref}=7.7$~ms (coloured rails mark the stored patterns;
    right: the ratio $\tau_{\rm ref}/\tau_s$). Sequences are retrieved only when
    $\tau_{\rm ref}/\tau_s$ is large enough (bottom); for too-small ratios the dynamics
    deteriorates into a superposition (top).}
    \label{fig:nmda_refractory}
\end{figure*}

\begin{figure*}[h!]
    \centering
    \includegraphics[width=1\linewidth]{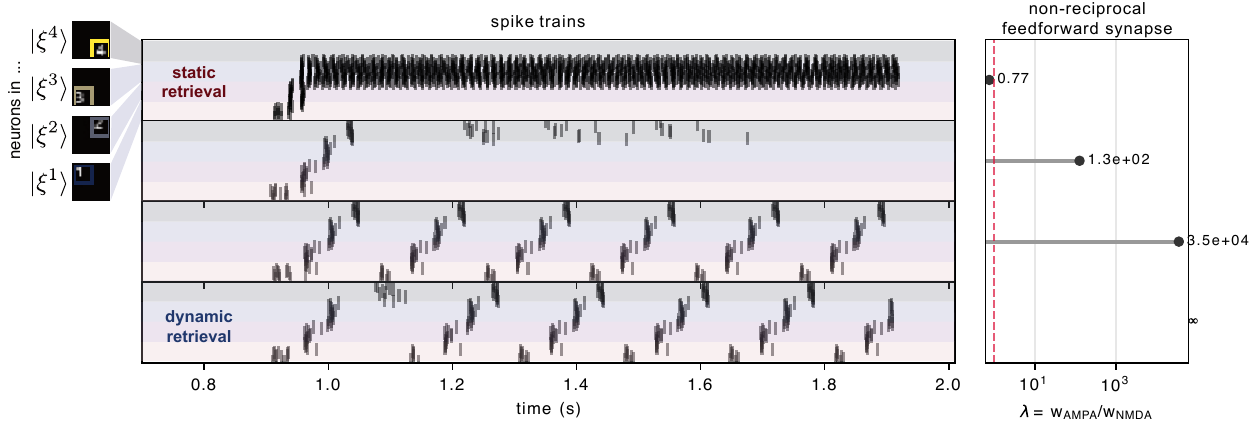}
    \caption{\textbf{Experimentally crossing the static--dynamic retrieval phase transition
    on the neuromorphic chip.} Spike rasters (coloured rails mark the stored patterns) as
    the recurrent NMDA weight $w_{\rm NMDA}$ (e.g. Hebbian $J$) is lowered from top to bottom at
    fixed AMPA feedforward weight $w_{\rm AMPA}$ ($\lambda$); right, the control ratio
    $w_{\rm AMPA}/w_{\rm NMDA}$. Strong $w_{\rm NMDA}$ (top) freezes the network in a static
    superposition; lowering it (bottom) releases sequential retrieval.}
    \label{fig:neuromorph_static_dynamic}
\end{figure*}

\medskip
\noindent{\bf Neuromorphic experiment.}
We implement the two-species model of Eq.~\eqref{eq:model} on the mixed-signal
neuromorphic DYNAP-SE chip, whose analog/digital circuits emulate spiking neurons and
synapses in continuous time. Species $x$ and $y$ occupy two cores (core $0=x$, core
$1=y$), each of 256 leaky integrate-and-fire neurons with programmable weights and time
constants. Writing the membrane potential of neuron $i$ in species $a\in\{x,y\}$ as
$V_i^a$,
\begin{subequations}\label{eq:methods_membrane}
    \begin{align}
        \tau_m\,\dot V_i^{x} &= -V_i^{x} + \sum_j J_{ij}\,(\kappa_s * s_j^{x})
                                + \lambda\,(\kappa_f * s_i^{y}) - \phi_i^{x}(t),
        \label{eq:methods_membrane_a}\\
        \tau_m\,\dot V_i^{y} &= -V_i^{y} + \sum_j K_{ij}\,(\kappa_s * s_j^{x})
                                - \phi_i^{y}(t),
        \label{eq:methods_membrane_b}
    \end{align}
\end{subequations}
where $s_j^{a}(t)=\sum_k\delta(t-t_j^{a,k})$ is the spike train of neuron $j$, $\kappa_s$
and $\kappa_f$ are the slow (NMDA, time constant $\tau_s$) and fast (AMPA) synaptic
kernels, and a neuron integrates until $V_i^{a}$ crosses threshold, spikes, and resets.
This realises the threshold ($\mathrm{sign}$) update of Eq.~\eqref{eq:model} in continuous
time, with $x_j, y_j$ replaced by the low-pass-filtered spike trains and the noise
$\eta^{a}$ provided by the intrinsic analog stochasticity of the neurons. The term
$\phi_i^{a}$ is the neuron's intrinsic spike-triggered self-inhibition, comprising a fast
(refractory, $\tau_{\rm ref}$) and a slow (adaptation, $\tau_a$) component.

The three couplings map onto distinct receptor classes: the symmetric intra-species memory
$J$ ($x\!\to\!x$) is carried by slow excitatory NMDA synapses; the asymmetric
cross-species coupling $K$ ($x\!\to\!y$) by NMDA where $K_{ij}>0$ and by
$\mathrm{GABA_B}$ where $K_{ij}<0$; and the non-reciprocal feedback $\lambda$ (local,
$y_i\!\to\!x_i$) by fast excitatory AMPA synapses. The inhibitory ($\mathrm{GABA_B}$)
entries of $K$ are what release the currently active pattern; if they are too weak the
patterns cannot switch and the dynamics arrests at a static superposition of the loaded
patterns. Sweeping these hardware parameters reproduces the phase diagram of the
model (Fig.~\ref{fig:hardware_demo}). We further investigated the effect of the refractory period relative to
the synaptic time scale, and found that sequences can be retrieved successfully only when
this ratio $\tau_{\rm ref}/\tau_s$ is sufficiently large (Extended Data
Fig.~\ref{fig:nmda_refractory}).

\begin{figure*}[hbt!]
    \centering
    \includegraphics[]{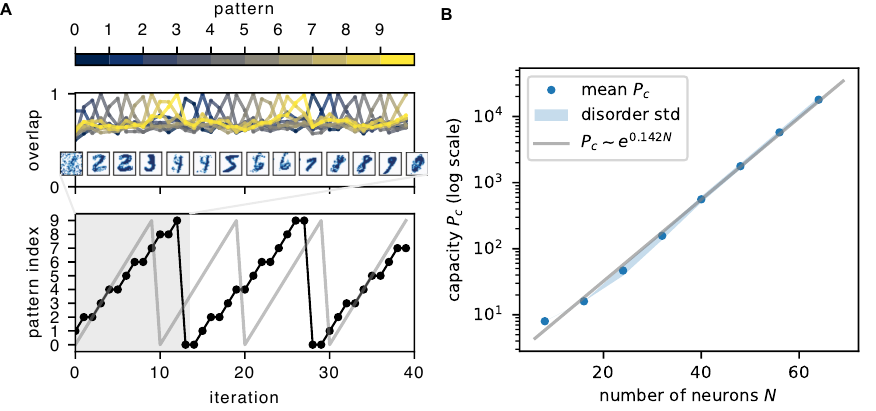}
    \caption{{\bf Asynchronous modern Hopfield network. 
    }
    {\bf A} We load a sequence of $P=10$ frames from a natural sequence, where each frame is a hand-written digit. Despite the strong overlaps between frames, the network with non-linearity successfully retrieves the entire sequence upon stimulation with a noisy partial cue, where the evolution overlaps between the network state and each stored pattern reveal clear sequential transitions between frames.
    {\bf B} For uncorrelated patterns, this memory capacity of this model scales exponentially with the number of neurons $N$.
        \label{fig:modern}
    }
\end{figure*}

\medskip
\noindent{\bf Modern (dense) non-reciprocal memory.}
We now lift the model to dense, correlated memories. The Hebbian construction of
Eq.~\eqref{eq:model} limits the capacity to a fraction of $N$ and breaks down for
correlated patterns. Both limitations are lifted by
inserting a strong nonlinearity $f$ acting on the pattern overlaps $m^\mu=\tfrac1N\sum_j\xi_j^\mu x_j$,
$\sum_\mu \xi_i^{\mu+1} m^\mu \to \sum_\mu \xi_i^{\mu+1} f(m^\mu)$, which amplifies the
separation between partially overlapping memories. With a sufficiently sharp $f$ (a high
power, or a softmax) the static capacity becomes exponential in $N$. The non-reciprocal,
sequence-generating version of our model is obtained by passing both channels through
$f$:
\begin{subequations}
\label{eq:methods_modern}
    \begin{align}
        x_i^1 &\leftarrow \mathrm{sign}\!\left[\sum_{\mu=1}^P \xi_i^\mu\, f\!\left(\sum_j \xi_j^\mu x_j^1\right) + \lambda x_i^2 + \eta_i^1\right],\\[4pt]
        x_i^2 &\leftarrow \mathrm{sign}\!\left[\sum_{\mu=1}^P \xi_i^{\mu+1}\, f\!\left(\sum_j \xi_j^\mu x_j^1\right) + \eta_i^2\right].
    \end{align}
\end{subequations}

Provided $f$ is sufficiently nonlinear, this model retrieves long sequences
\emph{asynchronously}, overcoming the central limitation of the linear model: for
uncorrelated patterns the sequence capacity scales exponentially with $N$, and for
strongly correlated frames (large temporal and spatial overlap) the network still
threads the entire sequence at high load. This is what enables the correlated-sequence
denoising/generalisation computation reported in the main text (Fig.~\ref{fig:denoise}),
and is illustrated for an image sequence in Fig.~\ref{fig:modern}. A biologically
plausible realisation that avoids explicit many-body interactions (auxiliary hidden
neurons storing each overlap), and an unlearning/``dreaming'' variant, are given in the
Supplementary Information.

\medskip
\noindent{\bf Universal computation by cellular-automaton emulation.}
In this section we show that the same non-reciprocal architecture, made \emph{local},
computes cellular automata (CA). We restrict the dense couplings of
Eq.~\eqref{eq:methods_modern} to a radius-1 window $\mathcal R=\{-1,0,+1\}$ and reuse a
single rule table at every site (a convolution). For an elementary two-state CA the table
has $P=2^3=8$ entries, stored as key/value pairs $(\kappa^\mu,v^\mu)$---the neighbourhood
configuration and the centre state it produces---which gives the update
\begin{equation}\label{eq:methods_localK}
\begin{aligned}
    x_i(t+\Delta t) &= \mathrm{sign}\!\Big[\sum_{\mu} \kappa^\mu_0\, f\!\big(\beta\!\sum_{d\in\mathcal R}\kappa^\mu_d x_{i+d}(t)\big) + \lambda y_i(t) + \eta^x_i\Big],\\
    y_i(t+\Delta t) &= \mathrm{sign}\!\Big[\sum_{\mu} v^\mu f\!\big(\beta\!\sum_{d\in\mathcal R}\kappa^\mu_d x_{i+d}(t)\big) + \eta^y_i\Big].
\end{aligned}
\end{equation}
At large $\beta$ the softmax selects the single stored key matching the current
neighbourhood, $y_i$ acquires the exact rule output, and $\lambda$ feeds it back into
$x_i$ one update later, so one CA step is realised over two sweeps with $y$ acting as an
implicit delay buffer (no explicit state cloning). This reproduces all 256 elementary
rules. When iterated asynchronously, the network reproduces the synchronous CA exactly
for the 18 (of 256) rules whose update operator is commutative on adjacent cells, whose
per-cell histories are schedule-invariant; all 18 are computationally simple (class 1/2).
The Turing-complete, non-commutative rules (e.g.\ rule~110) instead fail, because
asynchrony creates ``phantom'' half-updated neighbourhoods that never occur along the
synchronous trajectory.

\begin{figure*}[hbt!]
    \centering
    \includegraphics[width=1\linewidth]{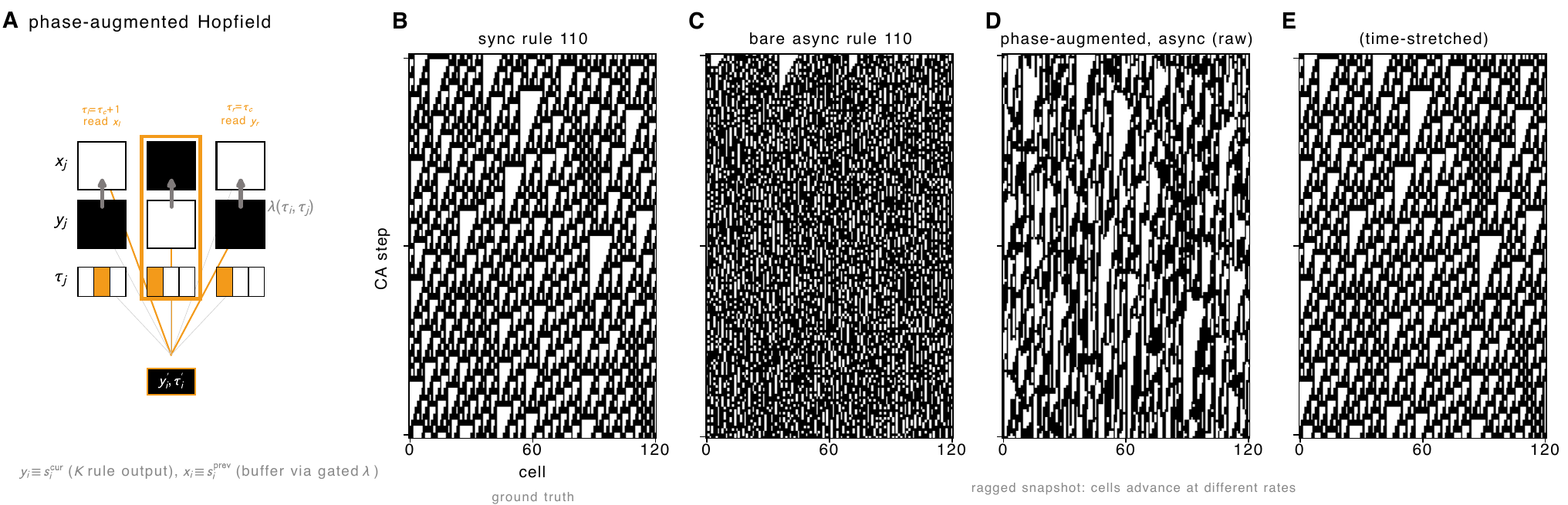}
    \caption{{\bf Universal computation under asynchrony.} We present a scheme that uses cellular automata with inertia and additional local clock degrees of freedom $\tau$.
    {\bf A}~Per-cell state $(x_i,y_i,\tau_i)$: the codebook $K$ writes the 
    current value $y_i'\!\equiv\! s^{\rm cur}_i$ and phase $\tau_i'$ from a 
    phase-selected read of the neighbourhood (orange: $x_l$ for a one-ahead 
    neighbour, $y_r$ for an in-phase one), and the $\tau$-gated coupling 
    $\lambda(\tau_i,\tau_j)$ relays $y\!\to\!x$ only on fire ($g_i\!=\!1$), so 
    $x_i\!\equiv\! s^{\rm prev}_i$ is buffered with no explicit clone. 
    {\bf B}~Synchronous rule~110 (ground truth). 
    {\bf C}~Bare async rule~110 disintegrates ($\sim\!50\%$ of cells wrong). 
    {\bf D}~The augmented model under async updates: the raw snapshot is 
    ragged as cells advance at different local rates. 
    {\bf E}~Re-indexed by each cell's logical time $\tau$, it matches the 
    synchronous trajectory exactly ($0\%$ mismatch).}
    \label{fig:methods_route_B}
\end{figure*}

Restoring universal computation under asynchrony requires the local clock that
non-reciprocity alone does not supply: $\lambda$ provides the one-step buffer but not the
decision of \emph{when} a cell may advance. The central construction is to wire in extra
clock degrees of freedom that realise the Nakamura phase-synchronisation scheme without
explicitly cloning state histories. We attach to each site a three-state Potts clock
$\boldsymbol\tau_i$ and promote the scalar $\lambda$ to a clock-dependent (Potts-coupled)
gate $\lambda(\boldsymbol\tau_i,\boldsymbol\tau_j)=\lambda\,\Lambda(\boldsymbol\tau_i,\boldsymbol\tau_j)$,
with the closed-form bilinear (in the one-hot clocks)
\begin{align}\label{eq:meth_gate}
&\Lambda(\boldsymbol\tau_i,\boldsymbol\tau_j)=
\begin{cases} 1 & (\tau_j-\tau_i)\bmod 3\in\{0,1\},\\ 0 & (\tau_j-\tau_i)\bmod 3 = 2,\end{cases}
\\
&g_i(t)=\!\!\prod_{j\in\{i-1,i+1\}}\!\!\Lambda\big(\boldsymbol\tau_i(t),\boldsymbol\tau_j(t)\big),
\end{align}
whose product gate $g_i$ is unity exactly when cell $i$ is fireable in Nakamura's sense
(no neighbour one phase behind). Identifying the queue $y_i$ with the next-step
prediction and $x_i$ with the one-step-stale state, the per-cell history buffer that
Nakamura stores as an explicit extra spin emerges here implicitly from the
non-reciprocity, while the clock gate supplies the missing synchronisation. This gate is the
computational counterpart of the neuronal refractory time invoked for asynchronous retrieval
above: both prevent a cell from advancing more than one step ahead of its neighbours---here
imposed exactly for universal computation, there supplied statistically by refractoriness. With the
augmented receptive field $\Phi_i(t)=(x,y,\boldsymbol\tau)_{i-1,i,i+1}$ and a codebook
$\{K^\mu,V^\mu\}$ over the $P_{\rm aug}$ augmented neighbourhoods, the gated dynamics reads
\begin{equation}\label{eq:meth_routeB}
\begin{aligned}
    y_i(t+\Delta t) &= \mathrm{sign}\!\Big[\textstyle\sum_{\mu} V^{\mu}_{y}\,
      f\!\big(\beta\, K^\mu\!\cdot\!\Phi_i(t)\big) + \eta^y_i(t)\Big],\\[2pt]
    \boldsymbol\tau_i(t+\Delta t) &= \mathrm{sign}\!\Big[\textstyle\sum_{\mu} V^{\mu}_{\tau}\,
      f\!\big(\beta\, K^\mu\!\cdot\!\Phi_i(t)\big) + \boldsymbol\eta^\tau_i(t)\Big],\\[2pt]
    x_i(t+\Delta t) &= \mathrm{sign}\!\Big[\big(1-g_i(t)\big)\,x_i(t) + g_i(t)\,\lambda\,y_i(t) + \eta^x_i(t)\Big],
\end{aligned}
\end{equation}
whose last line is the gated form of Eq.~\eqref{eq:model_a}: the queue reaches $x_i$ only
when the clock gate is open, otherwise $x_i$ holds its state. With this augmentation the
network reproduces the rule-110 trajectory exactly under any asynchronous schedule
(Fig.~\ref{fig:methods_route_B}); setting $\lambda=0$ while keeping
the codebook and gate breaks the construction. The full enumeration, the detailed failure
analysis of vanilla rule~110, and an alternative rule-engineered route are given in the
Supplementary Information.



\medskip
\noindent{\bf Acknowledgements.} D.E.G. is grateful to Junren Chen for helpful discussions about neuromorphic computing and establishing the initial connection to Neuromorphic Cognitive Systems group at the ETH Institute for Neuroinformatics. D.E.G acknowledges support by the NSF-Simons National Institute for Theory and Mathematics in Biology (NITMB) Fellowship supported via grants from the NSF (DMS-2235451) and Simons Foundation (MPS-NITMB-00005320). V.V. acknowledges partial support from the Army Research Office under grant W911NF-22-2-0109 and W911NF-23-1-0212. M.F. and V.V acknowledge partial support from the France Chicago center through a FACCTS grant. This research was partly supported from the National Science Foundation through the Center for Living Systems (grant no. 2317138), the Simons Foundation and the Chan Zuckerberg Foundation. This work was completed in part with resources provided by the University of Chicago’s Research Computing Center.

\bibliographystyle{unsrt}
\bibliography{bibliography}



\end{document}


\title{Supplementary Information for ``Inertial Asynchronous Computation''}

\author{Doruk Efe Gökmen}
\affiliation{NSF-Simons National Institute for Theory and Mathematics in Biology, Chicago, IL 60611, USA}
\affiliation{James Franck Institute, The University of Chicago, Chicago, IL 60637, USA}

\author{Michel Fruchart}
\affiliation{Gulliver, ESPCI Paris, Universit\'e PSL, CNRS, 75005 Paris, France}

\author{Dmitrii Zendrikov}
\affiliation{Institute of Neuroinformatics, University of Zurich and ETH Zurich, Zurich, Switzerland}

\author{Giacomo Indiveri}
\affiliation{Institute of Neuroinformatics, University of Zurich and ETH Zurich, Zurich, Switzerland}

\author{Giulio Biroli}
\affiliation{Laboratoire de Physique Statistique, \'Ecole normale sup\'erieure, PSL Research University, 24 rue Lhomond, 75005 Paris, France}

\author{Vincenzo Vitelli}
\affiliation{Leinweber Institute for Theoretical Physics, The University of Chicago, Chicago, IL 60637, USA}
\affiliation{James Franck Institute, The University of Chicago, Chicago, IL 60637, USA}

\maketitle                       


\clearpage
\onecolumngrid

\setcounter{section}{0}\renewcommand{\thesection}{S\arabic{section}}
\setcounter{equation}{0}\renewcommand{\theequation}{S\arabic{equation}}
\setcounter{figure}{0}\renewcommand{\thefigure}{S\arabic{figure}}\renewcommand{\theHfigure}{S\arabic{figure}}
\setcounter{table}{0}\renewcommand{\thetable}{S\arabic{table}}
\def\figurename{Figure}

\providecommand{\SIstandalone}{0}

\ifnum\SIstandalone=0
\begin{center}
  {\large\bfseries Supplementary Information for ``Inertial Asynchronous Computation''}\\[1.5em]
  {\normalsize Doruk Efe Gökmen,$^{1,2}$ Michel Fruchart,$^{3}$ Dmitrii Zendrikov,$^{4}$ Giacomo Indiveri,$^{4}$ Giulio Biroli,$^{5}$ and Vincenzo Vitelli$^{6,2}$}\\[0.8em]
  {\small\itshape
    $^{1}$NSF-Simons National Institute for Theory and Mathematics in Biology, Chicago, IL 60611, USA\\
    $^{2}$James Franck Institute, The University of Chicago, Chicago, IL 60637, USA\\
    $^{3}$Gulliver, ESPCI Paris, Université PSL, CNRS, 75005 Paris, France\\
    $^{4}$Institute of Neuroinformatics, University of Zurich and ETH Zurich, Zurich, Switzerland\\
    $^{5}$Laboratoire de Physique Statistique, École normale supérieure, PSL Research University, 24 rue Lhomond, 75005 Paris, France\\
    $^{6}$Leinweber Institute for Theoretical Physics, The University of Chicago, Chicago, IL 60637, USA}
\end{center}
\vspace{1.5em}
\fi

\titlecontents{section}[2.8em]
  {\addvspace{2pt}}
  {\contentslabel{2.6em}}
  {\hspace*{-2.6em}}
  {\titlerule*[0.6pc]{.}\contentspage}
\phantomsection
\section*{Contents}
\let\ORIGaddcontentsline\addcontentsline
\renewcommand{\addcontentsline}[3]{%
  \ifstrequal{#1}{toc}%
    {\ifstrequal{#2}{section}{\ORIGaddcontentsline{#1}{#2}{#3}}{}}%
    {\ORIGaddcontentsline{#1}{#2}{#3}}}
\startcontents[si]
\printcontents[si]{}{1}{}
\vspace{1em}

\section{Review of Hopfield networks}

\begin{figure}[ht!]
    \centering
    \includegraphics[width=0.25\linewidth]{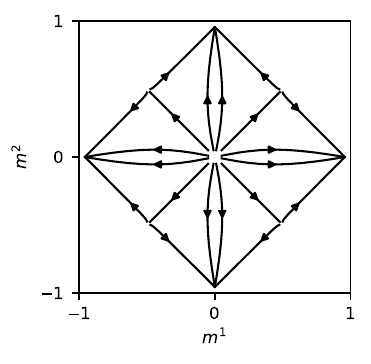}
    \caption{\label{fig:hm_simple_dynamics} Flow diagram of the Hopfield model in the simple limit
    $P/N\to 0$, under asynchronous updates. }
\end{figure}

A Hopfield network~\cite{LITTLE1978281, hopfield82} is a fully connected Ising model
with quenched disordered interactions,
\begin{equation}\label{eq:hopfield}
    H(\bm{x}) = -\frac{1}{2} \sum_{i,j} J_{ij} x_i x_j,
\end{equation}
whose spin configuration $\bm{x}=[x_i]_{i=1}^N$ is interpreted as the state of $N$
binary neurons. The symmetric couplings are trained by the Hebbian
rule~\cite{sompolinsky_beyond_hebb, Amit_1989}
\begin{equation}\label{eq:hebb}
    J_{ij} = \frac{1}{N} \sum_{\mu=1}^P \xi_i^\mu \xi_j^\mu
\end{equation}
---the auto-associative coupling $J$ of Eq.~\eqref{eq:model} in the main text---so that
$P$ random binary patterns $\boldsymbol{\xi}^\mu$ become attractors of the noisy
single-spin dynamics $x_i \leftarrow {\rm sign}(\sum_j J_{ij} x_j + \eta_i)$, see Fig.~\ref{fig:hm_simple_dynamics}.

The solution of Amit, Gutfreund and Sompolinsky maps out the phase
diagram: at
low temperature the memories are retrievable up to $P/N = 0.144$ (as global minima below
$P/N \approx 0.05$~\cite{Gardner_1986}, then as metastable states), beyond which the
network collapses discontinuously into a spin-glass phase with no
retrieval~\cite{PhysRevLett.55.1530}---the \emph{blackout catastrophe}. The capacity is
thus linear in $N$, and degrades further for correlated patterns~\cite{kinzel_85}.
\emph{Modern} Hopfield networks insert a strong nonlinearity $f$ that amplifies the
separation between partially overlapping memories~\cite{NIPS2016_eaae339c,
krotov_modern_hopfield}. These dense associative memories lift the linear capacity
scaling; a suitable choice of $f$ achieves exponentially many stored patterns while
retaining $\mathcal{O}(N)$ basins of attraction~\cite{modern_hopfield_capacity}.%

\subsection{How physically concrete is the Hopfield model?}
The Hopfield model is a mathematical abstraction, and its physical concreteness can be asked of many systems: beyond the brain, the same associative-memory dynamics has been invoked as an effective description of molecular self-assembly, DNA and chemical reaction networks, and gene-regulatory or epigenetic memory~\cite{Murugan2014, xiong_molecular_2022, Yampolskaya2025}, as well as analog and optical hardware. The crux in each case is the status of the two-state neuron: the physically concrete variable is not the discrete $S_i=\pm1$ but an underlying \emph{continuous} quantity---a mean firing rate, a concentration, a voltage, an optical amplitude---with its own deterministic dynamics, of which the binary description is the high-gain, time-averaged limit~\cite{continuous_hopfield, Amit_1989}. For neurons this is explicit: a cell's output is the short-time average of its firing rate $V_i=g(h_i)$, the gain $g(h)=\tfrac{1}{2}[1+\tanh(\beta h)]$ being exactly the Glauber form, and as $\beta\to\infty$ it saturates so that $S_i=\operatorname{sign}(h_i)$ records only whether the neuron fires near its maximum (``on'') or stays quiescent (``off''). The finite Glauber temperature is then not postulated but reappears on returning to spikes---a neuron fires in an interval $dt$ with probability $g(h_i)\,dt$, so the time-averaged ``on'' probability is again $g(h_i)$, with $\beta$ set by the spike statistics. In this sense the Hopfield model is physically implementable not as a literal microscopic dynamics but as the coarse-grained, high-gain limit of a continuous-rate description---a stochastic-to-deterministic shift in the level of description, of the kind that carries particle motion to hydrodynamics---in which the binary state and the temperature are emergent while only the couplings $J_{ij}$ are fundamental; the same saturation-and-noise logic governs the non-neural cases, with the firing-rate average replaced by the appropriate slow observable.

The two idealizations---reciprocal couplings $J_{ij}=J_{ji}$ and clocked, synchronous updating---are equally unphysical, but we set them aside since our architecture abandons both by construction. 
Indeed, the neuromorphic experiment (Fig.~\ref{fig:hardware_demo}) can be read as a direct demonstration of physical implementability in exactly this coarse-grained sense: the non-reciprocal Hopfield model emerges as the effective description of genuinely asynchronous, continuous-time spiking hardware---with the non-reciprocal couplings $K_{ij}\neq K_{ji}$ and the drive $\lambda$ built into the chip---even though here the reduction is exhibited empirically and no rigorous connection has yet been established.

\subsection{Sequential associative memory}
\begin{figure}[ht!]
    \centering
    \includegraphics[width=0.25\linewidth]{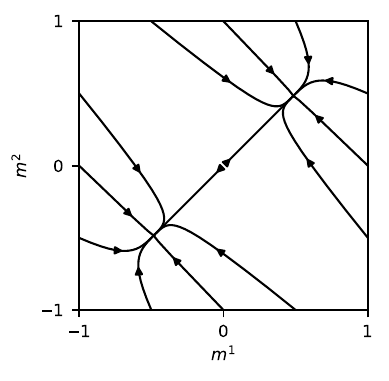}
    \caption{\label{fig:ahm_simple_dynamics} Flow diagram of the asymmetric sequence
    processing network \eqref{eq:naive} ($\lambda\to\infty$) in the simple limit
    $P/N\to 0$, under asynchronous updates. Instead of generating a limit cycle, the
    system is trapped in an attractor which is a mixture of the two patterns
    $\bm{\xi}^{1}+\bm{\xi}^{2}$.}
\end{figure}

Retrieving memories in a prescribed order
is harder than retrieving them individually, especially without a global clock. One can add a hetero-associative coupling, as the $K$ of Eq.~\eqref{eq:model} in the main text:
\begin{equation}\label{eq:naive}
    K_{ij} = \frac{1}{N} \sum_{\mu=1}^P \xi_i^{\mu+1} \xi_j^{\mu}, \quad x_i \leftarrow {\rm sign}\left(\sum_{j=1}^N J_{ij} x_j + \lambda \sum_{j=1}^N K_{ij} x_j\right).
\end{equation}
Under synchronous updates this model can cross between static and dynamic retrieval: for $\lambda<1$ retrieval is static; at finite $\lambda>1$ short sequences are retrieved with very low capacity; only the purely asymmetric limit
$\lambda \gg 1$ performs well, with $P_c/N \approx 0.269$~\cite{duringetal98}.
Dense versions of Eq.~\eqref{eq:naive} (see Sec.~\ref{si:dense}) improve the synchronous sequence capacity in the same
way as for static memories~\cite{NEURIPS2023_aa32ebcd}. But this cannot restore \emph{asynchronous} sequence retrieval, which is the problem addressed in this work.
Howe ver this model fails under asynchronous updates, as the network is
trapped in mixtures of consecutive patterns and $P_c/N = 0$~\cite{hopfield82,
synaptic_delay} (Fig.~\ref{fig:ahm_simple_dynamics}). 
Known remedies introduce finely tuned timescale separations (synaptic
delays)~\cite{Amit_1989, PhysRevLett.57.2861, sompolinsky_sequential}.

\subsection{Increasing capacity through dreaming}

It has been proposed that REM sleep facilitates learning by removing certain parasitic synapses~\cite{crick_unlearning_dream}.
Inspired by this hypothesis, the capacity of the network can be improved by including an \emph{unlearning} process (intentional forgetting)~\cite{hopfield_unlearning}, whereby synapses get updated by removing frustrated bonds.
In static memory retrieval setting, a recent result shows that the unlearning procedure approaches to a memory load $P_c/N\approx 0.6$, which is in fact optimal; it cannot beaten when $J_{ij}$ is learned by linear regression to encode a specific set of $P$ stable attractors~\cite{unlearning_optimal, unlearning_analysis, Derrida_1987}.
Concretely, the network is run at a \emph{dreaming mode}; updating the spins until converging to a stable fixed point, after which the synaptic matrix with an anti-Hebbian rule, \emph{i.e.} subtract the final state:
\begin{equation}
    J_{ij} \leftarrow J_{ij} -  \frac{\epsilon}{N}x_i x_j.
\end{equation}

In our model, the dreaming mode can be implemented as follows:
\begin{align}
    J_{ij} \leftarrow J_{ij} -  \theta(1-\lambda)\frac{\epsilon}{N} x^1_i x^1_j\\
    K_{ij} \leftarrow K_{ij} -  \theta(1-\lambda)\frac{\epsilon}{N} x_i^2 x_j^1,
\end{align}
where $\theta$ is the Heaviside step function, ensuring that the unlearning only occurs when the system is in the static retrieval phase, \emph{i.e.}, when it can converge to stable attractors.

\section{DMFT for Beginners}

\def\dd{\text{d}}
\def\ii{\text{i}}
\def\ee{\text{e}}

This section illustrates the main ideas of dynamical mean-field theory (DMFT), see Ref.~\cite{Cugliandolo2024} for more details and further references.

\subsection{The fully connected Ising model}
We start with the example of the mean-field Ising model, which can be seen as a toy model of ordering phenomena ranging from ferromagnetism \cite{Kardar2007} to decision making \cite{Redner2019}.

Let us consider (as in the main text) binary variables $s_i(t) = \pm 1$ ($i=1,\dots,N$) following the discrete-time dynamics\footnote{There is no substantial modification in the continuous-time case, which results in the DMFT equation as an ODE rather than a difference equation.}
\begin{equation}
        \label{hopfield_meth}
        s_i(t+1) = {\rm sign}\left(J_{ij} s_j(t) + \eta_i(t) \right)
\end{equation}
where a sum of the repeated index $j$ is implied, and in which $\eta_i(t)$ are independent Gaussian white noises satisfying $\langle \eta_i(t) \rangle = 0$ and $\langle \eta_i(s) \eta_j(t) \rangle = 2 T \delta_{ij} \delta_{st}$.
We focus on the case where $J_{ij} = J/N$ for all $i$ and $j$. 
In this case, the model describes the competition between interactions (which favor order by tending to make the $s_i$ identical to each other) and noise (which favors disorder by tending to make the $s_i$ randomly distributed).

We defined the stochastic magnetization
\begin{equation}
\hat{m}(t) \equiv \frac{1}{N}\sum_{i=1}^{N} s_i(t)
\end{equation}
in terms of which Eq.~\eqref{hopfield_meth} with our choice of $J_{ij} = J/N$ becomes
\begin{equation}
        \label{hopfield_meth_mfc}
        s_i(t+1) = {\rm sign}\left(J \hat{m}(t) + \eta_i(t) \right).
\end{equation}

We also define the average magnetization
\begin{equation}
    m(t) \equiv \langle \hat{m}(t)\rangle
\end{equation}
which characterizes the amount of order in the system.
Here, the average is performed on realizations of the noise process $\bm{\eta}(t)$ driving the system.
In particular the magnetization is an order parameter for the order-disorder (ferromagnet-paramagnet) phase transition: in the thermodynamic limit ($N\to \infty$), $m$ is nonzero in the ordered phase and it is zero in the disordered phase.

The fact that all spins are coupled together suggests that the random variable $\hat{m}(t)$ becomes deterministic as $N\to\infty$ (it is said to be self-averaging), meaning that
\begin{equation}
    \label{selfavg}
    \hat{m}(t) = m(t) + \text{(correction vanishing as $N \to \infty$)}.
\end{equation}
The underlying intuition is that $\hat{m}(t)$ is a sum of random variables with identical variances, so if these were independent with a fixed variance (independent of $N$), the central limit theorem would imply that $\hat{m}(t) \overset{?}{=} m(t) + \mathcal{O}(1/\sqrt{N})$.
However, the different spins are actually correlated, so this is not obvious.
In some cases (such as the present case), Eq.~\eqref{selfavg} can nevertheless be proved \cite{Spohn1980,Talagrand2011,Barra2008}.
Even when it is not the case, some progress can be made by using a replacement such as $\hat{m}(t) \to m(t)$ as an uncontrolled anzatz, which can be validated self-consistently later on (note that this is not a formal proof).

Here, the symmetry of the system under any permutation of the spins implies that in a stationary state, $\langle s_i(t) \rangle = \langle s_j(t) \rangle$ for any $i$ and $j$, and hence that $\langle s_i(t) \rangle = m(t)$.
Using the mean-field ansatz $\hat{m}(t) \simeq m(t)$ in Eq.~\eqref{hopfield_meth_mfc}, and taking the average, we end up with
\begin{equation}
	\label{dmft_magnetization}
    m(t+1) \simeq \langle {\rm sign}\left(J m(t) + \eta_i(t) \right) \rangle_{\eta}
\end{equation}
in which the statistical average now reduces to an average over the noise $\eta(t)$ (which can be seen as one of the $\eta_i(t)$).
The right-hand side of q.~\eqref{dmft_magnetization} is expressed as a statistical average of a quantity depending on the magnetization, which is itself defined as a statistical average. 
This is a defining feature of DMFT. 
In particular, in a steady-state where $m(t)=m(t+1)$, Eq.~\eqref{dmft_magnetization} becomes a self-consistency equation for the magnetization $m$, which allows us to determine its value.
Note however that this equation is dynamical: it allows us to do more and determine the evolution in time of the magnetization $m(t)$.
Expressing
\begin{equation}
   \langle f(\bm{\eta}) \rangle_{\bm{\eta}} \equiv \int f(\bm{\eta}) p(\bm{\eta}) d\bm{\eta}
\end{equation}
for any function $f$, with $p(\bm{\eta}) = 1/Z \ee^{-\lVert \bm{\eta} \rVert^2/2T^2}$ where $Z$ is a normalization constant (we have used the same notation $\bm{\eta}$ for the random variable and its possible values) allows us to evaluate
\begin{align}
	\label{gaussian_integral_dmft}
	\langle {\rm sign}\left(h + \eta \right) \rangle_\eta
    =
	\operatorname{erf}\left(\frac{h}{\sqrt{2} T}\right)
\end{align}
for any $h$, where $\operatorname{erf}$ is the error function (see \cite[\S~7.2.1]{NIST_DLMF}). 
Hence, the DMFT equation reads
\begin{equation}
    m(t+1) = \langle {\rm sign}\left(J m(t) + \eta_i(t) \right) \rangle_{\bm{\eta}} = \operatorname{erf}\left(\frac{J m(t)}{\sqrt{2} T}\right).
\end{equation}
The self-consistency equation obtained in the steady-state where $m$ does not depend on time leads to the expected qualitative picture of the mean-field Ising model, namely a pitchfork bifurcation whereby solutions $m_\pm \neq 0$ start existing as $J/T$ is varied.


\subsection{A limit cycle in nonreciprocal Ising DMFT}

We would like to now illustrate that DMFT also captures \emph{dynamical} phenomena.
Consider a mean-field version of the nonreciprocal Ising model of Ref.~\cite{Avni2025,Avni2025b}
\begin{align}
	s_i^a(t+1) &= {\rm sign}(J_{ij}^{ab} s^b_j(t) + \eta^a_i(t) )
\end{align}
where summations over repeated indices are implied, with $J_{ij}^{ab} = J_{ab}/N$ (for all $i$, $j$.
Defining a magnetization $m_a(t)$ for each species $a$, the same mean-field argument as above gives
\begin{align}
	m_a(t+1) &= \langle {\rm sign}(J_{ab} m_b(t) + \eta_a(t) )\rangle_{\bm{\eta}}.
\end{align}
using a mean-field approximation as before.
Using again Eq.~\eqref{gaussian_integral_dmft} to get an explicit expression, specializing to two species $a=1,2$, and using slightly different notations where $T$ is incorporated into the $J$ and $K$
\begin{align}
	m_1(t+1) &= \operatorname{erf}[J_1 m_1(t) + K_{12} m_2(t)] 
    \\
    m_2(t+1) &= \operatorname{erf}[J_2 m_2(t) + K_{21} m_1(t)] 
\end{align}
which takes the form $\bm{m}(t+1) = F[\bm{m}(t)]$ where $\bm{m}=(m_1, m_2)$.
We now set $J_1 = J_2 = J$ to simplify.
As $F[0] = 0$, the trivial state $\bm{m}=0$ is always a solution. 
The stability of a fixed point $\bm{m}$ can be assessed from the eigenvalues of the Jacobian $L_{ij} = \partial F_i/\partial m_j$ at $\bm{m}$, which reads
\begin{equation*}
    \mathbf L=
    \frac{2}{\sqrt{\pi}}
    \begin{pmatrix}
        J\, \ee^{-(J m_1+K_{12} m_2)^2} & K_{12}\, \ee^{-(J m_1+K_{12} m_2)^2} \\
        K_{21}\, \ee^{-(J m_2+K_{21} m_1)^2} & J\, \ee^{-(J m_2+K_{21} m_1)^2}
    \end{pmatrix}.
\end{equation*}
It is convenient to write $K_{12(21)}=K_+ \pm K_-$. 
The eigenvalues of $L|_{\bm{m}=0}$ then read $\lambda_{\pm} = \tfrac{2}{\sqrt{\pi}}[J \pm \sqrt{K_+^2 - K_-^2}\,]$.
The system is linearly stable when the eigenvalues are all inside the unit disk, and unstable when they get out, see \cite[ch.~4]{Kuznetsov2023} for details. 
In particular, when $K_{-} > K_{+}$ the eigenvalues become complex conjugate (with a finite imaginary part) and a Neimark-Sacker bifurcation (which can be seen as the discrete-time version of a Hopf bifurcation) to a limit cycle arises once $J$ is large enough.
While this example is \emph{autonomous}, i.e. the map $\mathbf F$ is the same at every step, in the next subsection we handle a simple example where the mean-field map itself periodically time-dependent, the setting of the main text.

\subsection{A more complicated toy example to challenge autonomy: periodically driven ferromagnet}

Both examples above are autonomous, in that the macroscopic map is the same at every
step. More generally, however, the evolution of the macroscopic order parameters can
carry an explicit time dependence---as exemplified by the periodic dependence
$\mathbf F(\cdot,t+\varpi)=\mathbf F(\cdot,t)$ that arises for the two-species
nonreciprocal Hopfield model and motivates the loop-space DMFT of the main text. 

It might be tempting to blame the loss of autonomy on asynchrony alone.
One might argue that the macroscopic map $\mathbf F$ should then depend on time
explicitly,
\begin{align}
    \mathbf F(\cdot, t) = \mathbf F(\cdot, t+2)\quad \text{but} \quad
    \mathbf F(\cdot, t) \neq \mathbf F(\cdot, t+1),
\end{align}
and indeed its single-sweep Jacobians at times $t$ and $t+1$ might not commute.
However when the system is ergodic, the fixed point is a stationary Gibbs measure, independently of the update order.
Microscopic asynchrony-induced non-commutativity therefore does \emph{not} by itself imply that the macroscopic map is non-autonomous.

We can, however, impose non-autonomy by hand, by driving the fully connected
ferromagnet of the first subsection with a periodic external field. Writing the
one-step map with its time argument shown explicitly,
\begin{equation}
    \label{driven_ferro}
    m(t+1) = F\big(m(t),\,t\big),
    \qquad
    F(m,t) \equiv \operatorname{erf}\!\left(\frac{J m + h(t)}{\sqrt{2}\,T}\right),
    \qquad h(t+\varpi) = h(t),
\end{equation}
the drive---and with it the map---is periodic, $F(\cdot,t+\varpi) = F(\cdot,t)$,
but changes from one step to the next within a period,
$F(\cdot,t) \neq F(\cdot,t+1)$. Labelling the phases by $k = t \bmod \varpi$, we
collect the field and the map at each phase,
\begin{equation}
    \label{driven_phases}
    h_k \equiv h(k),
    \qquad
    F_k(m) \equiv F(m,k) = \operatorname{erf}\!\left(\frac{J m + h_k}{\sqrt{2}\,T}\right),
    \qquad k = 0,\dots,\varpi-1 .
\end{equation}
For concreteness we take $\varpi = 3$ with $h_0 = a$, $h_1 = 0$, $h_2 = -a$, so
that $F_0, F_1, F_2$ are three distinct maps and the update at phase $k$ is
$m(t+1) = F_k\big(m(t)\big)$.

\begin{figure*}[t]
    \centering
    \includegraphics[width=0.95\linewidth]{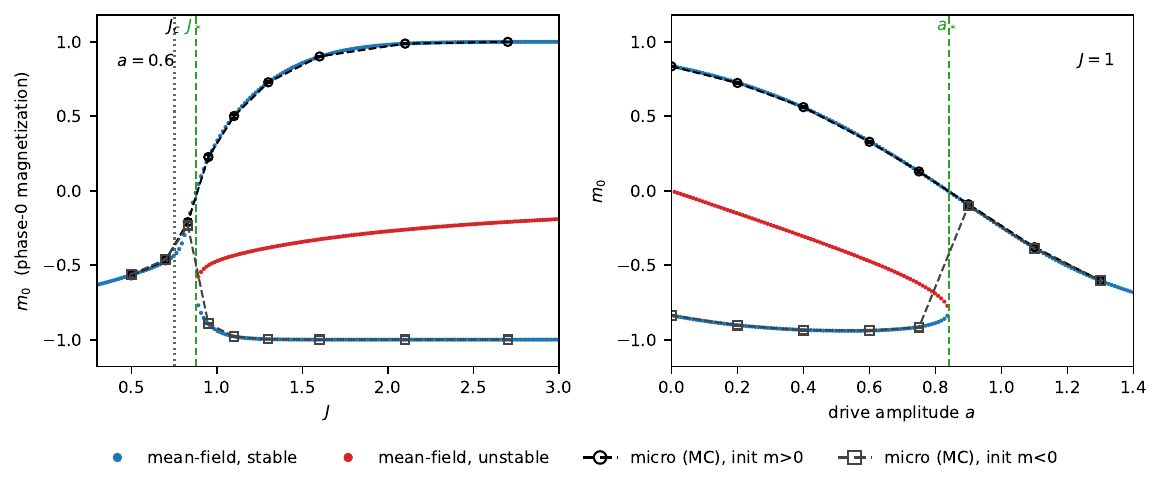}
    \caption{\label{fig:driven_cusp}%
    \textbf{Periodically driven ferromagnet ($\varpi = 3$, $T = 0.6$).} Phase-$0$
    magnetization $m_0$ of the period-$3$ orbits: mean-field fixed points of the
    stroboscopic map $\mathbf G$ (blue, stable; red, unstable) versus a direct
    simulation of $N = 1.2\times10^4$ Ising spins (markers, $\pm$ one standard
    deviation, dashed guides) started from positive (circles) and negative
    (squares) initial magnetization. \emph{Left:} varying $J$ at fixed drive
    amplitude $a = 0.6$; the undriven pitchfork at $J_c$ unfolds into a
    saddle-node at $J_\ast$, above which the dynamics is bistable. \emph{Right:}
    varying $a$ at fixed $J = 1 > J_c$; the two stable orbits annihilate at
    $a_\ast$. The simulation tracks the mean-field stable branches throughout.}
\end{figure*}
 
In the previous subsection, non-reciprocity produced a dynamically ordered limit
cycle by iterating an autonomous map $m(t+1) = F(m(t))$. Here the law itself
changes within the period, $F(\cdot,t) \neq F(\cdot,t+1)$, so successive phases
are related by the distinct maps $F_k$. A period-$\varpi$ orbit is then a fixed
point of the stroboscopic map $\mathbf G = F_2 \circ F_1 \circ F_0$, obtained by
closing the per-phase updates $m_{k+1} = F_k(m_k)$ into a ladder,
\begin{equation}
    \label{driven_ladder}
    m_1 = F_0(m_0), \qquad m_2 = F_1(m_1), \qquad m_0 = F_2(m_2),
\end{equation}
with $m_k \equiv m(t)$ for $t \equiv k \bmod \varpi$. The stability of such an
orbit is assessed, as in the previous subsection, from the Jacobian of the
stroboscopic map at the fixed point, $\Lambda = \mathbf G'(m_0)$ --- a single
number, since $m$ is scalar --- the orbit being stable when $|\Lambda| < 1$. By
the chain rule it is the ordered product of the single-step Jacobians along the
orbit,
\begin{equation}
    \label{driven_monodromy}
    \Lambda = \prod_{k=0}^{\varpi-1} F_k'(m_k),
    \qquad
    F_k'(m) = \frac{2}{\sqrt{\pi}}\,\frac{J}{\sqrt{2}\,T}\,
              \exp\!\left[-\left(\frac{J m + h_k}{\sqrt{2}\,T}\right)^{\!2}\right],
\end{equation}
the scalar analogue of the ordered Jacobian product $\mathbf J = \mathbf L_2
\mathbf L_1 \mathbf L_0$ of the main text. Leaving $J$ free and using
$J m_k + h_k = \sqrt{2}\,T\,\operatorname{erf}^{-1}(m_{k+1})$ on the orbit, it
takes the closed form
\begin{equation}
    \label{driven_monodromy_closed}
    \Lambda = \left(\frac{J}{J_c}\right)^{\!\varpi}
              \exp\!\left(-\sum_{k}\big[\operatorname{erf}^{-1}(m_k)\big]^2\right),
    \qquad J_c \equiv \sqrt{\pi/2}\;T .
\end{equation}
 
When the drive is switched off, the orbit collapses to $m_k = 0$, so
$\Lambda = (J/J_c)^{\varpi}$ and the disordered state loses stability at
$J = J_c$, where $\Lambda = 1$ --- the pitchfork of the first subsection. The
periodic drive is an explicit symmetry-breaking field, so $m = 0$ is no longer a
fixed point and this pitchfork unfolds into an imperfect one. The drive-following
orbit persists for all $J$, while a second stable orbit and an unstable one
($\Lambda > 1$, separating their basins) appear together at a saddle-node
$J_\ast > J_c$, with $J_\ast \to J_c$ as the drive amplitude vanishes; above
$J_\ast$ the dynamics is bistable (Fig.~\ref{fig:driven_cusp}, left). For
$J = 1$, $T = 0.6$, $a = 0.6$ (so $J_\ast \simeq 0.88$) the two stable period-$3$
orbits sit at $m_0 = -0.94$ and $0.33$, flanking the separatrix at
$m_0 = -0.47$. Equivalently, in the plane of coupling and drive amplitude the
pitchfork at $(J_c, 0)$ is the cusp of a saddle-node wedge, crossed by tuning
either the coupling or the drive strength (Fig.~\ref{fig:driven_cusp}, right). In
both slices a direct simulation of the underlying $N$-spin model follows the
mean-field stable branches, confirming the reduction (Fig.~\ref{fig:driven_cusp}).

{\color{gray}

}

\section{Loop-space dynamic mean-field theory}\label{app:dmft_details}
\subsection{Derivation of \texorpdfstring{$k$}{k}-Periodic DMFT Equations}
\label{sec:kperiodic_dmft}

We derive the dynamical mean-field theory (DMFT) equations for the non-reciprocal Hopfield model with asynchronous updates, extending the statistical neurodynamics approach of Amari and Maginu to the block-asynchronous setting. The central result is a self-consistent system with \emph{cross-block coupling}: the noise variance for each neuronal subpopulation depends on the other subpopulation's statistics. Please note that henceforth we will refer to the $\bm{p}$ degrees of freedom in the main text as $\bm{y}$.

\subsubsection{Model and Notation}

We consider $N$ binary neurons $y_i \in \{-1, +1\}$ storing $P = \alpha N$ sequential patterns $\{\bm{\xi}^\mu\}_{\mu=1}^P$ with periodic boundaries $\bm{\xi}^{P+1} = \bm{\xi}^1$. Pattern components are i.i.d.\ with $\mathrm{Prob}[\xi_i^\mu = \pm 1] = 1/2$. The synaptic matrix encodes sequential transitions:
\begin{equation}
K_{ij} = \frac{1}{N} \sum_{\mu=1}^{P} \xi_i^{\mu+1} \xi_j^{\mu}
\end{equation}
Neurons update stochastically via Glauber dynamics at inverse temperature $\beta$:
\begin{equation}
\langle y_i(T+\omega) \rangle = \tanh(\beta h_i(T))
\end{equation}
where the local field is $h_i = \sum_j K_{ij} y_j$.
Note that this single-species description is the $\lambda\to\infty$ limit of the bipartite model, in which $x_i = \mathrm{sign}(\lambda y_i + \cdots) = y_i$, so the readout becomes a one-step-lagged copy of the queue and the dynamics closes on the $y$ neurons alone.

\subsubsection{Block-Asynchronous Approximation}

In asynchronous dynamics, neurons update sequentially within each sweep. We partition neurons into two blocks of size $N/2$: block $B_b$ contains neurons updating early in each sweep, while block $B_a$ contains neurons updating late. Define block-resolved overlaps:
\begin{equation}
m_I^\mu(T) = \frac{2}{N} \sum_{j \in B_I} \xi_j^\mu y_j(T), \quad I \in \{a, b\}.
\end{equation}

When a block $B_b$ neuron (early updater) computes its local field, all of block $B_a$ remains in its previous state, so the field is dominated by block $a$'s overlap at the previous time step $T-\omega$, where $\omega = 1/k$, the integer $k$ being the number of sweeps after which the macroscopic dynamics repeats ($k=3$ below, the condensed pattern advancing by two per period).
Conversely, when a block $B_a$ neuron (late updater) computes its field, all of block $B_b$ has already updated, so the field is dominated by block $b$'s current overlap. This gives the cross-block field structure:
\begin{align}
h_i^a(T + n\omega) &= \sum_{\mu=1}^P \xi_i^{\mu+1} m_b^\mu(T + n\omega) \label{eq:field_a}\\[0.3em]
h_i^b(T + n\omega) &= \sum_{\mu=1}^P \xi_i^{\mu+1} m_a^\mu\left(T + (n-1)\omega\right), 
\label{eq:field_b}
\end{align}
for $n=0,1,\ldots,k-1$. This structure allows us to capture time-translation symmetry breaking local fields with a $k$-periodic ansatz.

\subsubsection{Signal-Noise Decomposition}

For successful sequence retrieval, at any given time, each block of the network tracks one condensed pattern $\mu_*$. This makes it natural to separate the field into signal and crosstalk noise contributions:
\begin{align}
h_i^a(T) &= \underbrace{\xi_i^{\mu_*+1} m_b^{\mu_*}(T)}_{\text{signal}} + \underbrace{\sum_{\nu \neq \mu_*} \xi_i^{\nu+1} m_b^{\nu}(T)}_{=:z_i^a(T)}, \label{eq:signal_noise_a}\\[0.5em]
h_i^b(T) &= \underbrace{\xi_i^{\mu_*+1} m_a^{\mu_*}(T-\omega)}_{\text{signal}} + \underbrace{\sum_{\nu \neq \mu_*} \xi_i^{\nu+1} m_a^{\nu}(T-\omega)}_{=:z_i^b(T)} \label{eq:signal_noise_b}.
\end{align}
The crosstalk noise $z_i^a$ is a sum over block $b$'s uncondensed overlaps, while $z_i^b$ is a sum over block $a$'s uncondensed overlaps.

Note that we have ignored the contribution of block $a$'s own overlaps to its field (and similarly for block $b$). This approximation is justified as follows: when a block $a$ neuron (late updater) computes its field, it sees \emph{all} of block $b$ in coherent, updated states, but only a \emph{fraction} of block $a$ in coherent states---the rest of block $a$ is a mixture of neurons that have and have not yet updated within the current sweep. For a typical late updater, roughly half of block $a$ contributes old states and half contributes new states; this incoherent mixture partially averages out, making the coherent block $b$ contribution dominant. The approximation becomes exact for neurons at the block boundary (the first neuron in block $a$ sees all of block $b$ updated and none of block $a$ updated).

The main assumption of statistical neurodynamics is that in the thermodynamic limit, the crosstalk noise is Gaussian:
\begin{equation}
z_i^a \sim \mathcal{N}(0, \alpha r^a), \quad z_i^b \sim \mathcal{N}(0, \alpha r^b),
\end{equation}
where the variances are given by the second moments of the uncondensed overlaps:
\begin{equation}
\alpha r^a = \sum_{\nu \neq \mu_*} (m_b^{\nu})^2, \quad \alpha r^b = \sum_{\nu \neq \mu_*} (m_a^{\nu})^2.
\label{eq:variance_def}
\end{equation}
Note that $r^a$ depends on block $b$'s overlaps, and $r^b$ depends on block $a$'s overlaps.

\subsubsection{Dynamics of Uncondensed Overlaps}

To derive the self-consistent relations between the variances and overlap order parameters, we need to understand how the uncondensed overlaps propagate under asynchronous updates.
Consider an uncondensed pattern $\nu \neq \mu_*$. The local field for neuron $i \in B_a$ can be written as:
\begin{equation}
h_i^a(T) = h_i^{a,(\nu)}(T) + \xi_i^{\nu+1} m_b^{\nu}(T),
\label{eq:field_decomp}
\end{equation}
where $h_i^{a,(\nu)}(T) = \sum_{\mu \neq \nu} \xi_i^{\mu+1} m_b^{\mu}(T)$ is the field with pattern $\nu$'s contribution removed.

Since $m_b^\nu = O(1/\sqrt{N})$ for uncondensed patterns, we expand $\tanh(\beta h_i^a)$ to first order:
\begin{align}
\tanh(\beta h_i^a) &= \tanh\left[\beta h_i^{a,(\nu)} + \beta \xi_i^{\nu+1} m_b^{\nu}\right] \notag\\[0.3em]
&= \tanh(\beta h_i^{a,(\nu)}) + \beta \xi_i^{\nu+1} m_b^{\nu} \, \frac{d}{dx}\tanh(x)\Big|_{x=\beta h_i^{a,(\nu)}} + O((m_b^\nu)^2) \notag\\[0.3em]
&= \tanh(\beta h_i^{a,(\nu)}) + \beta \xi_i^{\nu+1} m_b^{\nu} \, \mathrm{sech}^2(\beta h_i^{a,(\nu)}) + O(N^{-1}).
\label{eq:tanh_expansion}
\end{align}

The overlap of block $a$ with pattern $\nu+1$ after a Glauber update is:
\begin{align}
m_a^{\nu+1}(T') &= \frac{2}{N} \sum_{i \in B_a} \xi_i^{\nu+1} \langle y_i(T') \rangle \notag\\[0.3em]
&= \frac{2}{N} \sum_{i \in B_a} \xi_i^{\nu+1} \tanh\left(\beta h_i^a(T)\right).
\label{eq:overlap_update}
\end{align}

Substituting the expansion (\ref{eq:tanh_expansion}):
\begin{align}
m_a^{\nu+1}(T') &= \frac{2}{N} \sum_{i \in B_a} \xi_i^{\nu+1} \tanh\left(\beta h_i^{a,(\nu)}\right) + \frac{2}{N} \sum_{i \in B_a} \xi_i^{\nu+1} \cdot \beta \xi_i^{\nu+1} m_b^{\nu} \cdot \mathrm{sech}^2\left(\beta h_i^{a,(\nu)}\right) \notag\\[0.5em]
&= \underbrace{\frac{2}{N} \sum_{i \in B_a} \xi_i^{\nu+1} \tanh\left(\beta h_i^{a,(\nu)}\right)}_{=:\tilde{m}_a^{\nu+1}} + m_b^{\nu} \cdot \underbrace{\frac{2\beta}{N} \sum_{i \in B_a} (\xi_i^{\nu+1})^2 \, \mathrm{sech}^2\left(\beta h_i^{a,(\nu)}\right)}_{=:U^a}.
\label{eq:overlap_recursion_detailed}
\end{align}

Since $(\xi_i^{\nu+1})^2 = 1$ for all $i$, and using $\mathrm{sech}^2(x) = 1 - \tanh^2(x)$:
\begin{equation}
U^a = \frac{2\beta}{N} \sum_{i \in B_a} \left[1 - \tanh^2(\beta h_i^{a,(\nu)})\right] = \beta \left(1 - \frac{2}{N} \sum_{i \in B_a} \tanh^2(\beta h_i^{a,(\nu)})\right).
\label{eq:susceptibility_sum}
\end{equation}

In the thermodynamic limit, replacing the sum by an average over the Gaussian noise distribution:
\begin{equation}
U^a = \beta \left\{1 - \int \mathrm{D}z\, \tanh^2\left[\beta\left(\Omega^a + z\sqrt{\alpha r^a},\right)\right]\right\},
\label{eq:susceptibility}
\end{equation}
where $\mathrm{D}z = (2\pi)^{-1/2} e^{-z^2/2} dz$, $\Omega^a$ is the condensed overlap (signal), and $r^a$ is the noise variance.
Thus:
\begin{equation}
m_a^{\nu+1} = \tilde{m}_a^{\nu+1} + U^a \cdot m_b^{\nu}.
\label{eq:recursion_a}
\end{equation}
The identical analysis for block $b$ yields:
\begin{equation}
m_b^{\nu+1} = \tilde{m}_b^{\nu+1} + U^b \cdot m_a^{\nu}.
\label{eq:recursion_b}
\end{equation}

Note that the terms $\tilde{m}_{a,b}^{\nu+1}$ in (\ref{eq:recursion_a}, \ref{eq:recursion_b}) are $O(1/\sqrt{N})$ because they are sums of $N/2$ terms, each $O(1/N)$, with random signs $\xi_i^{\nu+1}$ that are statistically independent of $h_i^{a,(\nu)}$ since pattern $\nu$ was excluded from the field.

\subsubsection{Derivation of Variance Recursion}

We now derive the recursion for the noise variance $r^a$. Substituting the recursion (\ref{eq:recursion_b})
\begin{equation}
(m_b^{\nu})^2 = (\tilde{m}_b^{\nu})^2 + 2 \tilde{m}_b^{\nu} \cdot U^b \cdot m_a^{\nu-1} + (U^b)^2 (m_a^{\nu-1})^2,
\label{eq:mb_squared}
\end{equation}
and using (\ref{eq:variance_def}), we have:
\begin{equation}
\alpha r^a = \sum_\nu (m_b^{\nu})^2 = \sum_\nu (\tilde{m}_b^{\nu})^2 + 2 U^b \sum_\nu \tilde{m}_b^{\nu} \cdot m_a^{\nu-1} + (U^b)^2 \sum_\nu (m_a^{\nu-1})^2.
\label{eq:variance_expansion}
\end{equation}

We evaluate each term:

\paragraph{Term 1: Self-averaging of noise.}
Each $\tilde{m}_b^{\nu}$ is:
\begin{equation}
\tilde{m}_b^{\nu} = \frac{2}{N} \sum_{j \in B_b} \xi_j^{\nu} \tanh(\beta h_j^{b,(\nu)}).
\end{equation}

Since $\xi_j^{\nu}$ is independent of $h_j^{b,(\nu)}$ (pattern $\nu$ was excluded), $\langle \xi_j^{\nu} \rangle = 0$ and $\langle \xi_i^{\nu} \xi_j^{\nu} \rangle = \delta_{ij}$:
\begin{align}
\langle \tilde{m}_b^{\nu} \rangle = 0, \quad \langle (\tilde{m}_b^{\nu})^2 \rangle &= \frac{4}{N^2} \sum_{i, j \in B_b}  \underbrace{\langle \xi_i^{\nu} \xi_j^{\nu} \rangle}_{=\delta_{ij}} \langle \tanh\left(\beta h_i^{b,(\nu)}\right) \tanh\left(\beta h_j^{b,(\nu)}\right) \rangle \notag \\
&= \frac{4}{N^2} \sum_{i \in B_b}  \langle\tanh^2\left(\beta h_i^{b,(\nu)}\right)\rangle = \frac{2 q_b}{N}, \qquad q_b := \left\langle \tanh^2\!\big(\beta h^{b}\big)\right\rangle \leq 1.
\label{eq:mtilde_variance}
\end{align} 
The factor of $2$ follows from the block normalisation $m_I^\nu=(2/N)\sum_{j\in B_I}$: each
block enters the physical field with weight $1/2$, so the relevant quantity
is the variance of the \emph{full} overlap $m^\nu=\tfrac12(m_a^\nu+m_b^\nu)$. The blocks being
disjoint, $\langle\tilde m_a^\nu\tilde m_b^\nu\rangle=0$, hence
$\langle(\tilde m^\nu)^2\rangle=\tfrac14\big(\tfrac{2q_a}{N}+\tfrac{2q_b}{N}\big)=q/N$ with
$q=\tfrac12(q_a+q_b)$, and $\sum_{\nu=1}^{P}\langle(\tilde m^\nu)^2\rangle\to\alpha q$. The same
full-overlap normalisation applies to the signal (at block symmetry
$\tfrac12(m_a^{\mu_*}+m_b^{\mu_*})\approx m_b^{\mu_*}$). 
At low temperature $q\to1$ [$q=1-U/\beta$, cf.~\eqref{eq:susceptibility}] and we will set the new-noise constant to $1$ below.%


\paragraph{Term 2: Cross-terms vanish.}
Consider:
\begin{equation}
\sum_\nu \tilde{m}_b^{\nu} \cdot m_a^{\nu-1} = \sum_\nu \left(\frac{2}{N} \sum_{j \in B_b} \xi_j^{\nu} \tanh(\beta h_j^{b,(\nu)})\right) \cdot m_a^{\nu-1}
\label{eq:cross_term}
\end{equation}

The key observation is that $\tilde{m}_b^{\nu}$ involves $\xi_j^{\nu}$, while $m_a^{\nu-1}$ involves $\xi_i^{\nu-1}$. Since patterns are independent:
\begin{equation}
\langle \xi_j^{\nu} \xi_i^{\nu-1} \rangle = \langle \xi_j^{\nu} \rangle \langle \xi_i^{\nu-1} \rangle = 0
\end{equation}

Furthermore, $\tanh\left(\beta h_j^{b,(\nu)}\right)$ is independent of $\xi_j^{\nu}$ by construction. Therefore:
\begin{equation}
\langle \tilde{m}_b^{\nu} \cdot m_a^{\nu-1} \rangle = \frac{2}{N} \sum_{j \in B_b} \underbrace{\langle \xi_j^{\nu} \rangle}_{=0} \langle \tanh\left(\beta h_j^{b,(\nu)}\right) \cdot m_a^{\nu-1} \rangle = 0
\label{eq:cross_vanish}
\end{equation}

\paragraph{Term 3: Propagated variance.}
The third term in (\ref{eq:variance_expansion}) is:
\begin{equation}
(U^b)^2 \sum_\nu (m_a^{\nu-1})^2 = (U^b)^2 \cdot \alpha r^b
\label{eq:propagated}
\end{equation}
where we used $\sum_\nu (m_a^{\nu})^2 = \alpha r^b$ from (\ref{eq:variance_def}).

\paragraph{Complete variance recursion.}
Collecting terms from (\ref{eq:variance_expansion}):
\begin{align}
    \alpha r^a &= \alpha + 0 + (U^b)^2 \cdot \alpha r^b \notag \\
    \label{eq:variance_recursion_general}
    r^a &= 1 + (U^b)^2 \cdot r^b
\end{align}
This is the cross-block variance coupling: the variance $r^a$ (noise seen by block $a$, coming from block $b$'s overlaps) equals new noise (the constant term) plus propagated variance from block $a$'s overlaps (the $r^b$ term), amplified by block $b$'s susceptibility squaredpropagated coefficient.
By identical reasoning:
\begin{equation}
    r^b = 1 + (U^a)^2 \cdot r^a.
\label{eq:variance_recursion_b}
\end{equation}

\subsubsection{Three-Phase Structure}

Monte Carlo simulations reveal period-3 dynamics, with both blocks updating at each phase. We index phases by $k \in \{0, 1, 2\}$ within each macro time-step $\tau$. Define:
\begin{itemize}
\item $r_k^I$ = noise variance in block $I$'s field at phase $k$
\item $U_k^I$ = susceptibility of block $I$ at phase $k$
\item $\Omega_k^I$ = condensed overlap of block $I$ at phase $k$
\end{itemize}

At each phase $k$, block $a$'s noise comes from block $b$'s overlaps (and vice versa). The overlaps that block $a$ sees at phase $k$ were produced by block $b$ at phase $k-1$, using block $b$'s susceptibility $U_{k-1}^b$ acting on the noise variance $r_{k-1}^b$.

Following the same logic as (\ref{eq:variance_recursion_general}), but tracking phases:
\begin{equation}
    r_k^a = 1 + (U_{k-1}^b)^2 \cdot r_{k-1}^b
\label{eq:rka_phase}
\end{equation}
where indices are cyclic modulo 3.
The variance $r_{k-1}^b$ that appears here is the variance of block $b$'s \emph{output} overlaps after phase $k-1$. This output variance is determined by block $b$'s field at phase $k-1$, which had noise variance $r_{k-1}^b$. The output variance is precisely what we computed in (\ref{eq:variance_recursion_general}).

Tracing through the full cycle with both blocks updating at each phase, the recursions are:
\begin{alignat}{2}
r_0^a &= 1 + (U_2^b)^2 \cdot r_2^b, \quad & r_0^b &= 1 + (U_2^a)^2 \cdot r_2^a \label{eq:r0}\\[0.3em]
r_1^a &= 1 + (U_0^b)^2 \cdot r_0^b, \quad & r_1^b &= 1 + (U_0^a)^2 \cdot r_0^a \label{eq:r1}\\[0.3em]
r_2^a &= 1 + (U_1^b)^2 \cdot r_1^b, \quad & r_2^b &= 1  + (U_1^a)^2 \cdot r_1^a \label{eq:r2}
\end{alignat}

The susceptibilities are:
\begin{equation}
U_k^I = \beta \left\{1 - \int \mathrm{D}z\, \tanh^2\left[\beta\left(\Omega_k^I + z\sqrt{\alpha r_k^I}\right)\right]\right\}
\label{eq:Uk_full}
\end{equation}

The condensed overlaps satisfy:
\begin{equation}
\Omega_k^I = \int \mathrm{D}z\, \tanh\left[\beta\left(\Omega_{k-1}^{\bar{I}} + z\sqrt{\alpha r_k^I}\right)\right]
\label{eq:Omega_full}
\end{equation}
where $\bar{I}$ denotes the opposite block.

\begin{figure}[h!]
    \centering
    \includegraphics[]{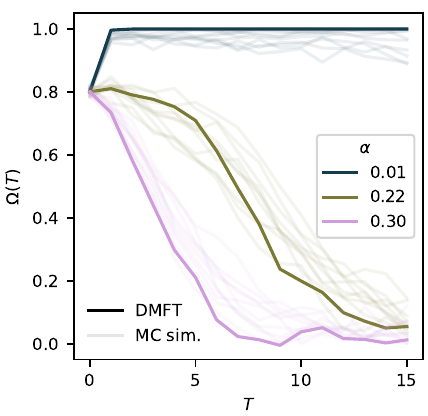}
    \caption{{\bf Macroscopic dynamics.} 
    Solid lines show the evolution of the order parameter vector $\bm{\theta}$ updated via the nonlinear DMFT map $\mathbf{F}$.
    Starting from an initial overlap of $0.8$, the order parameters $\Omega$ (left) and $\Psi$ (right) stabilise at a fixed point with finite overlap or 0 overlap depending on the memory load $\alpha$.
    These results are in agreement with finite-size Monte Carlo simulations (transparent curves).
    \label{fig:dmft_vs_mc}}
\end{figure}

\subsection{Dynamical mean-field theory as a discrete iterative dynamical map}
\label{sec:dmft_stability}

The loop-space self-consistency relations summarised in the Methods inherit the
periodic structure of the asynchronous dynamics: written sweep by sweep they are the
\emph{non-autonomous}, $\varpi$-periodic map $\mathbf F(\cdot,t)$ of Eq.~3, with
$\mathbf F(\cdot,t+\varpi)=\mathbf F(\cdot,t)$ and $\varpi=3$. Its three phases are the
distinct single-sweep maps $\mathbf F_0,\mathbf F_1,\mathbf F_2$. We close the problem
by working with the operator that advances the system by one \emph{full} period,
\begin{equation}\label{eq:period_map}
    \mathbf G \;:=\; \mathbf F_2\circ\mathbf F_1\circ\mathbf F_0,
\end{equation}
acting on the state vector that stacks the order parameters and variances of all
$\varpi$ phases,
\begin{equation}\label{eq:loop_state}
    \bm\theta=(\Theta_a,\Theta_b,\Phi_a,\Phi_b,\Psi_a,\Psi_b,\Omega_a,\Omega_b,
    r_0^a,r_0^b,r_1^a,r_1^b,r_2^a,r_2^b)^{\!\top}.
\end{equation}
By construction $\mathbf G$ is autonomous---the explicit time dependence of
$\mathbf F(\cdot,t)$ has been absorbed into the phase-resolved components of
$\bm\theta$. The pay-off is conceptual: a \emph{sequence-advancing} retrieval state,
which is a limit cycle in the laboratory frame, is nothing but a \emph{fixed point}
$\bm\theta^*=\mathbf G(\bm\theta^*)$ of the period map. Stationarity in loop space
\emph{is} periodic sequencing in real time. We obtain $\bm\theta^*$ by iterating
$\mathbf G$ to convergence (Fig.~\ref{fig:dmft_vs_mc}), at load $\alpha=P/N$ and
inverse noise strength $\beta$.

\begin{figure}[hbt!]
    \centering
    \includegraphics[]{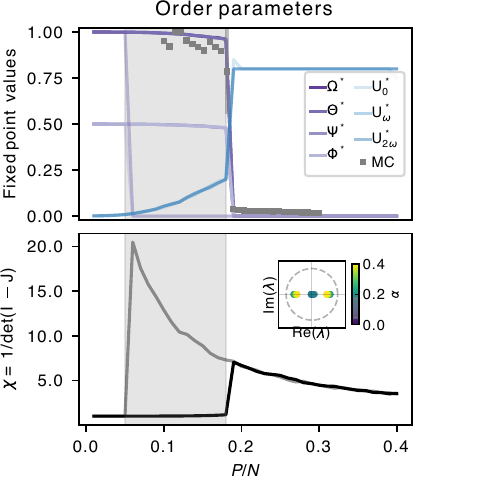}
    \caption{{\bf Spectral signature of the blackout transition.}
    Eigenvalues of the loop-space Jacobian $\mathbf J$ in the complex plane, coloured
    by memory load $P/N$ (inset; dashed line, unit circle). As $P/N$ increases the
    leading complex-conjugate pair \emph{approaches} the unit circle near
    $P/N\approx0.18$ but does not cross it, the stability indicator $\chi=1/\det(\mathbf I-\mathbf J)$ (lower panel) stays
    finite through the transition, peaking inside the coexistence region (grey).
    \label{fig:loop_spectrum}}
\end{figure}

The stability of such a state then follows from the single Jacobian of $\mathbf G$,
which by the chain rule is the ordered product of the per-sweep Jacobians,
\begin{equation}\label{eq:jac_product}
    \mathbf J=\mathbf L_2\,\mathbf L_1\,\mathbf L_0,
    \qquad
    \mathbf L_k=\frac{\partial\mathbf F_k}{\partial\bm\theta}\bigg|_{\bm\theta_k^*},
\end{equation}
each factor evaluated at the corresponding phase of the orbit. 

\paragraph*{Linear stability and the blackout transition.}
Linearising the period map $\mathbf G$ about a fixed point yields the Jacobian
$\mathbf J$ of Eq.~\eqref{eq:jac_product}, a discrete-time stability operator whose
eigenvalues govern stability through the unit-circle criterion $\rho(\mathbf J)<1$. We
solve the self-consistency by iterating $\mathbf G$ to convergence, and track the
leading eigenvalue pair as a function of the memory load (Fig.~\ref{fig:loop_spectrum}).
Their motion is \emph{non-monotonic}: at low load the multipliers sit close to the
origin, $|\lambda|\approx0$, and the retrieval fixed point is strongly stable; as $P/N$
enters the metastable (coexistence) window the leading pair---\emph{complex conjugate},
signalling an oscillatory mode---swells outward and approaches the unit circle as the
retrieval branch nears its Neimark--Sacker (discrete-time Hopf) threshold; beyond the
jump the multipliers relax back inward.

Consistently, the susceptibility indicator
$\chi := 1/\det(\mathbf I-\mathbf J) = \prod_i (1-\lambda_i)^{-1}$
(Fig.~\ref{fig:loop_spectrum}, lower panel)---which diverges precisely when an
eigenvalue of $\mathbf J$ reaches $+1$, where the static linear response
$(\mathbf I - \mathbf J)^{-1}$ of the fixed point diverges as well---peaks
inside the coexistence window yet stays finite throughout: it does not diverge, because
the leading eigenvalues crowd the unit circle at $e^{\pm i\phi}\neq1$ rather than
passing through $+1$. In other words, the eigenvalues are never observed to cross $|\lambda|=1$, as the iterative simulation follows the physical dynamics and cannot enter the unstable branches ($\rho(\mathbf J)>1$). 

An alternative way to obtain the stability operator, which we have not pursued here, is to evaluate $\mathbf J$ by automatic differentiation of the self-consistency map (e.g.\ in JAX), fixing a common set of Gaussian samples for each expectation and averaging over independent draws to control sampling noise.


\subsection{Autonomy and the two-species map}
\label{sec:two_species_autonomy}

The explicit period-$3$ time dependence of the macroscopic map is an artefact of
following a single species. Along the retrieval orbit the readout visits the cycle (Fig.~\ref{fig2}D)
\begin{equation}
    \text{pure state} \;\to\; \text{mixed state} \;\to\;
    \text{pure state} \;\to\; \text{pure state},
\end{equation}
so the same-looking single-species
state must be mapped sometimes to a pure and sometimes to a mixed state: no
time-independent deterministic map of one species' overlaps alone can generate this
dynamics, and the reduced map necessarily carries the explicit periodic time dependence
of $\mathbf F(\cdot,t)$. This is the generic consequence of projecting a deterministic
system onto a subset of its variables. Tracking \emph{both} species restores autonomy:
the joint state follows
\begin{equation}
    (\mathrm{pure},\mathrm{pure}) \;\to\; (\mathrm{mixed},\mathrm{pure}) \;\to\;
    (\mathrm{pure},\mathrm{mixed}) \;\to\; (\mathrm{pure},\mathrm{pure}),
\end{equation}
advancing the condensed pattern by two every three steps (effective speed $2/3$), with
the same joint input always mapped to the same output. The staggered cycle can then be thought of as an ordinary period-$3$ orbit of an \emph{autonomous} joint map---a fixed point of its third
iterate, in precisely the same way that the laboratory-frame limit cycle is a fixed
point of the loop-space period map $\mathbf G$ of Eq.~\eqref{eq:period_map}.

\section{Path Integrals for Synchronous Sequence-Processing Networks}
\label{sec:intro}

Here we consider a couple of related models in the synchronous mode of operation. In the case of synchronous updates, path integrals can be used as it is easier to identify the relevant macroscopic variables at any given time step.
Using the techniques developed by Düring, Coolen, and Sherrington~\cite{duringetal98}, we derive the storage capacity $\alpha_c = p_c/N$ in the thermodynamic limit, where $p$ is the number of stored patterns and $N\to \infty$ is the number of neurons.

For symmetric Hopfield networks storing static patterns, replica methods give $\alpha_c \approx 0.139$. For purely asymmetric networks storing sequences, path integral methods yield $\alpha_c \approx 0.269$---nearly double the symmetric capacity. This enhancement arises from the vanishing of the retarded self-interaction in asymmetric networks.

We first consider synaptic matrices of the form
\begin{equation}\label{eq:J_general}
    J_{ij} = \frac{1}{N}\sum_{\mu=1}^{p} (\xi_i^\mu + \lambda\xi_i^{\mu+1})\xi_j^\mu
\end{equation}
where $\bm{\xi}^\mu \in \{-1,+1\}^N$ are random patterns. The parameter $\lambda$ interpolates between purely symmetric ($\lambda = 0$) and purely asymmetric ($\lambda \to \infty$) limits.

\subsection{Synchronous single-species sequence processing network}
\label{sec:single_species}

\paragraph{Model definition}

Consider $N$ Ising neurons $\sigma_i = \pm 1$ with synaptic weights~\eqref{eq:J_general}. In matrix notation,
\begin{equation}\label{eq:J_matrix}
    \bm{J} = \frac{1}{N}\bm{\xi}^{\mathrm{T}}(\bm{1} + \lambda\bm{S})\bm{\xi}
\end{equation}
where $\bm{\xi}$ is the $p \times N$ pattern matrix with entries $\xi_{\mu i} = \xi_i^\mu$, and $S_{\mu\nu} = \delta_{\mu,(\nu+1)\mod p}$ is the cyclic shift matrix.

We will first consider the case where the neurons evolve via synchronous stochastic dynamics: at each discrete timestep, all neurons update simultaneously according to
\begin{equation}\label{eq:dynamics}
    \mathrm{Prob}[\sigma_i(s+1) = \pm 1] = \frac{1}{2}\left[1 \pm \tanh\beta h_i(s)\right]
\end{equation}
where $h_i(s) = \sum_j J_{ij}\sigma_j(s) + \theta_i(s)$ is the local field and $\beta = 1/T$ is the inverse temperature. The variables $\theta_i(s)$ are external fields.

\subsubsection{Purely asymmetric limit (\texorpdfstring{$\lambda \to \infty$}{lambda to infinity})}
\label{sec:purely_asymm}

As a warm-up, we first treat the purely asymmetric case where $J_{ij} = \frac{1}{N}\sum_\mu \xi_i^{\mu+1}\xi_j^\mu$, following Düring, Coolen, and Sherrington~\cite{duringetal98}. This corresponds to taking $\lambda \to \infty$ while rescaling the synaptic weights appropriately.


The generating functional for spin trajectories is
\begin{equation}\label{eq:Z_def}
    Z[\bm{\psi}] = \sum_{\{\bm{\sigma}(s)\}} p[\bm{\sigma}(0)] \prod_{s=0}^{t-1} W[\bm{\sigma}(s+1)|\bm{\sigma}(s)] \, e^{-\ii \sum_{s<t} \bm{\sigma}(s)\cdot\bm{\psi}(s)}
\end{equation}
where
\begin{equation}
    W[\bm{\sigma}(s+1)|\bm{\sigma}(s)] = \prod_{i=1}^N \frac{e^{\beta\sigma_i(s+1) h_i(s)}}{2\cosh[\beta h_i(s)]}
\end{equation}
is the transition probability. Correlation and response functions are obtained by differentiation:
\begin{align}
    \langle \sigma_i(s) \rangle &= \ii \lim_{\psi \to 0} \frac{\partial Z}{\partial \psi_i(s)}, \\
    \langle \sigma_i(s)\sigma_j(s') \rangle &= -\lim_{\psi \to 0} \frac{\partial^2 Z}{\partial \psi_i(s) \partial \psi_j(s')}, \\
    \frac{\partial \langle \sigma_i(s) \rangle}{\partial \theta_j(s')} &= \ii \lim_{\psi \to 0} \frac{\partial^2 Z}{\partial \psi_i(s) \partial \theta_j(s')}.
\end{align}


To decouple the spin-spin interactions, we introduce auxiliary fields $h_i(s)$ and their Fourier conjugates $\hat{h}_i(s)$ via the identity
\begin{equation}
    1 = \int \prod_{i,s} \frac{\dd h_i(s)\, \dd\hat{h}_i(s)}{2\pi} \exp\left[\ii\sum_{i,s} \hat{h}_i(s)\left(h_i(s) - \sum_j J_{ij}\sigma_j(s) - \theta_i(s)\right)\right].
\end{equation}
Inserting this into~\eqref{eq:Z_def} and using $\bm{J} = \frac{1}{N}\bm{\xi}^{\mathrm{T}}\bm{S}\bm{\xi}$, the generating functional becomes
\begin{align}\label{eq:Z_aux}
    Z[\bm{\psi}] = \sum_{\{\bm{\sigma}\}} p[\bm{\sigma}(0)] \int &\{\dd\bm{h}\,\dd\hat{\bm{h}}\} \prod_{s<t} e^{\beta\bm{\sigma}(s+1)\cdot\bm{h}(s) - \sum_i \ln 2\cosh[\beta h_i(s)]} \nonumber\\
    &\times e^{\ii\hat{\bm{h}}(s)\cdot[\bm{h}(s) - \bm{\theta}(s)] - \ii\bm{\psi}(s)\cdot\bm{\sigma}(s)} \times e^{-\ii N^{-1}\hat{\bm{h}}(s)\cdot(\bm{\xi}^{\mathrm{T}}\bm{S}\bm{\xi})\bm{\sigma}(s)}.
\end{align}


We assume that at time $s$, the system has $\mathcal{O}(1)$ overlap with pattern $\bm{\xi}^s$ (the ``condensed'' pattern), while overlaps with other patterns are $\mathcal{O}(N^{-1/2})$. This is the appropriate ansatz for describing sequence retrieval.

Introduce the non-condensed overlaps:
\begin{align}
    x_\mu(s) &= \frac{1}{\sqrt{N}}\sum_i \xi_i^{\mu+1}\hat{h}_i(s), \quad \mu \neq s, \\
    y_\mu(s) &= \frac{1}{\sqrt{N}}\sum_i \xi_i^\mu \sigma_i(s), \quad \mu \neq s.
\end{align}
Using Fourier representations with conjugate variables $\hat{x}_\mu(s)$ and $\hat{y}_\mu(s)$, the disorder-averaged generating functional becomes
\begin{align}\label{eq:Z_disorder}
    \overline{Z[\bm{\psi}]} = \sum_{\{\bm{\sigma}\}} p[\bm{\sigma}(0)] &\int \{\dd\bm{h}\,\dd\hat{\bm{h}}\} \, e^{\sum_{s<t}[\beta\bm{\sigma}(s+1)\cdot\bm{h}(s) - \sum_i \ln 2\cosh[\beta h_i(s)] + \ii\hat{\bm{h}}(s)\cdot(\bm{h}(s)-\bm{\theta}(s)) - \ii\bm{\psi}(s)\cdot\bm{\sigma}(s)]} \nonumber\\
    &\times e^{-\ii N^{-1}\sum_{s<t}[\hat{\bm{h}}(s)\cdot\bm{\xi}^{s+1}][\bm{\sigma}(s)\cdot\bm{\xi}^s]} \nonumber\\
    &\times \int \frac{\dd\bm{x}\,\dd\hat{\bm{x}}\,\dd\bm{y}\,\dd\hat{\bm{y}}}{(2\pi)^{2(p-1)t}} e^{\ii\sum_{s<t}\sum_{\mu \neq s}[\hat{x}_\mu(s)x_\mu(s) + \hat{y}_\mu(s)y_\mu(s) - x_\mu(s)y_\mu(s)]} \nonumber\\
    &\times \overline{\left[e^{-\ii N^{-1/2}\sum_{s<t}\sum_{\mu \neq s}[\hat{x}_\mu(s)\hat{\bm{h}}(s)\cdot\bm{\xi}^{\mu+1} + \hat{y}_\mu(s)\bm{\sigma}(s)\cdot\bm{\xi}^\mu]}\right]}.
\end{align}

\paragraph{Disorder average.}
\label{sec:disorder_average}

The disorder average is the central technical step that converts the quenched disorder problem into an annealed one with effective interactions. We provide a detailed derivation here.

From~\eqref{eq:Z_disorder}, the disorder-dependent factor involves products over non-condensed pattern indices $\mu > t$:
\begin{equation}\label{eq:disorder_factor}
    \overline{\left[e^{-\ii N^{-1/2}\sum_{s<t}\sum_{\mu \neq s}[\hat{x}_\mu(s)\hat{\bm{h}}(s)\cdot\bm{\xi}^{\mu+1} + \hat{y}_\mu(s)\bm{\sigma}(s)\cdot\bm{\xi}^\mu]}\right]}.
\end{equation}
The key observation is that the patterns $\bm{\xi}^\mu$ for $\mu > t$ are statistically independent of those appearing in the condensed terms ($\mu \leq t$). This allows us to average over them separately.

Since each $\xi_i^\mu$ is an independent random variable with $\mathrm{Prob}[\xi_i^\mu = \pm 1] = 1/2$, the average factorizes:
\begin{align}
    \overline{[\ldots]}_{\mu>t} &= \prod_{\mu > t} \overline{\exp\left[-\frac{\ii}{\sqrt{N}}\sum_{s<t}\sum_i \xi_i^\mu \phi_i^{\mu}(s)\right]}_{\xi^\mu} \nonumber\\
    &= \prod_{\mu > t} \prod_i \overline{\exp\left[-\frac{\ii}{\sqrt{N}} \xi_i^\mu \Phi_i^\mu\right]}_{\xi_i^\mu}
\end{align}
where we defined the ``charge'' coupled to pattern $\mu$ at site $i$:
\begin{equation}\label{eq:charge_def}
    \Phi_i^\mu = \sum_{s<t}\left[\hat{x}_{\mu-1}(s)\hat{h}_i(s) + \hat{y}_\mu(s)\sigma_i(s)\right].
\end{equation}
Note the index shift: $\xi_i^\mu$ couples to both $\hat{x}_{\mu-1}$ (from the $\bm{\xi}^{\mu+1}$ term with $\mu \to \mu-1$) and $\hat{y}_\mu$ (from the $\bm{\xi}^\mu$ term).

For a single Ising random variable $\xi = \pm 1$ with equal probability:
\begin{equation}
    \overline{e^{-\ii\varepsilon\xi\Phi}} = \frac{1}{2}\left(e^{-\ii\varepsilon\Phi} + e^{\ii\varepsilon\Phi}\right) = \cos(\varepsilon\Phi).
\end{equation}
With $\varepsilon = N^{-1/2}$, this becomes
\begin{equation}
    \overline{e^{-\ii N^{-1/2}\xi_i^\mu\Phi_i^\mu}} = \cos\left(\frac{\Phi_i^\mu}{\sqrt{N}}\right).
\end{equation}

\paragraph{Large-$N$ expansion.}
For $|\Phi_i^\mu| = \mathcal{O}(1)$ (which holds since $\hat{h}_i$, $\sigma_i$, $\hat{x}_\mu$, $\hat{y}_\mu$ are all $\mathcal{O}(1)$ and the time sum has $t = \mathcal{O}(1)$ terms), we expand:
\begin{equation}
    \cos\left(\frac{\Phi}{\sqrt{N}}\right) = 1 - \frac{\Phi^2}{2N} + \frac{\Phi^4}{24N^2} + \mathcal{O}(N^{-3}) = \exp\left[-\frac{\Phi^2}{2N} + \mathcal{O}(N^{-2})\right].
\end{equation}
The last equality uses $\ln(1-x) = -x - x^2/2 + \ldots$ and the fact that the $\Phi^4$ term is subleading.

\paragraph{Combining sites and patterns.}
Taking the product over all sites $i = 1, \ldots, N$ and non-condensed patterns $\mu > t$:
\begin{align}
    \prod_{\mu>t}\prod_i \cos\left(\frac{\Phi_i^\mu}{\sqrt{N}}\right) &= \exp\left[-\frac{1}{2N}\sum_{\mu>t}\sum_i (\Phi_i^\mu)^2 + \mathcal{O}(N^{-1})\right].
\end{align}
The sum over $i$ gives a factor of $N$, so the exponent is $\mathcal{O}(p-t) = \mathcal{O}(\alpha N)$ in total, which is $\mathcal{O}(N)$.

Substituting the definition~\eqref{eq:charge_def}:
\begin{align}
    (\Phi_i^\mu)^2 &= \left[\sum_{s<t}(\hat{x}_{\mu-1}(s)\hat{h}_i(s) + \hat{y}_\mu(s)\sigma_i(s))\right]^2 \nonumber\\
    &= \sum_{s,s'<t} \Big[\hat{x}_{\mu-1}(s)\hat{x}_{\mu-1}(s')\hat{h}_i(s)\hat{h}_i(s') + \hat{y}_\mu(s)\hat{y}_\mu(s')\sigma_i(s)\sigma_i(s') \nonumber\\
    &\qquad\qquad + 2\hat{x}_{\mu-1}(s)\hat{y}_\mu(s')\hat{h}_i(s)\sigma_i(s')\Big].
\end{align}

The site sums naturally define intensive order parameters:
\begin{align}
    Q(s,s') &= \frac{1}{N}\sum_i \hat{h}_i(s)\hat{h}_i(s'), \\
    q(s,s') &= \frac{1}{N}\sum_i \sigma_i(s)\sigma_i(s'), \\
    K(s,s') &= \frac{1}{N}\sum_i \hat{h}_i(s)\sigma_i(s').
\end{align}
The disorder-averaged exponent becomes:
\begin{align}\label{eq:disorder_avg_result}
    -\frac{1}{2N}\sum_{\mu>t}\sum_i (\Phi_i^\mu)^2 &= -\frac{1}{2}\sum_{\mu>t}\sum_{s,s'<t}\Big[\hat{x}_{\mu-1}(s)\hat{x}_{\mu-1}(s')Q(s,s') \nonumber\\
    &\qquad + \hat{y}_\mu(s)\hat{y}_\mu(s')q(s,s') + 2\hat{x}_{\mu-1}(s)\hat{y}_\mu(s')K(s,s')\Big].
\end{align}

The disorder average over the non-condensed patterns yields
\begin{align}
    \overline{[\ldots]} &= e^{\mathcal{O}(N^{1/2})} \prod_{\mu > t} \prod_i \cos\left[N^{-1/2}\sum_{s<t}(\hat{x}_{\mu-1}(s)\hat{h}_i(s) + \hat{y}_\mu(s)\sigma_i(s))\right] \nonumber\\
    &= e^{\mathcal{O}(N^{1/2})} \exp\left[-\frac{1}{2N}\sum_{\mu>t}\sum_{s,s'<t}\sum_i (\hat{x}_{\mu-1}(s)\hat{h}_i(s) + \hat{y}_\mu(s)\sigma_i(s))(\hat{x}_{\mu-1}(s')\hat{h}_i(s') + \hat{y}_\mu(s')\sigma_i(s'))\right]
\end{align}
The $\mathcal{O}(N^{1/2})$ prefactor accounts for the condensed patterns ($\mu \leq t$), which contribute terms that remain bounded as $N \to \infty$. The key result is that the disorder average converts the quenched randomness into Gaussian fluctuations with covariance determined by the order parameters $Q$, $q$, $K$.

\paragraph{Order parameters.}
We introduce relevant macroscopic order parameters via delta-function insertions:
\begin{align}
    m(s) &= \frac{1}{N}\sum_i \xi_i^s \sigma_i(s) & &\text{(sequence overlap)} \\
    k(s) &= \frac{1}{N}\sum_i \xi_i^{s+1} \hat{h}_i(s) & &\text{(conjugate field overlap)} \\
    q(s,s') &= \frac{1}{N}\sum_i \sigma_i(s)\sigma_i(s') & &\text{(spin correlation)} \\
    Q(s,s') &= \frac{1}{N}\sum_i \hat{h}_i(s)\hat{h}_i(s') & &\text{(conjugate field correlation)} \\
    K(s,s') &= \frac{1}{N}\sum_i \sigma_i(s)\hat{h}_i(s') & &\text{(spin-field correlation)}
\end{align}
The generating functional takes the saddle-point form
\begin{equation}\label{eq:Z_saddle}
    \overline{Z[\bm{\psi}]} = \int \dd\bm{m}\,\dd\hat{\bm{m}}\,\dd\bm{k}\,\dd\hat{\bm{k}}\,\dd\bm{q}\,\dd\hat{\bm{q}}\,\dd\bm{Q}\,\dd\hat{\bm{Q}}\,\dd\bm{K}\,\dd\hat{\bm{K}} \; e^{N[\Psi + \Phi + \Omega] + \mathcal{O}(N^{1/2})}
\end{equation}
where $\Psi$, $\Phi$, $\Omega$ are $\mathcal{O}(1)$ functionals.

The functional $\Psi$ enforces the order parameter definitions:
\begin{equation}
    \Psi = \ii\sum_{s<t}[\hat{m}(s)m(s) + \hat{k}(s)k(s) - m(s)k(s)] + \ii\sum_{s,s'<t}[\hat{q}(s,s')q(s,s') + \hat{Q}(s,s')Q(s,s') + \hat{K}(s,s')K(s,s')].
\end{equation}

The functional $\Phi$ contains the single-site dynamics:
\begin{align}
    \Phi = \frac{1}{N}\sum_i \ln\Bigg\{ &\sum_{\{\sigma(s)\}} \int\{\dd h\,\dd\hat{h}\} \, p_i(\sigma(0)) \, e^{\sum_{s<t}[\beta\sigma(s+1)h(s) - \ln 2\cosh(\beta h(s))]} \nonumber\\
    &\times e^{-\ii\sum_{s,s'<t}[\hat{q}(s,s')\sigma(s)\sigma(s') + \hat{Q}(s,s')\hat{h}(s)\hat{h}(s') + \hat{K}(s,s')\sigma(s)\hat{h}(s')]} \nonumber\\
    &\times e^{\ii\sum_{s<t}\hat{h}(s)[h(s) - \theta_i(s) - \hat{k}(s)\xi_i^{s+1}] - \ii\sum_{s<t}\sigma(s)[\hat{m}(s)\xi_i^s + \psi_i(s)]} \Bigg\}.
\end{align}

The functional $\Omega$ arises from integrating out the non-condensed overlaps:
\begin{align}\label{eq:Omega}
    \Omega[q,Q,K] = \frac{1}{N}\ln \int &\frac{\dd\bm{u}\,\dd\bm{v}}{(2\pi)^{(p-t)t}} e^{\ii\sum_{\mu>t}\sum_{s<t} u_{\mu+1}(s)v_\mu(s)} \nonumber\\
    &\times e^{-\frac{1}{2}\sum_{\mu>t}\sum_{s,s'<t}[u_\mu(s)Q(s,s')u_\mu(s') + u_\mu(s)K(s',s)v_\mu(s') + v_\mu(s)K(s,s')u_\mu(s') + v_\mu(s)q(s,s')v_\mu(s')]}
\end{align}
where we renamed variables $\hat{x}_{\mu-1} \to u_\mu$ and $\hat{y}_\mu \to v_\mu$.


Extremizing $\Psi + \Phi + \Omega$ yields the saddle-point equations. Variations with respect to conjugate fields give
\begin{equation}
    k(s) = \hat{m}(s) = Q(s,s') = 0
\end{equation}
and
\begin{align}
    \hat{k}(s) &= m(s) = \lim_{N\to\infty}\frac{1}{N}\sum_i \langle\sigma(s)\rangle_i \xi_i^s, \\
    q(s,s') &= C(s,s') = \lim_{N\to\infty}\frac{1}{N}\sum_i \langle\sigma(s)\sigma(s')\rangle_i, \\
    K(s,s') &= \ii G(s,s') = \lim_{N\to\infty}\frac{1}{N}\sum_i \langle\sigma(s)\hat{h}(s')\rangle_i. \label{eq:K_saddle}
\end{align}

Variations with respect to the order parameters themselves involve derivatives of $\Omega$:
\begin{align}
    \hat{q}(s,s') &= \ii\frac{\partial\Omega}{\partial C(s,s')}, \\
    \hat{Q}(s,s') &= \ii\frac{\partial\Omega}{\partial Q(s,s')}, \\
    \hat{K}(s,s') &= \ii\frac{\partial\Omega}{\partial K(s,s')}.
\end{align}

\paragraph{Evaluation of the $\Omega$ functional.}
To evaluate~\eqref{eq:Omega}, we work in the product space of pattern indices $\mu$ and time indices $s$. Define $\bm{\Gamma} = \bm{S} \otimes \bm{R}$ with matrix elements $\Gamma_{\mu\mu'}(s,s') = S_{\mu\mu'}R(s,s')$. The Gaussian integral gives
\begin{equation}
    \lim_{N\to\infty} \Omega[C,Q,\ii G] = -\frac{1}{2N}\left[\ln\det(\bm{1}\otimes\bm{C}) + \ln\det\left\{\bm{1}\otimes\bm{Q} + [\bm{S}\otimes\bm{1} - \bm{1}\otimes\bm{G}]^\dagger[\bm{1}\otimes\bm{C}]^{-1}[\bm{S}\otimes\bm{1} - \bm{1}\otimes\bm{G}]\right\}\right].
\end{equation}

The derivative $\hat{q}(s,s') = \ii\partial\Omega/\partial C(s,s')$ involves only $[\bm{S}\otimes\bm{1} - \bm{1}\otimes\bm{G}]^\dagger[\bm{S}\otimes\bm{1} - \bm{1}\otimes\bm{G}]$, which is independent of $C$. Hence $\hat{q}(s,s') = 0$.

For $\hat{Q}(s,s')$, setting $Q = 0$ and using the matrix identity $\ln\det(\bm{M} + \bm{Q}) = \ln\det\bm{M} + \Tr[\bm{M}^{-1}\bm{Q}] + \mathcal{O}(Q^2)$, one obtains
\begin{equation}
    \hat{Q}(s,s') = -\frac{\alpha\ii}{2}\sum_{n\geq 0}[(\bm{G}^\dagger)^n\bm{C}\bm{G}^n](s,s').
\end{equation}

The crucial calculation is for $\hat{K}(s,s')$. Using the eigenvalues $s_\mu = e^{2\pi\ii\mu/p}$ of the shift matrix $\bm{S}$, we have
\begin{align}\label{eq:Khat_calc}
    \hat{K}(s,s') &= \frac{\partial\Omega}{\partial G(s,s')}\bigg|_{Q=0} \nonumber\\
    &= -\frac{\alpha}{2}\frac{\partial}{\partial G(s,s')} \lim_{p\to\infty}\frac{1}{p}\sum_{\mu=1}^{p}\left[\ln\det(\bm{1} - e^{-2\pi\ii\mu/p}\bm{G}^\dagger) + \ln\det(\bm{1} - e^{2\pi\ii\mu/p}\bm{G})\right] \nonumber\\
    &= -\frac{\alpha}{2}\frac{\partial}{\partial G(s,s')} \Tr\sum_{n>0}\frac{1}{n}\int_{-\pi}^{\pi}\frac{\dd\omega}{2\pi}\left[e^{-n\ii\omega}(\bm{G}^\dagger)^n + e^{n\ii\omega}\bm{G}^n\right].
\end{align}
The integral $\int_{-\pi}^{\pi} e^{\pm n\ii\omega}\dd\omega = 0$ for all $n > 0$. Therefore,
\begin{equation}\label{eq:Khat_zero}
    \hat{K}(s,s') = 0
\end{equation}

This is the central result: the retarded self-interaction vanishes identically for purely asymmetric synapses. The phase cancellation arises from the shift structure $\bm{S}$ of the synaptic matrix.

The term $\hat{K}(s,s')$ is called the \textit{retarded self-interaction} because it couples a neuron's current state $\sigma(s)$ to its own past auxiliary field $\hat{h}(s')$ (for $s > s'$). To see this, examine the effective single-neuron action. In the generating functional, the term
\begin{equation}
    -\hat{K}(s,s')\sigma(s)\hat{h}(s')
\end{equation}
appears in the exponent. Integrating out $\hat{h}(s')$ (which enforces the local field constraint), this term generates a contribution to the effective field acting on $\sigma(s)$ that depends on the spin's own past trajectory. Specifically, since $\hat{h}(s')$ is conjugate to $h(s')$, which in turn depends on $\sigma(s'-1)$, the $\hat{K}$ term creates a coupling $\sigma(s) \leftrightarrow \sigma(s'-1)$---a self-feedback through time.

In systems with symmetric couplings (standard Hopfield model), $\hat{K} \neq 0$ and contributes additional noise to the effective dynamics. The remarkable property of purely asymmetric sequence-storing networks is that $\hat{K} = 0$: the phase structure of the shift matrix $\bm{S}$ causes exact cancellation, eliminating the retarded self-interaction entirely.

\paragraph{Effective single-neuron problem.}
With $\hat{K} = 0$, choosing staggered external fields $\theta_i(s) = \theta(s)\xi_i^{s+1}$, the effective single-neuron measure becomes site-independent after a gauge transformation $\sigma(s) \to \sigma(s)\xi_i^s$, $h(s) \to h(s)\xi_i^{s+1}$:
\begin{align}\label{eq:single_spin}
    \langle f[\{\sigma\}]\rangle_\star = \sum_{\{\sigma(s)\}} &\int \{\dd h\,\dd\hat{h}\} \, p(\sigma(0)) \, f[\{\sigma\}] \, e^{\sum_{s<t}[\beta\sigma(s+1)h(s) - \ln 2\cosh(\beta h(s))]} \nonumber\\
    &\times e^{\ii\sum_{s<t}\hat{h}(s)[h(s) - \theta(s) - m(s)] - \frac{\alpha}{2}\sum_{s,s'<t}R(s,s')\hat{h}(s)\hat{h}(s')}
\end{align}
where
\begin{equation}\label{eq:R_def}
    R(s,s') = \sum_{n\geq 0}[(\bm{G}^\dagger)^n\bm{C}\bm{G}^n](s,s').
\end{equation}

This describes an effective single spin with local field $h(s) = m(s) + \theta(s) + \sqrt{\alpha}\phi(s)$, where $\phi(s)$ is zero-mean Gaussian noise with covariance $\langle\phi(s)\phi(s')\rangle = R(s,s')$. Crucially, there is no retarded self-interaction term because $\hat{K} = 0$.

The self-consistent equations are
\begin{align}
    m(s) &= \langle\sigma(s)\rangle_\star, \\
    C(s,s') &= \langle\sigma(s)\sigma(s')\rangle_\star, \\
    G(s,s') &= \frac{\partial}{\partial\theta(s')}\langle\sigma(s)\rangle_\star.
\end{align}

Since the measure~\eqref{eq:single_spin} factorizes over time, the spin sums can be performed:
\begin{align}
    m(s) &= \int \{\dd v\,\dd w\} e^{\ii\bm{v}\cdot\bm{w} - \frac{1}{2}\bm{w}\cdot\bm{R}\bm{w}} \tanh\beta[m(s-1) + \theta(s-1) + \sqrt{\alpha}v(s-1)], \\
    G(s,s') &= \beta\delta_{s,s'+1}\left\{1 - \int \{\dd v\,\dd w\} e^{\ii\bm{v}\cdot\bm{w} - \frac{1}{2}\bm{w}\cdot\bm{R}\bm{w}} \tanh^2\beta[m(s-1) + \theta(s-1) + \sqrt{\alpha}v(s-1)]\right\}.
\end{align}
The response $G(s,s')$ is non-zero only for $s = s' + 1$, indicating that stationarity is reached on finite timescales.

\paragraph{Stationary state.}
For time-translation invariant solutions with constant external field $\theta(s) = \theta$, we have $m(s) = m$, $C(s,s') = C(s-s')$, $G(s,s') = G(s-s')$. The matrices $\bm{C}$ and $\bm{G}$ become Toeplitz and commute, so
\begin{equation}
    \bm{R} = \bm{C}[\bm{1} - \bm{G}^\dagger\bm{G}]^{-1}.
\end{equation}

Separating persistent and non-persistent parts, $C(\tau) = q + \tilde{C}(\tau)$ with $\tilde{C}(\tau) \to 0$ as $|\tau| \to \infty$, and writing $G(\tau) = \beta\delta_{\tau,1}(1-\tilde{q})$, the stationary equations become
\begin{align}
    \rho &= [1 - \beta^2(1-\tilde{q})^2]^{-1}, \label{eq:rho}\\
    m &= \int Dz \, \tanh\beta[m + \theta + z\sqrt{\alpha\rho}], \label{eq:m_stat}\\
    \tilde{q} &= \int Dz \, \tanh^2\beta[m + \theta + z\sqrt{\alpha\rho}], \label{eq:qtilde}\\
    q &= \int Dz \left[\int Dx \, \tanh\beta[m + \theta + z\sqrt{\alpha q\rho} + x\sqrt{\alpha(1-q)\rho}]\right]^2 \label{eq:q}
\end{align}
where $Dz = (2\pi)^{-1/2}e^{-z^2/2}\dd z$. Equations~\eqref{eq:rho}--\eqref{eq:qtilde} form a closed set; $q$ follows from~\eqref{eq:q}.

\begin{figure}
    \centering
    \includegraphics[]{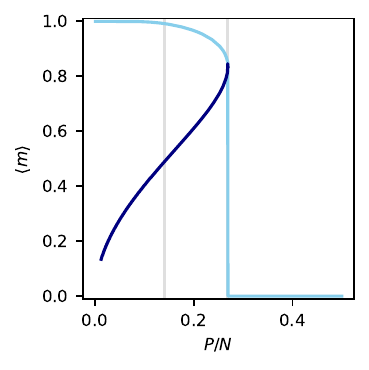}
    \caption{Phase diagram at zero temperature ($T = 0$) for purely asymmetric sequence-storing networks ($\lambda \to \infty$). The retrieval overlap $m$ (color scale) is plotted as a function of storage load $\alpha$. The light-blue curve indicates the retrieval phase, terminating at the critical capacity $\alpha_c \approx 0.269$ (equation~\eqref{eq:alpha_c}), and the dark-blue curve shows the metastable solution.}
    \label{fig:phase_diagram_infinite_lambda}
\end{figure}

\paragraph{Zero temperature and storage capacity.}
At $T = 0$ ($\beta \to \infty$), $\tilde{q} \to 1$ and $q \to 1$. Using
\begin{align}
    \lim_{\beta\to\infty} \int Dz \, \tanh\beta[m + z\sqrt{\alpha\rho}] &= \mathrm{erf}\left(\frac{m}{\sqrt{2\alpha\rho}}\right), \\
    \lim_{\beta\to\infty} \beta(1-\tilde{q}) &= \sqrt{\frac{2}{\pi\alpha\rho}} \exp\left(-\frac{m^2}{2\alpha\rho}\right),
\end{align}
and defining $x = m/\sqrt{2\alpha\rho}$ so that $m = \mathrm{erf}(x)$, the stationary equation becomes
\begin{equation}\label{eq:T0_eq}
    x\sqrt{2\alpha} = \pm\sqrt{\mathrm{erf}^2(x) - \frac{4x^2}{\pi}e^{-2x^2}}.
\end{equation}
The critical capacity $\alpha_c$ is the largest $\alpha$ for which~\eqref{eq:T0_eq} has non-trivial solutions:
\begin{equation}\label{eq:alpha_c}
    \alpha_c \approx 0.269.
\end{equation}

\subsubsection{Finite \texorpdfstring{$\lambda$}{lambda}: interpolating between static and dynamic synapses}
\label{sec:finite_lambda}
For finite $\lambda$, the synaptic matrix~\eqref{eq:J_matrix} contains both symmetric ($\bm{1}$) and asymmetric ($\lambda\bm{S}$) components. The phase diagram shows two distinct phases:
\begin{itemize}
    \item For $\lambda < 1$: static retrieval phase, where the network converges to a fixed pattern
    \item For $\lambda > 1$: dynamic sequence phase, where the network cycles through patterns
\end{itemize}
The transition point $\lambda = 1$ is special: both retrieval phases (dynamic and static) are destroyed. The coupling matrix becomes
\begin{equation}
    J_{ij} = \frac{1}{N}\sum_\mu \xi_i^\mu \underbrace{\left(\xi_j^\mu + \xi_j^{\mu-1}\right)}_{\approx\text{orthogonal to all patterns}},
\end{equation}
where the Hopfield (symmetric) and sequence (asymmetric) components have equal weight. This balance leads to destructive interference: the asymmetric component disrupts the static attractors of the Hopfield term, while the symmetric component interferes with clean sequence propagation.
 
To formulate the generating functional, we start by rewriting the synaptic matrix as
\begin{equation}
    J_{ij} = \frac{1}{N(1+\lambda)}\sum_{\mu=1}^{p} \xi_i^\mu \xi_j^\mu + \frac{\lambda}{N(1+\lambda)}\sum_{\mu=1}^{p} \xi_i^{\mu+1}\xi_j^\mu = \frac{1}{N(1+\lambda)}\bm{\xi}^{\mathrm{T}}(\bm{1} + \lambda\bm{S})\bm{\xi}.
\end{equation}
The disorder-dependent term in the generating functional becomes
\begin{equation}
    -\frac{\ii}{N}\hat{\bm{h}}(s)\cdot[\bm{\xi}^{\mathrm{T}}(\bm{1} + \lambda\bm{S})\bm{\xi}]\bm{\sigma}(s) = -\frac{\ii}{N}\sum_\mu [(\hat{\bm{h}}(s)\cdot\bm{\xi}^\mu) + \lambda(\hat{\bm{h}}(s)\cdot\bm{\xi}^{\mu+1})][\bm{\sigma}(s)\cdot\bm{\xi}^\mu].
\end{equation}
 
Define the non-condensed overlaps ($\mu \neq s$):
\begin{align}
    x_\mu(s) &= \frac{1}{\sqrt{N}}\sum_i \xi_i^{\mu+1} \hat{h}_i(s), \\
    y_\mu(s) &= \frac{1}{\sqrt{N}}\sum_i \xi_i^\mu \sigma_i(s),
\end{align}
and $x_s(s)=y_s(s)=0$. The interaction term becomes
\begin{equation}
    -\ii\sum_\mu [x_{\mu-1}(s) + \lambda x_\mu(s)] y_\mu(s) = -\ii\sum_\mu x_\mu(s) [y_{\mu+1}(s) + \lambda y_\mu(s)]
\end{equation}
where we shifted the summation index. In matrix notation,
\begin{equation}
    -\ii \bm{x}(s) \cdot \frac{\bm{S}^{\mathrm{T}} + \lambda\bm{1}}{\lambda + 1} \bm{y}(s).
\end{equation}
 
\paragraph{Structure of the modified generating functional.}
Upon writing the $\delta$-distributions in Fourier form, the disorder-averaged generating functional has the same structure as before, with the key difference that the interaction between the non-condensed overlaps $\bm{w}_\mu$ and $\bm{y}_\mu$ is not diagonal, but has the kernel $(\bm{S}^{\mathrm{T}} + \lambda\bm{1})$:
\begin{align}\label{eq:Z_disorder_finite_lambda}
    \overline{Z[\bm{\psi}]} = \sum_{\{\bm{\sigma}\}} p[\bm{\sigma}(0)] &\int \{\dd\bm{h}\,\dd\hat{\bm{h}}\} \, e^{\sum_{s<t}[\beta\bm{\sigma}(s+1)\cdot\bm{h}(s) - \sum_i \ln 2\cosh[\beta h_i(s)] + \ii\hat{\bm{h}}(s)\cdot(\bm{h}(s)-\bm{\theta}(s)) - \ii\bm{\psi}(s)\cdot\bm{\sigma}(s)]} \nonumber\\
    &\times e^{-\ii \frac{1}{\lambda + 1} N^{-1}\sum_{s<t}\left(\lambda[\hat{\bm{h}}(s)\cdot\bm{\xi}^{s+1}][\bm{\sigma}(s)\cdot\bm{\xi}^s] +  [\hat{\bm{h}}(s)\cdot\bm{\xi}^{s}][\bm{\sigma}(s)\cdot\bm{\xi}^s]\right)} \nonumber\\
    &\times \int \frac{\dd\bm{x}\,\dd\hat{\bm{x}}\,\dd\bm{y}\,\dd\hat{\bm{y}}}{(2\pi)^{2(p-1)t}} e^{\ii\sum_{s<t}\sum_{\mu \neq s}[\hat{x}_\mu(s)x_\mu(s) + \hat{y}_\mu(s)y_\mu(s) - x_\mu(s)\sum_\nu\frac{(\bm{S}^{\mathrm{T}} + \lambda\bm{1})_{\mu\nu}}{\lambda + 1} y_\nu(s)]} \nonumber\\
    &\times \overline{\left[e^{-\ii N^{-1/2}\sum_{s<t}\sum_{\mu \neq s}[\hat{x}_\mu(s)\hat{\bm{h}}(s)\cdot\bm{\xi}^{\mu+1} + \hat{y}_\mu(s)\bm{\sigma}(s)\cdot\bm{\xi}^\mu]}\right]}.
\end{align}
Note that the second factor now includes both the asymmetric term $[\hat{h}(s)\cdot \bm{\xi}^{s+1}] [{\bm \sigma}(s)\cdot \bm{\xi}^s]$ and the symmetric term $[\hat{h}(s)\cdot \bm{\xi}^s] [{\bm \sigma}(s)\cdot \bm{\xi}^s]$. Both are $O(N)$ and must be tracked for finite $\lambda$. The symmetric term introduces a new condensed variable $l(s) = \frac{1}{N}\hat{\bm{h}}(s)\cdot\bm{\xi}^{s}$, which modifies the effective single-neuron problem and significantly affects the critical capacity at finite $\lambda$.

The disorder average in the last factor is performed as before, yielding Gaussian fluctuations:
\begin{align*}
    \overline{\left[\cdots\right]} &= e^{\mathcal{O}(N^{1/2})} \overline{\left[e^{-\ii N^{-1/2}\sum_{s<t}\sum_{\mu > t}[\hat{x}_\mu(s)\hat{\bm{h}}(s)\cdot\bm{\xi}^{\mu+1} + \hat{y}_\mu(s)\bm{\sigma}(s)\cdot\bm{\xi}^\mu]}\right]} \\
    &= e^{\mathcal{O}(N^{1/2})} \prod_{\mu > t}\prod_{i} \cos\left[ N^{-1/2}\sum_{s<t}[\hat{x}_{\mu-1}(s)\hat{h}_i(s) + \hat{y}_\mu(s)\sigma_i(s)]\right]\\
    &= e^{\mathcal{O}(N^{1/2})} \exp\left[-\frac{1}{2N}\sum_{\mu>t}\sum_{s,s'<t}\sum_i [\hat{x}_{\mu-1}(s)\hat{h}_i(s) + \hat{y}_\mu(s)\sigma_i(s)][\hat{x}_{\mu-1}(s')\hat{h}_i(s') + \hat{y}_\mu(s')\sigma_i(s')]\right]
\end{align*}
The relevant order parameters are still $Q(s,s')$, $q(s,s')$, $K(s,s')$ as before. Plugging these in, we get:
\begin{align*}
    \exp\left[-\frac{1}{2}\sum_{\mu>t}\sum_{s,s'<t}[\hat{x}_{\mu-1}(s)\hat{x}_{\mu-1}(s')Q(s,s') + \hat{y}_\mu(s)\hat{y}_\mu(s')q(s,s') + \hat{x}_{\mu-1}(s)\hat{y}_\mu(s')K(s',s) + \hat{x}_{\mu-1}(s')\hat{y}_\mu(s)K(s,s')]\right],
\end{align*}
which will be included in the $\Omega$ functional along with the 
\begin{equation*}
    \sum_{s<t}\sum_{\mu \neq s}[\hat{x}_\mu(s)x_\mu(s) + \hat{y}_\mu(s)y_\mu(s) - x_\mu(s)\sum_\nu\frac{(\bm{S}^{\mathrm{T}} + \lambda\bm{1})_{\mu\nu}}{\lambda + 1}y_\nu(s)]
\end{equation*}
term.
 
The generating functional for finite $\lambda$ has a similar structure as the purely asymmetric case~\eqref{eq:Z_saddle}, with the decomposition
\begin{equation}
    \overline{Z[\bm{\psi}]} = \int \{\dd\text{(order params)}\} \, e^{N[\Psi + \Phi + \Omega] + \mathcal{O}(N^{1/2})}.
\end{equation}
The key modification is that the $\Omega$ functional---which arises from integrating out the non-condensed overlaps---now involves the coupling matrix $(\bm{S}^{\mathrm{T}} + \lambda\bm{1})$.

The functionals $\Psi$ and $\Phi$ depend only on the local structure of the problem (order parameter definitions and single-site dynamics) and are unchanged from Section~\ref{sec:purely_asymm}. The crucial difference at finite $\lambda$ arises entirely from $\Omega$, which encodes how the non-condensed patterns contribute to the effective noise and self-interaction. Specifically, the retarded self-interaction $\hat{K}(s,s') = \partial\Omega/\partial K(s,s')$ is the quantity that distinguishes the finite-$\lambda$ physics from the $\lambda \to \infty$ limit. Therefore, we focus on deriving the modified $\Omega$ functional.
 
\paragraph{Modified $\Omega$ functional.}
Following the same steps as Section~\ref{sec:purely_asymm}, we introduce Fourier conjugates $\hat{x}_\mu(s)$ and $\hat{y}_\mu(s)$ for the non-condensed overlaps ($\mu \neq s, s+1$). 
\begin{align}
    &\Omega[\bm{q}, \bm{Q}, \bm{K}] = \lim_{N\to\infty}\frac{1}{N}\ln \int \frac{\dd\bm{x}\,\dd\hat{\bm{x}}\, \dd\bm{y}\, \dd\hat{\bm{y}}}{(2\pi)^{2(p-1)t}} \exp\left[\ii\sum_{\mu>t} \sum_{s<t}[\hat{x}_\mu(s)x_\mu(s) + \hat{y}_\mu(s)y_\mu(s) - x_\mu(s)\sum_\nu \frac{(\bm{S}^{\mathrm{T}} + \lambda\bm{1})_{\mu\nu}}{\lambda + 1} y_\nu(s)]\right] \nonumber\\
    &\times \exp\left[-\frac{1}{2}\sum_{\mu>t}\sum_{s,s'<t}[\hat{x}_\mu(s)Q(s,s')\hat{x}_\mu(s') + \hat{y}_\mu(s)q(s,s')\hat{y}_\mu(s') + \hat{x}_{\mu-1}(s)K(s',s)\hat{y}_\mu(s') + \hat{x}_{\mu-1}(s')K(s,s')\hat{y}_\mu(s)]\right].
\end{align}
In the first factor we can identify a $\delta$ distribution:
\begin{equation*}
    \int \frac{\dd\bm{x}}{(2\pi)^{(p-t)t}} \exp\left[\ii\sum_{\mu>t}\sum_{s<t} x_\mu(s)[\hat{x}_\mu(s) - \sum_\nu \frac{(\bm{S}^{\mathrm{T}} + \lambda\bm{1})_{\mu\nu}}{\lambda + 1} y_\nu(s)]\right] =  \prod_{\mu>t}\prod_{s<t} \delta\left[\hat{x}_\mu(s) - \sum_\nu \frac{(\bm{S}^{\mathrm{T}} + \lambda\bm{1})_{\mu\nu}}{\lambda + 1} y_\nu(s)\right],
\end{equation*}
which imposes the constraint
\begin{equation}
    y_\mu(s) = (\lambda + 1) \sum_\nu (\bm{S} + \lambda\bm{1})^{-1}_{\mu\nu} \hat{x}_\nu(s).
\end{equation}
Substituting this back into the expression for $\Omega$, we can integrate out $\bm{y}$, and renaming $\hat{x}_{\mu-1}(s) \to u_\mu(s)$ and $\hat{y}_\mu(s) \to v_\mu(s)$, we obtain
\begin{align*}
    \Omega &= \lim_{N\to\infty}\frac{1}{N}\ln \int \frac{\dd\bm{u}\,\dd\bm{v}}{(2\pi)^{(p-t)t}} \exp\left[\ii\sum_{\mu>t}\sum_{s<t} v_\mu(s)(\lambda + 1)\sum_\nu(\bm{S}^{\mathrm{T}} + \lambda\bm{1})^{-1}_{\mu\nu} u_{\nu+1}(s)\right] \nonumber\\
    &\quad \times \exp\left[-\frac{1}{2}\sum_{\mu>t}\sum_{s,s'<t}[u_\mu(s)Q(s,s')u_\mu(s') + v_\mu(s)q(s,s')v_\mu(s') + u_\mu(s)K(s',s)v_\mu(s') + u_\mu(s')K(s,s')v_\mu(s)]\right].
\end{align*}
The shift $u_{\nu+1}$ in the first factor can be written as a matrix multiplication: $u_{\nu+1} = [\bm{S}^{\mathrm{T}}\bm{u}]_\nu$, since $S^{\mathrm{T}}_{\nu\rho} = \delta_{\nu+1,\rho}$. Therefore,
\begin{align}\label{eq:Omega_with_M}
    \Omega &= \lim_{N\to\infty}\frac{1}{N}\ln \int \frac{\dd\bm{u}\,\dd\bm{v}}{(2\pi)^{(p-t)t}} \exp\left[\ii\sum_{\mu>t}\sum_{s<t} v_\mu(s)\sum_\nu\mathcal{M}_{\mu\nu} u_{\nu}(s)\right] \nonumber\\
    &\quad \times \exp\left[-\frac{1}{2}\sum_{\mu>t}\sum_{s,s'<t}[u_\mu(s)Q(s,s')u_\mu(s') + v_\mu(s)q(s,s')v_\mu(s') + u_\mu(s)K(s',s)v_\mu(s') + u_\mu(s')K(s,s')v_\mu(s)]\right]\nonumber\\
    &= \lim_{N\to\infty}\frac{1}{N}\ln \int \frac{\dd\bm{u}\,\dd\bm{v}}{(2\pi)^{(p-t)t}} \exp\left[ -\frac{1}{2}\bm{u}\cdot[\bm{1}\otimes \bm{Q}]\bm{u} -\frac{1}{2}\bm{v}\cdot[\bm{1}\otimes \bm{q}]\bm{v} + \ii \bm{v}\cdot[\bm{\mathcal{M}}\otimes\bm{1} + \ii \bm{1}\otimes\bm{K}]\bm{u} \right],
\end{align}
where the coupling matrix is
\begin{equation}\label{eq:M_def}
    \bm{\mathcal{M}} = (\lambda + 1)(\bm{S}^{\mathrm{T}} + \lambda\bm{1})^{-1}\bm{S}^{\mathrm{T}}.
\end{equation}
 
Furthermore, this can be written in terms of the physical order parameters at the saddle point, $q(s,s')=C(s,s')$ and $K(s,s')=\ii G(s,s')$.
\begin{equation}
    \Omega = \lim_{N\to\infty}\frac{1}{N}\ln \int \frac{\dd\bm{u}\,\dd\bm{v}}{(2\pi)^{(p-t)t}} \exp\left[ -\frac{1}{2}\bm{u}\cdot[\bm{1}\otimes \bm{Q}]\bm{u} -\frac{1}{2}\bm{v}\cdot[\bm{1}\otimes \bm{C}]\bm{v} + \ii \bm{v}\cdot[\bm{\mathcal{M}}\otimes\bm{1} - \bm{1}\otimes\bm{G}]\bm{u} \right].
\end{equation}

\paragraph{Diagonalization in pattern space.}
The shift matrix $\bm{S}$ is a circulant matrix, diagonalized by the discrete Fourier transform with eigenvalues $e^{2\pi\ii\mu/p}$ for $\mu = 0, 1, \ldots, p-1$. Consequently, $\bm{S}^{\mathrm{T}}$ has eigenvalues $e^{-2\pi\ii\mu/p}$. Since both $\bm{S}^{\mathrm{T}}$ and $(\bm{S}^{\mathrm{T}} + \lambda\bm{1})$ are functions of $\bm{S}^{\mathrm{T}}$, they share the same eigenvectors (the Fourier modes) and are simultaneously diagonalizable. The eigenvalues of the coupling matrix $\mathcal{M}$ are therefore
\begin{equation}\label{eq:M_eigenvalues}
    z_\mu = (\lambda + 1)\frac{e^{-2\pi\ii\mu/p}}{\lambda + e^{-2\pi\ii\mu/p}} = \frac{\lambda + 1}{\lambda e^{2\pi\ii\mu/p} + 1}, \quad \mu = 0, 1, \ldots, p-1.
\end{equation}
 
Transforming to the eigenbasis of $\mathcal{M}$, the $\Omega$ functional factorizes over pattern modes:
\begin{equation}\label{eq:Omega_factorized}
    \Omega = \lim_{p\to\infty}\frac{1}{p}\sum_{\mu=1}^{p} \Omega_\mu(z_\mu),
\end{equation}
where $\Omega_\mu(z)$ is the contribution from the mode with eigenvalue $z$.

\paragraph{Evaluation of $\Omega_\mu(z)$.}
For a single eigenvalue $z$, the contribution $\Omega_\mu(z)$ is a Gaussian integral over $t$-dimensional vectors $\bm{u}$ and $\bm{v}$. At the saddle point, $q = C$ and $K = \ii G$, and using $Q = 0$, the integral takes the form
\begin{equation}\label{eq:Omega_mu_Gaussian}
    \Omega_\mu(z) = \alpha \ln \int \frac{\dd\bm{u}_\mu\,\dd\bm{v}_\mu}{(2\pi)^{2t}} \exp\left[ -\frac{1}{2}\bm{v}_\mu^{\mathrm{T}}\bm{C}\bm{v}_\mu + \ii \bm{v}_\mu^{\mathrm{T}}(z\bm{1} - \bm{G})\bm{u}_\mu \right].
\end{equation}
We evaluate this by first integrating over $\bm{u}$, then over $\bm{v}$~\cite{duringetal98}. The integral over $\bm{u}$ produces a $\delta$-function, and the subsequent Gaussian integral over $\bm{v}$ yields (cf.\ their Eq.~32)
\begin{equation}
    \Omega_\mu(z) = -\frac{1}{2}\ln\det\left[(z^*\bm{1} - \bm{G}^{\mathrm{T}})\bm{C}^{-1}(z\bm{1} - \bm{G})\right] - \frac{1}{2} \ln\det\bm{C}.
\end{equation}
Note that since $\bm{G}$ is real, $\bm{G}^{\mathrm{T}} = \bm{G}^\dagger$. Using $\det(\bm{A}\bm{B}\bm{C}) = \det\bm{A}\det\bm{B}\det\bm{C}$:
\begin{equation*}\label{eq:detA}
    \det\left[(z^*\bm{1} - \bm{G}^{\mathrm{T}})\bm{C}^{-1}(z\bm{1} - \bm{G})\right] = \det(z^*\bm{1} - \bm{G}^{\mathrm{T}})\det(\bm{C}^{-1})\det(z\bm{1} - \bm{G}),
\end{equation*}
which leads to 
\begin{equation}\label{eq:Omega_mu_z_final}
    \Omega_\mu(z) = -\frac{\alpha}{2} \left[ \ln\det(z^*\bm{1} - \bm{G}^{\mathrm{T}}) + \ln\det(z\bm{1} - \bm{G}) \right].
\end{equation}

The total $\Omega$ functional is
\begin{equation}\label{eq:Omega_finite_lambda}
    \Omega[\bm{C}, \bm{Q}, \bm{G}] = - \lim_{p\to\infty}\frac{\alpha}{2p}\sum_{\mu=1}^{p} \ln\det\left[ Q + (z_\mu^*\bm{1} - \bm{G}^{\mathrm{T}}) \bm{C}^{-1} (z_\mu\bm{1} - \bm{G}) \right] - \frac{\alpha}{2p} \ln\det\bm{C}
\end{equation}
with $z_\mu$ given by~\eqref{eq:M_eigenvalues}.
Using this result we can already start working out the saddle-point equations for the order parameters.
 
The first of the three equations, $\hat{q}$ remains trivial since $\Omega$ does not depend on $\bm{C}$.
The second equation gives $\hat{Q}$ and the third gives $\hat{K}$. We derive both in the following.
 
\paragraph{Evaluation of $\hat{Q}$.}
The saddle-point equation for $\hat{Q}$ is
\begin{equation}\label{eq:Qhat_derivative}
    \hat{Q}(s,s') = -\lim_{Q\to 0} \frac{\partial}{\partial Q(s,s')} \Omega[\bm{C}, \bm{Q}, \bm{G}].
\end{equation}
Using $\frac{\partial}{\partial Q(s,s')}\ln\det[\bm{Q} + \bm{A}] = [\bm{Q} + \bm{A}]^{-1}(s',s)$ and evaluating at $\bm{Q} = 0$:
\begin{equation}\label{eq:Qhat_general}
    \hat{Q}(s,s') = \frac{\alpha}{2} \lim_{p\to\infty}\frac{1}{p} \sum_{\mu=1}^{p} \left[(z_\mu^*\bm{1} - \bm{G}^{\mathrm{T}})\bm{C}^{-1}(z_\mu\bm{1} - \bm{G})\right]^{-1}(s,s').
\end{equation}
This is the general result before any time-translation-invariant assumptions.
 
To evaluate the sum, we use the contour integral representation. As $p \to \infty$, the sum becomes
\begin{equation}
    \hat{Q}(s,s') = \frac{\alpha}{2} \int_{-\pi}^{\pi} \frac{\dd\omega}{2\pi} \left[(z^*\bm{1} - \bm{G}^{\mathrm{T}})\bm{C}^{-1}(z\bm{1} - \bm{G})\right]^{-1}(s,s')\bigg|_{z = \frac{\lambda+1}{\lambda e^{\ii\omega}+1}}.
\end{equation}
 
\textbf{Purely asymmetric limit: $\lambda \to \infty$.}
In this limit, $z_\mu \to e^{-2\pi\ii\mu/p}$ are uniformly distributed on the unit circle. By the orthogonality of Fourier modes, the sum over $\mu$ projects onto the ``zero mode.'' To see this, note that
\begin{equation}
    (z\bm{1} - \bm{G})^{-1} = z^{-1}(\bm{1} - z^{-1}\bm{G})^{-1} = \sum_{n=0}^{\infty} z^{-(n+1)} \bm{G}^n.
\end{equation}
The matrix product becomes
\begin{equation}
    \left[(z^*\bm{1} - \bm{G}^{\mathrm{T}})\bm{C}^{-1}(z\bm{1} - \bm{G})\right]^{-1} = (z\bm{1} - \bm{G})^{-1}\bm{C}(z^*\bm{1} - \bm{G}^{\mathrm{T}})^{-1} = \sum_{n,m=0}^{\infty} z^{-(n+1)}(z^*)^{-(m+1)} \bm{G}^n \bm{C} (\bm{G}^{\mathrm{T}})^m.
\end{equation}
When we average over $z = e^{-\ii\omega}$ on the unit circle:
\begin{equation}
    \int_{-\pi}^{\pi}\frac{\dd\omega}{2\pi} e^{\ii(n-m)\omega} = \delta_{n,m}.
\end{equation}
This kills all cross-terms, leaving
\begin{equation}\label{eq:Qhat_asymm}
    \hat{Q}(s,s')\big|_{\lambda\to\infty} = \frac{\alpha}{2} \sum_{n=0}^{\infty} \left[\bm{G}^n \bm{C}\, (\bm{G}^{\dagger})^n\right](s,s'). 
\end{equation}
This is precisely the result we got in the purely asymmetric limit.
 
\textbf{Symmetric limit: $\lambda = 0$.} From~\eqref{eq:Qhat_general} with $z_\mu = 1$ for all $\mu$, we obtain
$\hat{Q}(s,s') = \frac{\alpha}{2}[(\bm{1} - \bm{G})^{-1}\bm{C}(\bm{1} - \bm{G}^{\dagger})^{-1}](s,s')
= \frac{\alpha}{2}\sum_{n,m\geq 0} [\bm{G}^n \bm{C}\, (\bm{G}^{\dagger})^m](s,s')$:
here there is no Fourier projection, so the full double sum survives. It coincides with
the diagonal sum of the opposite limit~\eqref{eq:Qhat_asymm} only when $\bm C$ is
diagonal --- the approximation adopted below since we are interested in the regime that produces sequential dynamics, under which the two limits indeed give identical noise variances.
 
\textbf{General $\lambda$.}
For finite $\lambda$, the eigenvalues $z_\mu = (\lambda+1)/(\lambda e^{2\pi\ii\mu/p}+1)$ no longer lie on the unit circle, breaking the simple Fourier orthogonality. The contour integral involves products like $(1+\lambda e^{\ii\omega})^{n+1}(1+\lambda e^{-\ii\omega})^{m+1}$, and the $n \neq m$ terms no longer vanish.
 
For the diagonal elements $\hat{Q}(s,s)$ (which enter the stationary equations), keeping the $n=m$ (diagonal) terms of the double expansion gives
\begin{equation}
    \hat{Q}(s,s) \;\approx\; \frac{\alpha}{2} \sum_{n=0}^{\infty} \frac{\mathcal{J}_n(\lambda)}{(\lambda+1)^{2(n+1)}} \left[\bm{G}^n \bm{C}\, (\bm{G}^{\dagger})^n\right](s,s)
\end{equation}
where $\mathcal{J}_n(\lambda) = \int_{-\pi}^{\pi}\frac{\dd\omega}{2\pi}|1+\lambda e^{\ii\omega}|^{2(n+1)}$. Note that the $n\neq m$ cross terms do not vanish at finite $\lambda$ [e.g.\ for $(n,m)=(1,0)$ the frequency integral gives $1+2\lambda^2$]; together with the $\mathcal J_n$ corrections they are subdominant to the effect of $\hat K \neq 0$, as we show below. The full resummation, including all cross terms, is given in closed spectral form in Eq.~\eqref{eq:V_spectral} below, which is what we evaluate numerically. For $\lambda \to \infty$, $\mathcal{J}_n(\lambda)/(\lambda+1)^{2(n+1)} \to 1$, recovering~\eqref{eq:Qhat_asymm}.

\paragraph{Evaluation of retarded self interaction $\hat{K}$.}
 
The retarded self-interaction is 
\begin{equation}\label{eq:Khat_finite_lambda}
    \hat{K}(s,s') = \lim_{p\to\infty}\frac{1}{p}\sum_{\mu=1}^{p} \frac{\partial \Omega_\mu(z)}{\partial G(s,s')}\bigg|_{z = \frac{\lambda + 1}{\lambda e^{2\pi\ii\mu/p} + 1}}.
\end{equation}
 
To evaluate~\eqref{eq:Khat_finite_lambda}, note that as $p\to\infty$ the eigenvalues $e^{-2\pi\ii\mu/p}$ of the circulant shift $\bm{S}$ are uniformly distributed on the unit circle, so $\frac{1}{p}\sum_{\mu=1}^{p} f(e^{-2\pi\ii\mu/p}) \to \int_{0}^{2\pi}\frac{\dd\phi}{2\pi}\,f(e^{-\ii\phi})$.
Setting $\omega = 2\pi\mu/p$ and taking $p \to \infty$:
\begin{equation}
    \hat{K}(s,s') = \int_{-\pi}^{\pi} \frac{\dd\omega}{2\pi} \, \mathcal{K}\left(s,s'; \frac{\lambda + 1}{\lambda e^{\ii\omega} + 1}\right)
\end{equation}
where $\mathcal{K}(s,s';z) = \partial\Omega/\partial G(s,s')|_z$. The integration is over the unit circle parametrized by $\omega \in [-\pi, \pi]$.
 
Using $\ln \det \bm{M} = \Tr \ln \bm{M}$, we can write $\mathcal{K}(s,s';z)$ as
\begin{align}
    \mathcal{K}(s,s';z) &= -\frac{\alpha}{2} \frac{\partial}{\partial G(s,s')} \lim_{p\to\infty} [\ln\det(z^*\bm{1}-\bm{G}^{\mathrm{T}}) + \ln\det(z\bm{1}-\bm{G})]  \nonumber\\
    &=-\frac{\alpha}{2} \lim_{p\to\infty} \Tr\left[(z^*\bm{1}-\bm{G}^{\mathrm{T}})^{-1}\frac{\partial (z^*\bm{1}-\bm{G}^{\mathrm{T}})}{\partial G(s,s')} + (z\bm{1}-\bm{G})^{-1}\frac{\partial (z\bm{1}-\bm{G})}{\partial G(s,s')}\right] \nonumber\\
    &=+\frac{\alpha}{2} \lim_{p\to\infty} \Tr\left[(z^*\bm{1}-\bm{G}^{\mathrm{T}})^{-1}\bm{E}_{s',s} + (z\bm{1}-\bm{G})^{-1}\bm{E}_{s,s'}\right] \nonumber\\
    &=+\frac{\alpha}{2} \lim_{p\to\infty} \left[(z^*\bm{1}-\bm{G}^{\mathrm{T}})^{-1}(s,s') + (z\bm{1}-\bm{G})^{-1}(s',s)\right],
\end{align}
where $\bm{E}_{s,s'}$ is the matrix with a $1$ in position $(s,s')$ and zeros elsewhere, and $\partial(z\bm{1}-\bm{G})/\partial G(s,s')=-\bm{E}_{s,s'}$ carries the sign. The last line follows since $\Tr[\bm{A}\bm{E}_{s',s}] = A(s,s')$.

To evaluate the integral, we first rewrite the matrix inverses. Using
\begin{equation}
    z = \frac{\lambda + 1}{1 + \lambda e^{\ii\omega}}, \quad z^* = \frac{\lambda + 1}{1 + \lambda e^{-\ii\omega}},
\end{equation}
we have
\begin{equation}
    (z\bm{1} - \bm{G})^{-1} = \frac{1 + \lambda e^{\ii\omega}}{\lambda + 1}\left(\bm{1} - \frac{1 + \lambda e^{\ii\omega}}{\lambda + 1}\bm{G}\right)^{-1}.
\end{equation}
Expanding the matrix inverse as a geometric series:
\begin{equation}
    \left(\bm{1} - a\bm{G}\right)^{-1} = \sum_{n=0}^{\infty} a^n \bm{G}^n,
\end{equation}
we obtain
\begin{equation}
    (z\bm{1} - \bm{G})^{-1} = \sum_{n=0}^{\infty} \frac{(1 + \lambda e^{\ii\omega})^{n+1}}{(\lambda + 1)^{n+1}} \bm{G}^n.
\end{equation}
 
The retarded self-interaction therefore involves integrals of the form
\begin{equation}
    I_n^{+}(\lambda) = \int_{-\pi}^{\pi}\frac{\dd\omega}{2\pi} (1 + \lambda e^{\ii\omega})^{n+1}.
\end{equation}
Substituting $w = e^{\ii\omega}$, which maps $\omega \in [-\pi, \pi]$ to the unit circle $|w| = 1$ traversed counterclockwise, with $\dd\omega = \dd w/(\ii w)$:
\begin{equation}
    I_n^{+}(\lambda) = \oint_{|w|=1}\frac{\dd w}{2\pi\ii w}(1 + \lambda w)^{n+1}.
\end{equation}
The integrand $f(w) = (1 + \lambda w)^{n+1}/w$ has a simple pole at $w = 0$ and a zero of order $n+1$ at $w = -1/\lambda$. Since only the pole at $w = 0$ contributes, the residue theorem gives:
\begin{equation}
    I_n^{+}(\lambda) = \mathrm{Res}_{w=0}\left[\frac{(1 + \lambda w)^{n+1}}{w}\right] = (1 + \lambda \cdot 0)^{n+1} = 1.
\end{equation}
Similarly, for the conjugate integral with $e^{-\ii\omega}$:
\begin{equation}
    I_n^{-}(\lambda) = \int_{-\pi}^{\pi}\frac{\dd\omega}{2\pi} (1 + \lambda e^{-\ii\omega})^{n+1} = 1.
\end{equation}
 
Substituting back into the expression for $\hat{K}(s,s')$:
\begin{align}
    \hat{K}(s,s') &= +\frac{\alpha}{2} \sum_{n=0}^{\infty} \frac{1}{(\lambda+1)^{n+1}} \left[ (\bm{G}^{\mathrm{T}})^n(s,s') + \bm{G}^n(s',s) \right].
\end{align}
Since $\bm{G}$ is the response function with $G(s,s') = 0$ for $s \leq s'$ (causality), $(\bm{G}^{\mathrm{T}})^n(s,s') = \bm{G}^n(s',s)$; both are supported on $s' > s$, so the two terms coincide and add. For $s' > s$ (the $n=0$ term $\delta_{s',s}$ drops):
\begin{equation}\label{eq:Khat_result}
    \hat{K}(s,s') = \alpha \sum_{n=1}^{\infty} \frac{\bm{G}^n(s',s)}{(\lambda+1)^{n+1}} = \frac{\alpha}{\lambda+1} \left[\left(\bm{1} - \frac{\bm{G}}{\lambda+1}\right)^{-1}\right](s',s),
\end{equation}
consistent with the retarded kernel $R=\hat{K}/\alpha$ used in Eq.~\eqref{eq:R_finite_lambda}.
 
Limiting cases:
\begin{itemize}
    \item \textbf{Purely asymmetric limit} ($\lambda \to \infty$): $\hat{K}(s,s') \to 0$. This recovers the result of Section~\ref{sec:purely_asymm}, where the retarded self-interaction vanishes.
    \item \textbf{Symmetric limit} ($\lambda = 0$): $\hat{K}(s,s') = \alpha\,[(\bm{1} - \bm{G})^{-1}](s',s)$ (for $s'>s$). This gives a non-zero retarded self-interaction, as expected for symmetric networks.
\end{itemize}
 
The non-vanishing $\hat{K}$ for finite $\lambda$ has a clear physical origin. The symmetric component in the synaptic matrix creates correlations between a neuron's current state and its own past local field---the retarded self-interaction. In the purely asymmetric case ($\lambda \to \infty$), pattern $\mu$ only influences pattern $\mu+1$, creating a ``feedforward'' structure in pattern space that eliminates self-feedback. The symmetric component breaks this feedforward structure, reducing the storage capacity.
 
\paragraph{Modified $\Psi$ functional.}
Similar to the fully asymmetric case, we introduce Fourier representations for all condensed variables. Note that we have a new relevant condensed variable
\begin{equation}
    l(s) = \frac{1}{N}\sum_i \xi_i^s \hat{h}_i(s)
\end{equation}
with conjugate $\hat{l}(s)$. The $\Psi$ functional becomes
\begin{align}
    \Psi =& \ii\sum_{s<t}\left[\hat{m}(s)m(s) + \hat{k}(s)k(s) + \hat{l}(s)l(s) - \frac{m(s)}{\lambda+1}[l(s) + \lambda k(s)]\right] \nonumber\\
    &+ \ii\sum_{s,s'<t}[\hat{q}(s,s')q(s,s') + \hat{Q}(s,s')Q(s,s') + \hat{K}(s,s')K(s,s')].
\end{align}
 
\paragraph{Modified $\Phi$ functional.}
The $\Phi$ functional is the single-site contribution, where the dynamical terms coming from the Glauber kernel are unchanged, but from the Fourier representation of the condensed variables we have additional terms involving $l(s)$ and $\hat{l}(s)$. The modified $\Phi$ functional is
\begin{align}
    \Phi = \frac{1}{N}\sum_i \ln\Bigg\{ &\sum_{\{\sigma(s)\}} \int\{\dd h\,\dd\hat{h}\} \, p_i(\sigma(0)) \, e^{\sum_{s<t}[\beta\sigma(s+1)h(s) - \ln 2\cosh(\beta h(s))]} \nonumber\\
    &\times e^{-\ii\sum_{s,s'<t}[\hat{q}(s,s')\sigma(s)\sigma(s') + \hat{Q}(s,s')\hat{h}(s)\hat{h}(s') + \hat{K}(s,s')\sigma(s)\hat{h}(s')]} \nonumber\\
    &\times e^{\ii\sum_{s<t}\hat{h}(s)[h(s) - \theta_i(s) - \hat{k}(s)\xi_i^{s+1} - \hat{l}(s)\xi_i^{s}] - \ii\sum_{s<t}\sigma(s)[\hat{m}(s)\xi_i^s + \psi_i(s)]} \Bigg\}.
\end{align}
 
\paragraph{Evaluation of conjugate order parameters.}
At the saddle point, varying with respect to each variable, for the conjugate variables we get:
\begin{align}
    \frac{\partial \Psi}{\partial m(s)} = 0 \implies & \hat{m}(s) = \frac{1}{\lambda+1}[l(s) + \lambda k(s)] = 0,\quad \text{from $\Phi$ below}, \\
    \frac{\partial \Psi}{\partial k(s)} = 0 \implies & \hat{k}(s) = \frac{\lambda}{\lambda+1} m(s), \\
    \frac{\partial \Psi}{\partial l(s)} = 0 \implies & \hat{l}(s) = \frac{1}{\lambda+1} m(s),
\end{align}
and for the physical variables:
\begin{align}
    &\frac{\partial (\Psi+\Phi)}{\partial \hat{m}(s)} = \ii[m(s) - \frac{1}{N}\sum_i \xi_i^s \langle \sigma_i(s) \rangle] = 0\implies m(s) = \frac{1}{N}\sum_i \xi_i^s \langle \sigma_i(s) \rangle, \\
    &\frac{\partial (\Psi+\Phi)}{\partial \hat{k}(s)} = \ii[k(s) - \frac{1}{N}\sum_i \xi_i^{s+1} \underbrace{\langle \hat{h}_i(s) \rangle}_{=0}] = 0 \, \implies k(s)=0 \\
    &\frac{\partial (\Psi+\Phi)}{\partial \hat{l}(s)} = \ii[l(s) - \frac{1}{N}\sum_i \xi_i^{s} \underbrace{\langle \hat{h}_i(s) \rangle}_{=0}] = 0 \, \implies \, l(s) = 0.
\end{align}

\paragraph{Deriving the effective single-neuron problem.}
 
Having obtained $\hat{K}$~\eqref{eq:Khat_result} and the general form of $\hat{Q}$~\eqref{eq:Qhat_general}, we now derive the effective single-neuron measure following the approach of Coolen (equations 132--137 in the review). While the full evaluation of $\hat{Q}$ for general $\lambda$ involves the $\mathcal{J}_n(\lambda)$ integrals discussed above, the dominant finite-$\lambda$ correction comes from the retarded self-interaction $\hat{K} \neq 0$. We therefore proceed with the Hubbard-Stratonovich transformation and later specialize to the stationary regime where explicit results can be obtained.
 
The key insight is that both $\hat{Q}$ and $\hat{K}$ enter the effective measure, and the Hubbard-Stratonovich transformation converts the quadratic $\hat{h}$ terms into Gaussian noise.
 
\textbf{Step 1: Structure of the effective measure.}
 
At the saddle point, the effective single-neuron measure has the form
\begin{align}\label{eq:M_general}
    &M[\{\sigma\},\{h\},\{\hat{h}\}] = \pi_0(\sigma(0)) \prod_t \left\{[2\cosh(\beta h(t))]^{-1} \exp[\beta\sigma(t+1)h(t)]\right\} \nonumber\\
    &\quad \times \exp\left\{i\sum_t \hat{h}(t)\left[h(t) - \xi^{t+1}\underbrace{\hat{k}(t)}_{=\frac{\lambda}{\lambda+1}m(t)} - \xi^{t}\underbrace{\hat{l}(t)}_{=\frac{1}{\lambda+1}m(t)} - \theta(t) - \sum_{t'}\hat{K}(t,t')\sigma(t')\right] - i\sum_{t,t'}\hat{Q}(t,t')\hat{h}(t)\hat{h}(t')\right\}.
\end{align}
 
Define
\begin{align}
    A(t) &= \frac{1}{1+\lambda}m(t)\left(\xi^{t} + \lambda \xi^{t+1}\right) + \theta(t) + \sum_{t'} \hat{K}(t,t')\sigma(t')\nonumber\\
    &= \frac{1}{1+\lambda}\xi^{t+1}m(t)\left(\eta_i^{t} + \lambda \right) + \theta(t) + \sum_{t'} \hat{K}(t,t')\sigma(t'),
\end{align}
where we used $\xi^{t} = \eta_i^{t}\xi^{t+1}$ with $\eta_i^{t} = \pm 1$ being an uncorrelated random variable independent of the condensed pattern $\xi^{t+1}$, since the patterns are random and independent.
The measure becomes
\begin{equation}
    M[\ldots] \propto \exp\left\{i\sum_t \hat{h}(t)[h(t) - A(t)] - i\sum_{t,t'}\hat{Q}(t,t')\hat{h}(t)\hat{h}(t')\right\} \times \prod_t \frac{1}{2}[1 + \sigma(t+1)\tanh(\beta h(t))].
\end{equation}

\textbf{Step 2: Hubbard-Stratonovich transformation.}
The quadratic term in $\hat{h}$ can be linearized by introducing auxiliary Gaussian fields $\phi(t)$. For a matrix $\bm{V}$ with $\hat{Q} = -\frac{i\alpha}{2}\bm{V}$:
\begin{equation}
    \exp\left(-\frac{\alpha}{2}\hat{h}^{\mathrm{T}}\bm{V}\hat{h}\right) = \mathcal{N}[\phi] \int \mathcal{D}[\phi] \exp\left(-\frac{1}{2}\phi^{\mathrm{T}}\bm{V}^{-1}\phi - i\sqrt{\alpha}\,\hat{h}^{\mathrm{T}}\phi\right),
\end{equation}
where $\mathcal{N}[\phi] = [(2\pi)^{t_m+1}\det\bm{V}]^{-1/2}$ is a normalization factor.
 
The Gaussian measure for $\phi$ is
\begin{equation}
    P[\{\phi\}] = \mathcal{N}[\phi] \exp\left(-\frac{1}{2}\sum_{t,t'}\phi(t)\bm{V}^{-1}(t,t')\phi(t')\right),
\end{equation}
with covariance $\langle\phi(t)\phi(t')\rangle = \bm{V}(t,t')$.
 
\textbf{Step 3: Integrating out $\hat{h}$.}
After the Hubbard-Stratonovich transformation, the integral over $\hat{h}$ becomes
\begin{equation}
    \int \mathcal{D}[\hat{h}] \exp\left[i\sum_t \hat{h}(t)(h(t) - A(t) - \sqrt{\alpha}\phi(t))\right] = \prod_t \delta\left[h(t) - A(t) - \sqrt{\alpha}\phi(t)\right].
\end{equation}
This fixes the local field:
\begin{equation}\label{eq:h_effective}
    h(t) = A(t) + \sqrt{\alpha}\phi(t) = \frac{1}{1+\lambda}\xi^{t+1}m(t)\left(\eta_i^{t} + \lambda \right) + \theta(t) + \sum_{t'} \hat{K}(t,t')\sigma(t') + \sqrt{\alpha}\phi(t).
\end{equation}
 
\textbf{Step 4: The effective single-neuron problem.}
Combining all pieces, the probability distribution over spin paths becomes
\begin{equation}\label{eq:P_sigma}
    P[\{\sigma\}] = \pi_0(\sigma(0)) \int \mathcal{D}[\phi]\, P[\{\phi\}] \prod_t \frac{1}{2}\left[1 + \sigma(t+1)\tanh(\beta h(t|\{\sigma\},\{\phi\}))\right],
\end{equation}
where $h(t|\{\sigma\},\{\phi\})$ is given by~\eqref{eq:h_effective}, and $\pi_0(\sigma(0)) = \frac{1}{2}[1 + \sigma(0)m_0]$ is the initial condition.
 
This is the central result: the effective single-neuron problem involves an effective local field
\begin{equation}\label{eq:h_linear_response}
    h(t|\{\sigma\},\{\phi\}) = \frac{m(t)}{1+\lambda}\left(\eta^t + \lambda\right)\xi^{t+1} + \theta(t) + \alpha\sum_{t'<t} R(t,t')\sigma(t') + \sqrt{\alpha}\phi(t),
\end{equation}
where the retarded self-interaction kernel is $R(t,t') = \frac{1}{\alpha}\hat{K}(t,t')$, $\phi(t)$ is correlated Gaussian noise with covariance $\langle\phi(t)\phi(t')\rangle = \bm{V}(t,t') = \frac{2}{\alpha}\cdot\frac{1}{i}\hat{Q}(t,t')$.
The key observation is that the signal term $\frac{m(t)}{1+\lambda}(\eta^t + \lambda)\xi^{t+1}$ depends on the random sign $\eta^t$:
\begin{itemize}
    \item When $\eta^t = +1$ (subsequent patterns aligned): signal strength is $m(t)$
    \item When $\eta^t = -1$ (subsequent patterns anti-aligned): signal strength is $\frac{\lambda-1}{\lambda+1}m(t)$
\end{itemize}
This splitting is crucial for finite $\lambda$.

From our results for $\hat{K}$~\eqref{eq:Khat_result} and $\hat{Q}$~\eqref{eq:Qhat_general}, we identify two important limiting cases for the retarded self-interaction $R(t,t')$ and noise covariance $\langle\phi(t)\phi(t')\rangle$:
 
For the \textbf{purely asymmetric case} ($\lambda \to \infty$):
\begin{align}
    R(t,t') &= 0 \quad \text{(no retarded self-interaction)}, \\
    \langle\phi(t)\phi(t')\rangle &= \sum_{n\geq 0}\left[(\bm{G}^{\dagger})^n \bm{C}\, \bm{G}^n\right](t,t'),
\end{align}
which matches the known results.
 
For the \textbf{symmetric Hopfield case} ($\lambda = 0$):
\begin{align}
    R(t,t') &= \left[\bm{G}(\bm{1}-\bm{G})^{-1}\right](t,t'), \\
    \langle\phi(t)\phi(t')\rangle &= \left[(\bm{1}-\bm{G})^{-1}\bm{C}(\bm{1}-\bm{G}^{\dagger})^{-1}\right](t,t').
\end{align}

For \textbf{finite $\lambda$} (with the diagonal of $J$, i.e. the self-interactions removed, as in our simulations):
\begin{align}
    R(t,t') &= \sum_{n\geq 1}\frac{\bm{G}^n(t,t')}{(\lambda+1)^{n+1}}
    \;=\; \left[\frac{\bm{G}}{(\lambda+1)^2}\left(\bm{1} - \frac{\bm{G}}{\lambda+1}\right)^{-1}\right](t,t'),
    \label{eq:R_finite_lambda}\\
    \langle\phi(t)\phi(t')\rangle &= \int_{-\pi}^{\pi}\frac{\dd\omega}{2\pi}\;|a(\omega)|^2
    \left[\big(\bm{1} - a(\omega)\bm{G}\big)^{-1}\bm{C}\,\big(\bm{1} - a^*(\omega)\bm{G}^{\dagger}\big)^{-1}\right](t,t'),
    \qquad a(\omega) = \frac{1+\lambda e^{-\ii\omega}}{1+\lambda}.
    \label{eq:V_spectral}
\end{align}
Both expressions follow compactly from the pattern-mode representation of the separable
coupling $\bm\xi^{\rm T} A\, \bm\xi/N$ with the normal pattern-space matrix
$A = (\bm 1 + \lambda S)/(1+\lambda)$ ($S$ the cyclic shift, eigenvalues $a(\omega)$):
the kernel resums the self-feedback through the pattern modes,
$R = \sum_{n} \langle a^{n+1}\rangle_A\, \bm G^{n}$ with moments
$\langle a^{n+1}\rangle_A = (\lambda+1)^{-(n+1)}$ [consistent with the contour integrals
evaluated above], while the noise resums the response-dressed pattern modes. They
interpolate the two limits: $a \to e^{-\ii\omega}$ projects the noise onto the
equal-power terms and sends $R\to0$ (purely asymmetric case), while $a \to 1$ gives the
full double resolvent and $R \to \bm G(\bm 1-\bm G)^{-1}$ (symmetric case). If the
diagonal of $J$ is retained in the model, $R$ acquires the additional equal-time term
$(\lambda+1)^{-1}\delta_{t,t'}$ (the $n=0$ moment). These expressions are validated
against direct Monte Carlo simulations across the full range of $\lambda$ in
Fig.~\ref{fig:alpha_c_vs_lambda}.

\paragraph{Self-consistency equations.}
The order parameters $m(t)$, $C(t,t')$, and $G(t,t')$ are determined self-consistently from the effective single-neuron problem~\eqref{eq:P_sigma}. At zero temperature, $\sigma(t+1) = \text{sign}(h(t))$, so:
\begin{align}
    m(t+1) &= \frac{1}{N}\sum_i \xi_i^{t+1}\sigma_i(t+1) = \left\langle \xi^{t+1} \,\text{sign}(h(t)) \right\rangle_{\eta,\xi,\phi,\sigma_{\text{past}}}, \\
    C(t,t') &= \langle\sigma(t)\sigma(t')\rangle_\star, \\
    G(t,t') &= \frac{\partial\langle\sigma(t)\rangle_\star}{\partial\theta(t')}.
\end{align}
 
Using the effective field~\eqref{eq:h_linear_response} and noting that $\xi^{t+1}\,\text{sign}(\xi^{t+1} A + B) = \text{sign}(A + \xi^{t+1}B)$ for $\xi^{t+1} = \pm 1$:
\begin{equation}
    m(t+1) = \left\langle \text{sign}\left(\frac{m(t)(\eta^t + \lambda)}{1+\lambda} + \tilde{\phi}(t)\right) \right\rangle_{\eta^t, \tilde{\phi}},
\end{equation}
where $\tilde{\phi}$ absorbs the noise and retarded terms (which are independent of $\xi^{t+1}$).
 
Averaging over $\eta^t = \pm 1$ with equal probability:
\begin{equation}\label{eq:m_self_consistency_finite_lambda}
    \boxed{m(t+1) = \frac{1}{2}\left\langle \text{sign}\left(m(t) + \tilde{\phi}\right) \right\rangle + \frac{1}{2}\left\langle \text{sign}\left(\frac{(\lambda-1)m(t)}{\lambda+1} + \tilde{\phi}\right) \right\rangle}
\end{equation}
This is the central modification for finite $\lambda$: the self-consistency equation splits into two terms corresponding to aligned ($\eta = +1$) and anti-aligned ($\eta = -1$) consecutive patterns.

\begin{figure}[hbt!]
    \centering
    \includegraphics[]{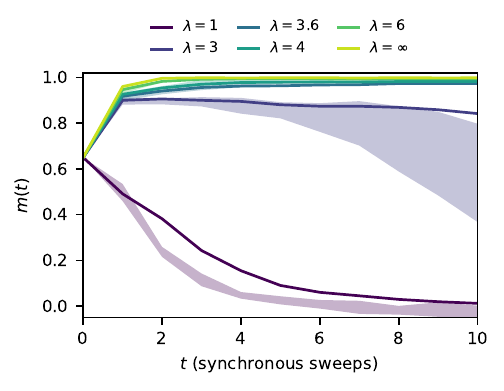}
    \caption{
        {\bf The numerical iteration of the DMFT equations.}
        Time evolution of the dynamical overlap $m(t)$ for different symmetry parameters
        $\lambda$, at zero temperature and load $\alpha = 0.1$, starting from
        $m_0 = 0.65$. Solid lines: numerically exact DMFT iteration of the effective
        single-neuron problem with the retarded kernel \eqref{eq:R_finite_lambda} and
        the spectral noise covariance \eqref{eq:V_spectral}, capturing the full
        non-Gaussian dynamics. Shaded bands: Monte Carlo simulations of the full
        microscopic network ($N = 3000$; mean $\pm$ one standard deviation over five
        independent disorder realisations). For $\lambda \gtrsim 3.6$ the load lies
        below the critical capacity and the sequence is retrieved ($m \approx 1$);
        at $\lambda = 3$ the load sits marginally above $\alpha_c(3) \approx 0.10$ and
        the overlap decays slowly beyond the window shown, with large
        sample-to-sample fluctuations; at $\lambda = 1$ retrieval fails within a few
        sweeps.
    }\label{fig:finite_lambda_dmft_iteration}
\end{figure}

\begin{figure}
    \centering
    \includegraphics[]{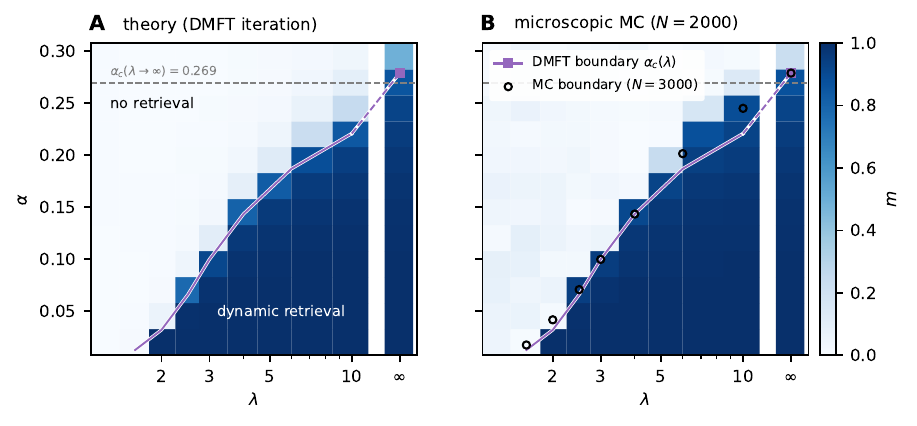}
    \caption{
        {\bf The synchronous phase diagram of the single-species sequence network at
        finite $\lambda$: theory versus microscopic dynamics.}
        The two panels compare the effective single-neuron theory (left) with the true
        many-body dynamics of the microscopic network (right) on the same grid.
        {\bf A.} Final sequence overlap $m$ in the $(\lambda, \alpha)$ plane at zero
        temperature, from the numerically exact DMFT iteration of the effective
        single-neuron problem. The dynamic-retrieval phase (blue) is bounded
        by the critical capacity $\alpha_c(\lambda)$ (purple line); retrieval deteriorates rapidly as
        $\lambda \to 1$, where the symmetric channel traps the network in mixtures of
        consecutive patterns.
        {\bf B.} The corresponding microscopic many-body phase diagram: the same observable from
        Monte Carlo simulations of the full network ($N = 2000$, diagonal of $J$
        removed) on the identical grid, with the DMFT boundary of panel A overlaid and
        the MC bisection estimates of $\alpha_c(\lambda)$ ($N = 3000$; open circles).
    }
    \label{fig:alpha_c_vs_lambda}
\end{figure}

\paragraph{Stationary state.}
In the stationary state we assume time-translation invariance, $m(t) = m$ and
$C(t,t') = C(\tau)$, with the one-step response
$G(\tau) = \beta(1-\tilde{q})\,\delta_{\tau,1}$, where
$\tilde{q} = \lim_{\tau\to\infty}C(\tau)$ is the Edwards--Anderson parameter; at zero
temperature $\beta(1-\tilde q) \to \gamma$ remains finite. In the purely asymmetric
limit the noise variance then resums to $\langle\phi^2\rangle = (1-\tilde q)\,\rho$ with
$\rho = [1-\beta^2(1-\tilde q)^2]^{-1}$. At finite $\lambda$, the retarded kernel
couples the field to the spin's own past: the effective field is no longer Gaussian,
and the one-step response ansatz itself becomes uncontrolled. The quantitative results
of Figs.~\ref{fig:finite_lambda_dmft_iteration} and \ref{fig:alpha_c_vs_lambda} are
therefore obtained from the exact iteration above; the closed stationary equations
below remain instructive as a rough analytical picture.

\paragraph{Numerically exact iterative solution.}
The DMFT equations can be solved by iterating forward in time. The key observation is that causality ensures $G(t,t') = 0$ for $t \leq t'$, which means:
\begin{itemize}
    \item The retarded kernel $R(t,t') = \sum_{\ell=1}^{t-t'} (\lambda+1)^{-(\ell+1)}(G^\ell)(t,t')$ involves only a \emph{finite} sum.
    \item At each time step $t$, the quantities $m(t)$, $C(t,t')$, and $G(t,t')$ depend only on previously computed values at times $t' < t$.
\end{itemize}
Starting from initial conditions $m(0) = m_0$ and $C(0,0) = 1$, one computes the full time-dependent order parameters by sampling paths from the effective single-neuron problem~\eqref{eq:P_sigma} at each time step. This approach tracks the full non-Gaussian structure of the local fields and is exact up to numerical precision.
 
In practice the noise is drawn with the covariance
\eqref{eq:V_spectral} (evaluated by quadrature over $\omega$), and the response matrix
is estimated by linear-response probes: a small field $\pm\varepsilon$ is applied at a
single time $t'$ with common random numbers, and
$G(t,t') = [\langle\sigma(t)\rangle_{+\varepsilon} -
\langle\sigma(t)\rangle_{-\varepsilon}]/2\varepsilon$. Alternatively, the linear response can be taken
in closed form: at zero temperature the single-step susceptibility is the field density
at the origin, $G(t{+}1,t) = \langle \delta(h(t)) \rangle$, averaged over the two signal
branches $\eta_t = \pm 1$, which is noise-free and avoids the linear-response sampling
used for the transient dynamics.
The iteration reproduces the
exact scalar recursion of the purely asymmetric limit (including the critical capacity
$\alpha_c = 0.269$) and tracks Monte Carlo simulations of the microscopic model across
the full range of $\lambda$, both for the transient overlap dynamics and for the phase
boundary $\alpha_c(\lambda)$ (Fig.~\ref{fig:alpha_c_vs_lambda}). The agreement is
quantitative for $\lambda \lesssim 4$ and in the $\lambda \to \infty$ limit; in the
intermediate window $\lambda \sim 6$--$10$ the single-condensate closure underestimates
$\alpha_c$ by $\sim 10\%$.
 
The number of order parameters grows as $\mathcal{O}(t^2)$ (the correlation and response matrices), which is computationally manageable for moderate times. This iterative approach is well-suited for studying transient dynamics and verifying the accuracy of approximate stationary-state solutions.

\paragraph{Gaussian ansatz to approximate the stationary state.}
For analytical tractability in the stationary state, we make a Gaussian ansatz: we approximate the non-Gaussian retarded self-interaction by absorbing its effect into a renormalized noise variance. The idea is to complete the square over the spin-field coupling term $\hat{K}\sigma\hat{h}$ in the effective measure~\eqref{eq:M_general}.
This closure replaces the full spectral covariance
\eqref{eq:V_spectral} by a single renormalised variance and should be regarded as a
rough analytical estimate: it captures the qualitative shape of the phase boundary but
substantially underestimates $\alpha_c$ at moderate $\lambda$. 
 
Completing the square over the $\hat K\sigma\hat h$ coupling and
treating the induced spin--spin term as additional Gaussian noise renormalises
$\rho \to \rho_{\rm eff} = \rho\,[1-\kappa\,\beta(1-\tilde q)]^{-2}$, with the geometric
factor $\kappa = \sum_{n}[\beta(1-\tilde q)]^{n}(\lambda+1)^{-(n+1)}$ inherited from the
kernel \eqref{eq:R_finite_lambda} (whether the sum starts at $n=0$ or $n=1$ depends on
the diagonal convention chosen for the self-interactions in $J$ matrix).

In the TTI stationary state with Gaussian noise of variance $\alpha\rho_{\mathrm{eff}}$, the self-consistency equation~\eqref{eq:m_self_consistency_finite_lambda} becomes:
\begin{equation}\label{eq:m_TTI_finite_lambda}
    m = \frac{1}{2}\mathrm{erf}\left(\frac{m}{\sqrt{2\alpha\rho_{\mathrm{eff}}}}\right) + \frac{1}{2}\mathrm{erf}\left(\frac{(\lambda-1)m}{(\lambda+1)\sqrt{2\alpha\rho_{\mathrm{eff}}}}\right).
\end{equation}
The two terms correspond to averaging over the random sign $\eta = \xi^t\xi^{t+1} = \pm 1$.
In the limit $\lambda \to \infty$, both terms contribute equally with argument $m/\sqrt{2\alpha\rho_{\mathrm{eff}}}$, recovering the standard result $m = \mathrm{erf}(m/\sqrt{2\alpha\rho_{\mathrm{eff}}})$.
For finite $\lambda$, the second term (corresponding to anti-aligned consecutive patterns, $\eta = -1$) has a reduced signal $(\lambda-1)/(\lambda+1) < 1$. This competition between the two cases reduces the effective signal strength and lowers the critical capacity.

At zero temperature ($\beta \to \infty$, $\tilde q \to 1$ with
$\beta(1-\tilde q) \to \gamma$ determined self-consistently), the closure gives
\begin{equation}\label{eq:rho_eff_T0}
    \rho_{\mathrm{eff}}(\lambda) = \frac{1}{1 - \gamma^2} \cdot \left(\frac{(\lambda+1) - \gamma}{(\lambda+1) - 2\gamma}\right)^{2}.
\end{equation}
Note that the factor $((\lambda+1) - \gamma)^2/((\lambda+1) - 2\gamma)^2$ captures the finite-$\lambda$ correction to the $\lambda \to \infty$ result. For large $\lambda$, this factor approaches 1.
 
Within the Gaussian ansatz, the $T=0$ self-consistency relations use the modified equation~\eqref{eq:m_TTI_finite_lambda} with the renormalized $\rho_{\mathrm{eff}}$ from~\eqref{eq:rho_eff_T0}:
\begin{equation}
    \gamma = \frac{1}{2}\sqrt{\frac{2}{\pi\alpha\rho_{\mathrm{eff}}}} \left[ e^{-m^2/(2\alpha\rho_{\mathrm{eff}})} + e^{-(\lambda-1)^2 m^2/((\lambda+1)^2\, 2\alpha\rho_{\mathrm{eff}})} \right]; \quad m = \frac{1}{2}\mathrm{erf}\left(\frac{m}{\sqrt{2\alpha\rho_{\mathrm{eff}}}}\right) + \frac{1}{2}\mathrm{erf}\left(\frac{(\lambda-1)m}{(\lambda+1)\sqrt{2\alpha\rho_{\mathrm{eff}}}}\right).
\end{equation}
The response, like the overlap, averages the two signal branches
$\eta = \pm1$; keeping only the $\eta = +1$ exponent is exact at $m=0$ or
$\lambda\to\infty$ but biases the finite-$\lambda$ capacity.
In the limit $\lambda \to \infty$, this reduces to $m = \mathrm{erf}(m/\sqrt{2\alpha\rho_{\mathrm{eff}}})$, $\gamma = \sqrt{2/(\pi\alpha\rho_{\rm eff})}\,e^{-m^2/(2\alpha\rho_{\rm eff})}$.

Solving this closed system for given $\lambda > 1$, the critical
capacity $\alpha_c(\lambda)$ can be estimated as the load at which a nontrivial
fixed point ceases to exist.
As $\lambda \to \infty$, we recover the purely asymmetric result $\rho = (1-\gamma^2)^{-1}$. The leading correction is $\mathcal{O}(1/\lambda)$, indicating that finite-$\lambda$ effects become significant when $\lambda \sim \gamma$.

The critical capacity $\alpha_c(\lambda)$ can be determined numerically by solving the self-consistent equations~\eqref{eq:rho_eff_T0} and the $T=0$ relations. As $\lambda$ decreases from infinity toward 1, $\alpha_c$ decreases monotonically from its asymptotic value $\alpha_c(\infty) \approx 0.269$, vanishing at $\lambda = 1$.

\paragraph{Estimating critical capacity via linearization.}
The critical capacity can be estimated analytically by linearizing the self-consistency equation~\eqref{eq:m_TTI_finite_lambda} around $m = 0$. Using $\mathrm{erf}(x) \approx \frac{2x}{\sqrt{\pi}}$ for small $x$:
\begin{equation}
    m \approx \frac{1}{2}\cdot\frac{2m}{\sqrt{2\pi\alpha\rho_{\mathrm{eff}}}} + \frac{1}{2}\cdot\frac{2(\lambda-1)m}{(\lambda+1)\sqrt{2\pi\alpha\rho_{\mathrm{eff}}}} = \frac{m}{\sqrt{2\pi\alpha\rho_{\mathrm{eff}}}}\left[1 + \frac{\lambda-1}{\lambda+1}\right] = \frac{2\lambda m}{(\lambda+1)\sqrt{2\pi\alpha\rho_{\mathrm{eff}}}}.
\end{equation}
For a non-trivial solution ($m \neq 0$) to exist, we require:
\begin{equation}
    \frac{2\lambda}{(\lambda+1)\sqrt{2\pi\alpha\rho_{\mathrm{eff}}}} \geq 1 \quad \Rightarrow \quad \sqrt{2\pi\alpha\rho_{\mathrm{eff}}} \leq \frac{2\lambda}{\lambda+1}.
\end{equation}
 
In the purely asymmetric limit $\lambda \to \infty$, this gives $\sqrt{2\pi\alpha_c\rho_c} = 2$, consistent with the known result. For finite $\lambda$:
\begin{equation}
    \sqrt{2\pi\alpha_c(\lambda)\rho_c(\lambda)} = \frac{2\lambda}{\lambda+1}.
\end{equation}
 
Since the response function $\gamma_0 = \sqrt{2/(\pi\alpha\rho_{\mathrm{eff}})}$ is unchanged at $m=0$, the critical $\rho_c$ is approximately independent of $\lambda$. Therefore:
\begin{equation}\label{eq:alpha_c_scaling}
    \boxed{\frac{\alpha_c(\lambda)}{\alpha_c(\infty)} = \left(\frac{\lambda}{\lambda+1}\right)^2}
\end{equation}
This predicts a significant reduction in critical capacity at finite $\lambda$.
While this estimate captures the origin of the reduction (the incoherent
$\eta$-split signal), it neglects the growth of the effective noise towards
$\lambda \to 1$ and the breakdown of the one-step response ansatz, and accordingly overestimates $\alpha_c$ at small $\lambda$ relative to the exact iteration and the microscopic simulations.

\subsection{Synchronous Two-Species Bipartite Model}
\label{sec:two_species}

We now consider a bipartite network with two species of neurons: visible neurons $\bm{x}^1 = (x_1^1, \ldots, x_N^1) \in \{-1,+1\}^N$ and hidden neurons $\bm{x}^2 = (x_1^2, \ldots, x_N^2) \in \{-1,+1\}^N$. The hidden layer provides an additional pathway for sequence encoding.


The local fields are
\begin{align}
    h_i^1(s) &= \frac{1}{1+\lambda} \sum_j J_{ij}^{\mathrm{S}} x_j^1(s) + \frac{\lambda}{1+\lambda} x_i^2(s) + \theta_i^1(s), \label{eq:h1_def}\\
    h_i^2(s) &= \sum_j J_{ij}^{\mathrm{T}} x_j^1(s) + \theta_i^2(s), \label{eq:h2_def}
\end{align}
where the synaptic matrices are:
\begin{align}
    \bm{J}^{\mathrm{T}} &= \frac{1}{N}\bm{\xi}^{\mathrm{T}}\bm{S}\bm{\xi}, \\
    \bm{J}^{\mathrm{S}} &= \frac{1}{N}\bm{\xi}^{\mathrm{T}}\bm{\xi}
\end{align}
where $\bm{\xi}$ is a $p \times N$ matrix with entries $\xi_{\mu i} = \xi_i^\mu$ and $\bm{S}$ is the cyclic shift matrix.

We consider the case where both species update synchronously according to
\begin{equation}\label{eq:W_2bit}
    W[\bm{x}(s+1)|\bm{x}(s)] = \prod_{i=1}^N \prod_{a=1}^2 \frac{e^{\beta x_i^a(s+1) h_i^a(s)}}{2\cosh[\beta h_i^a(s)]}.
\end{equation}
We develop the full path integral DMFT for this model by further extending the technique from the previous sections.


The generating functional for trajectories of both species is
\begin{equation}\label{eq:Z_2bit}
    Z[\bm{\psi}^1, \bm{\psi}^2] = \sum_{\{\bm{x}^1(s), \bm{x}^2(s)\}} p[\bm{x}(0)] \prod_{s=0}^{t-1} W[\bm{x}(s+1)|\bm{x}(s)] \, e^{-\ii\sum_{s<t}\sum_{a=1}^2\bm{x}^a(s)\cdot\bm{\psi}^a(s)}
\end{equation}
where $p[\bm{x}(0)]$ is the initial distribution and the source fields $\bm{\psi}^a(s)$ generate correlation functions:
\begin{align}
    \langle x_i^a(s) \rangle &= \ii\frac{\partial \ln Z}{\partial \psi_i^a(s)}\bigg|_{\psi=0}, \\
    \langle x_i^a(s) x_j^b(s') \rangle &= -\frac{\partial^2 \ln Z}{\partial \psi_i^a(s) \partial \psi_j^b(s')}\bigg|_{\psi=0}.
\end{align}


To decouple the synaptic interactions, we introduce auxiliary fields for each species:
\begin{align}
    1 &= \int \prod_{i,s} \frac{\dd h_i^1(s)\,\dd\hat{h}_i^1(s)}{2\pi} \exp\left[\ii\sum_{i,s} \hat{h}_i^1(s)\left(h_i^1(s) - \frac{1}{1+\lambda}\sum_j J_{ij}^{\mathrm{S}} x_j^1(s) - \frac{\lambda}{1+\lambda} x_i^2(s) - \theta_i^1(s)\right)\right],\\
    1 &= \int \prod_{i,s} \frac{\dd h_i^2(s)\,\dd\hat{h}_i^2(s)}{2\pi} \exp\left[\ii\sum_{i,s} \hat{h}_i^2(s)\left(h_i^2(s) - \sum_j J_{ij}^{\mathrm{T}} x_j^1(s) - \theta_i^2(s)\right)\right].
\end{align}

Inserting these into~\eqref{eq:Z_2bit} and using~\eqref{eq:W_2bit}:
\begin{align}\label{eq:Z_2bit_aux}
    Z = \sum_{\{x^1, x^2\}} p[x(0)] &\int \{\dd h^1\,\dd\hat{h}^1\,\dd h^2\,\dd\hat{h}^2\} \nonumber\\
    &\times \prod_{s<t} e^{\beta\bm{x}^1(s+1)\cdot\bm{h}^1(s) - \sum_i \ln 2\cosh(\beta h_i^1(s))} \nonumber\\
    &\times e^{\beta\bm{x}^2(s+1)\cdot\bm{h}^2(s) - \sum_i \ln 2\cosh(\beta h_i^2(s))} \nonumber\\
    &\times e^{\ii\hat{\bm{h}}^1(s)\cdot[\bm{h}^1(s) - \bm{\theta}^1(s)] + \ii\hat{\bm{h}}^2(s)\cdot[\bm{h}^2(s) - \bm{\theta}^2(s)]} \nonumber\\
    &\times e^{-\ii\bm{\psi}^1(s)\cdot\bm{x}^1(s) - \ii\bm{\psi}^2(s)\cdot\bm{x}^2(s)} \nonumber\\
    &\times e^{-\ii N^{-1}\frac{1}{1+\lambda}\hat{\bm{h}}^1(s)\cdot(\bm{\xi}^{\mathrm{T}}\bm{\xi})\bm{x}^1(s) - \ii\frac{\lambda}{1+\lambda}\hat{\bm{h}}^1(s)\cdot\bm{x}^2(s)} \nonumber\\
    &\times e^{-\ii N^{-1}\hat{\bm{h}}^2(s)\cdot(\bm{\xi}^{\mathrm{T}}\bm{S}\bm{\xi})\bm{x}^1(s)}.
\end{align}

Under successful sequence retrieval, the system is expected to have one condensed pattern $\mu$ in each subsystem. For example, for time $t=0, 1, 2, 3, 4, \dots$, the neurons $\bm{x}^1$ will have $\mathcal{O}(1)$ overlap with patterns $\bm{\xi}^0, \bm{\xi}^1, \bm{\xi}^1, \bm{\xi}^2, \bm{\xi}^2, \dots$. For convenience we reindex time using $\tau$ to track odd ($t=2\tau -1$) and even ($t=2\tau$) time steps. The condensed overlaps then become:
\begin{align}
    a(\tau) &= \frac{1}{N} \bm{\xi}^\tau \cdot \bm{x}^1(2\tau-1), \label{eq:a_def}\\
    b(\tau) &= \frac{1}{N} \bm{\xi}^\tau \cdot \bm{x}^1(2\tau). \label{eq:b_def}
\end{align}
Likewise, for the auxiliary fields, the relevant condensed overlaps are
\begin{align}
    l(\tau) &= \frac{1}{N} \bm{\xi}^\tau \cdot \hat{\bm{h}}^1(2\tau), \label{eq:l_def}\\
    p(\tau) &= \frac{1}{N} \bm{\xi}^{\tau} \cdot \hat{\bm{h}}^1(2\tau-1), \label{eq:r_def}\\
    k(\tau) &= \frac{1}{N} \bm{\xi}^{\tau+1} \cdot \hat{\bm{h}}^2(2\tau) \label{eq:k_def}\\
    r(\tau) &= \frac{1}{N} \bm{\xi}^{\tau+1} \cdot \hat{\bm{h}}^2(2\tau-1). \label{eq:p_def}
\end{align}

The non-condensed overlaps (crosstalk) can be defined similarly. For $\mu \neq \tau:=\left\lceil s/2 \right\rceil$, we have:
\begin{align}
    y_\mu(s) &=\frac{1}{\sqrt{N}}\begin{cases}
          \bm{\xi}^\mu \cdot \bm{x}^1(s=2\tau) \\
          \bm{\xi}^\mu \cdot \bm{x}^1(s=2\tau-1)
    \end{cases}, \\
    w_\mu^1(s) &=\frac{1}{\sqrt{N}}\begin{cases} 
         \bm{\xi}^\mu \cdot \hat{\bm{h}}^1(s=2\tau) \\
         \bm{\xi}^\mu \cdot \hat{\bm{h}}^1(s=2\tau-1)
    \end{cases}, \\
    w_\mu^2(s) &= \frac{1}{\sqrt{N}} \begin{cases}
         \bm{\xi}^{\mu+1} \cdot \hat{\bm{h}}^2(s=2\tau) \\
         \bm{\xi}^{\mu+1} \cdot \hat{\bm{h}}^2(s=2\tau-1)
        \end{cases}.
\end{align}
The non-condensed overlaps (crosstalk) are defined via the Fourier integrals:
\begin{align*}
    1 &= \int \prod_{\mu \neq \left\lceil s/2 \right\rceil} \frac{\dd y_\mu(s)\,\dd\hat{y}_\mu(s)}{2\pi} \exp\left[\ii \hat{y}_\mu(s)\left(\sqrt{N} y_\mu(s) - \bm{\xi}^\mu \cdot \bm{x}^1(s)\right)\right], \\
    1 &= \int \prod_{\mu \neq \left\lceil s/2 \right\rceil} \frac{\dd w_\mu^1(s)\,\dd\hat{w}_\mu^1(s)}{2\pi} \exp\left[\ii \hat{w}_\mu^1(s)\left(\sqrt{N} w_\mu^1(s) - \bm{\xi}^\mu \cdot \hat{\bm{h}}^1(s)\right)\right], \\
    1 &= \int \prod_{\mu \neq \left\lceil s/2 \right\rceil} \frac{\dd w_\mu^2(s)\,\dd\hat{w}_\mu^2(s)}{2\pi} \exp\left[\ii \hat{w}_\mu^2(s)\left(\sqrt{N} w_\mu^2(s) - \bm{\xi}^{\mu+1} \cdot \hat{\bm{h}}^2(s)\right)\right].
\end{align*}

The disorder-dependent terms in~\eqref{eq:Z_2bit_aux} separate into condensed and non-condensed contributions:
\begin{align}
    \frac{1}{N}\sum_{s<t} \hat{\bm{h}}^2(s)\cdot(\bm{\xi}^{\mathrm{T}}\bm{S}\bm{\xi})\bm{x}^1(s) &= \frac{1}{N}\sum_{s<t,\mu} [\hat{\bm{h}}^2(s)\cdot\bm{\xi}^{\mu+1}][\bm{x}^1(s)\cdot\bm{\xi}^\mu] = \sum_{\tau<t} \left[k(\tau) b(\tau) + r(\tau) a(\tau)\right] + \frac{1}{\sqrt{N}}\sum_{\substack{s<t,\\\mu \neq \left\lceil s/2 \right\rceil}} w_\mu^{2}(s) y_\mu(s),\\ 
    \frac{1}{N}\sum_{s<t} \hat{\bm{h}}^1(s)\cdot(\bm{\xi}^{\mathrm{T}}\bm{\xi})\bm{x}^1(s) &= \frac{1}{N}\sum_{s<t, \mu} [\hat{\bm{h}}^1(s)\cdot\bm{\xi}^\mu][\bm{x}^1(s)\cdot\bm{\xi}^\mu] = \sum_{\tau<t} \left[l(\tau) b(\tau) + p(\tau) a(\tau)\right] + \frac{1}{\sqrt{N}}\sum_{\substack{s<t,\\\mu \neq \left\lceil s/2 \right\rceil}} w_\mu^1(s) y_\mu(s).
\end{align}
The visible field~\eqref{eq:h1_def} carries $\bm{J}^{\mathrm S}$ with weight $1/(1+\lambda)$, so the
$\hat{\bm h}^1$--$\bm x^1$ interaction is really
$\tfrac{1}{1+\lambda}\tfrac1N\hat{\bm h}^1(s)\!\cdot\!(\bm\xi^{\mathrm T}\bm\xi)\bm x^1(s)$, i.e.\ the
visible--visible crosstalk channel enters with prefactor $(1+\lambda)^{-1}$~\footnote{
This factor propagates to the visible noise covariance as a
$(1+\lambda)^{-2}$ weight on the symmetric contribution,
${R^{11}\big|_{\mathrm{sym}}=(1+\lambda)^{-2}\,[(\bm 1-\tfrac{1}{1+\lambda}\bm G^{11})^{-1}\bm C^{11}(\bm 1-\tfrac{1}{1+\lambda}\bm G^{11})^{-\dagger}]}$,
and is what makes the visible self-crosstalk (and its Onsager term $\hat K^{11}$) fade as $\lambda$ grows.}.
The disorder average for the two-species model follows the same procedure as before, but with additional structure:
\begin{align}
    \overline{\exp\left[-\frac{\ii}{\sqrt{N}} \bm{\xi}^\mu \cdot \bm{\zeta}^\mu \right]} \exp{\mathcal{O}(N^{1/2})}
\end{align}
where for convenience, we define the ``charge'' $\zeta_i^\mu$ coupled to pattern $\mu$ at site $i$:
\begin{align}\label{eq:Z_2bit_disordercharge}
    \zeta_i^\mu = \sum_{s<t}\Big[&\hat{w}_\mu^1(s)\hat{h}_i^1(s) + \hat{w}_{\mu-1}^2(s)\hat{h}_i^2(s) + \hat{y}_\mu(s)x_i^1(s)\Big].
\end{align}

Following the same steps as Section~\ref{sec:disorder_average}, the disorder average over non-condensed pattern components $\xi_i^{\mu>t}$ gives:
\begin{equation*}
    \overline{\exp\left[-\frac{\ii}{\sqrt{N}}\sum_i \xi_i^\mu \zeta_i^\mu\right]}_\xi = \prod_i \cos\left(\frac{\zeta_i^\mu}{\sqrt{N}}\right) = \exp\left[-\frac{1}{2N}\sum_i (\zeta_i^\mu)^2 \right],
\end{equation*}
which completes the disorder average:
\begin{align}
    \exp\Bigg\{-\frac{1}{2}\sum_{\mu>t}\sum_{s,s'<t}\Big[&\hat{w}_\mu^{1}(s)\hat{w}_\mu^{1}(s')Q^{11}(s,s') + \hat{w}_{\mu-1}^{2}(s)\hat{w}_{\mu-1}^{2}(s')Q^{22}(s,s') + 2\hat{w}_\mu^{1}(s)\hat{w}_{\mu-1}^{2}(s')Q^{12}(s,s') \nonumber\\
    &+ \hat{y}_\mu(s)\hat{y}_\mu(s')C^{11}(s,s') + 2\hat{w}_{\mu}^{1}(s)\hat{y}_\mu(s')K^{11}(s,s') + 2\hat{w}_{\mu-1}^{2}(s)\hat{y}_\mu(s')K^{21}(s,s')\Big]\Bigg\}.
\end{align}
where
\begin{align}
    C^{11}(s,s') &= \frac{1}{N} \bm{x}^1(s) \cdot \bm{x}^1(s'), \\
    Q^{ab}(s,s') &= \frac{1}{N}\hat{\bm{h}}^a(s) \cdot \hat{\bm{h}}^b(s'), \\
    K^{a1}(s,s') &= \frac{1}{N}\hat{\bm{h}}^a(s') \cdot \bm{x}^1(s) .
\end{align}

We insert delta functions to define macroscopic two-time order parameters, using Fourier representations of the identity with conjugate fields $\hat{C}^{ab}$, $\hat{Q}^{ab}$, $\hat{K}^{ab}$:
\begin{align}
    1 &= \int \dd Q^{ab}(s,s')\,\dd\hat{Q}^{ab}(s,s') \exp\left[\ii N\hat{Q}^{ab}(s,s')\left(Q^{ab}(s,s') - \frac{1}{N}\hat{\bm{h}}^a(s) \cdot \hat{\bm{h}}^b(s')\right)\right], \quad a,b=1,2,\\
    1 &= \int \dd K^{a1}(s,s')\,\dd\hat{K}^{a1}(s,s') \exp\left[\ii N\hat{K}^{a1}(s,s')\left(K^{a1}(s,s') - \frac{1}{N}\hat{\bm{h}}^a(s') \cdot \bm{x}^1(s)\right)\right], \quad a=1,2,\\
    1 &= \int \dd C^{11}(s,s')\,\dd\hat{C}^{11}(s,s') \exp\left[\ii N\hat{C}^{11}(s,s')\left(C^{11}(s,s') - \frac{1}{N}\bm{x}^1(s) \cdot \bm{x}^1(s')\right)\right].
\end{align}

We now isolate the relevant single-time macroscopic observables using Fourier representations:
\begin{align*}
    1 &= \int \frac{\dd a(\tau)\,\dd\hat{a}(\tau)}{(2\pi/N)^{t/2}} \exp\left[\ii \sum_{\tau < t/2} \hat{a}(\tau)\left(N a(\tau) - \bm{\xi}^\tau \cdot \bm{x}^1(2\tau-1)\right)\right], \\
    1 &= \int \frac{\dd b(\tau)\,\dd\hat{b}(\tau)}{(2\pi/N)^{t/2}} \exp\left[\ii \sum_{\tau < t/2} \hat{b}(\tau)\left(N b(\tau) - \bm{\xi}^\tau \cdot \bm{x}^1(2\tau)\right)\right], \quad \cdots
\end{align*}
and similarly for $l(\tau), k(\tau), p(\tau)$ and $r(\tau)$.

The generating functional takes the form:
\begin{equation}
    \overline{Z[\bm{\psi}]} = \int \{\dd\text{(order params)}\} \, e^{N[\Psi + \Phi + \Omega] + \mathcal{O}(N^{1/2})},
\end{equation}
where the functionals $\Psi$, $\Phi$, and $\Omega$ are defined below. The saddle-point equations are obtained by extremizing the exponent with respect to all order parameters and their conjugates.

The $\Psi$ functional enforcing order parameter definitions is:
\begin{align}
    \Psi = &\ii\sum_{s<t}\Big\{\hat{k}(s)k(s) + \hat{l}(s)l(s) + \hat{b}(s)b(s) + \hat{r}(s)r(s) + \hat{a}(s)a(s) +  \hat{p}(s)p(s)  \nonumber\\
    &\quad - \left[k(s) + l(s)\right]b(s) - \left[r(s) + p(s)\right]a(s)\Big\} \nonumber\\
    &+ \ii\sum_{s,s'<t}\Big\{\hat{C}^{11}(s,s')C^{11}(s,s')+\sum_{a}\left[\hat{K}^{a1}(s,s')K^{a1}(s,s') + \sum_{b}\hat{Q}^{ab}(s,s')Q^{ab}(s,s')\right]\Big\}.
\end{align}


The $\Phi$ functional which will determine the structure of the single-site measure is:
\begin{align}\label{eq:Phi_2species}
    \Phi = &\frac{1}{N}\sum_i \ln\Bigg\{ \sum_{\{x^1(s),x^2(s)\}} \int\{\dd h^1\,\dd\hat{h}^1\} \int\{\dd h^2\,\dd\hat{h}^2\} \, p_i(x^1(0), x^2(0)) \nonumber\\
    &\times \exp\left\{\sum_{s<t}[\beta x^1(s+1)h^1(s) - \ln 2\cosh(\beta h^1(s)) + \beta x^2(s+1)h^2(s) - \ln 2\cosh(\beta h^2(s))]\right\} \nonumber\\
    &\times \exp\left\{\ii\sum_{\tau<t/2}\hat{h}^1(2\tau)\left[h^1(2\tau) - \theta_i^1(2\tau) - \frac{1}{1+\lambda}\hat{l}(\tau)\xi_i^{\tau} + \frac{\lambda}{1+\lambda} x^2(2\tau)\right]\right\} \nonumber\\
    &\times \exp\left\{\ii\sum_{\tau<t/2}\hat{h}^1(2\tau-1)\left[h^1(2\tau-1) - \theta_i^1(2\tau-1) - \frac{1}{1+\lambda}\hat{p}(\tau)\xi_i^{\tau} + \frac{\lambda}{1+\lambda} x^2(2\tau-1)\right]\right\} \nonumber\\
    &\times \exp\left\{\ii\sum_{\tau<t/2}\hat{h}^2(2\tau)[h^2(2\tau) - \theta_i^2(2\tau) - \hat{k}(\tau)\xi_i^{\tau+1}]\right\} \nonumber\\
    &\times \exp\left\{\ii\sum_{\tau<t/2}\hat{h}^2(2\tau-1)[h^2(2\tau-1) - \theta_i^2(2\tau-1) - \hat{r}(\tau)\xi_i^{\tau+1}]\right\} \nonumber\\
    &\times \exp\left\{-\ii\sum_{\tau<t/2}x^1(2\tau)\left[\hat{b}(\tau)\xi_i^\tau + \psi_i^1(2\tau)\right]\right\} \nonumber\\
    &\times \exp\left\{-\ii\sum_{\tau<t/2}x^1(2\tau-1)\left[\hat{a}(\tau)\xi_i^\tau + \psi_i^1(2\tau-1)\right]\right\} \nonumber\\
    &\times \exp\left\{-\sum_{s,s'<t}[\hat{C}^{11}(s,s')x^1(s)x^1(s') + \hat{Q}^{11}(s,s')\hat{h}^1(s)\hat{h}^1(s') + \hat{Q}^{12}(s,s')\hat{h}^1(s)\hat{h}^2(s')  + \hat{K}^{11}(s,s')x^1(s)\hat{h}^1(s')]\right\} \nonumber\\
    &\times \exp\left\{-\sum_{s,s'<t}[ \hat{Q}^{22}(s,s')\hat{h}^2(s)\hat{h}^2(s') + \hat{K}^{21}(s,s')\hat{h}^2(s')x^1(s)] \right\} \Bigg\}.
\end{align}
The order-parameter set contains only $C^{11}$, $Q^{ab}$ and $K^{a1}$. This follows from the
structure of the disorder charge in~\eqref{eq:Z_2bit_disordercharge}: the patterns couple to the
visible spins $x^1$ (never to $x^2$) and to both auxiliary fields $\hat{h}^1,\hat{h}^2$, so
the only $\bm x$--$\hat{\bm h}$ correlators that survive the average are $K^{11}$ and
$K^{21}$, and the only spin autocorrelator is $C^{11}$ (there is no $C^{22}$, $C^{12}$, or
$K^{22}$). 
The cross term $\hat{Q}^{12}\hat{h}^1\hat{h}^2$ is kept but it is shown below to vanish.

The $\Omega$ functional which arises from the disorder average and determine the retarded self-interactions and the effective noise covariances is:
\begin{align}\label{eq:Omega_2species_raw}
    \Omega[\bm{C}, \bm{Q},& \bm{K}] = \lim_{p\to\infty}\frac{1}{p}\ln \int \prod_{\mu>t}\frac{\dd\bm{w}^1\,\dd\bm{\hat{w}}^1\, \dd\bm{w}^2\,\dd\bm{\hat{w}}^2\, \dd\bm{y}\, \dd\bm{\hat{y}}}{(2\pi)^{3(p-t)t}}\nonumber\\ 
    &\times \exp\left\{-\ii\sum_{\mu>t}\sum_{s<t}\left[\hat{w}_\mu^{1}(s)w_\mu^{1}(s) + \hat{w}_\mu^{2}(s)w_\mu^{2}(s) + \hat{y}_\mu(s)y_\mu(s) - w_\mu^2(s) y_\mu(s) - \frac{1}{1+\lambda}\, w_\mu^1(s)y_\mu(s) \right]\right\} \nonumber\\
    &\times\exp\Bigg\{-\frac{1}{2}\sum_{\mu>t}\sum_{s,s'<t}\Big[\hat{w}_\mu^{1}(s)\hat{w}_\mu^{1}(s')Q^{11}(s,s') + \hat{w}_{\mu-1}^{2}(s)\hat{w}_{\mu-1}^{2}(s')Q^{22}(s,s') + 2\hat{w}_\mu^{1}(s)\hat{w}_{\mu-1}^{2}(s')Q^{12}(s,s') \nonumber\\
    &+ \hat{y}_\mu(s)\hat{y}_\mu(s')C^{11}(s,s') + 2\hat{w}_{\mu}^{1}(s)\hat{y}_\mu(s')K^{11}(s,s') + 2\hat{w}_{\mu-1}^{2}(s)\hat{y}_\mu(s')K^{21}(s,s')\Big]\Bigg\}.
\end{align}
To simplify, we integrate out the $w$ and $y$ variables. Integrating $\dd w_\mu^2$ gives $\delta(\hat{w}_\mu^2 - y_\mu)$, and integrating $\dd w_\mu^1$ now gives $\delta(\hat{w}_\mu^1 - \tfrac{1}{1+\lambda}y_\mu)$, since the visible coupling carries the $(1+\lambda)^{-1}$ weight reinstated above. This pins the two channels, $\hat{w}_\mu^2 = y_\mu$ and $\hat{w}_\mu^1 = \tfrac{1}{1+\lambda}y_\mu$. Keeping $y_\mu$ as the surviving lane and eliminating $\hat{w}^1_\mu,\hat{w}^2_\mu$ through the deltas, every visible-channel factor $\hat w^1\!\to\!\tfrac{1}{1+\lambda}y$ picks up one power of $(1+\lambda)^{-1}$, while the hidden channel $\hat w^2\!\to\! y$ is unweighted.
We thus have:
\begin{align}\label{eq:Omega_simplified}
    \Omega[\bm{C}, \bm{Q}, \bm{K}] = \lim_{p\to\infty}\frac{1}{p}\ln  \int \prod_{\mu>t}&\frac{\dd\bm{y}_\mu\,\dd\bm{\hat{y}}_\mu}{(2\pi)^{(p-t)t}} \exp\left[-\ii\sum_{\mu>t}\sum_{s<t}\hat{y}_\mu(s) y_{\mu}(s)\right] \nonumber\\
    &\times \exp\Bigg\{-\frac{1}{2}\sum_{\mu>t}\sum_{s,s'<t}\Big[\tfrac{1}{(1+\lambda)^2}\,y_\mu(s)y_\mu(s')Q^{11}(s,s') + y_{\mu-1}(s)y_{\mu-1}(s')Q^{22}(s,s') \nonumber\\
    &\quad + \tfrac{2}{1+\lambda}\,y_\mu(s)y_{\mu-1}(s')Q^{12}(s,s') + \hat{y}_\mu(s)\hat{y}_\mu(s')C^{11}(s,s') \nonumber\\
    &\quad + \tfrac{2}{1+\lambda}\,y_{\mu}(s)\hat{y}_\mu(s')K^{11}(s,s') + 2\,y_{\mu-1}(s)\hat{y}_\mu(s')K^{21}(s,s')\Big]\Bigg\}.
\end{align}
The key structural features are: (i) the first exponent has $\hat{y}_\mu y_\mu$ with \emph{no shift}, and (ii) the $Q^{12}$ and $K^{21}$ terms have cross-pattern couplings ($\mu$ with $\mu-1$) inherited from the $\hat{w}_{\mu-1}^2$ structure in the disorder average. 

\paragraph{Derivation of saddle-point equations.}
To evaluate~\eqref{eq:Omega_simplified}, we work in the product space of vectors labelled by both time indices $s$ and pattern indices $\mu$. Define the matrix $\bm{\Gamma} = \bm{S} \otimes \bm{R}$ with elements $\Gamma_{\mu\mu'}(s,s') = S_{\mu\mu'}R(s,s')$, where $S_{\mu\nu} = \delta_{\mu,(\nu+1)\mod p}$ is the shift matrix. Using the shorthand $\bm{u} = \bm{y}$ and $\bm{v} = \hat{\bm{y}}$, the Gaussian integral gives:
\begin{align}\label{eq:Omega_2species_evaluated}
    \lim_{N\to\infty}\Omega[\bm{C}, \bm{Q}, \ii\bm{G}] &= -\lim_{N\to\infty}\frac{\alpha}{2N}\Big\{\ln\det[\bm{1}\otimes\bm{C}^{11}] \nonumber\\
    &\quad + \ln\det\Big[\bm{1}\otimes\Big(\tfrac{1}{(1+\lambda)^2}\bm{Q}^{11} + \bm{Q}^{22}\Big) + \tfrac{1}{1+\lambda}(\bm{S}\otimes\bm{Q}^{12} + \bm{S}^{\dagger}\otimes\bm{Q}^{12}) \nonumber\\
    &\qquad\qquad + \Big(\bm{1}\otimes\bm{1} - \tfrac{1}{1+\lambda}\bm{1}\otimes\bm{G}^{11} - \bm{S}\otimes\bm{G}^{21}\Big)^{\dagger}[\bm{1}\otimes\bm{C}^{11}]^{-1}\Big(\bm{1}\otimes\bm{1} - \tfrac{1}{1+\lambda}\bm{1}\otimes\bm{G}^{11} - \bm{S}\otimes\bm{G}^{21}\Big)\Big]\Big\}.
\end{align}
The key structural difference from the single-species case is that the shift matrix $\bm{S}$ only multiplies $\bm{G}^{21}$ and $\bm{Q}^{12}$, while $\bm{G}^{11}$ and $\bm{Q}^{11}, \bm{Q}^{22}$ have no shift. 
The response kernels below are organized through the combination $\bm{G}_{\mathrm{eff}}(\omega)=\tfrac{1}{1+\lambda}\bm{G}^{11}+e^{-\ii\omega}\bm{G}^{21}$; the noise spectrum, by contrast, is dressed by $\bm{G}^{11}$ alone.

We use~\eqref{eq:Omega_2species_evaluated} to work out the saddle-point equations. The saddle-point conditions are:
\begin{align}
    \text{for all } s, s': \quad &\hat{C}^{11}(s,s') = \ii\frac{\partial\Omega}{\partial C^{11}(s,s')}\bigg|_{\mathrm{saddle}} = 0, \label{eq:qhat_2species}\\
    \text{for all } s, s': \quad &\hat{Q}^{ab}(s,s') = \ii\frac{\partial\Omega}{\partial Q^{ab}(s,s')}\bigg|_{\mathrm{saddle}}, \label{eq:Qhat_2species}\\
    \text{for all } s, s': \quad &\hat{K}^{ab}(s,s') = \ii\frac{\partial\Omega}{\partial K^{ab}(s,s')}\bigg|_{\mathrm{saddle}}. \label{eq:Khat_2species}
\end{align}
Note that the first equation follows from the fact that at the saddle point
\begin{equation}
    Q^{ab}\sim \langle \hat{h}^a \hat{h}^b \rangle = \frac{\partial^2 \bar{Z}[0]}{\partial h^a \partial h^b} = 0.
\end{equation}

\paragraph{Evaluation of $\hat{Q}^{ab}$.}
Using the matrix identity $\ln\det[\bm{M} + \bm{Q}] = \ln\det\bm{M} + \Tr[\bm{M}^{-1}\bm{Q}] + \mathcal{O}(Q^2)$ and the unitarity of $\bm{S}$ (with eigenvalues $s_\mu = e^{-2\pi\ii\mu/p}$), equation~\eqref{eq:Qhat_2species} reduces to a single-site noise kernel. Because the hidden spin $x^2$ is retained explicitly, the relay is carried by the sampled $x^2$ and the shift-carrying relay response $\bm{G}^{21}$ does not dress the visible crosstalk noise; the visible channel is dressed by its shift-free response $\bm{G}^{11}$ alone, and the pattern index decouples.

The visible-noise spectrum is therefore $\omega$-independent (since $\bm{G}^{11}$ carries no shift),
\begin{equation}\label{eq:Sigma_def}
    \bm{\Sigma}\;\equiv\;\Big[\bm{1}-\tfrac{1}{1+\lambda}\bm{G}^{11}\Big]^{-1}\,\bm{C}^{11}\,\Big[\bm{1}-\tfrac{1}{1+\lambda}\bm{G}^{11}\Big]^{-\dagger},
\end{equation}
while the hidden channel is dressed by the asymmetric relay response $\bm{G}^{21}$, which carries the shift $\bm{S}$. The conjugate noise parameters are then
\begin{equation}\label{eq:Qhat11_final}
    \boxed{\hat{Q}^{11} = -\frac{\ii\alpha}{2}\frac{1}{(1+\lambda)^2}\,\bm{\Sigma},\qquad \hat{Q}^{22} = -\frac{\ii\alpha}{2}\,\big[\bm{1}-(\bm{G}^{21})^{\dagger}\bm{G}^{21}\big]^{-1}\bm{C}^{11}.}
\end{equation}
The visible-layer noise $\hat{Q}^{11}$ is the symmetric, Onsager-dressed channel weighted by $(1+\lambda)^{-2}$; the hidden-layer noise $\hat{Q}^{22}$ is the asymmetric channel. In the scalar stationary form these are $\rho^1=(1+\lambda)^{-2}(1-G^{11})^{-2}$ and $\rho^2=(1-(G^{21})^2)^{-1}$.

For the cross-term, the $Q^{12}$ block enters the determinant through
$\bm{S}\otimes\bm{Q}^{12}+\bm{S}^{\dagger}\otimes\bm{Q}^{12}$, so its saddle insertion carries the
bare shift phase $e^{\ii\omega}+e^{-\ii\omega}=2\cos\omega$:
\begin{equation}\label{eq:Qhat12_final}
    \hat{Q}^{12} = \hat{Q}^{21} = -\frac{\ii\alpha}{2}\frac{1}{1+\lambda}\int_{-\pi}^{\pi}\frac{\dd\omega}{2\pi}\,2\cos\omega\,\bm{\Sigma}.
\end{equation}
Since $\bm{\Sigma}$ is independent of $\omega$, the angular average of the bare
shift phase vanishes, $\int_{-\pi}^{\pi}\tfrac{\dd\omega}{2\pi}\,2\cos\omega\,\bm{\Sigma}=\bm{0}$, so
\begin{equation}
    \boxed{\hat{Q}^{12}=\hat{Q}^{21}=0,}
\end{equation}
exactly as $\hat{K}^{21}=0$ below: a shift-coupled kernel (net shift power $\pm1$) is annihilated
by the angular average. The two species' non-condensed overlaps are sampled one pattern apart
($\bm{\xi}^\mu$ vs $\bm{\xi}^{\mu+1}$) and are uncorrelated; the inter-layer coupling is carried by
the explicit relay $x^2$ in the visible field.

\paragraph{Evaluation of $\hat{K}^{11}$ (retarded self-interaction from visible layer).}
Taking the derivative with respect to $G^{11}(s,s')$:
\begin{align}
    \hat{K}^{11}(s,s') &= \frac{\partial}{\partial G^{11}(s,s')}\Omega[\bm{C}, \bm{0}, \ii\bm{G}] \nonumber\\
    &= -\frac{\alpha}{2}\frac{\partial}{\partial G^{11}(s,s')}\lim_{p\to\infty}\frac{1}{p}\sum_{\mu=1}^{p}\left\{\ln\det\Big[\bm{1} - \tfrac{1}{1+\lambda}(\bm{G}^{11})^{\dagger} - s_\mu^*(\bm{G}^{21})^{\dagger}\Big] + \ln\det\Big[\bm{1} - \tfrac{1}{1+\lambda}\bm{G}^{11} - s_\mu\bm{G}^{21}\Big]\right\} \nonumber\\ 
    &= -\frac{\alpha}{2}\frac{1}{1+\lambda}\lim_{p\to\infty}\frac{1}{p}\sum_{\mu=1}^{p}\left\{[(\bm{1} - \bm{G}_{\mathrm{eff}}(\omega_\mu))^{-1}](s',s) + [(\bm{1} - \bm{G}_{\mathrm{eff}}(\omega_\mu)^{\dagger})^{-1}](s',s)\right\} \nonumber\\
    &= -\frac{\alpha}{2}\frac{1}{1+\lambda}\int_{-\pi}^{\pi}\frac{\dd\omega}{2\pi}\left\{[(\bm{1} - \bm{G}_{\mathrm{eff}}(\omega))^{-1}] + [(\bm{1} - \bm{G}_{\mathrm{eff}}(\omega)^{\dagger})^{-1}]\right\}(s',s).
\end{align}
The extra $(1+\lambda)^{-1}$ prefactor is the chain-rule factor from $\partial\bm{G}_{\mathrm{eff}}/\partial\bm{G}^{11}=(1+\lambda)^{-1}\bm 1$, since $\bm{G}^{11}$ enters $\Omega$ only through $\bm{G}_{\mathrm{eff}}=(1+\lambda)^{-1}\bm{G}^{11}+e^{-\ii\omega}\bm{G}^{21}$. Crucially, there is no phase factor $e^{\pm\ii\omega}$ in front of the integrand. Expanding in powers of $\bm{G}_{\mathrm{eff}}$:
\begin{align}
    \hat{K}^{11}(s,s') &= -\frac{\alpha}{2}\frac{1}{1+\lambda}\sum_{n\geq 0}\int_{-\pi}^{\pi}\frac{\dd\omega}{2\pi}\left\{\Big[\tfrac{1}{1+\lambda}\bm{G}^{11} + e^{-\ii\omega}\bm{G}^{21}\Big]^n + \Big[\tfrac{1}{1+\lambda}\bm{G}^{11} + e^{\ii\omega}\bm{G}^{21}\Big]^n\right\}(s',s).
\end{align}
For each $n$, the binomial expansion gives terms $\binom{n}{k}\Big(\tfrac{1}{1+\lambda}\bm{G}^{11}\Big)^{n-k}(\bm{G}^{21})^k e^{\mp\ii k\omega}$. The $\omega$-integral gives $\delta_{k,0}$, selecting only the $k=0$ term. Thus:
\begin{equation}\label{eq:Khat11_2species}
    \boxed{\hat{K}^{11}(s,s') = -\frac{\alpha}{1+\lambda}\sum_{n\geq 0}\left[\left(\tfrac{1}{1+\lambda}\bm{G}^{11}\right)^n\right](s',s) = -\frac{\alpha}{1+\lambda}\left[\left(\bm{1} - \tfrac{1}{1+\lambda}\bm{G}^{11}\right)^{-1}\right](s',s) \neq 0}
\end{equation}

The non-vanishing retarded self-interaction (it carries no shift, hence no phase to cancel) from the visible layer is the
fingerprint of the \emph{symmetric} coupling $\bm{J}^{\mathrm S}=\tfrac1N\bm{\xi}^{\mathrm T}\bm{\xi}$
that feeds $x^1$. It appears with an overall $(1+\lambda)^{-1}$ weight and a resolvent in $(1+\lambda)^{-1}\bm{G}^{11}$, so it fades as $\lambda\to\infty$ (see the recovery paragraph below) and reduces to the full symmetric Hopfield reaction $-\alpha(\bm 1-\bm{G}^{11})^{-1}$ at $\lambda=0$~\footnote{Note that we have a factor $\alpha$
(rather than $\alpha/2$) because in~\eqref{eq:Khat11_2species} the resolvent from
$\ln\det[\bm 1-\bm{G}_{\mathrm{eff}}]$ and its conjugate from $\ln\det[\bm 1-\bm{G}_{\mathrm{eff}}^{\dagger}]$
\emph{both} contribute, and once the $\omega$-integral projects onto the $\bm{G}^{21}$-free
sector they collapse onto the \emph{same real} matrix $(\bm 1-\tfrac{1}{1+\lambda}\bm{G}^{11})^{-1}(s',s)$ and add.
(Checked numerically against~\eqref{eq:Omega_2species_evaluated}.)}.

\paragraph{Evaluation of $\hat{K}^{21}$ (retarded self-interaction from hidden layer).}
Taking the derivative with respect to $G^{21}(s,s')$, the key difference is that $\bm{G}^{21}$ is multiplied by $\bm{S}$, so the derivative brings down a phase factor:
\begin{align}
    \hat{K}^{21}(s,s') &= \frac{\partial}{\partial G^{21}(s,s')}\Omega[\bm{C}, \bm{0}, \ii\bm{G}] \nonumber\\
    &= -\frac{\alpha}{2}\int_{-\pi}^{\pi}\frac{\dd\omega}{2\pi}\left\{e^{-\ii\omega}[(\bm{1} - \bm{G}_{\mathrm{eff}}(\omega))^{-1}] + e^{\ii\omega}[(\bm{1} - \bm{G}_{\mathrm{eff}}(\omega)^{\dagger})^{-1}]\right\}(s',s) \nonumber\\
    &= -\frac{\alpha}{2}\sum_{n\geq 0}\int_{-\pi}^{\pi}\frac{\dd\omega}{2\pi}\left\{e^{-\ii\omega}\Big[\tfrac{1}{1+\lambda}\bm{G}^{11} + e^{-\ii\omega}\bm{G}^{21}\Big]^n + e^{\ii\omega}\Big[\tfrac{1}{1+\lambda}\bm{G}^{11} + e^{\ii\omega}\bm{G}^{21}\Big]^n\right\}(s',s).
\end{align}
Each term carries an overall phase $e^{\pm\ii\omega}$. After binomial expansion, we get $e^{\pm\ii(k+1)\omega}$, and the integral $\int_{-\pi}^{\pi}\frac{\dd\omega}{2\pi}e^{\pm\ii(k+1)\omega} = \delta_{k,-1} = 0$ for all $k \geq 0$.
Therefore:
\begin{equation}\label{eq:Khat21_2species}
    \boxed{\hat{K}^{21}(s,s') = 0}
\end{equation}
The retarded self-interaction from the hidden layer pathway vanishes by phase cancellation, exactly as in the single-species purely asymmetric case.

\paragraph{Summary.}
At the physical saddle point, most macroscopic integration variables vanish: $k(s) = \hat{m}(s) = Q^{ab}(s,s') = \hat{q}(s,s') = 0$. The other conjugate order parameters which we just derived are:
\begin{align}
    \hat{Q}^{11} &= -\frac{\ii\alpha}{2}\frac{1}{(1+\lambda)^2}\,\bm{\Sigma}, \quad \hat{Q}^{22} = -\frac{\ii\alpha}{2}\big[\bm{1}-(\bm{G}^{21})^{\dagger}\bm{G}^{21}\big]^{-1}\bm{C}^{11}, \quad \hat{Q}^{12} = \hat{Q}^{21} = 0, \label{eq:Qhat_summary}\\
    \hat{K}^{11} &= -\frac{\alpha}{1+\lambda}\left(\bm{1} - \tfrac{1}{1+\lambda}\bm{G}^{11}\right)^{-1} \neq 0, \label{eq:Khat11_summary}\\
    \hat{K}^{21} &= 0, \label{eq:Khat21_summary}
\end{align}
The remaining macroscopic observables are expressed in terms of three quantities: the overlaps $a(\tau), b(\tau)$, the single-site correlation $C^{11}(s,s')$, and response functions $G^{11}(s,s'), G^{21}(s,s')$.

\paragraph{Evaluation of conjugate order parameters.}
The remaining saddle-point equations for $\hat{a}, \hat{b}, \hat{k}, \hat{l}, \hat{p}, \hat{r}$ and $a, b, k, l, p, r$ can be derived from the $\Phi$ functional~\eqref{eq:Phi_2species} in the usual way. They give:
\begin{align}
    \hat{k}(\tau) &= b(\tau), \qquad \hat{l}(\tau) = b(\tau), \qquad \hat{p}(\tau) = a(\tau), \qquad \hat{r}(\tau) = a(\tau), \nonumber\\
    a(\tau) &= \langle x^1(2\tau-1)\xi^{\tau}\rangle_{\star}, \qquad b(\tau) = \langle x^1(2\tau)\xi^{\tau}\rangle_{\star}, \nonumber\\
    k(\tau) &= \underbrace{\langle h^2(2\tau)\rangle_{\star}}_{=0}\xi^{\tau+1}=0, \qquad l(\tau) = \underbrace{\langle h^1(2\tau)\rangle_{\star}}_{=0}\xi^{\tau}=0, \nonumber\\
    r(\tau) &= \underbrace{\langle h^2(2\tau-1)\rangle_{\star}}_{=0}\xi^{\tau+1}=0, \qquad p(\tau) = \underbrace{\langle h^1(2\tau-1)\rangle_{\star}}_{=0}\xi^{\tau}=0, \nonumber\\
    \hat{a}(\tau) &= r(\tau) + p(\tau)=0, \qquad \hat{b}(\tau) = k(\tau) + l(\tau)=0, \nonumber
\end{align}
where $\langle\cdots\rangle_{\star}$ denotes averaging with respect to the single-site measure defined by~\eqref{eq:Phi_2species}.


\begin{figure}[h!]
    \centering
    \includegraphics[width=2.8in]{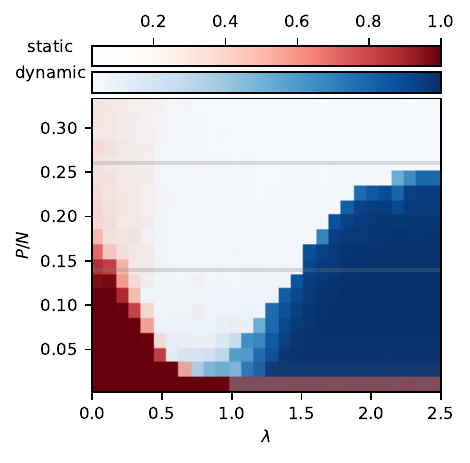}
    \caption{\label{fig:uncorrelated_dynamic_overlap} The Monte-Carlo phase diagram of the synchronous
    two-species model on the $\lambda$--$P/N$ plane, from simulations of the full microscopic network
    (synchronous updates, diagonal of $\bm{J}$ removed), probed by the static (red) and dynamic
    (blue) overlap order parameters. The dynamic-retrieval (blue) island opens near $\lambda\simeq1$
    and its capacity grows with $\lambda$ towards the purely asymmetric bound, reproducing the DMFT
    prediction of Fig.~\ref{fig:dmft_corrected_island}; the static-retrieval (red) region is confined
    to small $\lambda$, and both retrieval phases are suppressed near the balance point $\lambda=1$.
    Horizontal gray guides mark the single-species reference capacities $\alpha_c\simeq0.138$
    (static Hopfield) and $\alpha_c\simeq0.269$ (purely asymmetric sequence).}
\end{figure}

\paragraph{The effective single-neuron problem}

The effective single-neuron problem is defined via the $\Phi$ functional~\eqref{eq:Phi_2species} after substituting the saddle-point values of the macroscopic order parameters. For any function $f$ of the single-neuron trajectory $\{x^1(s), x^2(s), h^1(s), h^2(s), \hat{h}^1(s), \hat{h}^2(s)\}_{s<\tau}$, we define:
\begin{align}
    &\left\langle f\left[\left\{x^1, x^2, h^1, h^2, \hat{h}^1, \hat{h}^2\right\}\right] \right\rangle_{\star}\nonumber\\
    &\quad:=\frac{\sum_{\left\{x^1, x^2\right\}} \int\{\dd h^1\,\dd\hat{h}^1\}\int\{\dd h^2\,\dd\hat{h}^2\} \, W\left[\left\{x^1, x^2, h^1, h^2, \hat{h}^1, \hat{h}^2\right\}\right] \, f\left[\left\{x^1, x^2, h^1, h^2, \hat{h}^1, \hat{h}^2\right\}\right] }{\sum_{\left\{x^1, x^2\right\}} \int\{\dd h^1\,\dd\hat{h}^1\}\int\{\dd h^2\,\dd\hat{h}^2\}  \, W\left[\left\{x^1, x^2, h^1, h^2, \hat{h}^1, \hat{h}^2\right\}\right]}.
\end{align}

With the results~\eqref{eq:Qhat_summary}--\eqref{eq:Khat21_summary}, the single-site measure becomes:
\begin{align}
    W\left[\left\{x^1, x^2, h^1, h^2, \hat{h}^1, \hat{h}^2\right\}\right] = & p(x^1(0), x^2(0)) \, \prod_{s<t} \frac{e^{\beta x^1(s+1)h^1(s)}}{2\cosh(\beta h^1(s))} \cdot \frac{e^{\beta x^2(s+1)h^2(s)}}{2\cosh(\beta h^2(s))} \nonumber\\
    &\times e^{\ii\sum_{s<t}\hat{h}^1(s)\left[h^1(s) - \theta^1(s) - \frac{1}{1+\lambda}m^1(s) - \frac{\lambda}{1+\lambda} x^2(s)\right]} \nonumber\\
    &\times e^{\ii\sum_{s<t}\hat{h}^2(s)\left[h^2(s) - \theta^2(s) - m^2(s)\right]} \nonumber\\
    &\times e^{-\sum_{s,s'}\hat{K}^{11}(s,s')x^1(s)\hat{h}^1(s')} \nonumber\\
    &\times \exp\left\{-\frac{\alpha}{2}\sum_{s,s'}\Big[R^{11}(s,s')\hat{h}^1(s)\hat{h}^1(s') + R^{22}(s,s')\hat{h}^2(s)\hat{h}^2(s') + 2R^{12}(s,s')\hat{h}^1(s)\hat{h}^2(s')\Big]\right\}
\end{align}
where the noise kernels are set by the spectrum~\eqref{eq:Sigma_def} (related to the conjugate parameters by $\hat{Q}^{ab}=-\tfrac{\ii\alpha}{2}R^{ab}$),
\begin{equation}\label{eq:R_2species}
    R^{11}=\frac{1}{(1+\lambda)^2}\,\bm{\Sigma}, \quad R^{22}=\big[\bm{1}-(\bm{G}^{21})^{\dagger}\bm{G}^{21}\big]^{-1}\bm{C}^{11}, \quad R^{12}=R^{21}=0.
\end{equation}
The visible and hidden noise are therefore \emph{uncorrelated}, $R^{12}=0$: the two layers are coupled only through the explicit relay $x^2$ in the visible field, not through a noise cross-covariance. The retarded self-interaction $\hat{K}^{11}$ acts only within the visible layer ($\hat{K}^{21}=0$). The condensed signal terms are:
\begin{align}
    m^1(s) &= \xi^{\tau}\begin{cases}
        \hat{p}(\tau)  \\ 
        \hat{l}(\tau) 
    \end{cases} = \xi^{\tau}\begin{cases}
        a(\tau), & s = 2\tau-1 \text{ (odd timesteps)} \\ 
        b(\tau), & s = 2\tau \text{ (even timesteps)}
    \end{cases},\\
    m^2(s) &= \xi^{\tau+1}\begin{cases}
        \hat{r}(\tau)  \\ 
        \hat{k}(\tau)
    \end{cases} = \xi^{\tau+1}\begin{cases}
        a(\tau), & s = 2\tau-1 \text{ (odd timesteps)} \\ 
        b(\tau), & s = 2\tau \text{ (even timesteps)}
    \end{cases}.
\end{align}


Defining overlaps:
\begin{equation}
    a(\tau) = \langle x^1(2\tau-1)\xi^{\tau}\rangle_{\xi,\phi}, \qquad b(\tau) = \langle x^1(2\tau)\xi^{\tau}\rangle_{\xi,\phi},
\end{equation}
the effective single-neuron dynamics at finite temperature is:
\begin{equation}\label{eq:single_neuron_2species}
    \boxed{
    \begin{aligned}
    x^1(2\tau) &= \tanh\beta\left[\frac{1}{1+\lambda}a(\tau)\xi^{\tau} + \frac{\lambda}{1+\lambda} x^2(2\tau-1) + \sqrt{\alpha}\,\phi^1(2\tau-1) + \sum_{s'}K(2\tau-1, s')x^1(s')  \right], \\
    x^2(2\tau-1) &= \tanh\beta\left[b(\tau-1)\xi^{\tau} + \sqrt{\alpha}\,\phi^2(2\tau-2)\right], \\
    x^1(2\tau+1) &= \tanh\beta\left[\frac{1}{1+\lambda}b(\tau)\xi^{\tau} + \frac{\lambda}{1+\lambda} x^2(2\tau) + \sqrt{\alpha}\,\phi^1(2\tau) + \sum_{s'}K(2\tau, s')x^1(s')\right], \\
    x^2(2\tau) &= \tanh\beta\left[a(\tau)\xi^{\tau+1} + \sqrt{\alpha}\,\phi^2(2\tau-1)\right].
    \end{aligned}
    }
\end{equation}

The noise $\phi^1,\phi^2$ is jointly Gaussian with $\langle\phi^a(s)\phi^b(s')\rangle = R^{ab}(s,s')$;
since $R^{12}=0$ the visible readout noise and the relayed
hidden noise are \emph{independent}, so $c_a=c_b=0$ and the bivariate normal $\int\!D_c(z^1,z^2)$
below factorizes into independent Gaussians. Writing $\rho^a(s)=R^{aa}(s,s)$ and, for the two time pairs entering below, the
cross-correlation coefficients
\begin{equation}\label{eq:cross_corr_coeff}
    c_b(\tau)=\frac{R^{12}(2\tau-1,\,2\tau-2)}{\sqrt{\rho^1(2\tau-1)\,\rho^2(2\tau-2)}}, \qquad
    c_a(\tau)=\frac{R^{12}(2\tau,\,2\tau-1)}{\sqrt{\rho^1(2\tau)\,\rho^2(2\tau-1)}},
\end{equation}
we denote by $\int\! D_c(z^1,z^2)$ the average over a standardized bivariate normal with unit
variances and $\langle z^1z^2\rangle=c$. The order parameters then satisfy
\begin{align}
    b(\tau) = \int\! D_{c_b(\tau)}(z^1,z^2)\;\tanh\beta\Big[&\frac{1}{1+\lambda}\left(a(\tau) + \lambda\tanh\beta\!\left(b(\tau-1) + z^2\sqrt{\alpha \rho^2(2\tau-2)}\right)\right)\nonumber \\ &+ z^1\sqrt{\alpha \rho^1(2\tau-1)} + \sum_{s'}K(2\tau-1, s')x^1(s')  \Big],
\end{align}
\begin{align}
    a(\tau+1) =\;& \frac{1}{2}\int\! D_{c_a(\tau)}(z^1,z^2)\;\tanh\beta\Big[\frac{1}{1+\lambda}\left(b(\tau) + \lambda\tanh\beta\!\left(a(\tau) + z^2\sqrt{\alpha \rho^2(2\tau-1)}\right)\right)\nonumber\\ & \qquad\qquad\qquad\qquad\qquad\qquad + z^1\sqrt{\alpha \rho^{1}(2\tau)} + \sum_{s'}K(2\tau, s')x^1(s')  \Big] \nonumber\\
    &+ \frac{1}{2}\int\! D_{c_a(\tau)}(z^1,z^2)\;\tanh\beta\Big[\frac{1}{1+\lambda}\left(-b(\tau) + \lambda\tanh\beta\!\left(a(\tau) + z^2\sqrt{\alpha \rho^2(2\tau-1)}\right)\right)\nonumber\\  &\qquad\qquad\qquad\qquad\qquad\qquad  + z^1\sqrt{\alpha \rho^{1}(2\tau)} + \sum_{s'}K(2\tau, s')x^1(s')  \Big],
\end{align}
the two branches in $a(\tau+1)$ arising from the $\eta^\tau=\xi^\tau\xi^{\tau+1}=\pm1$ average
(each with weight $\tfrac12$). Since $R^{12}=0$, $c_a=c_b=0$ and these are
independent-noise equations: the $D_c$ integrals reduce to independent Gaussian averages over
$z^1,z^2$.

One approximation enters these scalar equations. The cross-species noise
correlation is zero, $R^{12}=0$, so $c_a=c_b=0$ and the readout noise $z^1$ and the relayed hidden
noise $z^2$ are independent. The binary $x^2$ has been
replaced inside the outer $\tanh$ by its conditional mean given its field. This replacement is exact
at zero temperature, where $x^2$ is deterministic given its field; at finite temperature the
exact average over $x^2=\pm1$ instead yields a two-branch form---a $\pm\lambda/(1+\lambda)$
shift of the readout field with weights $\tfrac12(1\pm\tanh\beta[\,\cdot\,])$.


\begin{figure}[hbt!]
    \centering
    \includegraphics[width=3.2in]{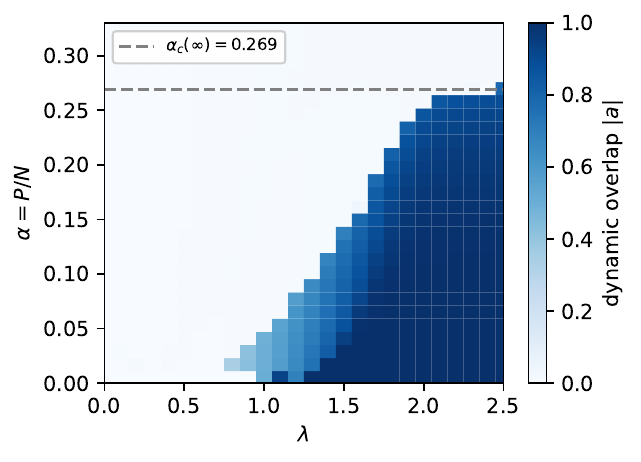}
    \caption{\label{fig:dmft_corrected_island} Dynamic-retrieval phase diagram of the synchronous
    two-species model from the scalar DMFT reduction
    (Eqs.~\eqref{eq:b_solver}--\eqref{eq:a_solver}), colour-coded by the dynamic overlap $|a|$ on
    the $\lambda$--$\alpha$ plane. The dynamic-retrieval island opens near $\lambda\simeq1$ and its
    capacity saturates to the purely asymmetric bound $\alpha_c(\infty)\simeq0.269$ (gray dashed)
    already at $\lambda\simeq2.5$; compare the Monte-Carlo phase diagram in
    Fig.~\ref{fig:uncorrelated_dynamic_overlap}.}
\end{figure}

\paragraph{Recovery of single-species asymmetric limit ($\lambda \to \infty$).}
The effective single-neuron dynamics~\eqref{eq:single_neuron_2species} has signal terms $\propto 1/(1+\lambda)$ and coupling terms $\propto \lambda/(1+\lambda)$. In the limit $\lambda \to \infty$:
\begin{itemize}
    \item The signal terms $\frac{a}{1+\lambda}\xi, \frac{b}{1+\lambda}\xi \to 0$,
    \item The coupling terms $\frac{\lambda}{1+\lambda}x^2 \to x^2$.
\end{itemize}
Thus the visible layer simply copies the hidden layer: $x^1(s) \to x^2(s-1)$. The retarded
self-interaction $\hat{K}^{11}=-\alpha(\bm 1-\bm{G}^{11})^{-1}$ originates entirely from the
\emph{symmetric} coupling $\bm{J}^{\mathrm S}$, whose weight in the visible field is
$1/(1+\lambda)\to0$; once it is switched off, $x^1$ relays the purely asymmetric hidden layer,
whose pathway carries the shift $\bm{S}$ and therefore generates no retarded self-interaction
(the same phase cancellation that gives $\hat{K}^{21}=0$). The effective self-interaction thus
\emph{vanishes} and we recover the purely asymmetric single-species result with critical
capacity $\alpha_c \approx 0.269$. 

\paragraph{Zero-temperature reduction and phase diagram.}
In the effective single-site process~\eqref{eq:single_neuron_2species} the hidden spin is kept
explicit as a sampled $\pm1$ variable, so the relayed signal and noise ride on $x^2$ itself. The
scalar zero-temperature reduction dresses the visible crosstalk by the visible self-response
alone, with $R^{12}=0$: the shift-coupled cross-noise ($\bm\xi^\mu$ vs
$\bm\xi^{\mu+1}$, one pattern apart) averages to zero.

With $\chi^a=\langle 2\,\delta(h^a)\rangle$, the zero-temperature limit of the scalar DMFT equations is
\begin{equation}
    \rho^1=\frac{(1+\lambda)^{-2}}{(1-G^{11})^2},\quad G^{11}=\frac{\chi^1}{1+\lambda};\qquad
    \rho^2=\frac{1}{1-(G^{21})^2},\quad G^{21}=\chi^2;\qquad R^{12}\simeq0,
\end{equation}
the relay being carried by the explicit $x^2=\mathrm{sign}(\cdot)$,
\begin{align}
    b &= \Big\langle \mathrm{erf}\!\Big[\frac{a+\lambda\,\mathrm{sign}\!\big(b+\sqrt{\alpha\rho^2}\,z\big)}{(1+\lambda)\sqrt{2\alpha\rho^1}}\Big]\Big\rangle_z,\label{eq:b_solver}\\
    a &= \Big\langle \tfrac12\!\!\sum_{\eta=\pm1}\mathrm{erf}\!\Big[\frac{\eta\,b+\lambda\,\mathrm{sign}\!\big(a+\sqrt{\alpha\rho^2}\,z\big)}{(1+\lambda)\sqrt{2\alpha\rho^1}}\Big]\Big\rangle_z,\label{eq:a_solver}
\end{align}
with $z\sim\mathcal N(0,1)$ and $\mathrm{erf}$ the analytic average of $\mathrm{sign}$ over the
visible noise. These collapse to the static Hopfield variance $1/(1-G^{11})^2$ as $\lambda\to0$
(relay off; the $\eta=\pm1$ branches cancel and the transition overlap $a\to0$), and to the purely
asymmetric sequence theory $\rho\to1/(1-(G^{21})^2)$, $\alpha_c\to0.269$ as $\lambda\to\infty$.

\section{Path integral for the scheduled asynchronous bipartite dynamics}
\label{sec:async_bipartite}

We now construct the generating-functional theory of the \emph{asynchronous}
two-species model---the dynamics actually simulated in the main text---and show
that it admits an \emph{autonomous} DMFT: the sequence-advancing retrieval
state, a limit cycle in the laboratory frame, becomes a fixed point of a
\emph{twisted stationarity} condition on one period cell of the order-parameter
kernels. The synchronous machinery of Sec.~\ref{sec:two_species} is reused
wholesale; asynchrony enters only through the \emph{time-index structure}
(substep supports, staircase plateaus, block weights) of otherwise identical
objects.

The functional, its disorder average, and the conjugate
kernels below can be obtained analytically at any $\lambda$. The sharpest
quantitative results (the $\varpi=3$ orbit and the capacity
\eqref{eq:ab_capacity}) are obtained in the pure-relay limit
$\lambda\to\infty$, where the retarded self-interaction vanishes by the
hop-counting rule of Stage~4 and the theory sits in the same closable class
as the purely asymmetric sequence network. At finite $\lambda$ the same
cell solver reproduces the measured retrieval speed at all $\lambda\ge0.5$
and the island boundary for $\lambda\gtrsim1.25$ (Stage~5), with the
remaining discrepancies quantified there.

\begin{figure}[hbt!]
    \centering
    \includegraphics[width=0.8\linewidth]{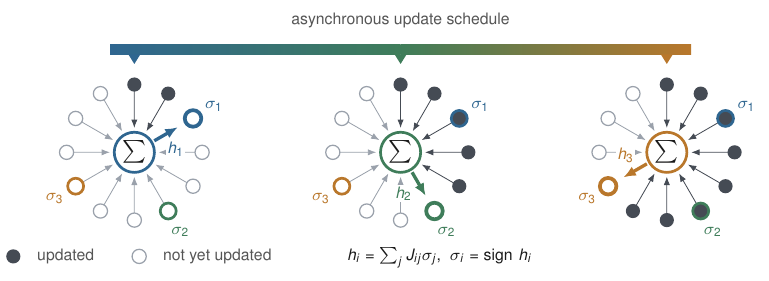}
    \caption{\label{fig:async_field_env}{\bf Asynchronous updating gives every
    neuron a different local-field environment.} As the sweep advances through
    the update schedule (colour bar), each neuron fires at a different point: when
    neuron $i$ updates, its local field $h_i=\sum_j J_{ij}\sigma_j$ mixes
    already-updated neighbours (filled) with not-yet-updated ones
    (open). The fraction of already-updated neurons---and hence the composition
    of $h_i$---differs for every neuron.}
\end{figure}

\subsection{Stage 0: model and schedule}
\label{sec:ab_stage0}

\paragraph{Units and fields.}
The hardware consists of $N$ \emph{units}, unit $i$ carrying the spin pair
$(x_i,y_i)\in\{-1,1\}^2$. With $\bm J=\tfrac1N\bm\xi^{\mathrm T}\bm\xi$ and
$\bm K=\tfrac1N\bm\xi^{\mathrm T}\bm S\bm\xi$ as before, the local fields are
those of Sec.~\ref{sec:two_species} [normalised as in \eqref{eq:h1_def}--\eqref{eq:h2_def}],
\begin{equation}\label{eq:ab_fields}
    h^1_i(n)=\frac{1}{1+\lambda}\sum_j J_{ij}x_j(n)+\frac{\lambda}{1+\lambda}\,y_i(n)+\theta^1_i(n),
    \qquad
    h^2_i(n)=\sum_j K_{ij}x_j(n)+\theta^2_i(n).
\end{equation}

At $\lambda\to\infty$ the readout field \eqref{eq:ab_fields} is dominated by
the relay, $x_i(n+1)=\mathrm{sign}\,y_i(n)=y_i(n)$ at every $x$-update
substep: the readout becomes a deterministic, one-substep-delayed copy of the
queue. The $J$-channel then decouples from the disorder---its weight
$(1+\lambda)^{-1}$ kills both its signal and its crosstalk---and the only
pattern coupling left is the $K$-channel. In this limit the entire
construction below collapses onto a \emph{single} disorder channel: of the
auxiliary fields only $\hat h^2$ survives, of the kernel families only
$C^J$, $Q^{22,J}$, $K^{21,J}$, and the effective single-unit problem closes
on the queue alone [Eq.~\eqref{eq:ab_single_unit_relay} below]. The reader
interested only in the solvable limit can apply this reduction at every
stage. We nonetheless carry general $\lambda$ through Stages~1--4, at the
price of heavier notation, because the finite-$\lambda$ conjugate-kernel
structure (the gate \eqref{eq:ab_gate}, the reaction kernel
\eqref{eq:ab_Khat11}, the cross-channel noise) is exact, connects this
section to Sec.~\ref{sec:two_species}, and is what locates the obstruction
to a finite-$\lambda$ closure (end of Stage~5).

The scheduled sweep---visit unit $i$, update $x_i$
reading its \emph{stale} $y_i$, then update $y_i$ reading the current
$\bm x$---reproduces the phenomenology of the main text exactly: the joint
$(x,y)$ condensed state cycles
\begin{equation}\label{eq:ab_cycle}
    (\mathrm{pure},\mathrm{pure})\;\to\;(\mathrm{pure},\mathrm{mixed})\;\to\;
    (\mathrm{mixed},\mathrm{pure})\;\to\;(\mathrm{pure},\mathrm{pure}),
\end{equation}
with period $\varpi=3$ sweeps and condensed-pattern advance $2$ per period
(speed $2/3$), the mixed member alternating between the species.

\begin{figure}[hbt!]
    \centering
    \includegraphics[width=0.99\linewidth]{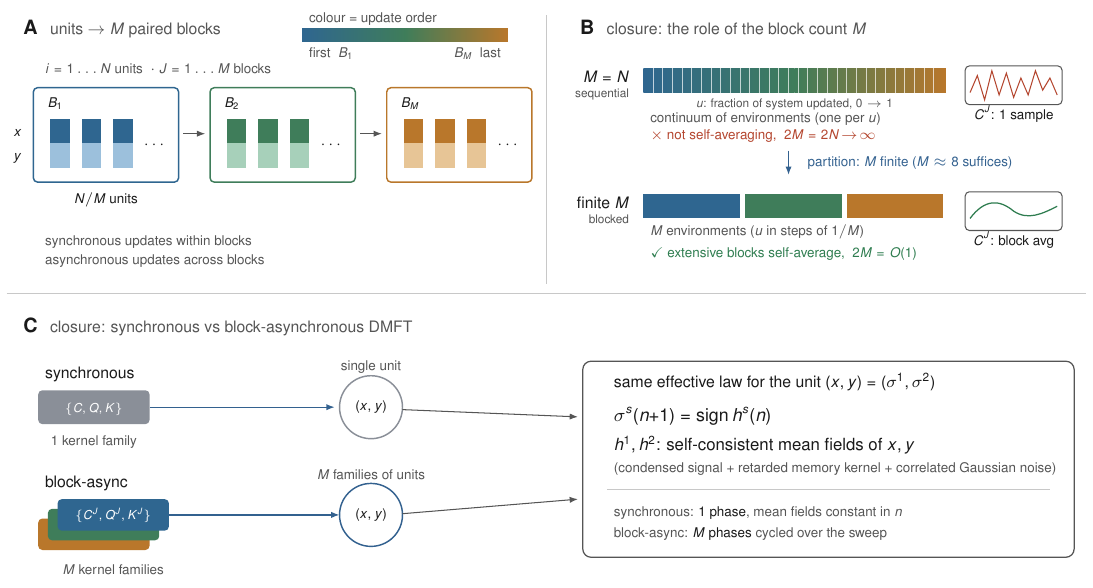}
    \caption{\label{fig:blocked_async} 
    {\bf Block construction for the generating-functional DMFT of the scheduled asynchronous dynamics.} 
    {\bf A} The $N$ units, each a spin pair $(x_i,y_i)$, are split into $M$ blocks of $N/M$ units in update order (colour): synchronous within a block, sequential across blocks ($x$ then $y$).
    {\bf B} A unit's environment is set by the fraction $u\in[0,1]$ of the system updated when it fires. Sequentially ($M=N$) $u$ is continuous, giving a continuum of single-unit environments whose kernels $C^J$ never self-average and whose number $2M=2N$ diverges; blocking samples $u$ at $M$ values, so for finite $M$ ($M\approx8$ suffices) the kernels are self-averaging block averages with $2M=O(1)$.
    {\bf C} Synchronous ($\{C,Q,K\}$) and block-asynchronous ($\{C^J,Q^J,K^J\}$) schedules share the effective single-unit law $\sigma^s(n{+}1)=\operatorname{sign}h^s(n)$, with $h^1,h^2$ the self-consistent mean fields of $x,y$, differing only in whether they are constant or cycled through $M$ phases per sweep.
    }
\end{figure}

\subsubsection{The blocked kernel ($M$ paired blocks).}
The entire effect of an update schedule on the dynamics lies in the pattern
of which spins have already updated within the current sweep at the moment a
given unit reads its field. In the sequential paired-unit sweep this pattern
varies continuously with the unit's position $u\in[0,1]$ in the sweep---a
mid-sweep unit sees, even among its own block of recent updaters, a mixture
of current and previous-sweep values---so the local-field environments are
heterogeneous at the level of single sites, and no finite set of
self-averaging order parameters describes them. The blocked kernel
\emph{organises} this heterogeneity: the units are partitioned into $M$
paired blocks $B_1,\dots,B_M$ in update order, $|B_J|=N/M$, with $x_i$ and
$y_i$ inheriting unit $i$'s block. Time is counted in substeps
$n=0,\dots,t-1$ with the period-$2M$ schedule defined by its update supports
\begin{equation}\label{eq:ab_schedule}
    \mathcal T^1_J=\big\{n:\ n\equiv 2(J{-}1)\ \mathrm{mod}\ 2M\big\},\qquad
    \mathcal T^2_J=\big\{n:\ n\equiv 2J{-}1\ \mathrm{mod}\ 2M\big\},\qquad
    J=1,\dots,M:
\end{equation}
at $n\in\mathcal T^1_J$ species $x$ of block $B_J$ updates, at
$n\in\mathcal T^2_J$ species $y$ of block $B_J$; each substep is a
\emph{parallel} Glauber update of that species-block at inverse temperature
$\beta$, all other spins frozen:
\begin{equation}\label{eq:ab_kernel}
    W_n[\bm\sigma(n+1)|\bm\sigma(n)]
    =\!\!\prod_{i\in B_{J(n)}}\!\!
    \frac{e^{\beta \sigma^{s(n)}_i(n+1)\,h^{s(n)}_i(n)}}{2\cosh\beta h^{s(n)}_i(n)}
    \times\!\!\prod_{\text{others}}\!\!\delta_{\sigma(n+1),\sigma(n)},
    \qquad \sigma^{1}\equiv x,\ \ \sigma^{2}\equiv y,
\end{equation}
where $s(n)\in\{1,2\}$ and $J(n)\in\{1,\dots,M\}$ are read off
\eqref{eq:ab_schedule} (i.e.\ $n\in\mathcal T^{s(n)}_{J(n)}$), the fields are
built from the \emph{current} state $\bm\sigma(n)=(\bm x(n),\bm y(n))$, and
$\mathcal T^s=\bigcup_J\mathcal T^s_J$. 

At any substep, the network seen by
an updating unit now consists of \emph{whole} blocks, each either fully
updated this sweep or still fully holding its previous-sweep values: the local
fields remain heterogeneous, but they take only $M$ distinct macroscopic
forms, one per block class, each shared by an extensive set of $N/M$ units.
This is precisely what a saddle-point theory requires---the block-resolved
kernels introduced below are averages over extensive classes and
self-average, and the time lattice per sweep has $2M=O(1)$ substeps, so the
saddle is over finitely many order parameters. The blocks thereby discretise
the update position $u$ in steps of $1/M$; the sequential sweep ($M=N$, one
unit per block) is recovered in the limit $M\to\infty$ taken \emph{after}
$N\to\infty$ (converged by $M\approx8$ in MC), while $M=2$ is the minimal
member, the direct analogue of the two-block loop-space closure. Two further
structural points: (i) at $n\in\mathcal T^2_J$ the
field $h^2$ already contains block $J$'s \emph{fresh} $x$ values (updated at
$n-1$)---the pair order of the sequential sweep is carried by the substep
chain, no skew bookkeeping needed; (ii) both species are staircases, $x_i$
changing only at $n\in\mathcal T^1_{J(i)}$ and $y_i$ at
$n\in\mathcal T^2_{J(i)}$, so all correlators inherit plateau structure. This
staircase is the \emph{exact} form of the staleness that App.~F's
statistical-neurodynamics closure modelled by the position mixture
$f\,m(T)+(1-f)\,m(T-\omega)$.

\paragraph{Macroscopic periods are included in the saddle ansatz, not input.}
Zero-temperature MC of the kernel family \eqref{eq:ab_kernel} retrieves
stably with periodically recurring mixed states at every $M$ tested, and
converges to the sequential schedule's $\varpi=3$ cell and speed $2/3$ by
$M\approx8$, at all $(\alpha,\lambda)$ tested
\footnote{Small $M$ can lock elsewhere: at $M=2$ the orbit sits on a \emph{swap-twisted ten-substep} cell---invariant under $(n\to n+10)\circ(B_1\leftrightarrow B_2)\circ(\mu\to\mu+2)$, i.e.\ laboratory period $5$ sweeps with pattern advance $4$ (speed $4/5$).}.
Nothing in Stages 1--4 below depends on any of this: the disorder average and the conjugate kernels hold for arbitrary trajectories, arbitrary $M$, and arbitrary fixed blocked schedules. 
The period enters only at Stage 5, as a
choice of stationarity ansatz, exactly as a nonzero overlap is an ansatz in
the static theory, and the same cell equations cover any $(q,\Delta)$~\footnote{This
is reminiscent of the heuristic loop-space theory, which evaluates two-block kernels on
the $\varpi=3$ cell of the physical schedule.}.

\paragraph{Assumptions.}
(A1) $t=O(1)$ as $N\to\infty$ (thermodynamic limit first).
(A2) Condensed ansatz: at every substep each species-block has $O(1)$ overlap
with at most two adjacent channels; all other overlaps are $O(N^{-1/2})$.
Imposed at the saddle, not in the functional.
(A3) Twisted stationarity (Stage 5 only).

\subsection{Stages 1--3: generating functional and disorder average}
\label{sec:ab_stage13}

\subsubsection{Generating functional and auxiliary fields.}
With sources for both species,
\begin{equation}\label{eq:ab_Z}
    Z[\bm\psi]=\sum_{\{\bm x(n),\bm y(n)\}}p[\bm\sigma(0)]
    \prod_{n=0}^{t-1}W_n[\bm\sigma(n+1)|\bm\sigma(n)]\,
    e^{-\ii\sum_{n<t}[\bm\psi^1(n)\cdot\bm x(n)+\bm\psi^2(n)\cdot\bm y(n)]},
\end{equation}
$Z[0]=1$, so the disorder average acts directly. Auxiliary fields are inserted
\emph{only where a field acts}: $(h^1_i(n),\hat h^1_i(n))$ for
$n\in\mathcal T^1$, $i\in B_{J(n)}$, and $(h^2_i(n),\hat h^2_i(n))$ for
$n\in\mathcal T^2$, $i\in B_{J(n)}$. The disorder then appears in exactly two
bilinears,
\begin{align}
    E_J&=-\frac{\ii}{1+\lambda}\,\frac1N\sum_{n\in\mathcal T^1}\sum_\mu
    \Big(\sum_{i\in B_{J(n)}}\xi_i^{\mu}\hat h^1_i(n)\Big)
    \Big(\sum_{j}\xi_j^{\mu}x_j(n)\Big), \label{eq:ab_EJ}\\
    E_K&=-\ii\,\frac1N\sum_{n\in\mathcal T^2}\sum_\mu
    \Big(\sum_{i\in B_{J(n)}}\xi_i^{\mu+1}\hat h^2_i(n)\Big)
    \Big(\sum_{j}\xi_j^{\mu}x_j(n)\Big), \label{eq:ab_EK}
\end{align}
while the relay term
$-\ii\tfrac{\lambda}{1+\lambda}\sum_{n\in\mathcal T^1}\hat h^1_i(n)\,y_i(n)$
is \emph{local} and disorder-free: it stays in the single-site measure, with
$y_i(n)$ at its current staircase value. The spin $y$ never couples to the
patterns. Note the now-familiar asymmetry: the $\hat h$-sums run over the
updating block only, the $x$-sums over all units (the field reads the frozen
block at its stale value). The $(1+\lambda)^{-1}$ weight of the symmetric
channel is carried explicitly from here on (cf.\ the normalisation remark in
Sec.~\ref{sec:two_species}).


\paragraph{Condensed sector.}
The physical content of the condensed ansatz (A3) is the assumption that at
any given time there exists an $O(1)$ set $\mathcal C$ of patterns carrying
macroscopic overlap: sequence retrieval confines the dynamics to a
travelling $O(1)$ window of the $p=\alpha N$ stored patterns, all others
remaining at $O(N^{-1/2})$. For channels $\mu\in\mathcal C$ ($\le2$ per
species-block per substep by (A3)) define block-resolved overlaps and the
channel multipliers on the updating block,
\begin{equation}\label{eq:ab_cond_ops}
    m_J^\mu(n)=\frac MN\sum_{j\in B_J}\xi_j^\mu x_j(n),\qquad
    m^\mu(n)=\frac1M\sum_{J=1}^{M} m_J^\mu(n),\qquad
    k^1_\mu(n)\ (n\in\mathcal T^1),\quad k^2_\mu(n)\ (n\in\mathcal T^2),
\end{equation}
with $k^1_\mu(n)=\tfrac MN\sum_{i\in B_{J(n)}}\xi_i^{\mu}\hat h^1_i(n)$ and
$k^2_\mu(n)=\tfrac MN\sum_{i\in B_{J(n)}}\xi_i^{\mu+1}\hat h^2_i(n)$. The
condensed parts of \eqref{eq:ab_EJ}--\eqref{eq:ab_EK} are extensive,
\begin{equation}\label{eq:ab_cond_bilinear}
    E^{\mathrm{cond}}=-\frac{\ii N}{M}\sum_{\mu\in\mathcal C}\Big[
    \frac{1}{1+\lambda}\sum_{n\in\mathcal T^1}k^1_\mu(n)\,m^\mu(n)
    +\sum_{n\in\mathcal T^2}k^2_\mu(n)\,m^\mu(n)\Big]:
\end{equation}
\emph{all} blocks enter each signal at weight $1/M$, the frozen blocks at
their stale staircase values---the exact field of the loop-space theory, here
built in rather than posited. As in the single-channel-pair situation of the
synchronous solver, two simultaneously condensed channels prevent gauging the
pattern labels away; the quenched pair $(\xi_i^{\nu},\xi_i^{\nu+1})$, i.e.\
$\eta_i=\xi_i^\nu\xi_i^{\nu+1}=\pm1$, survives into the single-site measure.

\begin{figure}[h!]
    \centering
    \includegraphics[width=3.4in]{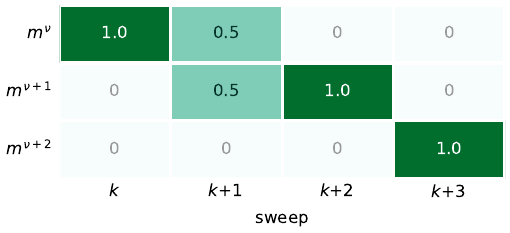}
    \caption{\label{fig:ab_window} {\bf The condensed window and its
    occupancy along the retrieval orbit.} The channels shown constitute an example of the
    condensed set $\mathcal C$ in asynchronous dynamics: the travelling $O(1)$ family of patterns
    assumed in (A3) to carry all macroscopic overlap at any given time, every
    other pattern remaining at $O(N^{-1/2})$ (the non-condensed set
    $\mathcal N$, whose collective effect is the crosstalk noise).
    Block-averaged $x$-overlaps with the window channels at sweep ends
    ($\lambda\gg1$, schematic values): a diagonal band---pure, mixed,
    pure---advancing two channels per three sweeps. The window definitions
    reserve $O(1)$ slots for all channels at all substeps; the band, i.e.\
    the time-locking of channel to sweep, is an output of the saddle (and the
    $y$-species band is the same staggered by one substep).}
\end{figure}

Note that, in contrast to the synchronous calculations---where the
time-locking $\mu=\tau$ could be hardwired into the definitions because the
orbit was known and trivial---here \emph{all} channels of the window
$\mathcal C$ are tracked at \emph{all} substeps. Which slot is macroscopic at
which $(n,J)$, including the half-occupied mixed entries and the stagger
between blocks, is the content of the orbit (period, speed, mixed-state
structure) and is filled in by the saddle: the occupancy forms a diagonal
band travelling through the window (Fig.~\ref{fig:ab_window}), inactive slots
sitting at $\approx0$ at no cost to the saddle-point bookkeeping. The
time-locking returns at Stage 5 as the twisted-stationarity ansatz
\eqref{eq:ab_twisted_kernels}, imposed on the saddle rather than on the
definitions.

\paragraph{Non-condensed sector and the index shift.}
Now we turn to the complementary set $\mathcal N=\{1,\dots,p\}\setminus\mathcal C$ which will be disorder average: its $\alpha N-O(1)$ overlaps are individually
$O(N^{-1/2})$, but their extensive number makes their collective effect---the
crosstalk---the noise of the theory. While the $\xi_i^\mu$ with
$\mu\in\mathcal C$ remain explicit quenched variables in the single-unit
measure, the $\mu\in\mathcal N$ components are averaged out below, and the
kernels of the disorder average exist precisely to describe their
statistics. (Since the window travels, a channel that has recently left
$\mathcal C$ retains an anomalously large, decaying overlap for a few sweeps;
it is kept in the window until relaxed, and the $\mathcal C$/$\mathcal N$
boundary contributes only sub-extensive corrections to the exponent.)
For $\mu\in\mathcal N$ define, with $N^{-1/2}$ scaling,
\begin{equation}\label{eq:ab_noncond_ops}
    Y^J_\mu(n)=\frac{1}{\sqrt N}\sum_{j\in B_J}\xi_j^\mu x_j(n)\ \ \forall n,\qquad
    u_\mu(n)=\frac{1}{\sqrt N}\!\!\sum_{i\in B_{J(n)}}\!\!\xi_i^{\mu}\hat h^1_i(n)\ \ n\in\mathcal T^1,\qquad
    v_\mu(n)=\frac{1}{\sqrt N}\!\!\sum_{i\in B_{J(n)}}\!\!\xi_i^{\mu+1}\hat h^2_i(n)\ \ n\in\mathcal T^2,
\end{equation}
so the non-condensed bilinear is
$-\ii\sum_\mu\{\tfrac{1}{1+\lambda}\sum_{\mathcal T^1}u_\mu(n)+\sum_{\mathcal T^2}v_\mu(n)\}\,\sum_J Y^J_\mu(n)$
(the factor in braces evaluated on its own support). Note that
$k^s_\mu,u_\mu,v_\mu$ carry no explicit block label $J$. This is because $\hat h(n)$ exists only on
the block $B_{J(n)}$ updating at $n$, so their block is fixed by the substep, $J=J(n)$, and the $J$-resolution of the conjugate sector reappears below in the kernels' supports. 

Enforcing each definition in \eqref{eq:ab_noncond_ops} by a Fourier conjugate
($\hat u_\mu,\hat v_\mu,\hat Y^J_\mu$) renders the $\mu\in\mathcal N$ pattern
dependence of the functional \emph{linear}: the bilinears
\eqref{eq:ab_EJ}--\eqref{eq:ab_EK} are rewritten in terms of the free
variables $u,v,Y^J$, which carry no $\xi$, and the non-condensed patterns
survive only in the enforcement exponent
$-\tfrac{\ii}{\sqrt N}\sum_{\mu,n}[\hat u_\mu\,\xi^\mu\hat h^1
+\hat v_\mu\,\xi^{\mu+1}\hat h^2+\hat Y^J_\mu\,\xi^\mu x]$. The site-by-site
average needs the total coefficient of each single $\xi_i^\mu$, which is
collected by shifting the dummy index on the periodic pattern ring in the
$\hat v$-term,
\begin{equation}\label{eq:ab_ringshift}
    \sum_{\mu\in\mathcal N}\hat v_\mu(n)\,\xi_i^{\mu+1}\,\hat h^2_i(n)
    \;=\;\sum_{\mu\in\mathcal N}\hat v_{\mu-1}(n)\,\xi_i^{\mu}\,\hat h^2_i(n).
\end{equation}
Consequently every product involving $\hat v_{\mu-1}$ in the squared charge
below ties pattern $\mu$ to pattern $\mu-1$: a \emph{one-step hop on the
pattern ring}, and such hops occur only at $\mathcal T^2$ ($K$-channel)
substeps. The schedule thus reschedules the hop \emph{times} but never the
hop \emph{length}---the pattern-index structure is that of the synchronous
calculation, and this is the single structural fact from which the $\hat K$
hop-counting rule below follows.

\subsubsection{Disorder average.}
Fix $i\in B_J$ and $\mu\in\mathcal N$. The total charge multiplying $\xi_i^\mu$ is
\begin{equation}\label{eq:ab_charge}
    \zeta_i^\mu=\sum_{n\in\mathcal T^1_J}\hat u_\mu(n)\,\hat h^1_i(n)
    +\sum_{n\in\mathcal T^2_J}\hat v_{\mu-1}(n)\,\hat h^2_i(n)
    +\sum_{n<t}\hat Y^J_\mu(n)\,x_i(n),
\end{equation}
and the Rademacher disorder average gives
$\overline{e^{-\ii N^{-1/2}\xi_i^\mu\zeta_i^\mu}}=\exp[-(\zeta_i^\mu)^2/2N+O(N^{-2})]$.
Expanding the square and summing over $i\in B_J$ generates six kernel families
per block---and only these (sites of different blocks never pair; $y$ never
appears in $\zeta_i^\mu$, so there is no $C^{22}$, $C^{12}$, $K^{12}$ or $K^{22}$):
\begin{equation}\label{eq:ab_kernels}
\begin{aligned}
    C^J(n,n')&=\tfrac MN\textstyle\sum_{i\in B_J}x_i(n)x_i(n') &&\text{all }n,n',\\
    Q^{11,J}(n,n')&=\tfrac MN\textstyle\sum_{i\in B_J}\hat h^1_i(n)\hat h^1_i(n') && n,n'\in\mathcal T^1_J,
    \qquad Q^{22,J}:\ n,n'\in\mathcal T^2_J,\qquad Q^{12,J}:\ n\in\mathcal T^1_J,\ n'\in\mathcal T^2_J,\\
    K^{11,J}(n,n')&=\tfrac MN\textstyle\sum_{i\in B_J}x_i(n)\hat h^1_i(n') && n'\in\mathcal T^1_J,\ n\ \text{arbitrary},
    \qquad K^{21,J}:\ n'\in\mathcal T^2_J,\ n\ \text{arbitrary},
\end{aligned}
\end{equation}
with
\begin{align}\label{eq:ab_disorder_factor}
    -\frac{1}{2N}\sum_{\mu\in\mathcal N}\sum_{i}(\zeta_i^\mu)^2
    =-\sum_{\mu\in\mathcal N}\sum_{J}\bigg[&
    \frac1{2M}\,\hat u_\mu Q^{11,J}\hat u_\mu
    +\frac1{2M}\,\hat v_{\mu-1}Q^{22,J}\hat v_{\mu-1}
    +\frac1M\,\hat u_\mu Q^{12,J}\hat v_{\mu-1}
    +\frac1{2M}\,\hat Y^J_\mu C^J\hat Y^J_\mu \nonumber\\
    &+\frac1M\,\hat Y^J_\mu K^{11,J}\hat u_\mu
    +\frac1M\,\hat Y^J_\mu K^{21,J}\hat v_{\mu-1}\bigg],
\end{align}
where, here and throughout, contracted bilinears abbreviate two-time sums
over the kernel's support,
\begin{equation}\label{eq:ab_contraction}
    \hat u_\mu Q^{11,J}\hat u_\mu
    \equiv\!\!\sum_{n,n'\in\mathcal T^1_J}\!\!\hat u_\mu(n)\,Q^{11,J}(n,n')\,\hat u_\mu(n'),
    \qquad
    \hat Y^J_\mu K^{11,J}\hat u_\mu
    \equiv\sum_{n<t}\,\sum_{n'\in\mathcal T^1_J}\hat Y^J_\mu(n)\,K^{11,J}(n,n')\,\hat u_\mu(n'),
\end{equation}
and likewise for the remaining terms, with supports as listed in
\eqref{eq:ab_kernels}. Note the
structural contrast with the synchronous calculation, where a single kernel
set $C,Q^{ab},K^{a1}$ lives on the full time lattice: the exponent
\eqref{eq:ab_disorder_factor} decomposes as a sum over blocks of quadratic
forms, each built from block $J$'s kernels on block $J$'s supports. As a
quadratic form in the conjugate variables it is block-diagonal in the
$\hat Y^J$---the average cannot generate $\hat Y^J$--$\hat Y^{J'}$ couplings,
since $(\zeta_i^\mu)^2$ is a single-site square and sites of different blocks
never pair---while $\hat u_\mu,\hat v_\mu$ are shared by all summands: (besides the condensed signals $m^\mu$), they
are the only lane [pinned in \eqref{eq:ab_collapse} below] through which the blocks communicate at all.

Enforcing
\eqref{eq:ab_cond_ops} and \eqref{eq:ab_kernels} with conjugate multipliers
yields the saddle form
$\overline{Z[\bm\psi]}=\int\dd\{\text{order params}\}\,e^{N[\Psi+\Phi+\Omega]+O(N^{1/2})}$,
with $\Phi$ the single-\emph{unit} measure and $\Omega$ the non-condensed
Gaussian, as we will shortly define below, and finally $\Psi$ collecting the
multiplier products and the condensed bilinear \eqref{eq:ab_cond_bilinear}.
Normalising each insertion as
$\exp\{\ii\tfrac NM\hat\Lambda[\Lambda-\tfrac MN\sum_{i\in B_J}(\cdot)]\}$,
\begin{align}\label{eq:ab_Psi}
    \Psi=\frac{\ii}{M}\Bigg\{&\sum_{\mu\in\mathcal C}\bigg[
    \sum_{n<t}\sum_J\hat m_J^\mu(n)\,m_J^\mu(n)
    +\frac{1}{1+\lambda}\!\!\sum_{n\in\mathcal T^1}\!\!\hat k^1_\mu(n)\,k^1_\mu(n)
    +\!\!\sum_{n\in\mathcal T^2}\!\!\hat k^2_\mu(n)\,k^2_\mu(n)\nonumber\\
    &\qquad
    -\frac{1}{1+\lambda}\!\!\sum_{n\in\mathcal T^1}\!\!k^1_\mu(n)\,m^\mu(n)
    -\!\!\sum_{n\in\mathcal T^2}\!\!k^2_\mu(n)\,m^\mu(n)\bigg]
    +\sum_J\Big[\hat q^J C^J+\hat Q^{ab,J}Q^{ab,J}
    +\hat K^{a1,J}K^{a1,J}\Big]\Bigg\},
\end{align}
with the contractions of \eqref{eq:ab_contraction}, summation over repeated
species indices, and the per-unit deposits of the same multipliers appearing
in \eqref{eq:ab_M}. The architecture is that of the synchronous $\Psi$ of
Sec.~\ref{sec:two_species}---multiplier products, the condensed bilinear in
order-parameter language, kernel multipliers---with block resolution,
supports and $1/M$ weights as the only changes.


\subsubsection{$\Phi$ and the single-unit measure.}
Let us start with the functional $\Phi$, whose origin is mechanical. After the disorder average, every
coupling between \emph{different} units $(x_i, y_i)$ has been traded for an order
parameter; what remains of the microscopic variables in the integrand is
(i) each unit's Glauber weights and field constraints, with the condensed
multipliers $\hat k^s_\mu$ attached to its quenched window components
$\xi_i^\mu$, (ii) the local relay tying $x_i$ to $y_i$, and (iii) the
per-unit terms $-\ii\hat q\,xx-\ii\hat Q\,\hat h\hat h-\ii\hat K\,x\hat h$
deposited by the Fourier multipliers that enforce the kernel definitions
\eqref{eq:ab_kernels} [contractions as in \eqref{eq:ab_contraction}]. The
interactions between units having been traded away, these conjugate kernels
are the \emph{effective two-time self-couplings} along one unit's own
trajectory that replace them; at the saddle, $\hat q$ will vanish, $\hat Q$
will become the Gaussian noise covariance and $\hat K$ the retarded
self-interaction of the effective dynamics. All of this is local in the unit index, so the
integrand is a product over units of factors of identical form; naming the
total path weight of unit $i$'s pair $\mathcal M_i$, the dynamic part of the
exponent is, \emph{by definition},
$\Phi=\tfrac1N\sum_i\ln\mathcal M_i$.
$\Phi$ thus factorises over the \emph{units} $(x_i,y_i)$, organised in $M$ block classes,
\begin{equation}\label{eq:ab_Phi_split}
    \Phi=\frac1M\sum_{J=1}^{M}\big\langle \ln\mathcal M_i\big\rangle_{i\in B_J},
    \qquad
    \big\langle\,\cdot\,\big\rangle_{i\in B_J}=\frac MN\sum_{i\in B_J}(\,\cdot\,).
\end{equation}
For $i\in B_J$ the path measure $\mathcal M_i$ couples the spin pair $(x,y)$
of one unit through (i) the local relay $\tfrac{\lambda}{1+\lambda}y(n)$
inside $h^1$, (ii) the shared quenched components
$\{\xi_i^\mu\}_{\mu\in\mathcal C}$, and---after the saddle---(iii) correlated
noise and the retarded kernel. Explicitly,
\begin{align}\label{eq:ab_M}
    \mathcal M_i=\sum_{\{x,y\}}\int\!\!\prod_{n\in\mathcal T^1_J}\!\!\frac{\dd h^1\dd\hat h^1}{2\pi}
    \prod_{n\in\mathcal T^2_J}\!\!\frac{\dd h^2\dd\hat h^2}{2\pi}\;p_i(x(0),y(0))\,
    &\exp\Big\{\sum_{s=1,2}\sum_{n\in\mathcal T^s_J}\big[\beta \sigma^s(n+1)h^s(n)-\ln2\cosh\beta h^s(n)\big]\Big\}\nonumber\\
    \times\;&\exp\Big\{\ii\!\!\sum_{n\in\mathcal T^1_J}\!\!\hat h^1(n)\Big[h^1(n)-\theta^1(n)
    -\tfrac{\lambda}{1+\lambda}y(n)-\tfrac{1}{1+\lambda}\!\sum_{\mu\in\mathcal C}\hat k^1_\mu(n)\xi_i^{\mu}\Big]\Big\}\nonumber\\
    \times\;&\exp\Big\{\ii\!\!\sum_{n\in\mathcal T^2_J}\!\!\hat h^2(n)\Big[h^2(n)-\theta^2(n)
    -\sum_{\mu\in\mathcal C}\hat k^2_\mu(n)\xi_i^{\mu+1}\Big]\Big\}\nonumber\\
    \times\;&\exp\Big\{-\ii\sum_n[\hat m\text{-terms}+\psi\text{-terms}]\Big\}\nonumber\\
    \times\;&\exp\Big\{-\ii\,\hat q^J\,xx
    -\ii\,\hat Q^{ab,J}\,\hat h^a\hat h^b
    -\ii\,\hat K^{a1,J}\,x\hat h^a\Big\},
\end{align}
with $\sigma^1=x$, $\sigma^2=y$, summation over repeated species indices
$a,b$, and the contraction convention of \eqref{eq:ab_contraction}. The
relay is a \emph{sampled binary spin} inside $h^1$, exactly as in the
synchronous two-species map.

\subsubsection{$\Omega$ and the shared noise source.}
The structure of $\Omega$ is likewise inherited from the network architecture: it is the receipt of the
non-condensed sector, as $\Phi$ was the receipt of the surviving microscopic
spins. That sector contributed three families of integration variables---the
overlaps $u_\mu,v_\mu,Y^J_\mu$ and their Fourier conjugates---and three kinds
of terms: the Fourier pairings
$\ii[\hat u_\mu u_\mu+\hat v_\mu v_\mu+\sum_J\hat Y^J_\mu Y^J_\mu]$, the
bilinear through which the crosstalk enters the fields [below
\eqref{eq:ab_noncond_ops}], and the disorder factor
\eqref{eq:ab_disorder_factor}, quadratic in the hatted variables with the
kernels as coefficients. Everything is at most quadratic, so the sector is
one Gaussian integral per pattern $\mu\in\mathcal N$ and, by definition,
$\Omega=\tfrac1N\sum_{\mu\in\mathcal N}\ln\int(\cdots)$: extensive because
$|\mathcal N|=\alpha N-O(1)$, and a circulant chain in $\mu$ because of the
one-step hops. Its role at the saddle is dual to $\Phi$'s: variations of
$\Phi$ reproduce the kernels as single-unit averages, while variations of
$\Omega$ with respect to the kernels produce the conjugates
$\hat q,\hat Q,\hat K$ appearing in $\mathcal M_i$---the noise covariances
and retarded self-interactions of the effective dynamics. Explicitly,
\begin{equation}\label{eq:ab_Omega_raw}
    \Omega=\frac1N\sum_{\mu\in\mathcal N}\ln\!\int\!
    \dd u_\mu\,\dd\hat u_\mu\,\dd v_\mu\,\dd\hat v_\mu
    \prod_J \dd Y^J_\mu\,\dd\hat Y^J_\mu\;
    e^{\,\mathcal A_\mu},
\end{equation}
with the per-pattern action
\begin{align}\label{eq:ab_Omega_action}
    \mathcal A_\mu=\;&\ii\!\!\sum_{n\in\mathcal T^1}\!\!\hat u_\mu(n)u_\mu(n)
    +\ii\!\!\sum_{n\in\mathcal T^2}\!\!\hat v_\mu(n)v_\mu(n)
    +\ii\sum_{n<t}\sum_J\hat Y^J_\mu(n)Y^J_\mu(n)\nonumber\\
    &-\;\ii\,\frac{1}{1+\lambda}\!\!\sum_{n\in\mathcal T^1}\!\!u_\mu(n)\,Y_\mu(n)
    \;-\;\ii\!\!\sum_{n\in\mathcal T^2}\!\!v_\mu(n)\,Y_\mu(n)
    \;+\;\big[\text{the $\mu$-summand of \eqref{eq:ab_disorder_factor}}\big],
\end{align}
where $Y_\mu(n):=\sum_{J=1}^{M}Y^J_\mu(n)$ is the total overlap fluctuation.

In this Gaussian, $u_\mu(n)$ and $v_\mu(n)$ appear linearly;
integrating them out pins their conjugates to the \emph{same} physical
overlap fluctuation, channel-weighted:
\begin{equation}\label{eq:ab_collapse}
    \hat u_\mu(n)=\frac{1}{1+\lambda}\,Y_\mu(n)\quad(n\in\mathcal T^1),\qquad
    \hat v_\mu(n)=Y_\mu(n)\quad(n\in\mathcal T^2).
\end{equation}
There is one crosstalk lane (i.e. the single aggregate overlap fluctuation $Y_\mu=\sum_J Y^J_\mu$), read by both channels and all blocks: all
cross-block and cross-species noise correlations are downstream consequences
of \eqref{eq:ab_collapse}. The evaluation of \eqref{eq:ab_Omega_raw} proceeds by exact
sequential elimination---one variable family is integrated out at a time, while the defining integral keeps its
value---and each display below is the integrand's action at the
corresponding stage.


Substituting \eqref{eq:ab_collapse} into the disorder factor reduces the
integrand to the $(Y^J_\mu,\hat Y^J_\mu)$ sector,
$\Omega=\tfrac1N\sum_{\mu}\ln\int\prod_J\dd Y^J_\mu\,\dd\hat Y^J_\mu\,
e^{\mathcal A'_\mu}$, with the reduced action
\begin{align}\label{eq:ab_Omega_reduced}
    \mathcal A'_\mu=\;\ii\sum_{n,J}\hat Y^J_\mu(n)Y^J_\mu(n)
    &-\frac1{2M}\sum_J\hat Y^J_\mu C^J\hat Y^J_\mu\nonumber\\
    &-\frac1M\sum_J\hat Y^J_\mu(n)\Big[\tfrac{1}{1+\lambda}K^{11,J}(n,n')\,Y_\mu(n')\big|_{n'\in\mathcal T^1_J}
    +K^{21,J}(n,n')\,Y_{\mu-1}(n')\big|_{n'\in\mathcal T^2_J}\Big]\nonumber\\
    &-\frac1{2M}\sum_J\Big[\tfrac{1}{(1+\lambda)^2}\,Y_\mu Q^{11,J}Y_\mu
    +Y_{\mu-1}Q^{22,J}Y_{\mu-1}
    +\tfrac{2}{1+\lambda}\,Y_\mu Q^{12,J}Y_{\mu-1}\Big],
\end{align}
a Gaussian chain, circulant in $\mu$. 

Naming the response coupling---the
bracket in the second term of \eqref{eq:ab_Omega_reduced}---
\begin{equation}\label{eq:ab_L}
    L^J_\mu(n):=\frac{1}{1+\lambda}\sum_{n'\in\mathcal T^1_J}K^{11,J}(n,n')\,Y_\mu(n')
    +\sum_{n'\in\mathcal T^2_J}K^{21,J}(n,n')\,Y_{\mu-1}(n'),
\end{equation}
the $\hat Y^J$ are integrated by completing the square
[$\int\dd\hat Y\,e^{-\frac{1}{2M}\hat Y C\hat Y+\hat Y\cdot b}\propto
e^{\frac M2\,b^{\top}C^{-1}b}$ with $b=\ii\,Y^J_\mu-\tfrac1M L^J_\mu$; the
blocks do not mix, by the block-diagonality noted above], leaving per block
\begin{equation}\label{eq:ab_yhat_int}
    -\frac M2\,Y^J_\mu\,(C^J)^{-1}Y^J_\mu
    \;-\;\ii\,Y^J_\mu\,(C^J)^{-1}L^J_\mu
    \;+\;\frac1{2M}\,L^J_\mu\,(C^J)^{-1}L^J_\mu,
\end{equation}
the three terms being, respectively, the entropic cost of an overlap
fluctuation, its propagation through the responses, and a contribution
quadratic in the kernels. At this stage
$\Omega=\tfrac1N\sum_\mu\ln\int\prod_J\dd Y^J_\mu\;e^{\mathcal A''_\mu}$,
where $\mathcal A''_\mu$ collects the per-block terms \eqref{eq:ab_yhat_int}
and the $Q$-terms of \eqref{eq:ab_Omega_reduced}, which the
$\hat Y$-integration left untouched; explicitly,
\begin{align}\label{eq:ab_App}
    \mathcal A''_\mu=\sum_{J=1}^{M}\Big[
    -\frac M2\,Y^J_\mu\,(C^J)^{-1}Y^J_\mu
    -\ii\,Y^J_\mu\,(C^J)^{-1}L^J_\mu
    +\frac1{2M}\,L^J_\mu\,(C^J)^{-1}L^J_\mu\Big]\nonumber\\
    -\frac1{2M}\sum_{J=1}^{M}\Big[\tfrac{1}{(1+\lambda)^2}\,Y_\mu Q^{11,J}Y_\mu
    +Y_{\mu-1}Q^{22,J}Y_{\mu-1}
    +\tfrac{2}{1+\lambda}\,Y_\mu Q^{12,J}Y_{\mu-1}\Big],
\end{align}
with $L^J_\mu$ as in \eqref{eq:ab_L}:
only the $Y^J$ remain, coupled across blocks
solely through $Y_\mu=\sum_J Y^J_\mu$, and \emph{every response insertion through
the $K$-channel carries exactly one unit of pattern shift} ($Y_{\mu-1}$), and unsurprisingly
the symmetric $J$-channel insertions none. 

The last elimination uses the structure of $\mathcal A''_\mu$ in the pattern
index: $\mu$ couples only to itself and to $\mu-1$ (the hops), with
coefficients independent of $\mu$---all non-condensed patterns are
statistically equivalent after the relabelling \eqref{eq:ab_ringshift}, and
the stored sequence is periodic, $\bm\xi^{p+1}=\bm\xi^1$. The quadratic form
over $\mu=1,\dots,p$ is therefore a \emph{circulant} chain, diagonalised by
the discrete Fourier transform
\begin{equation}\label{eq:ab_DFT}
    Y^J_\mu(n)=\frac{1}{\sqrt p}\sum_{\omega}e^{\ii\omega\mu}\,
    \tilde Y^J_\omega(n),\qquad
    \omega=\frac{2\pi k}{p},\ \ k=0,\dots,p-1,\qquad
    \tilde Y^J_{-\omega}=\big(\tilde Y^J_{\omega}\big)^{*},
\end{equation}
under which same-pattern couplings acquire no phase while each hop
contributes exactly one factor $e^{-\ii\omega}$,
\begin{equation}\label{eq:ab_hop_phase}
    \sum_\mu Y_\mu\,A\,Y_\mu\;\to\;\sum_\omega \tilde Y_\omega^{*}A\,\tilde Y_\omega,
    \qquad
    \sum_\mu Y_\mu\,B\,Y_{\mu-1}\;\to\;\sum_\omega e^{-\ii\omega}\,
    \tilde Y_\omega^{*}B\,\tilde Y_\omega .
\end{equation}
Note that, while this pattern-space ($\mu$) transform is available already at the level of the action, since all non-condensed patterns are exactly equivalent; the time direction ($n$) carries no such symmetry at this stage — sources, the initial condition and the transient live there. 


Since different modes no longer couple, the integral factorises over
$\omega$, and each mode is an ordinary finite-dimensional Gaussian. To write
its quadratic form, collect block $J$'s response kernels at mode $\omega$
into the matrix that the transform makes out of $L^J_\mu$
\eqref{eq:ab_L},
\begin{equation}\label{eq:ab_W}
    W^J(\omega):=\frac{1}{1+\lambda}\,K^{11,J}
    +e^{-\ii\omega}\,K^{21,J},
    \qquad\text{(columns on $\mathcal T^1_J$ and $\mathcal T^2_J$
    respectively)},
\end{equation}
so that $\tilde L^J_\omega=W^J(\omega)\,\tilde Y_\omega$ with
$\tilde Y_\omega=\sum_J\tilde Y^J_\omega$ the transformed lane. Stacking
$\bm{\mathcal Y}(\omega)=(\tilde Y^1_\omega,\dots,\tilde Y^M_\omega)$, a
vector of length $Mt$, the per-mode exponent of \eqref{eq:ab_App} reads
$-\tfrac12\,\bm{\mathcal Y}^{\dagger}\mathbb M(\omega)\,\bm{\mathcal Y}$
with the complex $(Mt)\times(Mt)$ matrix
\begin{equation}\label{eq:ab_Mmatrix}
    \mathbb M_{JJ'}(\omega)=\underbrace{M\,(C^J)^{-1}\,\delta_{JJ'}}_{\text{entropic core}}
    \;+\;\underbrace{\ii\,(C^J)^{-1}W^J(\omega)
    \;+\;\ii\,\big[(C^{J'})^{-1}W^{J'}(-\omega)\big]^{\!\top}}_{\text{propagation (acts on the lane: independent of the other block index)}}
    \;+\;\Lambda(\omega),
\end{equation}
where $\Lambda(\omega)$, the same for every pair $(J,J')$ because it acts
on the lane alone, collects the kernel-quadratic and noise insertions,
\begin{equation}\label{eq:ab_Lambda}
    \Lambda(\omega)=\frac1M\sum_{J''}\Big[
    \tfrac{1}{(1+\lambda)^2}Q^{11,J''}+Q^{22,J''}
    +\tfrac{1}{1+\lambda}\big(e^{-\ii\omega}Q^{12,J''}
    +e^{+\ii\omega}Q^{12,J''\,\top}\big)
    -W^{J''}(-\omega)^{\top}(C^{J''})^{-1}W^{J''}(\omega)\Big].
\end{equation}
The sequence-hop bookkeeping is now visible in the matrix itself: $e^{-\ii\omega}$
rides on $K^{21,J}$ inside $W^J$ and on $Q^{12,J}$ inside $\Lambda$, while
$Q^{22,J}$ is hop-neutral (it pairs $Y_{\mu-1}$ with $Y_{\mu-1}$). Gaussian
integration then gives $\ln[\det\mathbb M(\omega)]^{-1/2}$ per mode, and the
prefactor assembles as
$\tfrac1N\sum_\omega=\tfrac pN\cdot\tfrac1p\sum_\omega
\to\alpha\int\tfrac{\dd\omega}{2\pi}$---the only place where the load
$\alpha$ enters the theory, as the density of noise patterns per site:
\begin{equation}\label{eq:ab_Omega_logdet}
    \Omega=-\frac{\alpha}{2}\int_{-\pi}^{\pi}\frac{\dd\omega}{2\pi}\,
    \ln\det\mathbb M(\omega)\;+\;\text{const},
\end{equation}
exactly as $\bm S\otimes\bm G^{21}$
and $\bm S\otimes\bm Q^{12}$ did in \eqref{eq:Omega_2species_evaluated}.


\paragraph{Pure-relay limit of the mode matrix.}
The complexity of \eqref{eq:ab_Mmatrix} is that of the general-$\lambda$
theory carrying both channels; in the purely asymmetric limit it
simplifies. At $\lambda\to\infty$ the weight $(1+\lambda)^{-1}$ removes
$K^{11,J}$, $Q^{11,J}$ and $Q^{12,J}$ from
\eqref{eq:ab_W}--\eqref{eq:ab_Lambda}, so $W^J(\omega)\to
e^{-\ii\omega}K^{21,J}$, and the phases cancel inside
$W^{\dagger}(C)^{-1}W$; the mode matrix becomes a degree-one Laurent
polynomial in $e^{\ii\omega}$,
\begin{equation}\label{eq:ab_M_relay}
    \mathbb M(\omega)\;\xrightarrow{\ \lambda\to\infty\ }\;
    \mathbb D\;+\;e^{-\ii\omega}\,\mathbb P\;+\;e^{+\ii\omega}\,
    \mathbb P^{\top},
\end{equation}
with $\omega$-independent blocks (all supports now on $\mathcal T^2$)
\begin{equation}\label{eq:ab_DP_relay}
    \mathbb D_{JJ'}=M\,(C^J)^{-1}\delta_{JJ'}
    +\frac1M\sum_{J''}\Big[Q^{22,J''}
    -K^{21,J''\,\top}(C^{J''})^{-1}K^{21,J''}\Big],
    \qquad
    \mathbb P_{JJ'}=\ii\,(C^J)^{-1}K^{21,J}.
\end{equation}
Every appearance of the response now visibly carries exactly one phase, and
none of the $\omega$-independent blocks contains it unpaired---the
$\hat K$-gate of Stage~4 can be read off this display.

\paragraph{Consistency check: $M=1$ recovers the synchronous theory.}
At $M=1$ the we should get back to \emph{synchronous} dynamics \footnote{More precisely it is the species-alternating protocol ($x$ first, then $y$
reading the fresh $x$); at $\lambda\to\infty$ it closes on the queue as
$y(n{+}2)=\mathrm{sign}[K\,y(n)]$}. Setting
$M=1$ in \eqref{eq:ab_M_relay}--\eqref{eq:ab_DP_relay} and anticipating the
saddle identification $K^{21}=\ii\,G$ of \eqref{eq:ab_saddle_ids},
\begin{align}\label{eq:ab_M1_check}
    \mathbb M(\omega)
    &=Q^{22}+C^{-1}+G^{\top}C^{-1}G
    -e^{-\ii\omega}\,C^{-1}G-e^{+\ii\omega}\,G^{\top}C^{-1}\nonumber\\
    &=Q^{22}+\big(e^{-\ii\omega}\bm 1-G\big)^{\!\top}C^{-1}
    \big(e^{+\ii\omega}\bm 1-G\big)
\end{align}
which is exactly the perfect-square mode kernel of the synchronous asymmetric network
[Sec.~\ref{sec:disorder_average}], up to the immaterial ring relabelling
$\omega\to-\omega$.


\subsection{Stage 4: conjugate kernels and the hop-counting rule}
\label{sec:ab_stage4}

\subsubsection{Physical saddle.}
The normalisation identities ($Z[0]=1$ per realisation) give, per block $J$,
$\langle\hat h^a(n)\rangle_\star=0$, $\langle\hat h^a(n)\hat h^b(n')\rangle_\star=0$
and the response identification, so at the physical saddle
\begin{equation}\label{eq:ab_saddle_ids}
    k^s_\mu(n)=0,\qquad Q^{ab,J}(n,n')=0,\qquad
    K^{a1,J}(n,n')=\ii\,G^{a1,J}(n,n'),
\end{equation}
with $G^{11,J}(n,n')=\partial\langle x(n)\rangle/\partial\theta^1(n')$ and
$G^{21,J}(n,n')=\partial\langle x(n)\rangle/\partial\theta^2(n')$ the unit's
$x$-response to kicks of its two fields ($G^{21,J}$ is mediated by the
explicit relay: the kick shifts $y$ at its next update, which shifts $x$ one
update later; both kernels are causal staircases, $G(n,n')=0$ unless
$n\ge n'+1$, hence nilpotent). Variation of $\Psi$ gives the signal saddle
\begin{equation}\label{eq:ab_signal_saddle}
    \hat m_J^\mu(n)=0,\qquad
    \hat k^1_\mu(n)=m^\mu(n)\ \ (n\in\mathcal T^1),\qquad
    \hat k^2_\mu(n)=m^\mu(n)\ \ (n\in\mathcal T^2):
\end{equation}
each channel's signal is the $1/M$-weighted, staleness-carrying \emph{full}
overlap, evaluated at the channel's own substeps. (The synchronous
calculation contains the same statement, dispersed over its parity-resolved
multipliers: $\hat l=\hat k=b$, $\hat p=\hat r=a$.)


\subsubsection{The hop-counting rule.}
Each conjugate kernel is an $\Omega$-derivative, and $\Omega$ is a
log-determinant, so each is a trace formula evaluated at the physical
saddle,
\begin{equation}\label{eq:ab_trace_formula}
    \hat X=\ii\,\frac{\partial\Omega}{\partial X}\bigg|_{\rm saddle}
    =-\frac{\ii\alpha}{2}\int_{-\pi}^{\pi}\frac{\dd\omega}{2\pi}\;
    \Tr\Big[\mathbb M^{-1}(\omega)\,
    \frac{\partial\mathbb M(\omega)}{\partial X}\Big],
    \qquad X\in\{C^J,\,Q^{ab,J},\,K^{a1,J}\}.
\end{equation}
The two factors of the integrand are controlled separately. The insertion
$\partial\mathbb M/\partial X$ is read off from
\eqref{eq:ab_Mmatrix}--\eqref{eq:ab_Lambda}: each kernel enters
$\mathbb M$ linearly, in a known slot, with a known phase ($Q^{11}$,
$Q^{22}$, $K^{11}$ phase-free; $Q^{12}$ and $K^{21}$ with $e^{-\ii\omega}$,
their transposed partners with $e^{+\ii\omega}$). The resolvent
$\mathbb M^{-1}(\omega)$ is a \emph{finite} polynomial in $e^{\pm\ii\omega}$
(nilpotent responses). 

A term survives then the $\omega$-integral if and only
if its total phase balances to $e^{0}$---the hop-counting rule. 
Recall that the same counting criterion is the mechanism behind the solvability of
purely asymmetric sequence processing networks: with the phases written as
eigenvalues $e^{-\ii\omega}$ of the shift matrix $\bm S$ [as in
\eqref{eq:Omega_2species_evaluated}], it underlies $\hat K=0$ in the
single-species theory and $\hat K^{21}=0$ in \eqref{eq:Khat21_2species} for
the synchronous two-species theory. Applied row
by row:
\begin{center}
\begin{tabular}{l l l l}
 conjugate kernel & insertion attaches to & net shift & survives?\\
\hline
 $\hat q^J$ (spin-spin) & $C^J$, no shift & 0 & no --- $=0$ (cancellation at $Q=0$)\\
 $\hat Q^{11,J},\ \hat Q^{22,J}$ & $Y_\mu Y_\mu$, $Y_{\mu-1}Y_{\mu-1}$ & 0 & yes --- noise kernels\\
 $\hat Q^{12,J}$ & $Y_\mu Y_{\mu-1}$ & 1 & only via $G^{21}$ phase-matching\\
 $\hat K^{11,J}$ ($J$-channel) & $Y_\mu$, no shift & 0 & \textbf{yes}, weight $(1+\lambda)^{-1}$\\
 $\hat K^{21,J}$ ($K$-channel) & $Y_{\mu-1}$ & 1 & \textbf{no} --- $=0$ exactly\\
\end{tabular}
\end{center}

The two $\hat K$ statements deserve spelling out. For $\hat K^{21,J}$ one
must count with care: in the $Y$-eliminated representation
\eqref{eq:ab_Mmatrix} \emph{both} phases appear (the transposed propagation
branch carries $e^{+\ii\omega}$), so the vanishing is not term-by-term
there. It becomes term-by-term in the opposite elimination order:
integrating $Y$ \emph{first}, the saddle ($Q=0$) determinant factorises into
a $C$-block and a single factor in which every $K^{21}$ carries
$e^{-\ii\omega}$ and no $e^{+\ii\omega}$ appears anywhere. In that
representation the derivative is a single one-sided resolvent insertion,
every term carries $e^{-\ii(k+1)\omega}$ with $k\ge0$, and
$\int_{-\pi}^{\pi}\dd\omega\,e^{-\ii(k+1)\omega}=0$ term by term:
\begin{equation}\label{eq:ab_gate}
    \boxed{\;\hat K^{21,J}(n,n')=0\qquad\text{exactly, for all blocks and all time pairs.}\;}
\end{equation}
\emph{Scheduled asynchrony generates no retarded self-interaction on the
sequence pathway, at any $\lambda$}: the phase count is a property of pattern
space, which the schedule never touches. In particular, our heuristics loop-space
closure assumed exactly this---its recursions carry no
memory kernel---and this result turns that assumption into a theorem for such blocked models.

For $\hat K^{11,J}$ the insertion is phase-free, so the $\omega$-integral
\emph{projects} instead of killing: it deletes every word of the resolvent
containing an unpaired $G^{21}$ (a \emph{word}: one ordered product of
response factors in the expansion of the resolvent, carrying one factor
$e^{-\ii\omega}$ per $K^{21}$ letter, cf.\ the same usage below
\eqref{eq:Qhat12_final}), and what survives are the \emph{round trips
through the symmetric channel}---the unit perturbs the network through $J$
(weight $\tfrac{1}{1+\lambda}$), the network responds ($\bar{\bm G}^{11}$),
and the response returns through $J$ to the same unit; echoes through the
$K$-channel advance one pattern per leg and never return. Summing all round
trips gives a geometric series,
\begin{equation}\label{eq:ab_Khat11}
    \hat K^{11,J}(n,n')=-\frac{\alpha}{1+\lambda}\,
    \Big[\big(\bm 1-\tfrac{1}{1+\lambda}\bar{\bm G}^{11}\big)^{-1}\Big](n',n),
\end{equation}
where $\bar{\bm G}^{11}$ assembles the $M$ blocks' $x$-responses with their
$1/M$ weights on the $\mathcal T^1$ supports, the precise assembly being
fixed by \eqref{eq:ab_Mmatrix}. (We do not quote a closed scalar form; the
kernel's defining properties---the support $n'\ge n$, the $(1+\lambda)^{-1}$
weight, and both $\lambda$-limits below---are checked numerically in the
verification paragraph at the end of this subsection.) 

Physically, this is
the \emph{Onsager reaction term} of the model: in $\mathcal M_i$ it couples
$x(n)$ to $\hat h^1(n')$ with $n'\ge n$---the unit's spin now shifts its own
field later---rendering the effective single-unit process non-Markovian at
finite $\lambda$; its diagonal, $-\alpha/(1+\lambda)$, is the instantaneous
reaction field familiar from the TAP equations~\cite{}. As $\lambda\to\infty$ the
weight kills it, $\hat K^{11,J}\to0$ (no echo can return through the pure
relay), recovering the solvable class; at $\lambda=0$ it collapses onto the
full symmetric Hopfield reaction term $-\alpha(\bm 1-\bar{\bm G}^{11})^{-1}$. Asynchrony thus changes \emph{where} the
self-interaction lives (supports, block weights) but not \emph{whether} it
lives: that is decided, exactly as in the synchronous theory, by which
synaptic pathway carries the shift.

\subsubsection{Noise kernels.}
Differentiating at $Q=0$ gives the noise covariances as $\omega$-integrals of
the resolvent sandwich $\mathbb B(\omega)$ of \eqref{eq:ab_Omega_reduced}
[the block/parity-decorated analogue of $\bm\Sigma(\omega)$ in
\eqref{eq:Sigma_def}]:
\begin{equation}\label{eq:ab_noise}
    \hat Q^{11,J}=-\frac{\ii\alpha}{2}\,\frac{1}{(1+\lambda)^{2}}\!\int_{-\pi}^{\pi}\!\frac{\dd\omega}{2\pi}\,
    \mathbb B(\omega)\Big|_{\mathcal T^1_J\times\mathcal T^1_J},\qquad
    \hat Q^{22,J}=-\frac{\ii\alpha}{2}\!\int_{-\pi}^{\pi}\!\frac{\dd\omega}{2\pi}\,
    \mathbb B(\omega)\Big|_{\mathcal T^2_J\times\mathcal T^2_J},
\end{equation}
and a cross-channel kernel $\hat Q^{12,J}$ with weight
$\propto(1+\lambda)^{-1}e^{\pm\ii\omega}$. Two facts, both verified
numerically: (i) the channel weights sit on the \emph{bare} amplitudes, so the
visible ($J$-channel) crosstalk fades as $(1+\lambda)^{-2}$ while the
$K$-channel noise saturates---consistent with the saturation of the dynamic
phase at moderate $\lambda$ observed in MC; (ii) $\hat Q^{12,J}$ vanishes identically when
$G^{21}=0$ and is generated, at lowest order linearly, by the relay response.
Downstream these conjugates \emph{are} the noise of the effective process: in
$\mathcal M_i$ [last line of \eqref{eq:ab_M}] the term
$-\ii\,\hat Q^{ab,J}\hat h^a\hat h^b=-\tfrac{\alpha}{2}R^{ab,J}\hat h^a\hat h^b$
is undone by a Hubbard--Stratonovich transformation, replacing the quadratic
form by a Gaussian field $\sqrt{\alpha}\,\phi^a(n)$ added to $h^a(n)$ with
$\langle\phi^a(n)\phi^b(n')\rangle=R^{ab,J}(n,n')$---the coloured noise of the
single-unit process \eqref{eq:ab_single_unit} like in the previous sections.

\subsubsection{Numerical verification.}
All structural claims of this subsection were checked by finite differences
on the exact finite-ring $\Omega$ [the Gaussian \eqref{eq:ab_Omega_reduced}
with $t=8$ substeps (two sweeps), $p=16$, random causal kernels on the correct
supports; the mode
matrix \eqref{eq:ab_Mmatrix}--\eqref{eq:ab_Lambda} and its pure-relay limit
\eqref{eq:ab_M_relay}--\eqref{eq:ab_DP_relay} were additionally verified
entrywise against the pattern-space construction]:
$\max|\partial\Omega/\partial K^{21,J}|\sim10^{-36}$ (gate, machine zero);
$\partial\Omega/\partial K^{11,J}\ne0$ on $n'\ge n$ and $=0$ on $n'<n$;
$\partial\Omega/\partial C^J=0$ at $Q=0$ ($\hat q^J=0$);
$\partial\Omega/\partial Q^{11,J},\,\partial\Omega/\partial Q^{22,J}\ne0$;
$\partial\Omega/\partial Q^{12,J}=0$ when $G^{21}=0$ and $\ne0$ otherwise;
and $\partial\Omega/\partial K^{11,J}\to0$ at $(1+\lambda)^{-1}\to0$.

\subsection{Stage 5: twisted stationarity --- the autonomous period-cell DMFT}
\label{sec:ab_stage5}

\subsubsection{Effective single-unit process.}
Collecting \eqref{eq:ab_signal_saddle}, \eqref{eq:ab_gate},
\eqref{eq:ab_Khat11} and \eqref{eq:ab_noise} in \eqref{eq:ab_M}: a unit in
block $J$ carries quenched condensed components
$\{\xi^\mu\}_{\mu\in\mathcal C}$ (equivalently, per condensed window, the pair
$(\xi,\eta)$), its spins freeze between their substeps, and at update substeps
they redraw with $\mathrm{Prob}[\sigma^s(n+1)=\pm1]=\tfrac12[1\pm\tanh\beta h^s(n)]$ from
\begin{equation}\label{eq:ab_single_unit}
    \boxed{
    \begin{aligned}
    h^1(n)&=\frac{1}{1+\lambda}\sum_{\mu\in\mathcal C}\xi^{\mu}m^\mu(n)
    +\frac{\lambda}{1+\lambda}\,y(n)
    +\sum_{n'\in\mathcal T^1_J}\hat K^{11,J}_R(n,n')\,x(n')
    +\sqrt{\alpha}\,\phi^1(n), &&n\in\mathcal T^1_J,\\[0.2em]
    h^2(n)&=\sum_{\mu\in\mathcal C}\xi^{\mu+1}m^\mu(n)
    +\sqrt{\alpha}\,\phi^2(n), &&n\in\mathcal T^2_J,
    \end{aligned}}
\end{equation}
Here $m^\mu(n)$ is the \emph{all-block} staircase overlap of
\eqref{eq:ab_cond_ops}, $m^\mu(n)=\tfrac1M\sum_{J'}m^\mu_{J'}(n)$, each
$m^\mu_{J'}$ constant between block $J'$'s substeps: a unit updating
mid-sweep therefore reads the already-updated blocks at their current values
and the remaining blocks at their previous-sweep plateaus. The mixed states
of the period-$\varpi$ orbit enter the single-unit theory through precisely
this term. $\hat K^{11,J}_R$ is the (real) retarded kernel corresponding to
\eqref{eq:ab_Khat11}, and $(\phi^1,\phi^2)$ are zero-mean Gaussian with
covariances $R^{ab,J}\propto\hat Q^{ab,J}$ on the respective supports. The
relay enters as the sampled spin $y(n)$ at its staircase value---the
two-branch $\pm\tfrac{\lambda}{1+\lambda}$ structure of the synchronous map,
now with the previous-sweep $y$ read built into the substep chain. Self-consistency
closes in the familiar way:
$m^\mu_J(n)=\overline{\langle\xi^\mu x(n)\rangle}_J$,
$C^J(n,n')=\overline{\langle x(n)x(n')\rangle}_J$,
$G^{a1,J}(n,n')=\partial\overline{\langle x(n)\rangle}_J/\partial\theta^a(n')$,
fed back through \eqref{eq:ab_Khat11}--\eqref{eq:ab_noise}. The two unit
classes communicate only through the macroscopic kernels.

\paragraph{Pure-relay reduction of the single-unit process.}
In the $\lambda\to\infty$ limit announced at Stage~0 the process
\eqref{eq:ab_single_unit} collapses onto the queue: $x(n+1)=y(n)$ on
$\mathcal T^1_J$ (deterministic copy; no reaction term, no noise on $x$),
and the unit reduces to
\begin{equation}\label{eq:ab_single_unit_relay}
    \boxed{\;
    y(n+1)=\pm1\ \text{w.p.}\ \tfrac12\big[1\pm\tanh\beta h^2(n)\big],
    \qquad
    h^2(n)=\sum_{\mu\in\mathcal C}\xi^{\mu+1}m^\mu(n)+\sqrt{\alpha}\,\phi(n),
    \qquad n\in\mathcal T^2_J,\;}
\end{equation}
with $m^\mu(n)$ the all-block staircase overlaps, as above, of the copied $x$
(i.e.\ of $y$ one substep earlier) and
$\langle\phi(n)\phi(n')\rangle=\bar R^{22}(n,n')$ on
$\mathcal T^2_J\times\mathcal T^2_J$. All quantitative results below solve
\eqref{eq:ab_single_unit_relay}.

\subsubsection{Twisted stationarity and autonomy.}
The equations above are exact for the blocked schedule but live on the full
two-time lattice. The usual stationary ansatz for DMFT, $C(n,n')=C(n-n')$ with
$n$-independent overlaps, cannot describe retrieval: the condensed window
travels, so $m^\mu(n)$ is never $n$-independent, and the orbit breaks time
translation, pattern-ring rotation and block relabelling separately. What
survives, by the block exchangeability and the $\mu\to\mu+\Delta$ invariance
of the pattern measure, is their composition, the \emph{twist}
\begin{equation}\label{eq:ab_twist}
    \Sigma_{q,\Delta}^{(r)}:\qquad
    (n\to n+q)\ \circ\ (\mu\to\mu+\Delta)\ \circ\ \rho^{\,r},
    \qquad \rho(J)=J+1\ \mathrm{mod}\ M\ \text{the cyclic relabelling of the blocks},
\end{equation}
with $r\equiv q/2\bmod M$ fixed by the offset of $q$ within a sweep
($\rho^{\,r}=\mathrm{id}$ when $q$ is a whole number of sweeps). The
relabelling frees the cells from the schedule: a shift $q$ landing
mid-sweep is realigned by $\rho^{\,r}$, so periods incommensurate with the
sweep stay admissible---a non-symmorphic screw axis, whose primitive cell
is shorter than any laboratory period. The slower speeds $v=\Delta/q$
selected at smaller $\lambda$ (Fig.~\ref{fig:ab_staircase}) are of this
kind.

\begin{figure}[h!]
    \centering
    \includegraphics[width=3.2in]{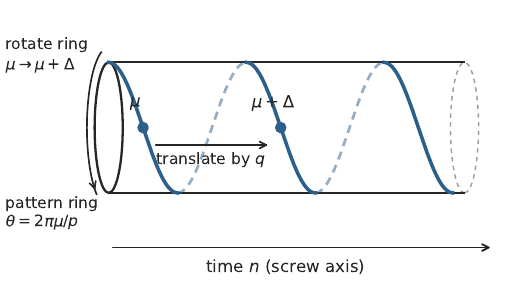}
    \caption{\label{fig:ab_helix} {\bf The retrieval orbit is a screw
    motion.} Projection of the orbit onto (time, condensed-window position):
    the window's position on the pattern ring (angular direction) against
    substep time $n$ (axis). Neither the time shift $n\to n+q$ nor the ring
    relabelling $\mu\to\mu+\Delta$ alone maps the trajectory onto itself;
    their composition, the twist \eqref{eq:ab_twist}, does---the geometry
    repeats while the pattern labels advance by $\Delta$ per cell. 
    frame.}
\end{figure}

The retrieval
orbit has a helical geometry with time running along the axis, and the pattern ring as the angular
direction. It is mapped to itself by neither translation nor rotation but by the
screw (Fig.~\ref{fig:ab_helix}). The stationarity ansatz (A4) is
$\Sigma$-invariance of the saddle:
\begin{equation}\label{eq:ab_twisted_kernels}
    m_J^\mu(n)=m_{\rho^r(J)}^{\mu+\Delta}(n+q),\qquad
    C^J(n,n')=C^{\rho^r(J)}(n+q,n'+q),\qquad
    G^{a1,J}(n,n')=G^{a1,\rho^r(J)}(n+q,n'+q),
\end{equation}
under which every kernel is determined by its values on one \emph{period
cell}, the fundamental domain $n\in\{0,\dots,q-1\}$ of the twist: as a unit
cell tiles a crystal, the cell regenerates the full two-time lattice under
powers of $\Sigma$, each application relabelling patterns by $\Delta$ and
blocks by $\rho^{\,r}$ (two-time kernels need one cell column each, since the
twist acts on both arguments and the staircase memory is finite at $T=0$).

Using this ansatz, the self-consistency becomes a \emph{closed, autonomous
fixed-point problem on the cell}: collecting the cell-restricted order
parameters into
\begin{equation}\label{eq:ab_cellmap}
    \bm\Theta\equiv\big(m^\mu_J,\;C^J,\;G^{a1,J}\big)\Big|_{\text{cell}},
    \qquad
    \boxed{\;\bm\Theta=\mathcal G[\bm\Theta]\;},
\end{equation}
In components the map is explicit:
\begin{equation}\label{eq:ab_cellmap_explicit}
    \mathcal G[\bm\Theta]
    =\Big(\,
    \overline{\big\langle \xi^{\mu}\,\sigma^1(n)\big\rangle}_{\bm\Theta,J}\,,\;
    \overline{\big\langle \sigma^1(n)\,\sigma^1(n')\big\rangle}_{\bm\Theta,J}\,,\;
    \partial\,\overline{\big\langle \sigma^1(n)\big\rangle}_{\bm\Theta,J}\big/\partial\theta^{a}(n')
    \Big)_{n,n'\in\mathrm{cell}},
\end{equation}
where $\langle\cdot\rangle_{\bm\Theta,J}$ denotes the expectation in the
single-unit process \eqref{eq:ab_single_unit} of block class $J$, whose
kernels $\hat K^{11,J}_R$ and $R^{ab,J}$ are evaluated from $\bm\Theta$
through \eqref{eq:ab_Khat11} and \eqref{eq:ab_noise}.

The ansatz
\eqref{eq:ab_twisted_kernels} simplifies two things. First, every
kernel argument outside the cell is pulled back into it, so the kernel
formulas consume only cell entries of $\bm\Theta$ and the
$\omega$-integrals of \eqref{eq:ab_noise} become finite matrix algebra on
the cell lattice. Second, the read-back on the left-hand side of
\eqref{eq:ab_cellmap_explicit} is relabelled---$\mu\to\mu+\Delta$,
$J\to\rho^{r}(J)$---before being identified with the input: retrieval is a
fixed point of evolution \emph{composed with} the twist, not of evolution
alone. $\mathcal G$ carries no explicit time dependence---the
generating-functional counterpart of the heuristic loop-space period map
$\mathbf G$, now derived rather than constructed, and carrying full
within-cell memory instead of one-step closure.
\emph{Twisted stationarity in loop space is periodic sequencing in real time}:
the laboratory-frame limit cycle, mixed states included, is the orbit of the
cell fixed point under \eqref{eq:ab_twist}, and the non-autonomous
single-species map may be recovered by projecting the autonomous
cell dynamics onto one species. At $M\to\infty$ the substep lattice becomes
continuous and $\Sigma$-invariance becomes a travelling-wave ansatz,
$m^\mu(t)=f(\mu-vt)$ with $v=\Delta/q$: the cell fixed point is the wave
profile $f$ itself, and the $\varpi=3$ cycle with its mixed states is $f$
sampled at sweep boundaries (Fig.~\ref{fig:ab_orbit}).

\begin{figure}[hbt!]
    \centering
    \includegraphics[]{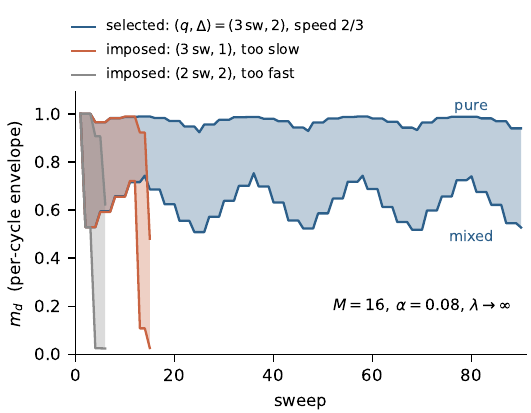}
    \caption{\label{fig:ab_twist_demo} {\bf The twist is selected, not
    chosen.} Per-cycle envelope (rolling min/max over three sweeps) of the
    peak window overlap $m_d$ in the cell population dynamics at $M=16$,
    $\alpha=0.08$, $\lambda\to\infty$. In the freely selected cell
    $(q,\Delta)=(3\ \text{sweeps},\,2)$ the band persists indefinitely, its
    edges tracing the pure and mixed substates of the $\varpi=3$ orbit.
    Imposing a too-slow $(3\ \text{sweeps},\,1)$ or too-fast
    $(2\ \text{sweeps},\,2)$ relabelling destroys retrieval within a few
    cells---the condensate climbs out of, or is relabelled off, the
    co-rotating window---and bisection finds $\alpha_c=0$: the wrong-cell
    retrieval fixed point does not exist at any load.}
\end{figure}

\paragraph{Selecting the twist.}
Nothing in the theory so far fixes the cell twist $(q,\Delta,r)$:
the ansatz \eqref{eq:ab_twisted_kernels} can be imposed for any admissible
triple, and the cell equations \eqref{eq:ab_cellmap} are identical in form
for all of them. Can the correct twist be determined self-consistently?
Yes---by the fixed point itself: a twist is realised when
\eqref{eq:ab_cellmap} admits a stable retrieval solution on its cell, and
among coexisting cells the dynamics selects by stability. The twist is thus
an output of the theory, on the same footing as the order parameters.

\begin{figure*}[hbt!]
    \centering
    \includegraphics[]{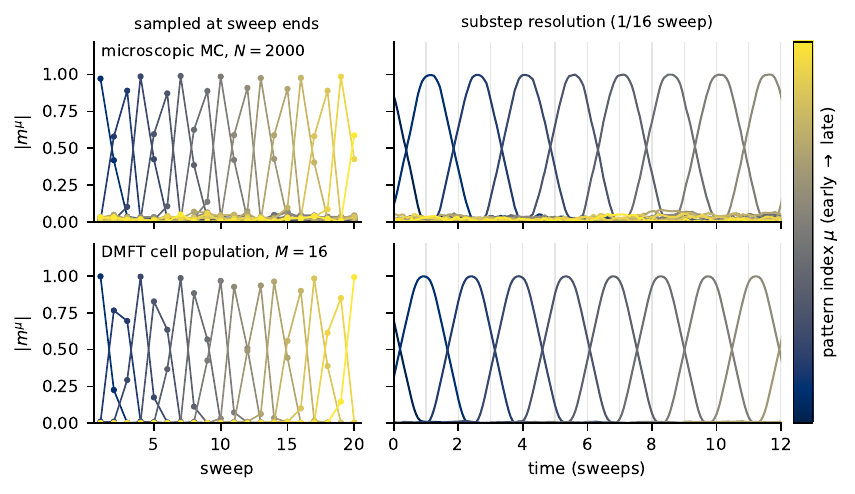}
    \caption{\label{fig:ab_orbit} {\bf The retrieval orbit is a smooth
    travelling wave.} Pattern overlaps $|m^\mu|$ of the same runs at two
    time resolutions (pure-relay limit, $\alpha=0.05$). Top row:
    microscopic MC of the sequential paired-unit schedule ($N=2000$;
    overlaps recorded every $N/16$ site updates). Bottom row: population
    dynamics of the effective single-unit process
    \eqref{eq:ab_single_unit} ($M=16$ blocks, cyclic noise closure, one
    point per block update); the condensed window follows the leading
    channel, so the speed is an \emph{output}. Left column: sampled once
    per sweep, the orbit appears as the $\varpi=3$ cycle pure $\to$ mixed
    $\to$ pure of the sweep-level description. Right column: at substep
    resolution ($1/16$ sweep, gridlines marking sweep ends) the same
    dynamics is a common smooth profile travelling at speed $2/3$,
    neighbouring arches crossing at overlap $\approx1/2$: the sweep-level
    cycle is this wave sampled stroboscopically, and the mixed state is
    the wave caught mid-crossing.}
\end{figure*}

Which twist does the dynamics actually select? For the physical sequential
schedule at large $M>8$, it is $q=3$ sweeps ($6M$ substeps), $\Delta=2$,
$r=0$: the condensed window advances two patterns every three sweeps
(retrieval speed $2/3$), no block rotation is needed because $q$ is a whole
number of sweeps, and the window content is the pure/mixed joint cycle
\eqref{eq:ab_cycle}.
This is precisely the structure we imposed to construct the heuristic loop-space closure, but now the period is not put by hand and instead derived self-consistently. 
Imposing a cell that is incompatible with the kernels suppresses
retrieval artificially. For example at $M=16$, forcing
$(q,\Delta)=(3\ \text{sweeps},\,1)$ or $(2\ \text{sweeps},\,2)$ in place of
the selected $(3\ \text{sweeps},\,2)$ gives $\alpha_c=0$
(Fig.~\ref{fig:ab_twist_demo}). The numerical
solver accordingly lets the condensed window follow the leading channel and
reads $(q,\Delta,r)$ off the converged orbit. Mapping which cells are
stable where in $(\alpha,\lambda,M)$ (the MC locking at small $M$ hints at
a devil's staircase across the blocked family) is left to Stage~6.

\begin{figure*}[hbt!]
    \centering
    \includegraphics[]{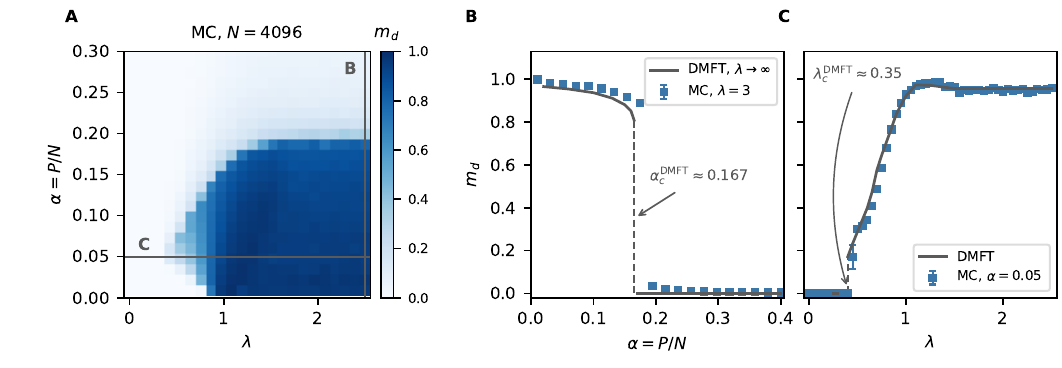}
    \caption{\label{fig:ab_pd} {\bf Asynchronous phase diagram: theory vs.\
    MC.} (A) Dynamic-retrieval island of the sequential paired-unit schedule
    (MC, $N=4096$, $T=100$ sweeps, $20$ disorder samples; dynamic overlap
    $m_d$). (B) Vertical slice at $\lambda=3$: $m_d(\alpha)$ from MC at
    $N=10^4$ ($2000$ sweeps, $100$ samples per point; SEM smaller than the
    symbols) against the $\lambda\to\infty$ DMFT curve. The theory's
    transition is discontinuous at $\alpha_c^{\rm DMFT}\approx0.167$; the MC
    drop occurs after $\alpha\approx0.174$. (C) Horizontal slice at
    $\alpha=0.05$: $m_d(\lambda)$ from MC ($N=4096$, $150$ sweeps, $10$
    samples) against the finite-$\lambda$ cell solver (perturbatively
    measured gain; mixed-state ignition); the dynamic phase opens at
    $\lambda_c^{\rm DMFT}\approx0.35$. Both slices use the same
    sequence-conditioned estimator for theory and MC; the gray lines in
    (A) locate the two slices (the B plane, $\lambda=3$, lies just beyond
    the plotted range).}
\end{figure*}

\paragraph{The $\lambda\to\infty$ theory: $\varpi=3$ dynamics and capacity.}
We solve the reduced process \eqref{eq:ab_single_unit_relay} by population
dynamics ($\sim10^4$ units per block class, each carrying its quenched
window components and spin pair). The condensed set $\mathcal C$ of
\eqref{eq:ab_single_unit} is realised as a travelling window of six
consecutive patterns, re-anchored on the leading channel as the condensate
advances (this implements the ansatz with twist symmetry \eqref{eq:ab_twist} realised by the orbit as needed); a channel scrolled out of the window is absorbed into the
Gaussian sector and its slot refilled with fresh i.i.d.\ components. 

\begin{figure}[hbt!]
    \centering
    \includegraphics[]{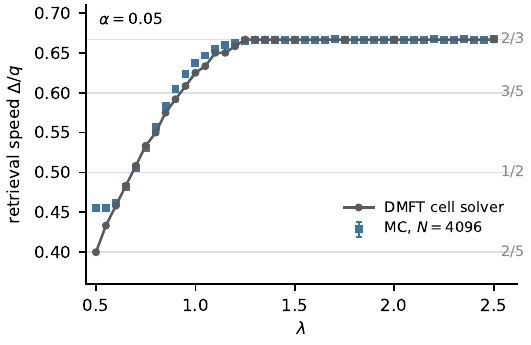}
    \caption{\label{fig:ab_staircase} {\bf The selected twist at finite
    $\lambda$.} Retrieval speed $\Delta/q$ at $\alpha=0.05$: the cell
    solver (gray; free window, synchronous gain convention) against
    microscopic MC ($N=4096$, $150$ sweeps, $10$ samples; speed-agnostic
    estimator, SEM bars). Both lock onto the $\varpi=3$ cell (speed $2/3$)
    for $\lambda\gtrsim1.3$ and descend the cell family together at smaller
    $\lambda$; gridlines mark simple rational speeds.}
\end{figure}

The noise variances close through the cross-block cyclic recursion and the advancing
gain is measured perturbatively on the orbit itself.
The dynamics locks onto the $\varpi=3$ cycle of the physical schedule
(Fig.~\ref{fig:ab_orbit}), with the mixed member alternating between the
species and the speed converging to $2/3$ under block refinement
($0.76,\,0.70,\,0.65$ for $M=2,4,8$), which interestingly also mirrors the convergence of a blocked microscopic MC simulation ($0.77\to0.68$). Driving the load to the boundary
gives
\begin{equation}\label{eq:ab_capacity}
    \alpha_c^{\rm async}(\lambda\to\infty)\;\approx\;0.167
\end{equation}
at converged horizon, within $\sim7\%$ of the microscopic MC plateau. The asynchronous capacity is thus $\sim\!66\%$ of the
synchronous one ($\alpha_c^{\rm sync}\simeq0.269$).

\paragraph{Finite $\lambda$.}
The same cell solver applies at finite $\lambda$. The comparison with the microscopic MC is quantitative on three
fronts. \emph{Kinematics}: the selected twist varies with $\lambda$---the
retrieval speed locks at $2/3$ for $\lambda\gtrsim1.3$ and descends the
cell family at smaller $\lambda$ (through $\simeq1/2$ near
$\lambda\simeq0.7$)---and the solver tracks the measured speed within
symbol size across the range (Fig.~\ref{fig:ab_staircase}).
\emph{Coherence}: theory and MC share the shallow maximum of $m_d$ near
$\lambda\simeq1.2$ (the bump in Fig.~\ref{fig:ab_pd}C; the darkest island
core in Fig.~\ref{fig:ab_pd}A). The visible field
\eqref{eq:ab_single_unit} carries the sequence state twice---attractor
copy and relay copy---and near their balance, $\lambda\approx m$, the two
reinforce on the units that need not flip, so the attractor \emph{vetoes}
relay-borne errors before they re-enter the crosstalk (the measured gain
drops from $0.63$ at large $\lambda$ to $\approx0.47$ at $\lambda=1$).
\emph{Capacity}: the converged solver boundary agrees with the MC island
edge for $\lambda\gtrsim1.25$ ($0.17$--$0.18$). Nevertheless, the closure cannot properly capture the blackout dynamics at high load (above capacity) near the balance point $\lambda=1$.

\section{Circular embedding: pseudo-inverse and fidelity metrics}\label{si:projection}

\subsection{Linear stability in the embedded plane: the sequence-mode multiplier}
\label{sec:embed_stability}

\begin{figure}[hbt!]
    \centering
    \includegraphics[width=0.8\linewidth]{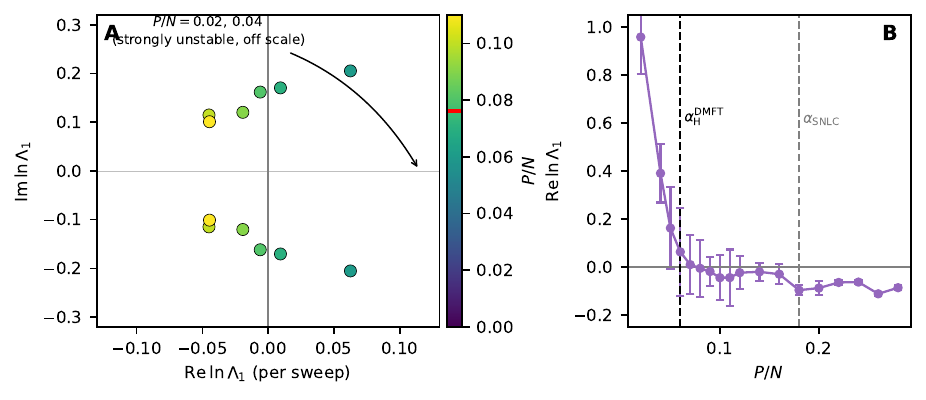}
    \caption{{\bf Oscillatory instability of the blackout state,
    measured in the embedded plane.}
    {\bf A.} The eigenvalue pair of the linearised embedded flow at the origin [Eq.~\eqref{eq:embed_linear}] estimated from the complex one-sweep multiplier $\Lambda_1$ of the fundamental sequence mode
    $c_1$ [Eq.~\eqref{eq:mode_multiplier}] for loads $P/N \le 0.11$ (colour bar).
    Each point averaged over 12 disorder realisations $\times$ 24 random initial conditions. 
    As the load increases the pair migrates across the stability boundary $\mathrm{Re}\ln\Lambda_1 = 0$ (grey line) at $P/N \approx 0.076$ (red tick on the colour bar). 
    The finite rotation rate $|\mathrm{Im}\ln\Lambda_1|$ 
    identifies the instability as oscillatory. 
    {\bf B.} Growth rate $\mathrm{Re}\ln\Lambda_1$ versus load: positive (blackout
    unstable, sequence recovered from zero overlap) below the crossing, negative
    beyond, and smoothly stable through the saddle-node of cycles at
    $(P/N)_{\rm SNLC}\simeq0.18$---the subcritical coexistence window of Fig.~2B.
    \label{fig:eig_blackout}}
\end{figure}

We now connect the circular embedding to the linear stability of the blackout state,
and use this connection to measure the spectral signature of the remembering--forgetting transition (Fig.~\ref{fig:eig_blackout}).


\paragraph{The embedding linearises onto the fundamental sequence mode.}
Around the blackout state all overlaps vanish, and a perturbation is described by the
overlap vector $\mathbf m = (m^1,\dots,m^P)$. Since the stored patterns are i.i.d.\ and
the sequence coupling is cyclic in the pattern index, the disorder-averaged
linearisation of one asynchronous sweep commutes with the shift $\mu\to\mu+1$; it is
therefore diagonalised by the Fourier modes of the overlap vector,
\begin{equation}
    c_q(t) \;=\; \sum_{\mu=1}^P m^\mu(t)\, e^{-2\pi i q \mu/P},
    \qquad
    c_q(t+1) \;=\; \Lambda_q\, c_q(t),
    \label{eq:mode_multiplier}
\end{equation}
with complex one-sweep multipliers $\Lambda_q$: the modulus $|\Lambda_q|$ is the
growth factor of the mode, and the phase is its rotation---the sequence advance. The
blackout state is linearly stable when $|\Lambda_q|<1$ for all $q$, i.e.\
$\mathrm{Re}\ln\Lambda_q<0$.

The circular embedding of Eq.~\eqref{eq:methods_proj} measures precisely the $q=1$
member of this family. Writing the embedded position as a complex number
$Z(\mathbf x) = \Pi_1 + i\,\Pi_2$ and expanding the softmax weights about
$\mathbf m = 0$ (where they flatten to $w_\mu = 1/P$, and
$\sum_\mu \mathbf p_\mu = 0$ for the equally spaced pattern positions),
\begin{equation}
    Z \;=\; \frac{\kappa}{P} \sum_{\mu=1}^P m^\mu\, e^{i\theta_\mu} \;+\; O(m^2)
    \;\;\propto\;\; \bar c_1,
    \label{eq:embed_linear}
\end{equation}
up to a constant phase: to linear order the embedded coordinate \emph{is} the (conjugate) fundamental mode amplitude. 
Here $\kappa = \tau_d$ for the pure softmax
weights of Eq.~\eqref{eq:methods_proj}, and $\kappa = 1$ for the overlap-weighted
variant ($w_\mu \to w_\mu m_\mu$) used in the figures; both linearise onto the same
mode. Consequently the linearised dynamics of the embedded point is the complex
multiplication $Z(t+1) = \bar\Lambda_1 Z(t)$: the Jacobian
$\mathbf J_\Pi$ of Eq.~\eqref{eq:proj_jacobian} at the origin is, in the thermodynamic
limit, the conformal matrix $|\Lambda_1| R(-\arg\Lambda_1)$, whose eigenvalue pair
$\{\Lambda_1,\bar\Lambda_1\}$ is shown in Fig.~\ref{fig:eig_blackout}. The
embedding-based stability analysis and the mode analysis are the same calculation.

\paragraph{Estimator and results.}
We can estimate eigenvalues of the embedded flow\footnote{One may equally fit the multiplier to the embedded coordinate $Z(t)$ of the
\emph{nonlinear} embedding. At finite $N$, however, the smallest overlaps accessible
from random states are $O(\sqrt{2\ln P/N})$, for which $\tau_d\, m$ is already of order
one: the softmax then compresses the mode and biases the multiplier low. In the zero-temperature limit of the embedding
($\tau_d \to 0$, uniform weights with overlap weighting) the identity
$Z \propto \bar c_1$ becomes exact at all amplitudes, and the embedded fit coincides
with the mode fit to machine precision. The mode-space estimator is therefore the
operationally robust form of the embedded linear stability analysis.} directly from $\Lambda_1$ in Eq.~\eqref{eq:mode_multiplier}: trajectories are started from random (zero-overlap) initial conditions, $c_1(t)$ is recorded once per
sweep, and the multiplier is obtained by least squares,
$\Lambda_1 = \sum_t \bar c_1(t)\,c_1(t{+}1) / \sum_t |c_1(t)|^2$, over the window between
the initial transient and saturation ($\max_\mu |m^\mu| > 0.35$) or dynamical arrest,
then averaged over initial conditions and disorder realisations. 
The measured multipliers of the modes $q=1,\dots,4$ agree within errors in modulus, with rotation
rates scaling as $q$; the embedding selects $q=1$.

At $N=400$ ($\lambda=2.5$, zero temperature) the pair
crosses the imaginary axis at $P/N \approx 0.076$ (Fig.~\ref{fig:eig_blackout}), the
blackout state destabilises at a finite load consistent with the DMFT Hopf point
$(P/N)_{\rm H} \simeq 0.06$ of Fig.~2B.

\subsection{Pseudo-inverse Embedding}\label{sec:inverse_projection}

To construct flow fields visualizing the network dynamics, we require an approximate inverse mapping from 2D points back to high-dimensional states. Given a point $\mathbf{z} \in \mathbb{R}^2$, we seek a state $\hat{\mathbf{x}} \in \{-1,+1\}^N$ such that $\Pi(\hat{\mathbf{x}}) \approx \mathbf{z}$. This inverse is fundamentally ambiguous, as the forward function is many-to-one.

We considered two approaches to approximate the inverse. The deterministic method computes weights based on 2D distances to pattern positions:
\begin{equation}
\tilde{w}_\mu = \frac{\exp(-\tau_d \, d_\mu)}{\sum_{\nu=1}^P \exp(-\tau_d \, d_\nu)}, \quad d_\mu = |\mathbf{z} - \mathbf{p}_\mu|_2,
\end{equation}
where $\tau_d$ is a distance-space temperature (distinct from the overlap-space temperature $\tau$ of the forward map). This forms a continuous-valued state as a weighted blend of patterns, $\tilde{\mathbf{x}} = \sum_{\mu=1}^P \tilde{w}_\mu \boldsymbol{\xi}^\mu$, which is subsequently binarized via the sign function to yield $\hat{\mathbf{x}} = \text{sign}(\tilde{\mathbf{x}})$. While computationally efficient and deterministic, this approach provides no guarantee that the reconstructed state's overlaps match the intended weights.

We can instead use a stochastic approximation for the inverse embedding to address this limitation through probabilistic sampling. Rather than blending patterns continuously, we construct the state $\hat{\mathbf{x}}$ by independently sampling each spin from the stored patterns according to the weight distribution $\{\tilde{w}_\mu\}$:
\begin{equation}
\hat{x}_i \sim \sum_{\mu=1}^P \tilde{w}_\mu \, \delta(\xi_i^\mu), \quad i = 1, \ldots, N,
\end{equation}
where each spin $\hat{x}_i$ takes the value from pattern $\mu$ with probability $\tilde{w}_\mu$. For patterns that are approximately orthogonal (as commonly occurs in Hopfield networks), this ensures that the expected overlap between $\hat{\mathbf{x}}$ and pattern $\mu$ equals $\tilde{w}_\mu$, providing statistical fidelity to the target weight distribution. The stochastic approach introduces variability—repeated inversions of the same point yield different states—but this diversity better represents the multiplicity of high-dimensional states embedded at that location.

\subsection{Quantifying Embedding Fidelity}

Unlike linear methods such as PCA, which naturally provide an explained variance ratio, our nonlinear embedding requires custom metrics to assess information preservation.

\paragraph{Overlap preservation.} We measure how well the embedding-reconstruction cycle preserves the overlap vector $\mathbf{m}(\mathbf{x}) = (m_1, \ldots, m_P)^T$ that characterizes each state's relationship to the stored patterns. The overlap preservation score is defined as:
\begin{equation}
\rho_{\text{overlap}} = 1 - \frac{|\mathbf{m}(\mathbf{x}) - \mathbf{m}(\hat{\mathbf{x}})|_2}{2} \in [0, 1],
\end{equation}
where $\hat{\mathbf{x}} = \Pi^{-1}(\Pi(\mathbf{x}))$ is the reconstructed state. Perfect preservation yields $\rho_{\text{overlap}} = 1$.

\paragraph{Explained overlap variance.} As a nonlinear analog to PCA's explained variance ratio, we quantify how much variance in the overlap space is preserved by the embedding. The total variance across all pattern overlaps is:
\begin{equation}
\sigma^2_{\text{total}} = \sum_{\mu=1}^P \frac{1}{T} \sum_{t=1}^T (m_\mu^{(t)} - \bar{m}_\mu)^2,
\end{equation}
while the residual variance after reconstruction is:
\begin{equation}
\sigma^2_{\text{residual}} = \sum_{\mu=1}^P \frac{1}{T} \sum_{t=1}^T (m_\mu^{(t)} - \hat{m}_\mu^{(t)})^2,
\end{equation}
yielding the explained variance:
\begin{equation}
R^2_{\text{overlap}} = 1 - \frac{\sigma^2_{\text{residual}}}{\sigma^2_{\text{total}}} \in [0, 1].
\end{equation}

\paragraph{Out-of-plane flow.} Finally, we quantify how much of the dynamics occurs in directions orthogonal to the 2D embedding plane. For each state transition, we compare the actual embedded displacement $\mathbf{z}^{(t+1)} - \mathbf{z}^{(t)}$ with the displacement obtained by inverse-embedding, evolving, and re-embedding: $\Pi(\mathcal{D}(\Pi^{-1}(\mathbf{z}^{(t)}))) - \mathbf{z}^{(t)}$. The out-of-plane flow ratio:
\begin{equation}
\gamma = \frac{1}{T-1} \sum_{t=1}^{T-1} \frac{|\mathbf{z}^{(t+1)} - \Pi(\mathcal{D}(\Pi^{-1}(\mathbf{z}^{(t)})))|}{|\mathbf{z}^{(t+1)} - \mathbf{z}^{(t)}|}
\end{equation}
quantifies this discrepancy. Small values of $\gamma$ indicate that most dynamics are well-captured in the 2D embedding, while large values reveal significant transverse flow that escapes the projected plane.


\section{Dense non-reciprocal memory: biological plausibility and cellular automata}\label{si:dense}
\subsection{Biologically plausible version of dense associative memories.}
It can be argued that the neurobiological plausiblity of dense associative memories is questionable. 
While synaptic junctions of real neurons connect exactly two neurons, the non-linear activations on pattern overlaps involve higher-than-quadratic order interactions between neurons (biological neurons exist only at two-body synaptic junctions):
\begin{equation}
    f\left(\sum_{j=1}^N\xi^\mu_j x^1_j\right) = \sum_n f^{(n)} \left(\sum_{j_1,\ldots,j_n} \xi^\mu_{j_1}\ldots \xi^\mu_{j_n} x^1_{j_1}\ldots x^1_{j_n}\right).
\end{equation}
This issue can be circumvented by considering an additional set of $P$ ``hidden'' neurons that store the overlaps with each pattern, and then the non-linear elements are only applied to the (individual) hidden neurons~\cite{modern_bio_plausible}:
\begin{equation}\label{eq:modern_bio_model}
    \begin{cases}
        x_i^1 & \leftarrow {\rm sign}\left[\sum_{\mu=1}^P \xi_i^\mu f\left(m^\mu\right) + \lambda x_i^2 + \eta^1_i \right]\\
        x_i^2 & \leftarrow {\rm sign}\left[\sum_{\mu=1}^P \xi_i^{\mu+1} f\left(m^\mu\right)  + \eta^2_i\right]\\
        m^\mu & \leftarrow \sum_{j=1}^N\xi^\mu_j x^1_j.
    \end{cases}
\end{equation}

\begin{figure}[hbt!]
    \centering
    \includegraphics[]{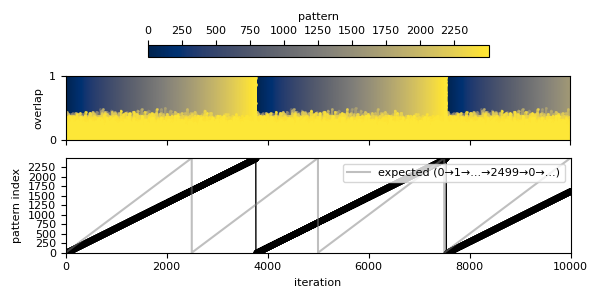}
    \caption{{\bf Retrieval of very long sequences of uncorrelated patterns using the modern non-reciprocal model.}
    Simulation of an $N=100$ neuron modern non-reciprocal Hopfield model successfully retrieving a sequence of $P=2500$ patterns. Such large memory load $\alpha=P/N=25$ is in contrast with the model without non-linearity which had a capacity given by only a fraction of the number of neurons $N$.
    }
    \label{fig:modern_uncorrelated}
\end{figure}

\subsection*{Convolutional modern Hopfield network as a local cellular automaton}
We specialise the bipartite non-reciprocal network of Eq.~(2), in its modern-Hopfield form (Eqs.~(4)--(5)), to a \emph{local} (convolutional) receptive field. Take a 1D elementary CA with two states mapped to the memory species $x_i\in\{-1,+1\}$, and a radius-$1$ neighbourhood set $\mathcal R=\{-1,0,+1\}$. The rule is a lookup table assigning to each of the $P=2^{|\mathcal R|}=8$ local configurations a key $\kappa^\mu\in\{\pm1\}^{|\mathcal R|}$ (the neighbourhood pattern, whose centre component is $\kappa^\mu_0$) and a value $v^\mu\in\{\pm1\}$ (the new centre state the rule produces). Storing configurations as keys and outputs as values, the two-species update of Eq.~(2) becomes
\begin{equation}
\begin{aligned}
x_i(t+\Delta t) &= \mathrm{sign}\!\Big[\,
  \sum_{\mu=1}^{P} \kappa^\mu_{0}\,
  f\!\big(\beta\!\!\sum_{d\in\mathcal R}\!\kappa^\mu_{d}\,x_{i+d}(t)\big)
  \;+\; \lambda\, y_i(t) \;+\; \eta^x_i(t)\,\Big],\\[4pt]
y_i(t+\Delta t) &= \mathrm{sign}\!\Big[\,
  \sum_{\mu=1}^{P} v^\mu\,
  f\!\big(\beta\!\!\sum_{d\in\mathcal R}\!\kappa^\mu_{d}\,x_{i+d}(t)\big)
  \;+\; \eta^y_i(t)\,\Big],
\end{aligned}
\label{eq:methods-localK}
\end{equation}
where $f$ is the nonlinearity of Eqs.~(4)--(5)---a polynomial $f(u)=u^n$, or $f=\exp$ normalised over $\mu$ (softmax). The same key/value set $\{\kappa^\mu,v^\mu\}$ is reused at every site $i$: this weight tying over the shifted window $\{x_{i+d}\}_{d\in\mathcal R}$ is what makes the layer \emph{convolutional} and the rule \emph{local}. The first line is the auto-associative ($J$) channel of Eq.~(2a) restricted to the local window---it holds $x_i$ at the centre state of the matched neighbourhood---while the second is the sequence-flow ($K$) channel of Eq.~(2b), in which $y_i$ stores the rule output for that neighbourhood. In the notation of Eqs.~(4)--(5) the keys $\kappa^\mu$ play the role of the stored patterns $\xi^\mu$ and the values $v^\mu$ that of the cyclic successor $\xi^{\mu+1}$, with the dense overlap restricted to the local window. At large $\beta$ the softmax concentrates on the single stored key $\kappa^{\mu^\star}$ matching the current receptive field, so $y_i$ acquires the exact CA output and the non-reciprocal coupling $\lambda$ feeds it back into $x_i$ one update later: one CA step is realised over two sweeps, with $y$ acting as an implicit $1$-step delay buffer at no extra cost (no explicit state cloning).
We verified this exact match for all $256$ elementary rules.

\begin{figure}[hbt!]
    \centering
    \includegraphics[width=0.7\linewidth]{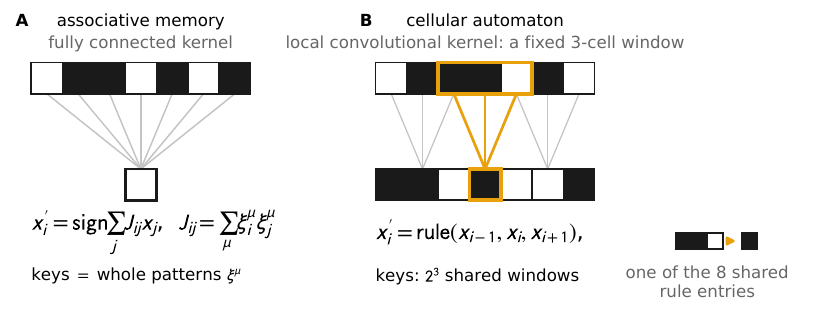}
    \caption{Implementing cellular automata as associative memory.}
    \label{fig:si_ca_vs_hopfield}
\end{figure}

\paragraph{Commutative rules are robust against update asynchrony.}

\begin{figure}[hbt!]
  \centering
  \includegraphics[width=0.95\columnwidth]{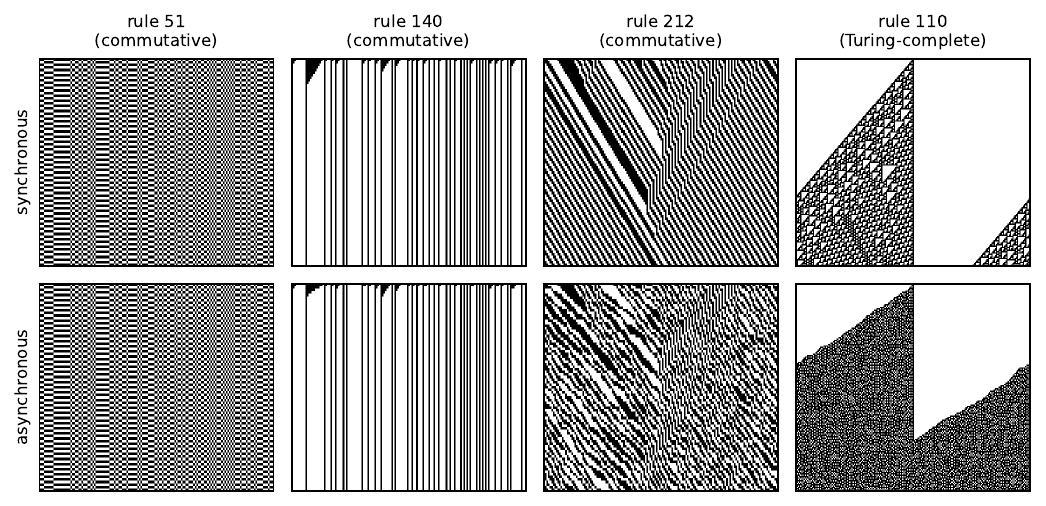}
  \caption{Space-time of elementary CAs on the local Hopfield network of
    Eq.~\eqref{eq:methods-localK}. Each column is one rule; top row
    synchronous, bottom row random-sequential asynchronous. Columns
    $1$--$3$: commutative rules $51$, $140$, $212$ from a common random
    initial condition ($N=151$ cells, $80$ sweeps; rules $51$ and $212$
    have rich space-time, most of the other $15$ of the $18$ collapse to
    a fixed point or short period). Under asynchrony the raw snapshot is
    ragged because cells advance at different local rates, but the
    per-cell update history is schedule-invariant and the computation stays coherent. Column $4$ shows the Turing-complete, non-commutative rule $110$; from a single active cell, the asynchronous trajectory disintegrates into structureless noise. }
  \label{fig:methods_commutative}
\end{figure}

When Eq.~(2) is iterated asynchronously, one neuron at a time in random sequential order, the network reproduces the synchronous CA computation if the update operator is commutative on adjacent cells~\cite{Gacs2001}. Algebraic enumeration over the $2^4=16$ four-cell window configurations identifies exactly $18$ commutative rules out of $256$; for each, the per-cell update history is schedule-invariant (G\'acs Thm.~7), so the computation stays coherent under any schedule even though a raw async snapshot is ragged because cells advance at different local rates (Fig.~\ref{fig:methods_commutative}, top rows). None of the $18$ is Turing-capable: all sit in class~1 or 2, with at most fixed-point or short-period attractors.

\paragraph{Failure of Turing-complete 1D rules under asynchronous updates.}

Storing the lookup of rule~$110$ (Turing-complete) in the same network and updating asynchronously corrupts ${\sim}35\%$ of the space-time cells relative to the synchronous CA trajectory
(Fig.~\ref{fig:methods_commutative}, bottom row). The failure is that asynchrony routinely produces \emph{phantom} neighbourhoods, combinations of half-updated and half-stale neighbour states that never appear along the synchronous trajectory, and rule~$110$'s non-commutative lookup answers them with values that drift from the synchronous reference. To rule out a locality artefact, we also tested an orthogonally scrambled basis $\tilde\xi^\mu = R\xi^\mu$ with $R\in O(N)$ random, which delocalises the rule across all $N$ neurons while preserving the stored lookup; the async failure rate is essentially unchanged. 

\subsubsection*{Routes to async-robust universal computation?}
\paragraph{Explicit delay buffer plus phase-tag gates.}
One strategy for achieving schedule invariance on non-commutative CA rules is the classical construction of Nakamura~\cite{Nakamura1974,Fates2014}.
The idea is to augment each cell of a $q$-state CA from $q$ states to $3q^2$ states, storing per-cell triples $(s^{\rm prev}_i,\, s^{\rm cur}_i,\, t_i)$ with
$t_i\in\mathbb Z_3$ a phase tag. A cell is allowed to fire only when none of its neighbours is more than one phase ahead; when reading a neighbour that \emph{is} one phase ahead, the cell consults the stored $s^{\rm prev}_j$ in place of $s^{\rm cur}_j$. The $1$-step history buffer $s^{\rm prev}$ plays exactly the role of the delayed term $s_j(t-\tau_d)$ in the Sompolinsky--Kanter sequence-storing Hopfield network~\cite{PhysRevLett.57.2861}; the modular phase tag is a separate ``logic'' apparatus that synchronises the parallel local clocks across
adjacent sites. The cost is multiplicative in the number of states ($3q^2$ per cell, e.g.\ $12$ states for $q=2$) and additive in the wiring (the phase-tag gates).

In contrast, our model realises emergent time delays dynamically: the
non-reciprocal coupling $\lambda$ of Eq.~(2) makes $x$ lag $y$ by one update step at every site, so the $1$-step short-term memory of $x$ relative to $y$ emerges here without explicit cloning. A scalar interaction $\lambda$ alone, however, supplies only the buffer; it does not tell a cell \emph{when} it is safe to advance. That decision is made by Nakamura's phase-tag gate, which compares a cell's modular phase against each neighbour's and suppresses the update unless they are compatible. Below we show that the gate, too, can be implemented physically---by promoting $\lambda$ to a $\boldsymbol\tau$-dependent (Potts-like coupling) form---so that no per-cell conditional branch is needed and the codebook needs no separate $s^{\rm prev}$ clone.

%

\paragraph{Nakamura's modern Hopfield model with gated non-reciprocity.}
Although the phase-tag gate is a conditional branch on discrete data, it can be implemented as a physical interaction if we consider the phase tags to be physical degrees of freedom.
The construction is to augment the receptive field with the phase tag: at every site $i$ introduce a $3$-state Potts variable $\boldsymbol\tau_i\in\{\pm1\}^3$ encoded one-hot in $\pm 1$, and enumerate the $12^3=1728$ augmented neighbourhoods $(s^{\rm cur}_j,s^{\rm prev}_j,\tau_j)_{j\in\{i-1,i,i+1\}}$ that can appear along a Nakamura trajectory.

At every site, Eq.~(2) writes the predicted next state into $y$ from the current state of $x$ ($K$-channel) and delivers $y$ back into $x$ one update later ($\lambda$-channel), so $x$ lags $y$ by exactly one update step. Identifying $y_i$ with Nakamura's $s^{\rm cur}_i$ (the next-step prediction, just written by $K$) and $x_i$ with $s^{\rm prev}_i$, the per-cell history buffer that Nakamura stores as an explicit extra spin is realised here implicitly.
 
The phase gate can be achieved by promoting $\lambda$ to a Potts-coupled form: i.e. replacing the scalar $\lambda$ of Eq.~(2a) with the pairwise function
\begin{equation}
\lambda(\boldsymbol\tau_i,\boldsymbol\tau_j) \;=\; \lambda\,\Lambda(\boldsymbol\tau_i,\boldsymbol\tau_j),
\qquad \Lambda(\boldsymbol\tau_i,\boldsymbol\tau_j) \;=\;
\begin{cases} 1 & (\tau_j-\tau_i)\bmod 3\in\{0,1\}\\ 0 & (\tau_j-\tau_i)\bmod 3 = 2\end{cases}
\label{eq:methods-lambdaPotts}
\end{equation}
($\Lambda$ is a closed-form bilinear in one-hot $\boldsymbol\tau$). The product over both neighbours
$g_i(t)\equiv\prod_{j\in\{i-1,i+1\}}\Lambda(\boldsymbol\tau_i(t),\boldsymbol\tau_j(t))$ is the gate: $g_i = 1$ iff cell $i$ is fireable in Nakamura's sense (no neighbour one phase behind, i.e.\ two ahead modulo $3$). With the augmented $15$-dim receptive field $\Phi_i(t)=\big(x_{i-1},y_{i-1},\boldsymbol\tau_{i-1}\,;\,x_i,y_i,\boldsymbol\tau_i\,;\,x_{i+1},y_{i+1},\boldsymbol\tau_{i+1}\big)$, the dynamics is
\begin{equation}
\begin{aligned}
y_i(t+\Delta t) &= \mathrm{sign}\!\Big[\sum_{\mu=1}^{P_\mathrm{aug}} V^{\mu}_{y}\,
  f\!\big(\beta\, K^\mu\!\cdot\!\Phi_i(t)\big) + \eta^y_i(t)\Big],\\[4pt]
\boldsymbol\tau_i(t+\Delta t) &= \mathrm{sign}\!\Big[\sum_{\mu=1}^{P_\mathrm{aug}} V^{\mu}_{\tau}\,
  f\!\big(\beta\, K^\mu\!\cdot\!\Phi_i(t)\big) + \boldsymbol\eta^\tau_i(t)\Big],\\[4pt]
x_i(t+\Delta t) &= \mathrm{sign}\!\Big[\,\big(1-g_i(t)\big)\,x_i(t) \;+\; g_i(t)\,\lambda\,y_i(t) \;+\; \eta^x_i(t)\Big],
\end{aligned}
\label{eq:methods-routeB}
\end{equation}
where the last line plays the role of Eq.~(2a) with the  $\lambda(\boldsymbol\tau_i,\boldsymbol\tau_j)$ of Eq.~\eqref{eq:methods-lambdaPotts} factored out, and the $(1-g_i)\,x_i$ hold term is the degenerate auto-associative channel that preserves $x_i$ when the gate is closed.

The codebook is built by enumerating each augmented neighbourhood and 
writing into $V^\mu_y$ the Nakamura centre update---a freeze if any 
neighbour is two phases ahead, otherwise rule~$110$ applied to a 
phase-dependent neighbourhood reading $y_j$ for in-phase neighbours and 
$x_j$ for one-ahead neighbours---and into $V^\mu_\tau$ either the held 
value or $\tau_i+1\bmod 3$. The only piece of Nakamura's logic that 
remains in $K$ is this phase-selected ``which neighbour value to read''; 
the storage is supplied by the non-reciprocal $\lambda$.

We verified Eq.~\eqref{eq:methods-routeB} numerically on rule~$110$ ($N=51$ cells, $20$ CA steps). Per-cell logical-time-indexed configurations match the correct rule~$110$ trajectory exactly under asyncrhonous updates. As a control, setting $\lambda=0$ while keeping the codebook and gate intact breaks the construction ($\sim\!43\%$ mismatch in random-sequential async).

\begin{figure}[hbt!]
    \centering
    \includegraphics[width=0.6\linewidth]{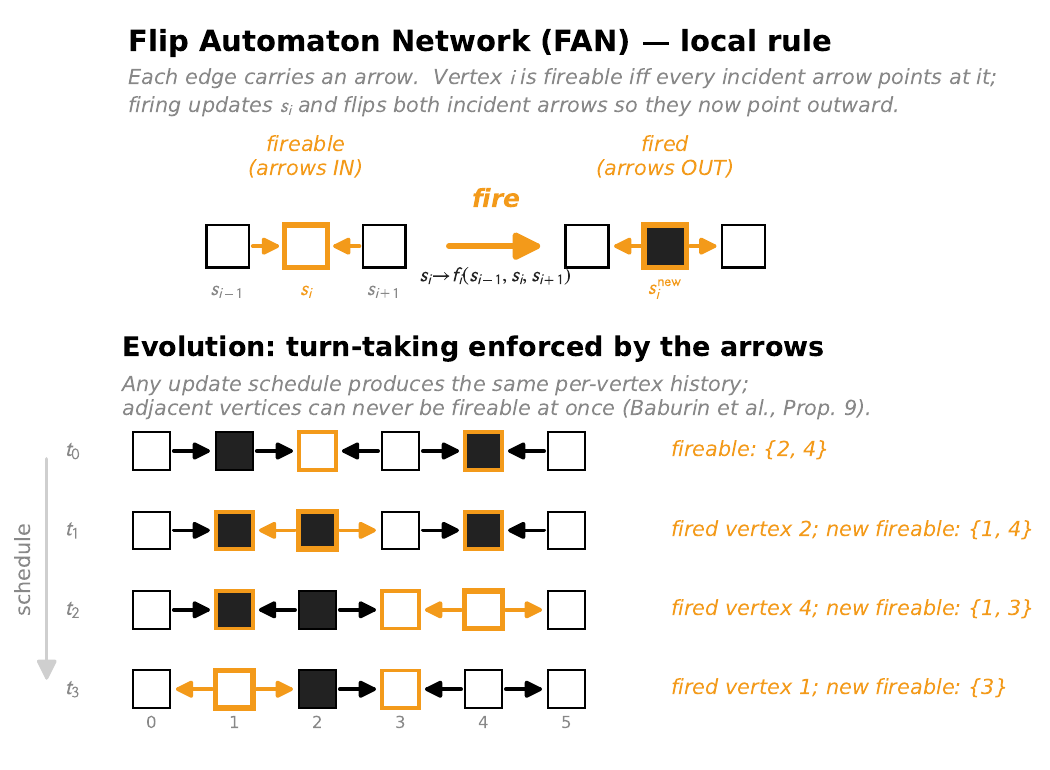}
    \caption{Flip automata networks.}
    \label{fig:si_fan}
\end{figure}

\medskip
\paragraph{Engineering more sophisticated update rules so logic gate is unnecessary.}
A different strategy is to dispense with the per-cell phase-tag gate entirely and design the local rule so that schedule invariance follows from its lookup-table structure. A framework is the \emph{flip automaton network} (FAN) of Baburin et al.~\cite{Baburin2025}: on a graph $T=(I,E)$, the configuration carries a vertex state $s_i\in\mathcal S$ at each $i\in I$ and a direction indicator $a_{ij}\in\{\pm1\}$ (with $a_{ij}=-a_{ji}$) at each edge $\{i,j\}\in E$. Vertex $i$ is \emph{fireable} iff every incident arrow points at $i$, i.e.\ $a_{ji}=-1$ for all $j\in N(i)$; when fired, its local flip function
\begin{equation}
f_i:\;\mathcal S^{N(i)\cup\{i\}}\;\longrightarrow\;\mathcal S\times\{\pm1\}^{N(i)},
\qquad \big(s_j\big)_{j\in N(i)\cup\{i\}}\;\longmapsto\;\big(s_i^{\rm new},\,\{a_{ij}^{\rm new}\}_{j\in N(i)}\big),
\label{eq:methods-FAN}
\end{equation}
updates the vertex state together with a specified subset of incident arrows, subject to the constraint that $a_{ji}^{\rm new}=+1$ for every $j$ (after a fire, every incident arrow points away from $i$). For a $1$D path graph with right-edge arrow $a_i\equiv a_{i,i+1}$, fireability reduces to $a_{i-1}=+1\wedge a_i=-1$ and a fire flips both incident arrows in addition to updating the state. As a result, FANs have invariant histories under any asynchronous schedule. 

\medskip
Non-reciprocity played a crucial role in making \emph{temporal-sequence retrieval} async-robust (main text) by creating implicit time delays.
This is because the computation to perform global associative recall only depends on macroscopic overlap order parameters without a local per-cell logic to keep in step. 
In contrast, for an intrinsically local Boolean computation the delay is not the sufficient mechanism. 
Robustness to update order either requires additional digital logic gates that suppress updates until the local neighbourhood is in a compatible state, or it must be engineered into the rule itself.



\stopcontents[si]
\let\addcontentsline\ORIGaddcontentsline   


\bibliographystyle{unsrt}
\bibliography{bibliography}